%% file: main.tex
\documentclass[11pt]{book} 
\usepackage[utf8]{inputenc} 

\usepackage[english]{babel}

\usepackage[T1]{fontenc} 
\usepackage[a4paper, margin=2cm, head=18.0pt]{geometry} 
\usepackage{graphicx}
\usepackage[onehalfspacing]{setspace}
\usepackage{tocbasic}
\usepackage{booktabs}
\usepackage{multicol}
\usepackage{multirow}
\usepackage[]{scrlayer-scrpage}
\usepackage[titletoc]{appendix}
\usepackage{comment}

\usepackage{dirtree} 
\usepackage{float} 
\usepackage{microtype} 
\usepackage{mathtools} 
\usepackage{amssymb} 
\usepackage{algorithm} 
\usepackage{algpseudocode} 

\usepackage{tikz} 

\usepackage{csquotes}
\usepackage[
    left = \flqq{},%
    right = \frqq{},%
    leftsub = \flq{},%
    rightsub = \frq{} %
]{dirtytalk}
\DeclareQuoteStyle{english}{\glqq}{\grqq}{\glq}{\grq}

\usepackage[backend=biber, style=numeric, sorting=none, defernumbers=true, language=american]{biblatex}
\DeclareBibliographyCategory{cited}
\AtEveryCitekey{\addtocategory{cited}{\thefield{entrykey}}}
\nocite{*} 

\usepackage{hyperref}
\usepackage{xurl}
\usepackage[acronym, toc]{glossaries} 
\makeglossaries
\newglossaryentry{memgraph}
{
  name=memory graph,
  description={A memory graph, or \textit{memgraph} is a graph representation of a memory dump. This graph can be a graph of blocks, where each node in the graph corresponds to a block of 8 bytes in the heap dump and each edge corresponds to a pointer from one block to another, or describes which blocks are part of a chunk whose root note is a Chunk Header Node. It can also be a graph of chunks (only CHNs), where each node in the graph corresponds to a chunk in heap dump and each edge corresponds to a pointer from one object to another.}
}

\newglossaryentry{nodes}
{
  name=nodes,
  description={A node is an entity in a graph, it can be a person, a place, a thing, or any other entity.}
}
\newglossaryentry{chn}{
  name={CHN},
  description={Chunk Header Node. This is a node whose bytes have been identified as a data structure header. In the graph, this node is the root node of an malloc-allocated memory chunk.}
}

\newglossaryentry{pn}{
  name={PN},
  description={Pointer Node. This is a node whose bytes have been identified as a pointer.}
}

\newglossaryentry{kn}{
  name={KN},
  description={Key Node. This is a node whose bytes have been identified as a key. This identification relies both on the annotations and some verification checks.}
}

\newglossaryentry{vn}{
  name={VN},
  description={Value Node. These are all blocks that have not been identified. It is the default node type.}
}

\newacronym{kg}{KG}{Knowledge Graph}
\newacronym{foss}{FOSS}{Free and Open Source Software}
\newacronym{rdf}{RDF}{Resource Description Framework}
\newacronym{rdfs}{RDFS}{Resource Description Framework Schema}
\newacronym{owl}{OWL}{Web Ontology Language}
\newacronym{ml}{ML}{Machine Learning}
\newacronym{dl}{DL}{Deep Learning}
\newacronym{fe}{FE}{Feature Evaluation}
\newacronym{nlp}{NLP}{Natural Language Processing}
\newacronym{ke}{KE}{Knowledge Engineering}
\newacronym{del}{DEL}{Directed Edge-labelled Graphs}
\newacronym{er}{ER}{Entity Resolution}
\newacronym{qa}{QA}{Quality Assurance}
\newacronym{sparql}{SPARQL}{SPARQL Protocol and RDF Query Language}
\newacronym{ssh}{SSH}{Secure Shell Protocol}
\newacronym{os}{OS}{Operating System}
\newacronym{vm}{VM}{Virtual Machine}
\newacronym{ddos}{DDoS}{Distributed Denial of Service}
\newacronym{ess}{ESS}{Estimated Security Strength}
\newacronym{vmi}{VMI}{Virtual Machine Introspection}
\newacronym{smote}{SMOTE}{Synthetic Minority Over-sampling Technique}
\newacronym{svm}{SVM}{Support Vector Machine}
\newacronym{knn}{KNN}{K-Nearest Neighbors}
\newacronym{rf}{RF}{Random Forest}
\newacronym{mlp}{MLP}{Multi-Layer Perceptron}
\newacronym{relu}{ReLU}{Rectified Linear Unit}
\newacronym{sgd}{SGD}{Stochastic Gradient Descent}
\newacronym{ai}{AI}{Artificial Intelligence}
\newacronym{pca}{PCA}{Principal Component Analysis}
\newacronym{lda}{LDA}{Linear Discriminant Analysis}
\newacronym{tsne}{t-SNE}{t-distributed Stochastic Neighbor Embedding}
\newacronym{msb}{MSB}{Most Significant Bit}
\newacronym{lsb}{LSB}{Least Significant Bit}

\newacronym{lstm}{LSTM}{Long Short-Term Memory}
\newacronym{gru}{GRU}{Gated Recurrent Units}
\newacronym{rnn}{RNN}{Recurrent Neural Networks}
\newacronym{cnn}{CNN}{Convolutional Neural Networks}
\newacronym{rcnn}{RCNN}{Recurrent Convolutional Neural Network}
\newacronym{gnn}{GNN}{Graph Neural Network}
\newacronym{gcn}{GCN}{Graph Convolutional Networks}
\newacronym{llm}{LLM}{Large Language Model}

\usepackage{titlesec}

\titleclass{\chapter}{straight} 
\titleformat{\chapter}
  {\normalfont\Huge\bfseries}{\thechapter}{1em}{}
\titlespacing*{\chapter}{0pt}{\parskip}{\parskip}

\usepackage{listings} 

\usepackage{bera}
\usepackage{listings}
\usepackage{xcolor}

\definecolor{eclipseStrings}{RGB}{42,0.0,255}
\definecolor{eclipseKeywords}{RGB}{127,0,85}
\definecolor{punctuationcolor}{rgb}{0.5,0,0}
\definecolor{delimcolor}{rgb}{0,0.5,0}
\definecolor{red}{rgb}{1,0,0}
\colorlet{numb}{magenta!60!black}

\lstdefinestyle{json}{
  basicstyle=\ttfamily\small,
  breaklines=true,
  postbreak=\mbox{\space},
  columns=fullflexible,
  showstringspaces=false,
  commentstyle=\color{gray},
  keywordstyle=\color{black},
  numberstyle=\tiny\color{gray},
  numbers=left,
  frame=single,
  captionpos=b
}

\lstdefinestyle{text}{
  basicstyle=\ttfamily\small,
  breaklines=true,
  postbreak=\mbox{\space},
  columns=fullflexible,
  showstringspaces=false,
  commentstyle=\color{gray},
  keywordstyle=\color{black},
  numberstyle=\tiny\color{gray},
  numbers=left,
  frame=single,
  captionpos=b
}

\lstdefinestyle{hexdump}{
  basicstyle=\ttfamily\small,
  breaklines=true,
  postbreak=\mbox{\space},
  columns=fullflexible,
  showstringspaces=false,
  commentstyle=\color{gray},
  keywordstyle=\color{black},
  numberstyle=\tiny\color{gray},
  numbers=left,
  frame=single,
  captionpos=b
}

\lstdefinestyle{rust}{
  basicstyle=\ttfamily\small,
  breaklines=true,
  postbreak=\mbox{\space},
  columns=fullflexible,
  showstringspaces=false,
  commentstyle=\color{gray},
  keywordstyle=\color{black},
  numberstyle=\tiny\color{gray},
  numbers=left,
  frame=single,
  captionpos=b
}

\usepackage{hyphenat} 
\graphicspath{{img/}}

\usepackage{scrlayer-scrpage}
\clearpairofpagestyles
\newcommand{\thetitle}{Predicting SSH keys in Open SSH Memory dumps}
\newcommand{\theauthor}{Rascoussier, Florian Guillaume Pierre}

\title{\thetitle}
\author{\theauthor}
\date{April-Mai 2023}

\begin{document}

\input{tex/title.tex}
\newpage


\section*{Abstract}
As the digital landscape evolves, cybersecurity has become an indispensable focus of IT systems. Its ever-escalating challenges have amplified the importance of digital forensics, particularly in the analysis of heap dumps from main memory. In this context, the Secure Shell protocol (\acrshort{ssh}) designed for encrypted communications, serves as both a safeguard and a potential veil for malicious activities. This research project focuses on predicting SSH keys in OpenSSH memory dumps, aiming to enhance protective measures against illicit access and enable the development of advanced security frameworks or tools like honeypots. 

This Masterarbeit is situated within the broader SmartVMI project, a collaborative research initiative with the objective to advance artificial intelligence-based mechanisms for attack detection and digital forensics. Specifically, this work seeks to build upon existing research on key prediction in OpenSSH heap dumps. Utilizing machine learning and deep learning models, the study aims to refine feature for embedding techniques and explore innovative methods for effective key detection. The objective is to accurately predict the presence and location of SSH keys within memory dumps. This work builds upon, and aims to enhance, the foundations laid by SSHkex \cite{SSHkex22} and SmartKex \cite{SmartKex22}, enriching both the methodology and the results of the original research while exploring the untapped potential of newly proposed approaches.

The current thesis dives into memory graph modelization from raw binary heap dump files. Each \gls{memgraph} can support a range of embeddings than can be used directly for model training, through the use of classic \acrshort{ml} models and graph neural network. The report encapsulates the progress of a year-long Master's thesis research project executed between October 2022 and October 2023. Conducted within the framework of the PhDTrack program between the University of Passau and INSA Lyon, the research has been supervised by Prof. Dr. Michael Granitzer and Christofer Fellicious from the University of Passau, as well as Prof. Dr. Pierre-Edouard Portier from INSA Lyon. It offers an in-depth discussion on the current state-of-the-art in key prediction for OpenSSH memory dumps, research questions, experimental setups, programs development, results as well as discussing potential future directions.

\newpage
\section*{Acknowledgements}
First acknowledgement goes to Christofer Fellicious, my supervisor at the University of Passau, for his guidance, support and feedback during the Masterarbeit. 

I want to express my sincere gratitude to my colleague and friend, Clément Lahoche, whose human and technical skills have been a great source of inspiration and motivation throughout this project;  especially considering that we have been working on closely related subjects. It has been a great pleasure to share our ideas and insights, and to collaborate on the development of several programs necessary for the experimentation. 

Another acknowledgements go to my esteemed supervisors Prof. Dr. Granitzer and Prof. Dr. Portier for their support and feedback during the Masterarbeit. 

I would also like to express my sincere gratitude to all the persons that have helped me, even punctually, during the Masterarbeit with their valuable help, insights, discussions and contributions as well as all the persons involved in the PhDTrack program that made this Masterarbeit possible, including but not limited to: 
\begin{itemize}
    \item Lionel Brunie, Director of CS Department at INSA Lyon, that makes this PhDTrack program possible from the French side.
    \item Harald Kosch, Head of the Chair of Distributed Information Systems at the University of Passau, that makes this PhDTrack program possible from the German side.
    \item Natalia Lucari, PhDTrack coordinator at INSA Lyon, for her support and help during the PhDTrack program.
    \item Ophelie Coueffe, PhDTrack coordinator at the University of Passau, for her patience, support and invaluable help during the PhDTrack program.
    \item Elöd Egyed-Zsigmond, PhDTrack coordinator at INSA Lyon, for the subject selection, administrative support and final proofreading.
    \item All the other PhDTrack students for the great atmosphere, mutual help and the interesting discussions during almost two years.
\end{itemize}

I cannot forget to mention my many friends from Lyon to Passau and beyond, for their encouragements during this Masterarbeit.

Finally, my last acknowledgements go to my family whose support has been precious throughout the two years of the PhDTrack program.

\newpage
\tableofcontents
\listoffigures
\listoftables
\lstlistoflistings
\newpage

\pagenumbering{arabic} 

\include{tex/chapters/introduction}

\include{tex/chapters/background}
\include{tex/chapters/methods}

\include{tex/chapters/results}
\include{tex/chapters/discussion}
\include{tex/chapters/conclusion}

\newpage
\begin{appendices}
    \input{tex/appendix.tex}

\end{appendices}

\newpage
\printglossary[type=\acronymtype]

\printglossary

\newpage
\printbibliography[
    heading=bibintoc,
    category=cited,
    title={References}
]

\printbibliography[
    notcategory=cited,
    heading=bibintoc,
    title={Additional bibliography},
]

\include{tex/german_affidavit}

\restoregeometry
\end{document}

%% file: tex/title.tex
\begin{titlepage}
    \centering
    \begin{onehalfspace}
    	
        	\includegraphics[width=7cm, height=1.5cm]{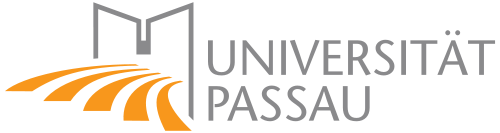}
			\hspace*{1.0cm}
			\includegraphics*[width=7cm, height=1.5cm]{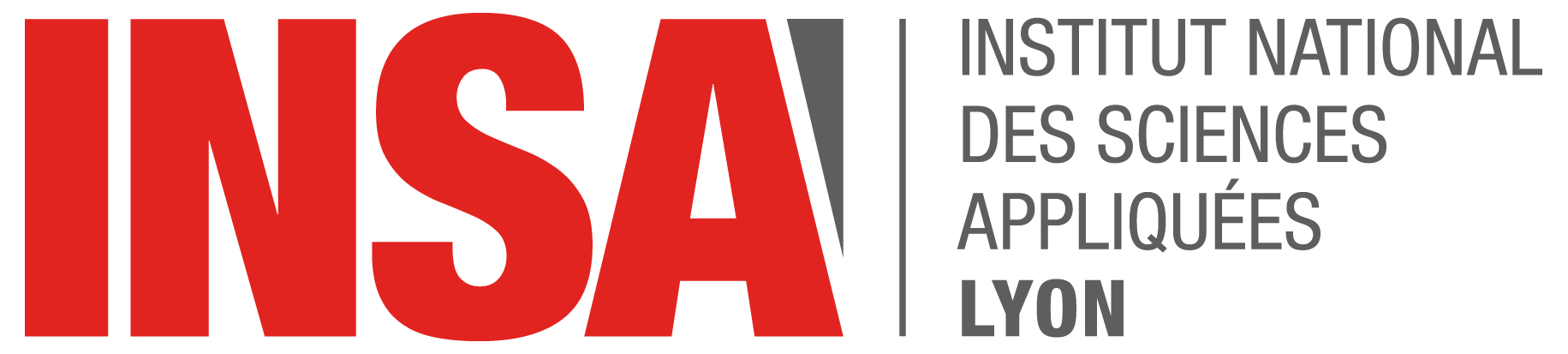}\\
        	\vspace{1.0cm}
        	{\Large \bfseries Masterarbeit}\\

        	\vspace{2.5cm}

            \begin{doublespace}
            	{\textsf{\Huge{\thetitle}}}
            \end{doublespace}

        	\vspace{2cm}

            {\Large A report by}\\

        	\vspace{1cm}

        	{\bfseries \large{\theauthor}} \\
			\vspace{0.5cm}
			Matrikelnummer (Passau): 112485 \\
			Matrikelnummer (INSA): 4018543 \\
			ORCID: 0009-0005-3253-9814 \\

        	\vfill

        	{\Large
                \textit{Erstpr\"ufer} \\
				Prof. Dr. Harald Kosch \\
				\textit{Zweitpr\"ufer} \\
				Prof. Dr. Michael Granitzer\\
				\vspace{0.5cm}
				\textit{Betreuer} \\
				Christofer Fellicious\\
				Prof. Dr. Pierre-Edouard Portier\\
				Prof. Dr. Elöd Egyed-Zsigmond\\
        	}

        	\vspace{1cm}

        	\parbox{\linewidth}{\hrule\strut}

            \vfill

			{\large \today}
    \end{onehalfspace}
\end{titlepage}

%% file: tex/chapters/introduction.tex
\chapter{Introduction}\label{chap:introduction}



The digital age has brought with it an unprecedented increase in the volume and complexity of data that is being generated, stored, and processed. This data is often sensitive in nature, and its security is of paramount importance, making cybersecurity a critical focus area. This evolving landscape is fraught with challenges that continue to amplify the importance of digital forensics in IT systems. One area that stands out for its widespread use and importance is the Secure Shell protocol (SSH) and its most popular implementation, OpenSSH. SSH is a cryptographic network protocol widely used for secure remote access to systems. It is also used for secure file transfer, and as a secure tunnel for other applications. SSH is a key component of IT systems whose encryption capabilities are critical to the security of IT systems. However, it also presents a unique set of challenges, most notably the concealment of malicious activities.

A common case is when an unauthorized actor gains access to SSH keys so as to get access to a system. This can happen through a malicious human actor, but more commonly through automated processes such as malwares and botnets. This situation presents a formidable and growing threat to cybersecurity, affecting a broad range of stakeholders from governments and financial institutions to individual users. In just 2019, the number of Command and Control (C\&C) servers for botnets increased by 71.5\%, leading to an estimated \$19 billion in advertising theft \cite{SSHBotnetInfect21}. Many malwares and botnets \say{have in common that they have used as attack vector the Secure Shell (SSH) remote access service} \cite{SSHBotnetInfect21}. 

At the heart of the issue lies the fact that SSH veils its communications through encryption, making it difficult to detect malicious activities. To be able to detect those potential malicious actors, it is possible to replace SSH by a honeypot that enables to monitor pseudo-SSH activities. There is a range of readily available honeypots, such as Kippo or Cowrie, which are designed to emulate a vulnerable SSH system and attract attackers \cite{ClassificationMalware21}. The problem lies that those honeypots are not able to mimic perfectly a real system, which makes them easy to detect by experienced attackers. As stated by \citetitle{SSHHoneypotEffectiveness23}: \say{The ability to collect meaningful malware from attackers depends on how the attackers receive the honeypot. Most attackers fingerprint targets before they launch their attack, so it would be very beneficial for security researchers to understand how to hide honeypots from fingerprinting and trick the attackers into depositing malware. [...] What is certain is that if a cautious attacker believes they are in a honeypot, they will leave without depositing malware onto the system, which reduces the effectiveness of the honeypot} \cite{SSHHoneypotEffectiveness23}. 

There are other approaches that allow to decrypt SSH connections without relying on a honeypot, like the \textit{man-in-the-middle} or \textit{binary manipulation} with their own set of challenges \cite{SSHkex22}. Instead of relying on softwares that mimics or modify a real system, it is possible to use a real unmodified system directly. The idea is to be able to decrypt SSH connection channels, which is possible if the SSH keys are known. Since SSH encryption keys are typically stored in the main memory of a system, it is possible for the administrators to extract them through the exploitation of memory dumps of a targeted system. In this context, the ability to detect SSH keys in memory dumps, and specifically OpenSSH keys, is critical to the development of effective SSH honeypot-like systems. The research introduced by the SmartVMI project with SSHKex, SmartKex, the present thesis and the future related work could be used to develop such a new type of system-monitoring tools. This new kind of tools would be very difficult to detect by attackers, increasing their effectiveness, and wouldn't require the alteration of the system. The present report is focused on the SSH key detection in memory dumps, which is a key component allowing to decode SSH communications such that it becomes possible to intercept malicious communications and to detect malicious activities.

\section{Research Questions}


At the very beginning of this thesis, first questions were:
\begin{itemize}
	\item What is the state of the art in the field of security key detection in heap dump memory?
	\item What are the challenges of security key detection in heap dump memory?
	\item How can the existing methods for detecting SSH keys in OpenSSH heap dumps be improved?
\end{itemize}

The SmartVMI project has already made significant progress in the detection of SSH keys in OpenSSH heap dumps. An open dataset of memory dumps has been created, and a simple yet effective method for detecting SSH keys has been developed. The dataset has been used to train and test simple machine learning algorithms, and the results have been promising. The research has been published in the form of two papers, SSHkex \cite{SSHkex22} and SmartKex \cite{SmartKex22}, which is the basis of this thesis. 

However, there is still room for improvement, particularly in the area of machine learning algorithms. This thesis seeks to build upon the existing research by refining feature extraction techniques and exploring innovative methods for effective key detection prediction. The objective is to accurately predict the presence and location of SSH keys within memory dumps. Rooted in this context, this Masterarbeit aims to address several key research questions:

\begin{itemize}
	\item \textbf{Memory graph:} How can we develop a memory graph representation to improve the prediction of SSH keys in memory dumps?
	\item \textbf{Memory graph embedding:} How can we develop a memory graph embedding representation to improve the prediction of SSH keys in memory dumps?
	\item \textbf{Feature importance:} What features are most indicative of SSH keys in memory dumps?
	\item \textbf{Feature extraction:} How can these features be extracted from memory dumps and used to train machine learning algorithms?
	\item \textbf{\acrshort{ml} for key prediction:} How can machine learning algorithms be optimized for the prediction of SSH keys in memory dumps? 
	\item \textbf{\acrlong{gcn} for key prediction:} How can \acrshort{gcn} be used to improve the prediction of SSH keys in memory dumps, and how does it compare to traditional machine learning algorithms?
\end{itemize}

By tackling these research questions, this thesis seeks not only to advance the academic understanding of SSH key prediction and digital forensics but also to provide practical insights that could lead to the development of more secure and effective systems.

\section{Commitment to Open Science and Reproducibility}

In alignment with the principles of Open Science, this thesis aims to be not just a scholarly report but also a comprehensive guide for anyone who wishes to understand, replicate, or extend the work presented. Open Science is a movement that advocates for the transparent and accessible sharing of scientific research, data, and dissemination processes \cite{WhyNotShareData22}. It is built on six fundamental principles \cite{WasIstOpenScience23}:

\begin{enumerate}
    \item \textbf{Open Methodology}: Detailed methodologies are provided to ensure that the experiments can be replicated.
    \item \textbf{Open Source}: All code used in this research is available for scrutiny and reuse. As such, all code including the \LaTeX{} code for the present report \footnote{The present report repository can be found here: \url{https://github.com/0nyr/masterarbeit\_report}} is properly documented and can be accessed on GitHub. 
    \item \textbf{Open Data}: Raw data and the data processing techniques are made publicly available.
    \item \textbf{Open Access}: The research is published in a manner that is free for all to read and download.
    \item \textbf{Open Peer Review}: The peer review process is transparent. In the case of this Masterarbeit, the research is reviewed by the supervisors of the project.
    \item \textbf{Open Educational Resources}: Any educational content produced is shared openly.
\end{enumerate}

To ensure the highest level of reproducibility and accessibility, this thesis includes what might seem like exhaustive details, such as hardware or software specifications, precise shell commands and some code implementations used during the research. These are included to provide a complete picture and to minimize the friction for those who wish to replicate the experiments, whatever their level of expertise may be. By adhering to the principles of Open Science, this thesis aims to contribute to a more transparent, collaborative, and efficient scientific community.
	
	\subsection{GitHub Repositories}

	In the context of the present Masterarbeit, a number of GitHub repositories have been created to facilitate the sharing of code and data. These repositories are listed below:

	\begin{itemize}

		\item \textbf{masterarbeit\_report\_onyr}: Repository containing the LaTeX code for the report as well as several scripts related to dataset exploration: \url{https://github.com/passau-masterarbeit-2023/masterarbeit_report_onyr}
		
		\item \textbf{mem2graph}: Memory graph creation utility built in Rust, featuring different graph creation and embedding strategies. Collaboration with Clément Lahoche: \url{https://github.com/passau-masterarbeit-2023/mem2graph}

		\item \textbf{research-base}: Custom Python framework for developing programs that include all the basics of a research project, such as logging, environment and argument loading, result keeping, and more. Collaboration with Clément Lahoche: \url{https://github.com/0nyr/research-base}

		\item \textbf{data\_processing}: Python program for data processing and machine learning for SSH key prediction. This repository contains tests on machine learning model training and evaluation for classical .csv based embedding files from \textit{mem2graph}: \url{https://github.com/passau-masterarbeit-2023/data_processing}
		
		\item \textbf{phdtrack\_project\_3}: Legacy repository containing the first version of the memory graph creation utility and the first version of the dataset creation script. Collaboration with Clément Lahoche. \url{https://github.com/0nyr/phdtrack_project_3}
		
		\item \textbf{memory\_graph\_gcn}: Main Python program and scripts around GCN for SSH key prediction. This program leverages the modified DOT file with embedding generated by \textit{mem2graph}: \textit{mem2graph}:
		\url{https://github.com/passau-masterarbeit-2023/memory_graph_gcn}

		\item \textbf{phdtrack\_server\_scripts}: Scripts for the servers used for computing experiments. This repository contains the scripts used to install the necessary tooling and run the experiments on the different servers we used. Collaboration with Clément Lahoche:
		\url{https://github.com/passau-masterarbeit-2023/phdtrack_server_scripts}
	\end{itemize}

	\subsection{Datasets}

	All datasets used in this research are publicly available and can be accessed on the Zenodo. The datasets are organized in the following manner:

	\begin{itemize}
		\item \textbf{Original Heap Dumps Dataset}: This is the raw dataset used for the research and produced by the SmartKex team \cite{SSHkex22}. It contains the original heap dumps in the form of \textit{-heap.raw} files with \textit{.json} annotation files. The dataset is available here: \url{https://zenodo.org/records/6537904}.
		\item \textbf{Cleaned Heap Dumps Dataset}: This dataset contains heap dumps with annotation files but has been parsed as described in section \ref{sec:background:kex:dataset}. The dataset is available here: \url{https://doi.org/10.5281/zenodo.10514199}.
	\end{itemize}

	As one can see, and considering the collaborative work effort that has been, it has been decided to regroup all repositories related to the OpenSSH heap dump exploration in a single GitHub organization, \textit{passau-masterarbeit-2023} \url{https://github.com/passau-masterarbeit-2023}.

	\section{Structure of the Thesis}


	The present thesis is organized in a manner that ensures a coherent and logical flow of information, following the standard structure of a Masterarbeit report. The structure is designed to gradually guide the reader from understanding the context and background of the research to the intricacies of the methods employed, and finally to the interpretation of the results. Below is a breakdown of each section:
	
	\begin{itemize}
		\item \textbf{Background Section:} This section serves as an introduction to the research context and establishes the foundation for the thesis. It outlines the previous work and state of the art, providing the reader with an understanding of existing knowledge and identifying gaps that this research aims to address. Key concepts, terminologies, and theories relevant to the study are introduced, setting the stage for the subsequent sections.
		
		
		\item \textbf{Methods Section:} This section meticulously describes the methods and approaches employed during the research. From the creation of the dataset to the selection and implementation of machine learning algorithms, this section ensures that the research process is transparent and reproducible.
		
		\item \textbf{Results Section:} The results' section presents the data obtained from the experiments conducted, outlining both the layout of programs used and the raw results. It provides a factual account of the findings without delving into interpretation or discussion.
		
		\item \textbf{Discussion Section:} This section offers an analysis and interpretation of the results obtained. It explores the implications of the findings, discusses the limitations of the study, and contextualizes the results within the broader research landscape.
		
		\item \textbf{Conclusion:} The concluding section succinctly recall the salient points of the thesis. It underscores the contributions made to the field and suggests avenues for future research, providing a fitting closure to the report.
	\end{itemize}
	
	In structuring the thesis in this manner, the intention is to provide the reader with a comprehensive yet accessible insight into the research undertaken all along this year-long project.

%% file: tex/chapters/background.tex
\chapter{Background}\label{sec:background}

%
%

This section is dedicated to the background information needed to understand the work developed in the thesis. It provides the necessary context for the research, including the problem being solved, why it's important, and the related state-of-the-art and background information. It also includes fundamental concepts and theories, terminology definitions, and a high-level overview of the problem domain. Likewise, it serves as a primer to the rest of the report, providing the necessary context for the research and is intended for readers who may not be experts in the specific area of the research but have some knowledge of the broader field.

\section{SSH and OpenSSH Implementation}\label{sec:background:ssh}

    \subsection{Basics of the Secure Shell Protocol (SSH)}
    
    The Secure Shell Protocol, commonly known as \acrshort{ssh}, is designed to facilitate secure remote login and other secure network services over insecure networks. \acrshort{ssh} has been designed since its inception with security in mind, as a successor of the Telnet protocol, which is not secure, and other \say{unsecured remote shell protocols such as rlogin, rsh and rexec} \cite{SSHkex22}. 
    
    \subsubsection{SSH design and origin}
    As stated by the authors of the \citetitle{SSHReport18}, \say{The founder of SSH, Tatu Ylönen, designed the first version of the SSH protocol after a password-sniffing attack at his university network. Tatu released his implementation as freeware in July 1995, and the tool quickly gained in popularity. Towards the end of 1995, the SSH user base had grown to 20,000 users in fifty countries. By 2000, there were an estimated 2,000,000 users of the protocol. Today, more than 95\% of the servers used to power the Internet have SSH installed in them. The SSH protocol is truly one of the cornerstones of a safe Internet.} \cite{SSHReport18}.

    \acrshort{ssh} is defined in \citetitle{RFC4251} \cite{RFC4251}. It is divided into three major components:

    \begin{itemize}
        \item \textbf{Transport Layer Protocol:} This provides server authentication, confidentiality, and integrity. It can also optionally provide compression. Typically, the transport layer runs over a TCP/IP connection but can also be used on top of any other reliable data stream.
        \item \textbf{User Authentication Protocol:} Running over the transport layer, this protocol authenticates the client-side user to the server. Multiple methods of authentication such as password and public key are supported.
        \item \textbf{Connection Protocol:} This multiplexes the encrypted tunnel established by the preceding layers into several logical channels. Channels can be used for various purposes, such as setting up secure interactive shell sessions or tunneling arbitrary TCP/IP ports.
    \end{itemize}

    \say{The client sends a service request once a secure transport layer connection has been established. A second service request is sent
    after user authentication is complete. This allows new protocols to be defined and coexist with the protocols listed above} \cite{RFC4251}.

    \subsubsection{SSH keys}\label{sec:background:ssh:ssh_keys}
    For the purposes of this Masterarbeit, a comprehensive understanding of SSH's key exchange and encryption mechanism is important. As outlined in SSHKex \cite{SSHkex22}, the SSH protocol utilizes a key exchange procedure that culminates in a derived master key \( K \) and a hash value \( h \). These components are pivotal for encrypting client-server communications and identifying sessions.

    During the key exchange process, Diffie-Hellman is employed to negotiate an ephemeral shared key between the client and the server \cite{RFC4251} \cite{StackExchangeSSHuseRASandDH23}. The Diffie-Hellman key exchange is a method of securely exchanging cryptographic keys over a public channel. Proposed by Whitfield Diffie and Martin Hellman in 1976, this protocol is one of the first practical implementations of public key exchange. The fundamental principle behind Diffie-Hellman is the difficulty of solving discrete logarithm problems \cite{ComOverInsecureChannels78}. This ephemeral key is then signed by the server using either RSA, DSA, or another suitable signature algorithm (see \ref{subsubsec:background:ssh:hashing}). The signed key confirms to the client that the negotiated key is indeed from the intended server and not an imposter or a middleman, thereby preventing man-in-the-middle (MITM) attacks.

    In addition, the host key of the server is used to sign the Diffie-Hellman parameters. This key is not to be confused with the client key listed in the server's \texttt{authorized\_keys} file, which is used later for client authentication.

    The key exchange process results in multiple session keys computed for various purposes:

    \begin{itemize}
        \item \textbf{Initialization Vectors:} Key A and Key B are designated for initialization vectors from the client to the server and vice versa.
        \item \textbf{Encryption Keys:} Key C and Key D act as encryption keys for client-to-server and server-to-client communications, respectively.
        \item \textbf{Integrity Keys:} Key E and Key F are utilized to preserve the integrity of data transmitted between the client and server.
    \end{itemize}

    This approach provides forward secrecy: if the private key is stolen, it does not compromise the encryption of old sessions. This is because the Diffie-Hellman parameters are ephemeral and discarded once they are no longer needed. Therefore, the only long-lasting keypair is used for authentication purposes.

    \subsubsection{SSH key encryption}
    These keys are computed using hash functions that take the master key \textit{K} and a hash value \textit{H}, a unique letter (A, B, C, D, E, or F), and the session ID as inputs. This is summarized in \citetitle{OpenSSHUnderHood07}:
    \say{
        The equations used for deriving the above vectors and keys are taken from RFC 4253 \cite{RFC4253}. In the following, the $||$ symbol stands for concatenation, K is encoded as mpint, $K$ is already a number (hash), $"A"$ as byte and $session_id$ as raw data. Any letter, such as the $"A"$ (in quotation marks) means the single character A, or ASCII 65.
        \begin{itemize}
            \item Initial IV client to server: $HASH(K || H || "A" || session_id)$.
            \item Initial IV server to client: $HASH(K || H || "B" || session_id)$.
            \item Encryption key client to server: $HASH(K || H || "C" || session_id)$.
            \item Encryption key server to client: $HASH(K || H || "D" || session_id)$.
            \item Integrity key client to server: $HASH(K || H || "E" || session_id)$.
            \item Integrity key server to client: $HASH(K || H || "F" || session_id)$.
        \end{itemize}
    } \cite{OpenSSHUnderHood07}. Details about the hash function are given in the next section.
    
    The most interesting keys are the encryption keys, as they are used to encrypt the communication between the client and the server. The other keys are used for integrity checks and initialization vectors. Decrypting encrypted SSH communication necessitates either to retrieve these session keys and variables so as to recompute the keys, or to retrieve those keys directly, which is the focus of this Masterarbeit. 

    \subsection{OpenSSH Implementation}
    OpenSSH (OpenBSD Secure Shell) is an open-source implementation written in C of the SSH protocol suite, and it is the most widely used SSH implementation \cite{OpenSSHUnderHood07}. It is the default SSH implementation on most Linux distributions, and it is also available for Windows. OpenSSH is used for a wide range of purposes, including remote command-line login and remote command execution. It is also used for port forwarding, tunneling, and transferring files via SCP and SFTP either manually or via automated processes, such as backup systems, configuration management tools, and automated software deployment tools. 

    \subsubsection{OpenSSH components}
    OpenSSH is composed of several tools and daemons, including client and server components \cite{PortableOpenSSHGitHub}:
    \begin{itemize}
        \item \textbf{ssh:} The basic client program that allows to log into and execute commands on a remote machine.
        \item \textbf{sftp:} An interactive file transfer program that uses SSH to secure the connection.
        \item \textbf{sshd:} This is the SSH daemon that runs on the server. This is used for connecting to a remote machine when using the SSH client from another system.
        \item \textbf{ssh-agent:} The program that holds private keys in memory, so one doesn't have to enter one's passphrase every time.
        \item \textbf{ssh-add:} A program for adding RSA or DSA identities to the authentication agent.
        \item \textbf{ssh-keygen:} A utility for creating and managing SSH keys.
        \item \textbf{ssh-keyscan:} A utility for gathering public SSH host keys from a number of hosts.
        \item \textbf{ssh-keychk:} A utility for checking the validity of SSH keys.
        \item Several other tools to support the SSH protocol and the OpenSSH implementation.
    \end{itemize}

    \subsubsection{OpenSSH hashing}\label{subsubsec:background:ssh:hashing}
    OpenSSH employs a variety of hash functions and algorithms to secure data, most commonly using SHA1. However, SHA1 is increasingly seen as weak due to its vulnerability to collision attacks \cite{OpenSSHUnderHood07}. In light of this, the contemporary standard leans towards SHA512. The hash functions are used alongside cipher algorithms like \say{Advance Encryption Standard (AES) Cipher Block Chaining (CBC), AES Counter (AES-CTR), and ChaCha20} \cite{SmartKex22}. The Message Authentication Code (MAC) typically uses either MD5 or SHA1 hash algorithms in combination with a secret key. Since cybersecurity and cryptography are constantly evolving, so do \acrshort{ssh} and OpenSSH. Depending on the version \cite{OpenSSHUnderHood07}, the available hash options include:
    
    \begin{itemize}
        \item \textbf{ssh-dss:} \textit{(disabled at run-time since OpenSSH 7.0 released in 2015)} SSH-1 version using Digital Signature Algorithm (DSA) from the Digital Signature Standard (DSS). Originally popular but phased out due to vulnerabilities to collision attacks for DSA Key in a 1024-bit modulus. As stated by \citetitle{RFC9142}: \say{These attacks are still computationally very difficult to perform, but it is desirable that any key exchange using SHA-1 be phased out as soon as possible} \cite{RFC9142} \cite{OpenSSHReleaseNotes7-0}.
        
        \item \textbf{ssh-rsa:} \textit{(disabled at run-time since OpenSSH 8.8 released in 2021)} It refers to the use of RSA (Rivest-Shamir-Adleman) encryption algorithm. In the context of SSH-1, this version had to be replaced due to the related to key size issue similar to DSS: \say{RSA 1024-bit keys have approximately 80 bits of security strength}... \say{which may not be sufficient for most users.} \cite{RFC9142} \cite{OpenSSHReleaseNotes8-8}.
        
        \item \textbf{ecdsa-sha2-nistp256:} \textit{(since OpenSSH 5.7 released in 2011)} Uses the SHA-2 family for hashing and the NIST P-256 curve. It is considered secure and efficient, with an \acrfull{ess} of 128 bits \cite{RFC9142} \cite{OpenSSHReleaseNotes5-7}.
        
        \item \textbf{ecdsa-sha2-nistp384:} \textit{(since OpenSSH 5.7)} Utilizes the SHA-2 family and the larger NIST P-384 curve for additional security at the cost of performance. It has an \acrshort{ess} of 192 bits \cite{RFC9142} \cite{OpenSSHReleaseNotes5-7}.
        
        \item \textbf{ecdsa-sha2-nistp521:} \textit{(since OpenSSH 5.7)} Employs SHA-2 and the even larger NIST P-521 curve for maximal security with an \acrshort{ess} of 256 bits \cite{RFC9142}. It is less commonly used due to performance considerations \cite{OpenSSHReleaseNotes5-7}. 
        
        \item \textbf{ssh-ed25519:} \textit{(since OpenSSH 6.5 released in 2014)} Known for high security and performance efficiency; employs the Ed25519 elliptic curve with an \acrshort{ess} of 128 bits \cite{RFC9142} which is similar to $ecdsa-sha2-nistp256$, and has been more prevalent following the 2013 suspicions of NSA backdoors in NIST curves \cite{Adamantiadis2013} following the Snowden revelations \cite{NSAFoilSafeguards2013} \cite{GuardianEncryption2013} \cite{OpenSSHReleaseNotes6-5}.
        
        \item \textbf{rsa-sha2-256:} \textit{(since OpenSSH 7.2 released in 2016)} An upgrade from ssh-rsa, using SHA-256 (with \acrshort{ess} of 128 bits) for hashing to improve security without major performance hits \cite{OpenSSHReleaseNotes7-2}.
         
        \item \textbf{rsa-sha2-512:} \textit{(since OpenSSH 7.2)} Similar to $rsa-sha2-256$ but employs SHA-512 for even stronger security, albeit with some performance cost \cite{OpenSSHReleaseNotes7-2}.
        
        \item \textbf{ecdsa-sk:} \textit{(since OpenSSH 8.2 released in 2020)} Security Key-enabled, uses NIST curves and is geared towards modern hardware-based authentication \cite{OpenSSHReleaseNotes8-2}.
        
        \item \textbf{ed25519-sk:} \textit{(since OpenSSH 8.2)} Similar to ssh-ed25519 but integrates hardware-based Security Keys for an additional layer of security \cite{OpenSSHReleaseNotes8-2}.
        
        \item \textbf{NTRU Prime-x25519:} \textit{(since OpenSSH 9.0)} A new, highly secure algorithm focused on post-quantum cryptography, providing future-proof security \cite{NTRUPostQuantum17} \cite{OpenSSHReleaseNotes9-0}.
    \end{itemize}
    
    These hashes have fixed lengths such that key lengths range between 12 and 64 bytes \cite{SmartKex22}. Since high-quality random number generation is crucial to ensure that those keys are secure and difficult to predict, it can thus be assumed that those key have a high entropy \cite{McLaren2019}. This is a crucial assumption as it is the basis for the use of both brute force and machine learning algorithms to predict the presence and location of SSH keys in memory dumps.

    The keys generated by these hash functions are pseudo-random numbers stored in the system's RAM. Following the Kerckhoffs' principle: that \say{a cryptosystem should be secure, even if everything about the system, except the key, is public knowledge}, the code for the OpenSSH implementation is open-source and available on GitHub \cite{PortableOpenSSHGitHub}. This allows for the analysis of the code and the identification of the memory structures where the keys are stored.

    \subsection{The state of SSH security}
    Since its origins, SSH has been developed with cybersecurity in mind, and is generally considered a secure method for remote login and other secure network services over an insecure network. However, as with any technology, it can be exploited if not configured or managed correctly. The protocol is used by system administrators to manage remote systems, and it is also used by automated processes to transfer data and perform other tasks. This makes SSH a valuable target for attackers. In fact, SSH has been a popular target for cyber-attacks. Due to being so prevalent, it is often used by threat actors either as a vector for initial access, as a means to move laterally across a network or as a covered exit for exfiltration of sensitive data \cite{APTTactics19}. The encrypted nature of its communications makes it an attractive option for attackers, as it can be difficult to detect malicious activity.
    
    \subsubsection{SSH security issues}
    Here are some cases where SSH can involve in cyber-attacks, although it's important to note that SSH itself is not inherently insecure:
    \begin{itemize}
        \item \textbf{SSH Brute-Force Attacks:} One of the most common types of attacks involving SSH is a brute-force attack, where an attacker tries to gain access by repeatedly attempting to log in with different username-password combinations. These attacks are not sophisticated but can be effective if strong authentication measures are not in place. For instance, the botnet \textit{Chabulo} was used to launch a large-scale brute-force attack \say{through compromised SSH servers and IoT devices} in 2018 \cite{SSHReport18}. 
        \item \textbf{SSH Key Theft:} In some advanced attacks, threat actors have stolen SSH keys to move laterally across a network after initial entry. This allows them to authenticate as a legitimate user and can make detection much more challenging. It can \say{ occur when users have their SSH password or unencrypted keys stolen through a variety of methods (sniffed via a key-logging console program, shoulder-surfed via bad security awareness, poor key management practices, etc.).} \cite{SSHIdentityTheft05}.
        \item \textbf{Man-in-the-Middle Attacks:} Although SSH is designed to be secure, it can be susceptible to man-in-the-middle attacks if proper verification of SSH keys is not done during the initial connection setup \cite{OpenSSHUnderHood07}.
        \item \textbf{Misconfiguration:} As with any technology, misconfiguration can lead to security issues. For example, leaving default passwords, using weak encryption algorithms, or enabling root login can all make an SSH-enabled system vulnerable \cite{SSHBotnetInfect21}.
    \end{itemize}

    \subsubsection{SSH vulnerabilities}
    In cybersecurity, it is generally considered that any system that is connected to the Internet will be attacked at some point. Similarly, it is a common saying that no system is 100\% secure. This is true for SSH as well. Although it is a secure protocol, it can be exploited if not configured or managed correctly. 
    
    Some vulnerabilities have also been discovered in the protocol itself, although these are rare.
    \begin{itemize}
        \item \textbf{SSH-1 Vulnerabilities:} A series of vulnerabilities in the first implementation of SSH were discovered from 1998 to 2001, with its subsequent fixes leading to unauthorized content insertion and arbitrary code execution. SSH-1 had many design flows and is now considered obsolete. \cite{CoreSecurity23}, \cite{SSH1Vulnerability01}. 
        \item \textbf{CBC Plaintext Recovery:} A theoretical vulnerability discovered in 2008 affecting all versions of SSH, allowing the recovery of up to 32 bits of plaintext from CBC-encrypted ciphertext \cite{USCERT2011}.
        \item \textbf{Suspected Decryption by NSA:} Leaked information in 2014 suggested that the NSA might be able to decrypt some SSH traffic, although the protocol itself was not confirmed to be compromised \cite{Spiegel14}.
    \end{itemize}
    
    \subsubsection{SSH and cyber-attacks}
    SSH has been used in many high-profile cyber-attacks and malwares, including the following:
    \begin{itemize}
        \item \textbf{Operation Windigo:} This was a large-scale campaign that infected over 25,000 UNIX servers. SSH was one of the vectors used for maintaining control over compromised servers. A report by ESET mentions that the  OpenSSH backdoor Linux/Ebury was first discovered in 2011 as a component of the aforementioned operation. \say{This operation has been ongoing since at least 2011 and has affected high profile servers and companies, including cPanel - the company behind the famous web hosting control panel - and Linux Foundation's kernel.org - the main repository of source code for the Linux kernel} \cite{ESETWindigo14}. 
        \item \textbf{Linux/Hydra:} Initially unleashed in 2008, this malware is a fast login cracker that targets a range of popular protocols including SSH. Hence, SSH is one of its primary vectors to gain initial access to Internet of Things (IoT) devices. Once a device is infected by Linux/Hydra, it joins an IRC channel and initiates a SYN Flood attack \cite{ClassificationMalware21}.
        \item \textbf{Psyb0t:} Discovered in early 2009, Psyb0t is an IRC-controlled malware specifically designed to target devices with MIPS architecture, such as routers and modems. Notably, it was responsible for orchestrating a DDoS attack against the DroneBL service, infecting up to 100,000 devices for this purpose. The malware is equipped to conduct UDP and ICMP flood attacks and employs a brute-force attack mechanism against Telnet and SSH ports. Remarkably, it uses a pre-configured list of 6,000 usernames and 13,000 passwords to perform these attacks \cite{ClassificationMalware21}.
        \item \textbf{Chuck Noris:} Similar to Psyb0t in its objectives and methods, Chuck Noris targets routers and DSL modems, focusing on SoHo (small office/home office) devices. However, unlike Psyb0t, which uses ICMP flood attacks, Chuck Noris deploys ACK flood attacks. The malware carries out brute-force attacks on Telnet and SSH open ports, drawing parallels to the tactics employed by Psyb0t but with the specific variation in flooding techniques \cite{ClassificationMalware21}.
    \end{itemize}

    It's worth noting that in many of these cases, SSH was not the initial attack vector but was used at some stage in the attack lifecycle. Properly configured and managed SSH is still considered a secure and robust protocol for remote access and data transfer. In all those situations, a tool monitoring the SSH traffic could have detected the malicious activities and prevented the attack.

    \subsection{The Imperative of SSH Honeypots in Cybersecurity Monitoring}
    SH (Secure Shell) has become an indispensable protocol for secure communication but can also conceal malicious agents. This reality underscores the urgency for robust monitoring mechanisms capable of identifying suspicious activities in real-time. Among various countermeasures, SSH honeypots have emerged as a particularly effective tool for monitoring and gathering intelligence on potential threats. 

    An SSH honeypot is a decoy server or service that mimics legitimate SSH services. The primary aim is to attract cybercriminals and study their tactics, thereby offering an active form of surveillance and data collection. Unlike traditional intrusion detection systems, honeypots do not merely identify an attack; they engage the attacker in a controlled environment, enabling detailed observation and logging of the intruder's actions. This allows for the collection of valuable information, such as the attacker's IP address, the tools used, and the techniques employed. This data can then be used to enhance security measures and develop more robust countermeasures \cite{ClassificationMalware21}. 

    SSH honeypots serve as an invaluable asset in the cybersecurity arsenal, providing not just a reactive but a proactive measure against evolving cyber threats. They can collect actionable intelligence on new hacking methods, malware, and exploitation scripts. This information can be crucial for proactively securing actual production environments. The data collected can also be used to trace back to the origin of the attack, facilitating legal pursuits against the perpetrators. By diverting attackers to decoy servers, honeypots also protect real assets from being targeted, saving both computational resources and administrative effort needed for post-incident recovery.

    Popular SSH honeypots include Kippo, Cowrie, and HoneySSH.  Cowrie is a fork of Kippo, with additional features such as logging of attacker's keystrokes and file transfer. 

    \begin{itemize}
        \item \textbf{Kippo:} Kippo is a medium-interaction honeypot that logs the attacker's shell interaction. It specializes in capturing brute force and Telnet-based attacks \cite{ClassificationMalware21}.
        
        \item \textbf{Cowrie:} Serving as Kippo's successor, Cowrie emulates various protocols including SSH, SFTP, and SCP. It logs events in JSON format, making it particularly useful for detecting brute force and Telnet-based attacks, as well as spoofing attacks \cite{ClassificationMalware21}.
        
        \item \textbf{IoTPOT:} This IoT-focused honeypot supports multiple CPU architectures and can detect a variety of attacks including brute force, DoS, and sniffing attacks on Telnet, SSH, and HTTP ports \cite{ClassificationMalware21}.
        
        \item \textbf{HoneySSH:} HoneySSH is a low-interaction honeypot that emulates an SSH server and logs the attacker's IP address, username, and password \cite{honeyssh17}.
        
        \item \textbf{Sarracenia (SSHKex):} Introduced in 2018, Sarracenia is a high-interaction SSH honeypot that has been enhanced by SSHKex. Instead of \say{requiring the VM to be paused for every incoming or outgoing packet, which degrades the server performance} \cite{SSHkex22}, SSHKex allows for the extraction of derived SSH session keys. This reduces the performance degradation significantly, as the VM is paused less frequently \cite{SSHkex22} \cite{SarraceniaSSHHoneypot18}.
    \end{itemize}

    These honeypots are useful tools for gathering intelligence on potential threats. However, they are not without their limitations.

    \subsection{Research context and motivation for this Masterarbeit}

    Security and malware detection are active areas of research, with SSH honeypots being a particularly promising tool for gathering intelligence on potential threats. However, they are not without their limitations. For instance, they are not able to perfectly mimic a real system, such that attackers might be able to detect them \citetitle{SSHHoneypotEffectiveness23}. 
    
    As explained in \citetitle{SSHHoneypotEffectiveness23}: \say{As attackers become more sophisticated with their ransomware and malware campaigns, there is a significant need for security researchers to assist the greater community by running vulnerable honeypot machines to collect malicious software}. \citeauthor{SSHHoneypotEffectiveness23} explain that \say{the ability to collect meaningful malware from attackers depends on how the attackers receive the honeypot. Most attackers fingerprint targets before they launch their attack, so it would be very beneficial for security researchers to understand how to hide honeypots from fingerprinting and trick the attackers into depositing malware.} They conclude that \say{What is certain is that if a cautious attacker believes they are in a honeypot, they will leave without depositing malware onto the system, which reduces the effectiveness of the honeypot for security research.} We can extrapolate this conclusion to SSH honeypots, which are also vulnerable to fingerprinting and detection by attackers. 
    
    Hence, the need for more advanced SSH honeypots-inspired tools that can leverage data forensic and machine learning techniques so as to be able to use directly a real server as a honeypot, without the need to emulate a system. The current master's thesis is aligned with this ongoing research (see \ref{sec:background:previous_work}), further enhancing the state of SSH honeypots. It aims to develop algorithms, proof of concepts and tools that can extract SSH keys from memory dumps of a real server, and use them to decrypt SSH traffic. This could lead to the development of new tools for SSH monitoring, as discussed in the future work section \ref{conclusion:sec:future_work}.

\section{Previous Work on OpenSSH key extraction}\label{sec:background:previous_work}

    Now that the necessary context has been established, this section will present the related work in the field of machine learning for memory forensics in the context of OpenSSH. It is divided into two parts. The first part will present the related work in the field of memory forensics, and the second part will present the related work in the field of machine learning for memory forensics.

    \subsection{SSHKex}\label{sec:background:kex:sshkex}
        
    SSHKex is a research project that aims to address the challenges of analyzing encrypted SSH traffic by leveraging \acrfull{vmi} techniques. Developed by \citeauthor{SSHkex22}, the project focuses on extracting SSH keys and decrypting SSH network traffic in a stealthy, non-intrusive manner while maintaining evidence integrity \cite{SSHkex22}. This paper is itself a continuation of the work presented in \citetitle{SarraceniaSSHHoneypot18} \cite{SarraceniaSSHHoneypot18}, which introduced Sarracenia, a high-interaction SSH honeypot. It is also related to a range of other research projects and papers \cite[section 5.6 and 6]{SSHkex22}.
    
    The SSHKex approach combines standard network traffic capturing methods with dynamic SSH session key extraction. It assumes that the SSH implementation running on the server is known, which is crucial for the key extraction process. The project employs VMI tools like LibVMI and Volatility to gain a complete and untainted view of all guest VM's state information. This allows to efficiently locate SSH session keys in the main memory of a Linux machine. 
    
    Here is a summary of the SSHKex methodology for key extraction:
    \begin{enumerate}
        \item \textbf{Data Structure Information:} The method leverages detailed knowledge about the data structures used to store the keys. Specific debugging symbols corresponding to the SSH implementation version on the target system provide essential offset values to facilitate the extraction of key material. The structures of interest include \texttt{struct ssh}, \texttt{struct session\_state}, \texttt{struct newkeys}, and \texttt{struct sshenc}. These structures store a range of information such as IP addresses, ports, session states, and encryption keys.
    
        \item \textbf{Tracing OpenSSH Functions:} Function tracing is employed to identify the precise locations of data structures and to extract keys at the right time. The focus is on two key functions: \texttt{kex\_derive\_keys} (which initiates key generation) and \texttt{do\_authentication2} (which kicks off user authentication).
    
        \item \textbf{Breakpoints Injection:} Software breakpoints are intentionally placed in the program execution to facilitate debugging. SSHKex utilizes Virtual Machine Introspection (VMI) to inject these breakpoints at the initial points of the two aforementioned key functions.
    
        \item \textbf{Key Extraction:} Upon calling the \textit{kex\_derive\_keys} function, SSHKex initially stores the address of the \textit{ssh struct}. The actual keys are extracted from memory when the \textit{do\_authentication2} function is subsequently called, adhering to the known structures. 
    
        \item \textbf{Key Indexing:} OpenSSH stores client-to-server and server-to-client keys in distinct indices of the \texttt{newkeys} structure. SSHKex extracts keys based on these specific indices.
    
        \item \textbf{Handling Multiple Connections:} To manage multiple SSH connections, OpenSSH spawns child processes. SSHKex extends its key extraction strategy to each child process by identifying them through their unique process IDs.
    \end{enumerate}
    
    One of the key strengths of SSHKex is its focus on stealthiness, preservation, and evidence integrity. The approach aims to be as unobtrusive as possible, avoiding any modifications to the system under investigation. This is particularly important in forensic contexts, where the integrity of the evidence is crucial \cite{SSHkex22}.
    
    \subsection{SmartKex}
    
    SmartKex is a direct followup project that focuses on the extraction of SSH keys from heap memory dumps. Its primary objective is to automate the process of SSH key extraction from heap memory dumps. The project introduces a machine learning-assisted methodology that significantly improves the efficiency and accuracy of key extraction compared to traditional brute-force methods. This method is also significantly more straightforward to implement compared to the previous SSHKex approach, which requires detailed knowledge of the SSH implementation and the ability to inject breakpoints into the program execution.
    
    SmartKex discusses two distinct methods for SSH key extraction:
    \begin{itemize}
        \item \textbf{Brute-Force Baseline Method:} This is a traditional approach that scans through the heap memory to identify potential keys based on known patterns.
        \item \textbf{Machine Learning-Assisted Method:} This  approach uses a Random Forest algorithm trained on a highly imbalanced dataset using \acrshort{smote} balancing. The machine learning model is designed to identify SSH keys with high precision and recall rates, but is not exact as compared to the brute-force method since it is based on a probabilistic model.
    \end{itemize}
    
        \subsubsection{Baseline brute-force method}
    
        Here is a summary of SmartKex's brute-force method for SSH key extraction from heap dumps \cite{SmartKex22}:
        \begin{enumerate}
            \item \textbf{Heap Dump Generation:} Heap dump binary files of OpenSSH server process have been generated (ASK HOW) and serves as the input for the key extraction process. The exact process and architecture is not described in SmartKex paper, but we suppose it was done on a \textit{linux-x86\_64} architecture.
            
            \item \textbf{Data Reduction:} To minimize the heap size, the method removes memory pages that are irrelevant (empty) based on Hamming distance.
            
            \item \textbf{Brute-force key search:} Starting from the first byte, a key length of 128 bytes is taken from the heap dump as the potential key. The algorithm iterates over the entire heap, continuously updating the potential key until the heap's end is reached.
            
            \item \textbf{Decryption Attempt:} For every potential key, an attempt is made to decrypt network packets. If decryption fails, the process is repeated with a new potential key.
        \end{enumerate}
        
        Although the brute-force approach is exact, it is computationally expensive. It performs poorly especially when keys are located at the end of the heap dump \cite[section 6.2]{SmartKex22}.
    
        \subsubsection{SmartKex machine-learning method}
    
        The real innovation of SmartKex is its machine learning-assisted methodology for SSH key extraction. At the cost fo exactness, this approach is significantly faster than the brute-force method and has a high degree of accuracy in identifying encryption keys. It also allows for the heap size to be reduced to less than 2\% of its original size, further optimizing the extraction process.
    
        Here is a summary of SmartKex's machine learning-assisted method for SSH key extraction from heap dumps \cite{SmartKex22}:
    
        \begin{enumerate}
            \item \textbf{Heap Dump inputs:} Similarly to the brute-force method, heap dump binary files of OpenSSH also serve as inputs for the key extraction process.
            \item \textbf{Preprocessing:} The raw heap dump is resized into an $N \times 8$ matrix. High entropy parts of the heap dump, which are likely to be encryption keys, are identified using the logical AND operation on the vertical and horizontal differences of adjacent bytes. This creates an array that flags potential key locations.
            \item \textbf{Training:} A Random Forest algorithm is trained on 128-byte slices of the preprocessed heap. The dataset is imbalanced, with the slices that contain keys being rare. A stacked classifier approach is used, comprising a high precision classifier and a high recall classifier.
            \item \textbf{Key Identification:} The machine learning model is used to predict which 128-byte slices of the heap dump are likely to contain encryption keys. These slices are then subjected to a brute-force method to actually extract the keys.
        \end{enumerate}
        
    SmartKex is significantly faster than the brute-force method alone and has a high degree of accuracy in identifying encryption keys. It also allows for the heap size to be reduced to less than 2\% of its original size, further optimizing the extraction process.
    
    SmartKex has broad applications in the field of cybersecurity, particularly in memory forensics. Its machine learning-assisted methodology can be adapted for other types of sensitive data extraction, making it a versatile tool for researchers and practitioners alike. The project is open-source, with the code available on GitHub\footnote{\url{https://github.com/smartvmi/Smart-and-Naive-SSH-Key-Extraction}}.
    
    \subsection{Objectives of the present work}
    This Masterarbeit can be seen as a direct followup to the paper \citetitle{SmartKex22}. The present work aims to improve the SmartKex methodology by exploring new machine learning architectures and algorithms. The goal is to improve the accuracy of the machine learning model and to reduce the computational complexity of the overall process.
    
    To do so, this work has significantly broadened the research area by exploring entirely new ways to deal with the dataset by leveraging memory graph representation, feature engineering, new machine learning and deep learning model architectures, and new training strategies. A range of different tools and script, with a focus on code quality and reproducibility with careful packaging using Nix ensure that the present research can be easily extended and reproduced by other researchers.

\section{Graph-based memory modelization}\label{sec:background:graph}

    In the following section, we present important concepts that will be used for the memory modelization of the heap dump.

    Because the dataset used is composed of RAW heap dump files from OpenSSH, it is a critical aspect to understand how memory works at a low level point of view. This section aims to provide an in-depth understanding of how memory is managed in C, the language used in the OpenSSH implementation of SSH, for a \textit{linux-x86\_64} architecture. We will explore the fundamental concepts of memory management, including the heap and the stack, memory allocation, and the role of pointers. These concepts will serve later as the foundation for our graph-based approach to memory modelization.

    This section will also introduce many graph theory and \acrfull{kg} concepts. We will explore the fundamentals of graph theory, including the definition of a graph, its components, and its properties. We will also discuss the concept of a \acrfull{kg}, which is a type of graph that stores information in the form of nodes and edges, and its applications in the field of machine learning.

    \subsection{Defining memory concepts and modelization}
    Memory management in C is a complex task that requires a deep understanding of the language's features, the operating system's capabilities and the compiler used. In C, memory is primarily managed through two built-in functions: \texttt{malloc} (memory allocation) and \texttt{free} (memory deallocation). These functions operate on two primary types of memory: the heap and the stack.

    \begin{itemize}
        \item \textbf{Heap vs Stack:} The heap is used for dynamic memory allocation, where variables are allocated and freed at runtime. In contrast, the stack is used for static memory allocation, where variables are allocated and deallocated automatically. The stack is faster but has a limited size, while the heap is more flexible but requires manual management to prevent memory leaks.
        
        \item \textbf{Heap Dump:} A heap dump is a snapshot of the heap's state at a given time. It provides valuable information about the memory layout, active pointers, and data stored in the heap. Analyzing heap dumps can help in debugging memory-related issues and understanding the program's behavior.
        
        \item \textbf{Memory Addresses:} Each location in memory is identified by a unique memory address. These addresses are usually represented in hexadecimal notation. Note that the address $0x0$ is reserved for the \textit{NULL} pointer, which is used to indicate that a pointer does not point to any memory location.

        \item \textbf{Pointer:} A pointer is a variable that stores the memory address of another variable. It is used to indirectly access the value of the variable it points to. Pointers are used extensively in C, particularly for dynamic memory allocation.
        
        \item \textbf{Data Structure:} A data structure is a collection of data values, the relationships among them, and the functions or operations that can be applied to the data. In the context of C programming, data structures are declared using the keyword \texttt{struct} and are byte aligned. This means that the size of the data structure is always a multiple of the size of the largest member of the structure. Structures can be nested within each other, and pointers can be used to indirectly access the members of a structure. Data structures are often stored in the heap using \texttt{malloc}.
        
        \item \textbf{Malloc headers:} When \texttt{malloc} is called, it allocates a block of memory in the heap and returns a pointer to the first byte of the block. The heap manager keeps track of these allocations through metadata, often stored in headers preceding the allocated blocks. These headers contain information such as the size of the allocated block and whether it is free or occupied. Note that the pointer returned by malloc in C points to the first byte of the block of memory that has been allocated for your use, not to the malloc header. The malloc header, is managed internally by the memory allocator and is not exposed to the programmer, but is visible in the heap dump.
        
    \end{itemize}

    \subsubsection{Endianness}
    
    Endianness refers to the byte order used to represent multibyte data types. In a \textit{little-endian} system, the least significant byte is stored first, while in a \textit{big-endian} system, the most significant byte is stored first. Knowing the endianness of the system is crucial for interpreting the content of memory \cite{InferenceEndianness17}.
    
    For instance, the hexadecimal value \textit{0x56343a198000} (taken from \textit{"HEAP\_START"} of \ref{lst:json-annotation-ex-1}) is represented as $550179058774$ ($\simeq 5.50e+11$) in decimal basis in a little-endian system, while it is represented as $94782313037824$ ($\simeq 9.48e+13$) in a big-endian system.

    \begin{minipage}{\dimexpr\linewidth-20pt}
        \paragraph{Little-Endian Conversion}
        The conversion of a hexadecimal number in \textit{little-endian} format to a decimal number is given by the following formula:
        \par 
        
        \vspace{2em}  
        \begin{center}
            $
            \text{Decimal} = \sum_{i=0}^{N-1} \left( \text{HexDigit}_{i} \times 16^{i} \right)
            $
        \end{center}
        \vspace{1em}
    \end{minipage}
    
    \begin{minipage}{\dimexpr\linewidth-20pt}
        \paragraph{Big-Endian conversion}
        And the conversion of a hexadecimal number in \textit{big-endian} format to a decimal number is given by following formula:
        \par 
        
        \vspace{2em}  
        \begin{center}
            $
            \text{Decimal} = \sum_{i=0}^{N-1} \left( \text{HexDigit}_{N-1-i} \times 16^{i} \right)
            $
        \end{center}
        \vspace{1em}
    \end{minipage}

    Here, $ \text{HexDigit}_{i} $ is the value of the \(i\)-th digit in the little-endian hexadecimal number, and \( N \) is the number of digits in the hexadecimal number. Note that $ \text{HexDigit}_{i} $ should be converted to its decimal equivalent ('A' becomes 10, 'B' becomes 11, etc.) before performing the calculation.

    These formulas will be used later to convert pointer addresses from hexadecimal to decimal format in \textit{mem2graph}.

    \subsubsection{The role of entropy in forensic analysis}

    Entropy plays a pivotal role in forensic analysis, particularly in the context of memory dumps analysis. It serves as a measure of uncertainty and randomness, which can be crucial for tasks such as endianness detection and identifying encrypted keys in memory.

    As defined by Shannon in the realm of information theory, it is a measure of the uncertainty or randomness associated with a set of possible outcomes \cite{InferenceEndianness17} \cite{TheoryOfCommunicationShannon1948}. In digital applications, when calculated using the logarithm to base 2, entropy represents the amount of bits of information in a message.

    \begin{minipage}{\dimexpr\linewidth-20pt}
        \paragraph{Entropy Formula}
        The entropy \( H \) of a message is calculated using the formula:
        \par 
        
        \vspace{2em}  
        \begin{center}
            $
            H = - \sum_{i=1}^{n} p_i \log_2 p_i
            $
        \end{center}
        \vspace{1em}

        Where $ p_1, p_2, \ldots, p_n $ are the probabilities of the set of all possible messages \cite{InferenceEndianness17}.
    \end{minipage}

    Endianness, as presented before, refers to the byte order used to represent multibyte data types in computer memory \cite{InferenceEndianness17}. It is necessary to know the endianness of the heap dump to correctly interpret the content of memory, and especially the addresses of potential pointers. In this context, entropy can be used to infer the endianness of a system by analyzing the distribution of byte values in a memory dump, as presented in \citetitle{InferenceEndianness17} \cite{InferenceEndianness17}.

    Entropy is also a key element for encryption key detection, since those should be random byte sequences with high entropy \cite{SmartKex22}. By examining 8-byte aligned data for high entropy, it is possible to detect potential keys in a heap dump. Techniques such as the calculation of discrete differences and logical operations can further refine this detection as described in \citetitle{SmartKex22} \cite{SmartKex22}.

    \subsection{Graphs and Knowledge Graphs}
    In this project, we (i.e. Clément Lahoche and the author) have built a program called \textit{mem2graph}, a generic tool developed in Rust for efficiency. It is used to convert a RAW memory dump into a graph representation. This graph is technically a directed Edge-labeled heterogeneous property graph, whose concepts are described later. As such, and while it is debatable whether or not \textit{mem2graph} can be seen as building Knowledge Graphs (KGs), it relies on many concepts from the domain, necessitating a proper introduction.

    \subsubsection{Defining Graph Theory concepts}
    Graph theory is a mathematical field concerned with the study of graphs. It has applications in various fields, including computer science, social sciences, or linguistics. Graphs are used to model pairwise relations between objects, and the study of graphs involves analyzing the properties of these relations. In a few words, a graph is just a collection of nodes and edges.
    
    A graph can be formally defined  as: \say{a pair \( G = (S, A) \) where:
    \begin{itemize}
        \item \( S \) is a finite set of vertices.
        \item \( A \) is a set of pairs of vertices \((s_i, s_j) \in S^2\).
    \end{itemize}
    Graphs can be either directed or undirected. In a directed graph, the pairs \((s_i, s_j)\) are ordered, representing arcs from \(s_i\) to \(s_j\). In an undirected graph, the pairs are unordered, representing edges between \(s_i\) and \(s_j\) } \cite{GraphTheorySolnon}. Let's introduce some vocabulary to describe graphs:

    \begin{itemize}
        \item \textbf{Node (or Vertex)}: A single entity in the graph, often represented as a circle.
        \item \textbf{Edge (or Arc)}: A connection between two nodes, often represented as a line or arrow.
        \item \textbf{Degree}: The number of edges connected to a node.
        \item \textbf{Path}: A sequence of edges that connect two nodes.
        \item \textbf{Cycle}: A path that starts and ends at the same node.
        \item \textbf{Order}: The number of vertices in the graph.
        \item \textbf{Adjacency}: Two nodes are adjacent if there is an edge between them.
        \item \textbf{Loop}: An edge that connects a node to itself.
        \item \textbf{Weight}: A value assigned to an edge.
        \item \textbf{Ancestors (parents) and Descendants (children)}: A node \(s_i\) is an ancestor of \(s_j\) if there is a path from \(s_i\) to \(s_j\). A node \(s_j\) is a descendant of \(s_i\) if there is a path from \(s_i\) to \(s_j\).
    \end{itemize}
    
    Various other terminologies and concepts exist in graph theory, but these are the most important ones for our purposes. For a more in-depth understanding of graph theory, the reader is encouraged to consult the work by \citeauthor{GraphTheorySolnon} as a quick introduction, \cite{GraphTheorySolnon} or a more in-depth one by \citeauthor{GraphTheoryIntro01} \cite{GraphTheoryIntro01}.

    \subsubsection{Graphs types}
    Graphs offer a flexible way to conceptualize, represent, and integrate diverse and incomplete data. Many graph models exist, each with its own advantages and disadvantages, as well as graph properties \footnote{Some parts of the following section are directly from a prior work for Seminar 5369S: Knowledge Graphs, written during summer 2023. As of the date of writting and to the best of my knowledge, this work has not been published, and as such, cannot be properly referenced}. Those different types of graphs include:

    \begin{itemize}
        \item \textbf{Directed Edge-labelled Graphs (DEL):} The classic graph, set of nodes and edges that connect the nodes with in certain way. \acrshort{rdf} is a popular \acrshort{del} data model.
        \item \textbf{Heterogeneous Graphs:} Each node and edge is assigned one type, allowing for partitioning nodes according to their type, which is useful for machine learning.
        \item \textbf{Property Graphs:} Allows a set of property-value pairs and a label to be associated with nodes and edges. This model is used in Neo4j and offers great flexibility but is harder to handle and query.
        \item \textbf{Graph Dataset:} A set of named graphs, with a default graph with no ID. Useful when working with different sources.
        \item \textbf{Hypergraphs:} Hypergraphs generalize the concept of graphs by allowing edges, known as hyperedges, to connect any number of nodes, rather than just pairs. This means that a single hyperedge can link together two or more nodes, forming a subset of the hypergraph's node set. This feature makes hypergraphs particularly useful for modeling relationships in complex systems where connections are not merely binary. They are widely used in areas such as database design, combinatorial optimization, and complex network analysis, where multi-way relationships are prevalent.

    \end{itemize}

    A given graph can have a range of properties that give some insights into its structure. Here are some important properties of graphs:

    \begin{itemize}
        \item \textbf{Connected Graph}: A graph is connected if there is a path between every pair of nodes.
        \item \textbf{Disconnected Graph}: A graph is disconnected if there is at least one pair of nodes that are not connected by a path.
        \item \textbf{Cyclic Graph}: A graph is cyclic if it contains at least one cycle.
        \item \textbf{Acyclic Graph}: A graph is acyclic if it does not contain any cycles.
        \item \textbf{Complete Graph}: A graph is complete if there is an edge between every pair of nodes.
    \end{itemize}

    \subsubsection{Graph vs Knowledge Graph}
    A \acrfull{kg} is a specialized form of graph intended to accumulate and convey real-world knowledge. As said before, we are not technically building KGs, but we rely on many concepts from the domain. Research on \acrshort{kg} has further accelerated in recent years, and introduced significant improvement to Graph Theory, especially in the practical applications of graph construction and use for Machine Learning or Deep Learning \cite{KG21}. They have a number of benefits when compared with a relational model or NoSQL alternatives, such as the ability for data to evolve in a more flexible manner, and the capacity to organize data in a way that is not hierarchical. They can represent incomplete information, and does not require a precise schema \cite[p.2]{KG21}, which is invaluable in the context of memory analysis, where the structure of the heap is not known in advance.

    While all KGs are graphs, not all graphs are KGs. However, the line between the two is often blurry, and the distinction is not always clear. The term "knowledge graph" first appeared in 1973, but really gained popularity through a 2012 blog post about Google's Knowledge Graph \cite{googleblog2023knowledgegraph}. Several definition attempts have been made, but none of them are universally accepted. Below are listed some of the most common definitions of Knowledge Graphs:
    \begin{itemize}
        \item \say{A graph of data intended to accumulate and convey knowledge of the real world, whose nodes represent entities of interest and whose edges represent potentially different relations between these entities.} \cite{KG21}
        \item \say{A graph of data consisting of semantically described entities and relations of different types that are integrated from different sources. Entities have a unique identifier. KG entities and relations are semantically described using an ontology or, more clearly, an ontological representation.} \cite{CKG23}
    \end{itemize}
    
    In KGs, edges are often labeled and may represent complex relationships like "is a subclass of" or "is married to", allowing for more expressive power. The very nature of KG makes any definition attempt difficult. Indeed, KG is a broad concept that can be applied to many domains, use cases and can have diverse implementations. The definition of KGs is thus very context-dependent, and it's debatable where the line between a graph and a KG is drawn. 
    
    For the purpose of this thesis, we won't focus on this distinction and just consider that we deal with memory \textit{graphs}, which are graphs that are not necessarily KGs, but that can be used to represent complex relationships between entities extracted from memory dumps and be leveraged using \acrshort{kg}-related techniques for advance tasks like feature engineering, automated embedding, inductive reasoning and learning.

    \subsubsection{Ontologies}\label{sec:background:ontology}
    An ontology is a formal representation of knowledge within a domain, providing a structured framework for organizing and interpreting information. It consists of a set of concepts, relationships, and rules that define how data is interconnected and how it can be reasoned about. Ontologies are often used to model a domain and support reasoning about entities and their relationships \cite{KG22}.

    Ontologies play a crucial role in the development and utility of knowledge graphs. They provide a semantic layer to the knowledge graph, enabling machines to understand the meaning and context of the data. The axioms and rules in an ontology enable automated reasoning, allowing the knowledge graph to infer new facts from existing data. Ontologies also help in maintaining the quality and consistency of data by enforcing constraints and validation rules. Finally, ontologies enable interoperability by providing a common vocabulary for data exchange and integration. Different popular ontologies exists, such as \acrfull{owl} or \acrfull{rdf}.

    By incorporating ontologies, knowledge graphs become more than just a collection of nodes and edges; they become a rich, interconnected web of semantically meaningful information that can be easily queried, analyzed, and extended. We will be referring to concepts that have been inspired by ontologies in the next sections, like \textit{rdf:Bag}.

    The \texttt{rdf:Bag} container is a part of the Resource Description Framework (RDF) used to represent collections where the order of elements is not significant. Unlike other containers such as \texttt{rdf:Seq} and \texttt{rdf:Alt}, \texttt{rdf:Bag} permits duplicate entries. It can be used to model unsorted collections of resources or literals. An instance of \texttt{rdf:Bag} can be represented as a graph where each node connected to the root node represents an item in the collection \cite{OrderedDataInRDF20}. For example, consider a root node named \texttt{"DataStructure"} with a property \texttt{Address} specifying its \texttt{malloc header address}.

    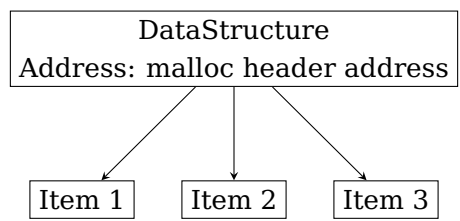
\begin{figure}[H]
        \centering
        \begin{tikzpicture}[>=stealth, every node/.style={shape=rectangle, draw, align=center}]
            \node (root) at (0,0) {DataStructure\\Address: malloc header address};
            \node (item1) at (-2,-2) {Item 1};
            \node (item2) at (0,-2) {Item 2};
            \node (item3) at (2,-2) {Item 3};
            \draw[->] (root) -- (item1);
            \draw[->] (root) -- (item2);
            \draw[->] (root) -- (item3);
        \end{tikzpicture}
        \caption{Graphical representation of an \texttt{rdf:Bag} container.}
    \end{figure}

    This small example illustrates how we can represent a container, or a relationship of belonging, using a graph. This is a concept that will be used later in the memory modelization process.

    \subsubsection{Inductive Reasoning and Learning}
    Inductive reasoning in Knowledge Graphs (KGs) involves techniques like embedding and Graph Convolutional Networks (GCNs) to learn the potential underlying structure of the graph. This is particularly useful for tasks like link prediction, node classification, and clustering.

    For the sake of clarity and grouping, we will present the concepts of graph embedding and GCNs in the next sections, but it is important to note that they are not mutually exclusive. In fact, GCNs can be used to generate embeddings, and embeddings can be used as input for GCNs. As such, they are often used together in the context of KGs, or in our case, memory graphs.

\section{Data preprocessing for Machine Learning}\label{sec:background:processing}
    \acrfull{ml} is a subfield of \acrfull{ai} that focuses on the development of algorithms that can learn from data and make predictions. It is a powerful tool that has been used to solve a wide range of problems, including image classification, speech recognition, and natural language processing. Before we can apply machine learning algorithms to a dataset, we must first prepare the data by performing various preprocessing steps. In this section, we will discuss the most common data preprocessing techniques, including data cleaning, feature engineering, and dataset splitting, as well as the importance of feature selection and dimensionality reduction. All those elements are crucial for the development of effective machine learning models for key extraction.

    \subsection{Feature engineering}
    In the realm of machine learning and data science, \textit{features} refer to individual measurable attributes or characteristics of the phenomena under study. Those features can be of different types, and can be used to predict the value of a target variable. 
    
    \subsubsection{Types of Features}
    Features can be of different types, depending on the nature of the data. The type of feature determines how it is processed and utilized by the model. The most common types of features include:

    \begin{itemize}
        \item \textbf{Numerical Features}: These are quantitative attributes representing measurements like height, weight, or age.
        \item \textbf{Categorical Features}: These are qualitative attributes representing discrete classes or labels, such as gender (Male, Female) or educational level (High School, Bachelor's, Master's).
        \item \textbf{Ordinal Features}: Similar to categorical features but with an inherent order, like ratings on a scale of 1 to 5.
        \item \textbf{Text Features}: These contain textual data and often require special preprocessing steps like tokenization and vectorization.
        \item \textbf{Temporal Features}: These are time-based attributes, such as timestamps, requiring special handling to capture time-dependent patterns.
        \item \textbf{Geospatial Features}: These attributes represent geographical or spatial coordinates.
    \end{itemize}

    Features are pivotal for the performance of machine learning models. The quality and pertinence of features can significantly influence the model's capability to discern underlying patterns in the data. Inadequately chosen, or irrelevant features can lead to a poorly performing model, while carefully selected, relevant features can result in a robust and accurate model.

    \subsubsection{The Curse of Dimensionality}
    The \textit{curse of dimensionality} refers to a set of challenges that arise when dealing with high-dimensional data. As the number of features, or dimensions, in a dataset increases, the volume of the feature space grows exponentially. This exponential growth leads to several issues:

    \begin{itemize}
        \item \textbf{Data Sparsity}: In high-dimensional spaces, data points tend to be sparse, making it difficult for algorithms to identify patterns. The sparsity also means that the notion of "distance" becomes less meaningful, which is problematic for algorithms that rely on distance metrics.
        
        \item \textbf{Computational Complexity}: The exponential increase in volume demands significantly more computational power and memory, making it challenging to process and analyze the data efficiently.
        
        \item \textbf{Overfitting}: High dimensionality increases the risk of overfitting, where a model learns the noise in the data rather than the actual pattern. Overfit models perform poorly on unseen data.
        
        \item \textbf{Statistical Significance}: As dimensions increase, the amount of data required to achieve statistical significance also increases exponentially, often making it impractical to collect sufficient data.
        
        \item \textbf{Visualization and Interpretability}: High-dimensional data are difficult to visualize and interpret, making it challenging to derive intuitive insights.
    \end{itemize}

    Due to these challenges, many dimensionality reduction techniques have been developed to transform high-dimensional data into a lower-dimensional form, aiming to preserve as much of the relevant information as possible. These techniques are discussed in the next section.

    \subsubsection{Feature Engineering techniques}
    The meticulous process of selecting the most relevant features, or constructing new features from existing ones, is known as \textit{feature engineering}. This step can encompass normalization, transformation, and the creation of interaction terms among features. This complex process requires a deep understanding of the domain of study and the datasets, and is often a crucial step in the development of machine learning models \cite{FeatureEngineeringMadeEasy18}.

    In the context of \acrlong{ml}, features essentially serve as the input variables $ X $ that a machine learning model employs to make predictions or inferences about the output variable $ Y $. But not all features are equally informative. Some features may be redundant, irrelevant, or even detrimental to the model's performance. Irrelevant features are those that do not contribute to the predictive power of the model, while redundant features are those that are highly correlated with other features. Feature engineering techniques, can be used to build, transform or eliminate features, thereby reducing the dimensionality of the data and enhancing model performance \cite{FeatureSelecExtract14}.

    \begin{itemize}
        \item \textbf{Scaling and Normalization:} Scaling and normalization are techniques used to transform the features to a similar scale. This is particularly important for algorithms that rely on distance metrics, such as \acrfull{knn} and \acrfull{svm}. Scaling and normalization can also help accelerate the training process by reducing the number of iterations required for the model to converge. Some common techniques include min-max scaling, z-score normalization, and log transformation.
        
        \item \textbf{Feature Extraction}: This technique involves transforming the original set of features into a new set of features, which is usually of lower dimensionality. The new features are often combinations of the original features and aim to capture as much of the information in the original data as possible. Many methods exits, like \acrfull{pca}, \acrfull{lda} and \acrfull{tsne} are often employed to transform high-dimensional data into a lower-dimensional form are commonly used for feature extraction \cite{FeatureSelecExtract14}.
        
        \item \textbf{Feature Selection}: Unlike feature extraction, feature selection aims to pick a subset of the most important features from the original set, without changing them \cite{FeatureSelecExtract14}. The goal is to remove irrelevant or redundant features that do not contribute significantly to the model's performance. Techniques like Recursive Feature Elimination (RFE) and using importance scores from tree-based algorithms like Random Forest are popular methods for feature selection.
    \end{itemize}

    Both techniques have their own advantages and disadvantages, and the choice between the two often depends on the specific requirements of the task at hand. Feature extraction is generally more suitable when the original features do not have much interpretability to begin with, or when transforming features can lead to a more compact and effective representation. On the other hand, feature selection is often preferred when it is important to maintain the interpretability of the features, or when computational efficiency is a concern \cite{FeatureEngineeringMadeEasy18}.

    \subsubsection{Evaluating Features with Correlation Tests}
    To assess the quality of features, various statistical measures can be employed. Correlation tests are statistical tests that measure the strength and direction of the relationship between two variables. Pearson, Kendall, and Spearman correlation coefficients are commonly used to quantify the linear or monotonic relationship between each feature and the target variable \cite{StatisticalMethodsInPractice09}. A high absolute value indicates a strong relationship, aiding in feature selection.

    \begin{itemize}
        \item \textbf{Pearson Correlation}: Measures the linear relationship between two variables. It ranges from -1 to 1, where -1 indicates a strong negative linear correlation, 1 indicates a strong positive linear correlation, and 0 indicates no linear correlation.
        \item \textbf{Kendall's Tau}: A non-parametric test that measures the strength and direction of a monotonic relationship between two variables.
        \item \textbf{Spearman's Rank}: Also a non-parametric test, it assesses how well an arbitrary monotonic function can describe the relationship between two variables without making any assumptions about the frequency distribution.
    \end{itemize}

    These techniques are useful for evaluating the relationship between each feature and allows generating correlation matrices, which can be used to identify redundant features. It's also possible to evaluate each feature independently through univariate feature selection techniques. In Python's Scikit-learn library \cite{ScikitLearn}, methods like \textit{F-test} and the \textit{p-value} are often used for this purpose.
"
    \begin{itemize}
        \item \textbf{F-test value}: Measures the linear dependency between the feature variable and the target. A higher F-test value indicates a more useful feature.
        \item \textbf{p-value}: Indicates the probability of an F-test value this large arising if the null hypothesis is true. A smaller p-value suggests rejecting the null hypothesis, making the feature significant.
    \end{itemize}

    In summary, features are the foundational elements of any machine learning model. The quality of these features, along with how they are processed and utilized, can markedly impact the model's performance. The significance of feature engineering cannot be overstated. Properly engineered features can drastically reduce modeling errors, leading to more accurate and reliable predictions. It serves as a bridge between raw data and predictive models, ensuring that the models are fed with the most relevant and informative features.

    \subsection{Embeddings}
    Embeddings are low-dimensional vector representations of high-dimensional objects. They are often used to capture complex relationships between objects and are particularly useful for machine learning tasks like clustering, classification, and link prediction. Embeddings are widely used in the field of natural language processing (NLP) to represent words, sentences, and documents \cite{UnderstandingWordEmbeddingsAndLM20}. In recent years, they have also been applied to graphs to learn node representations \cite{KG21}. In this section, we will discuss the concept of embeddings and explore some common techniques for creating them.
    
    \subsubsection{Embedding Creation vs Feature Engineering}    
    Both embedding creation and feature engineering are techniques used to transform data for machine learning models. However, they are different in terms of their goals and how they are achieved.

    \begin{itemize}
        \item \textbf{Embedding Creation}: Embedding creation is the process of learning a low-dimensional vector representation of a high-dimensional object. This involves mapping discrete objects, such as words in \acrfull{nlp} or nodes in a graph in our case, to vectors of continuous values in a lower-dimensional space \cite{UnderstandingWordEmbeddingsAndLM20}. The goal is to capture the semantic or structural relationships between these objects. Specialized algorithms like Word2Vec for word embeddings or Node2Vec for graph embeddings are often used.
        
        \item \textbf{Feature Engineering}: As discussed before, it is a more general practice that involves creating new features or modifying existing ones to improve the performance of a machine learning model. While feature engineering can include creating embeddings, it also encompasses a wide range of other techniques like normalization, transformation, outlier detection, and handling missing values.
    \end{itemize}

    In general, feature engineering is a more manual process, while embedding creation is more automated. Feature engineering requires domain knowledge and understanding of the problem, while embedding creation can be done using a variety of machine learning techniques.

    Thus, embedding creation is a specialized form of feature engineering aimed at mapping discrete objects to continuous vectors of numbers, usually for capturing complex relationships. Feature engineering, on the other hand, is a broader practice that can involve a variety of techniques, including but not limited to embedding creation.

    \subsubsection{Embeddings for graphs}\label{sec:background:processing:embedding:graph}
    Graph embedding techniques aim to map nodes and edges in a graph to vectors in a low-dimensional space \cite{KG21}. The primary goal is to preserve the graph's structural properties, such as node connectivity and community structure, in the embedded space. These vectors can then be used for various machine learning tasks like clustering, classification, and link prediction. A quite complete overview of the different techniques can be found in \citetitle{KG21} \cite[Section 4.2]{KG21}. Here are some common and advanced techniques:


    \paragraph{Translational Models}
    The first type of graph embeddings techniques is based around using transactional models that interpret edge labels as transformations from subject nodes to object nodes \citetitle{KG21}. 

    \begin{itemize}
        \item \textbf{TransE:} is one of the earliest and most straightforward translational models. It represents entities as points in a vector space and relations as translations between these points. The primary idea is that for a valid triple \((h, r, t)\), the equation \(h + r = t\) should hold, where \( h \) stands for the head entity, \( r \) represents the relation, and \( t \) is the tail entity. This model is simple and computationally efficient but is limited in its ability to capture complex relationships \cite{KG21}.
        \item \textbf{TransH:} TransH extends TransE by introducing relation-specific hyperplanes. This allows the model to capture more complex relationships by projecting the entity embeddings onto these hyperplanes before performing translations \cite{KG21}.
        \item \textbf{TransR:} TransR goes a step further by not only introducing relation-specific hyperplanes but also relation-specific translations. This allows for a more flexible representation of relations, accommodating various types of complexities \cite{KG21}.
        \item Other improvements include \textit{TransD} or \textit{MuRP}, which shows that research in this domain is still very active.
    \end{itemize}

    \paragraph{Tensor Decomposition Models}
    Tensor decomposition models represent entities and relations as vectors or matrices in a low-dimensional space. These models are based on the assumption that the relationship between entities can be represented by a bilinear function. They are computationally efficient and can capture complex relationships, but they are also limited in their ability to model asymmetric and reflexive relations.

    \begin{itemize}
        \item \textbf{RESCAL:} RESCAL employs a bilinear model where each relation is represented by a full-rank matrix. This allows for capturing asymmetric and reflexive relations but at the cost of increased computational complexity \cite{KG21}.
        \item \textbf{DistMult:} DistMult simplifies RESCAL by assuming that the relation matrices are diagonal. This reduces the number of parameters and computational complexity, making it more scalable \cite{KG21} \cite{KG22}.
        \item \textbf{ComplEx:} ComplEx extends DistMult by introducing complex-valued embeddings. This allows the model to capture asymmetric relations effectively while maintaining computational efficiency \cite{KG21} \cite{KG22}.
    \end{itemize}

    \paragraph{Neural Models}
    Neural models employ neural networks to learn the features of entities and relations. This provides a more flexible and adaptive approach to graph embeddings.

    \begin{itemize}
        \item \textbf{ConvKB:} ConvKB employs a convolutional neural network to automatically learn the features of entities and relations. It is technically a translational model as introduced before, but uses a convolutional layer to capture the interactions between entities and relations. This provides a more flexible and adaptive approach to graph embeddings \cite{ConvKB18}.
        \item \textbf{RotatE:} RotatE uses complex rotations in the embedding space to model relations. This neural model captures the semantics of relations in a more expressive manner \cite{KG22}.
        \item \textbf{SDNE (Structural Deep Network Embedding)}: SDNE employs a deep autoencoder to learn complex and non-linear node embeddings while preserving first-order and second-order proximities. It is particularly effective for capturing intricate patterns and structures in the graph \cite{SDNE16}.
        \item \textbf{R-GCN:} Relational Graph Convolutional Networks (R-GCNs) combine the strengths of GCNs and traditional embedding methods to capture both topological and semantic information \cite{RGCN18}. A more recent study by \citeauthor{RGCN22} argues that the main contribution of R-GCN lies in its message passing paradigm rather than the learned weights. This paper introduces a variant called Random R-GCN (RR-GCN) \cite{RGCN22}.
        \item \textbf{ConvE:} ConvE employs convolutional layers to capture local and global interactions between entities and relations, offering a more expressive representation \cite{KG22}.
    \end{itemize}

    \paragraph{Language Models}
    Language models utilize pre-trained language models to enrich the embeddings with contextual information. This approach leverages the recent developments of language models to capture the semantics of entities and relations.

    \begin{itemize}
        \item \textbf{BERT for KGE:} Utilizing pre-trained BERT models, this approach leverages the power of language models to enrich the embeddings with contextual information \cite{KGBERT19}.
        \item \textbf{BART KGE:} Bidirectional and Auto-Regressive Transformers (BART) is a denoising autoencoder that can be used for various NLP tasks. This approach utilizes BART to learn entity and relation embeddings. The paper introducing this also compares other \acrshort{llm} like GPT-2 \cite{KGBART21}.
        \item Due to the recent developments in NLP, this domain is still very active, and new approaches are being developed regularly.
    \end{itemize}

    \subsubsection{Word Embeddings}
    Word embeddings are vector representations of words in a low-dimensional space. They are often used as input for machine learning models in natural language processing (NLP) tasks like text classification, sentiment analysis, and machine translation. Word embeddings are typically learned from large text corpora using unsupervised learning techniques like Word2Vec and GloVe. These embeddings can then be used to capture semantic relationships between words and phrases, which is particularly useful for NLP tasks \cite{UnderstandingWordEmbeddingsAndLM20}.

    \begin{itemize}
        \item \textbf{Word2Vec}: Word2Vec is a popular algorithm for learning word embeddings from text data. It employs a shallow neural network to learn the embeddings and is often used as a pre-processing step for NLP tasks \cite{KG22}.
        \item \textbf{GloVe}: Global Vectors for Word Representation (GloVe) is another popular algorithm for learning word embeddings. It is based on the co-occurrence matrix of words and utilizes matrix factorization to learn the embeddings \cite{KG22}.
    \end{itemize}

    \paragraph{Graph Embeddings}
    Graph embeddings are vector representations of nodes in a graph.Graph embeddings are typically learned using unsupervised learning techniques like Node2Vec and DeepWalk. These embeddings can then be used to capture structural relationships between nodes, which is particularly useful for graph analytics tasks \cite{KG21}.
    
    \begin{itemize}
        \item \textbf{One-Hot Encoding}: This is a simple technique that represents each node as a vector of 0s and 1s, where the length of the vector is equal to the number of nodes in the graph. The vector contains a 1 at the index corresponding to the node and 0s everywhere else. As is, this method is not really suitable for large graphs as it results in a high-dimensional and sparse matrix representation \cite{KG22}.
        
        \item \textbf{Node2Vec}: This algorithm learns continuous feature representations for nodes by optimizing a neighborhood-preserving objective. It employs biased random walks and uses the Skip-gram model to generate embeddings. Node2Vec is particularly effective for capturing local structures and can be fine-tuned for specific tasks \cite{Node2vec16}.
        
        \item \textbf{DeepWalk}: Similar to Node2Vec, DeepWalk uses random walks to generate node sequences. It employs the Skip-gram model from natural language processing to learn embeddings. Unlike Node2Vec, it does not use biased walks, making it more suitable for capturing global structures \cite{Deepwalk14}.
        
        \item \textbf{Spectral Clustering}: This technique is based on the spectral theory of graphs. It utilizes the eigenvalues and eigenvectors of the Laplacian matrix of the graph to find an optimal embedding. Spectral Clustering is particularly useful for community detection and can capture the global structure of the graph \cite{SpectralNetworks13} \cite{SemiSupervisedClassificationSpectralConvGCN16}.
        
        \item \textbf{LINE}: Large-scale Information Network Embedding (LINE) aims to preserve both local and global network structures. It optimizes two objectives: first-order and second-order proximities between nodes. LINE is scalable and can handle large graphs efficiently \cite{LINEEmbedding15}.
        
        \item \textbf{Graph Factorization}: This method directly factorizes the adjacency matrix of the graph to learn node embeddings, making it computationally efficient but less capable of capturing complex structures. It is often used for large-scale graphs where computational resources are limited \cite{KG22}.
        
    \end{itemize}

    \subsection{Other preprocessing techniques for Machine Learning}
    When working with real-world data and datasets, it is common to find issues like missing values, outliers, and imbalanced classes that can adversely affect the performance of machine learning models. In this section, we will discuss some common preprocessing techniques used to improve the quality of data and ensure that the models can learn effectively.

    \subsubsection{Data Cleaning}
    Data cleaning is the process of detecting and correcting errors in the data. It is a crucial step in data preprocessing and involves various techniques like outlier detection, handling missing values, and data normalization. Data cleaning is necessary to ensure that the data is accurate and consistent, which is essential for machine learning models to learn effectively.

    \begin{itemize}
        \item \textbf{Outlier Detection}: Outliers are data points that deviate significantly from the rest of the dataset. They can be caused by errors in data collection or genuine variations. Outliers can skew the model's learning and should be identified and handled appropriately, either by removal or transformation.
        
        \item \textbf{Handling Missing Values}: Missing data can lead to biased or incorrect model training. Techniques for handling missing values include imputation, where missing values are replaced with statistical measures like mean, median, or mode, and deletion, where rows with missing values are removed.
        
        \item \textbf{Data Normalization}: Features with different scales can affect the performance of machine learning algorithms. Normalization rescales the features to a standard range, usually [0, 1], or transforms them to have a mean of 0 and a standard deviation of 1.
        
        \item \textbf{Encoding Categorical Variables}: Many machine learning algorithms require numerical input and output variables. Categorical variables are converted to numerical format through techniques like one-hot encoding or label encoding.
        
        \item \textbf{Text Cleaning}: In natural language processing tasks, text data may require cleaning to remove irrelevant characters, correct typos, or standardize text format.
        
        \item \textbf{Duplicate Removal}: Duplicate entries can bias the model and should be identified and removed from the dataset.
        
        \item \textbf{Feature Engineering}: While not strictly data cleaning, feature engineering involves transforming existing features or creating new ones to improve model performance.
    \end{itemize}

    \subsubsection{Dataset splitting and sampling}
    One other typical step is the division of the dataset into training and testing sets. This separation is crucial for evaluating the generalization performance of a model. The training set is used to train the model, while the testing set is used to evaluate its performance on unseen data. Failing to separate these sets can lead to overfitting, where the model performs well on the training data but poorly on new, unseen data.

    Various techniques exist for dataset splitting and sampling, each with its own advantages and disadvantages:

    \begin{itemize}
        \item \textbf{Random Split}: The dataset is randomly divided into training and testing sets based on a given ratio, such as 70\% for training and 30\% for testing. This method is simple but may result in imbalanced classes in the splits.
        
        \item \textbf{Stratified Split}: Similar to random split, but ensures that the distribution of classes is the same in both training and testing sets. This is particularly useful for imbalanced datasets.
        
        \item \textbf{k-Fold Cross-Validation}: The dataset is divided into 'k' subsets or "folds." The model is trained on k-1 folds and tested on the remaining fold. This process is repeated k times, each time with a different fold as the testing set. The average performance metric is used for evaluation.
        
        \item \textbf{Leave-One-Out Cross-Validation (LOOCV)}: A special case of k-Fold Cross-Validation where k is equal to the number of data points. Each data point is used once as the test set while the remaining points form the training set.
        
        \item \textbf{Bootstrapping}: Random samples are drawn with replacement from the dataset to create multiple training sets. The model is trained and tested on these sets, and the average performance is calculated.
    \end{itemize}

    \subsubsection{Dealing with Imbalanced Datasets}
    Imbalanced datasets are those where the classes are not represented equally. This is a common issue in machine learning, especially in classification problems. An imbalanced dataset can lead to a biased model that may not effectively predict the minority class. Therefore, it is crucial to address this issue during data preprocessing.

    \begin{itemize}
        \item \textbf{Why it is Necessary}: In imbalanced datasets, machine learning algorithms tend to be biased towards the majority class, ignoring the minority class. This results in poor classification performance for the minority class, which is often the class of interest in problems like fraud detection, medical diagnosis, etc.
        
        \item \textbf{Oversampling}: This involves adding more copies of the minority class. Oversampling can be random with replacement, or it can involve generating synthetic samples.
        
        \item \textbf{Undersampling}: This involves removing some of the samples of the majority class. This method is generally not preferred as it can lead to loss of data.
        
        \item \textbf{SMOTE (Synthetic Minority Over-sampling Technique)}: SMOTE is an oversampling method that creates synthetic samples in the feature space. It selects two or more similar instances (using a distance measure) and perturbing an instance one at a time by a random amount within the difference to the neighboring instances.
        
        \item \textbf{ADASYN (Adaptive Synthetic Sampling)}: Similar to SMOTE, but it uses a weighted distribution for different minority class examples according to their level of difficulty in learning. 
        
        \item \textbf{Cost-sensitive Learning}: This involves modifying the algorithm to increase the weight of the minority class during training, thereby making the algorithm more sensitive to it.
        
        \item \textbf{Ensemble Methods}: Methods like Random Forest and Gradient Boosting can be used with techniques like bagging and boosting to handle imbalanced datasets effectively.
        
        \item \textbf{Resampling Methods}: These involve randomly partitioning the data into subsets, balancing each subset, and then aggregating the results.
    \end{itemize}

    To conclude this section, data preprocessing is a crucial step in machine learning. The number of concepts and techniques discussed here is by no means exhaustive, which further highlights the importance and complexity of data preprocessing. 

\section{Machine Learning and Deep Learning}\label{sec:background:ml}
    In the preceding sections, we have discussed the importance of data preprocessing and feature engineering for machine learning. In this section, we will discuss the machine learning pipeline and explore some common machine learning algorithms. We will also discuss deep learning and neural networks especially in the context of graphs, since they have gained popularity in recent years due to their superior performance on many tasks.

    \subsection{Machine Learning}
    Machine learning is a cornerstone in the field of artificial intelligence and has been instrumental in driving many of today's technological and scientific breakthroughs. From natural language processing to computer vision, machine learning algorithms play a critical role in making sense of large and complex data sets.

        \subsubsection{What is Machine Learning}
        Machine learning is a subfield of artificial intelligence that provides systems the ability to automatically learn and improve from experience without being explicitly programmed. This learning process is based on the recognition of complex patterns in data and the making of intelligent decisions based on them \cite{ScienceMachineLearning15}.

        The machine learning pipeline typically involves several steps including some pre-steps like data collection, preprocessing, feature extraction that we have already discussed before, usually followed by model training, evaluation, and deployment. Algorithms are trained on a dataset, and the learned patterns are used to make predictions or decisions without human intervention. \acrshort{ml} algorithms can be broadly classified into three categories \cite{ScienceMachineLearning15}:

        \begin{itemize}
            \item \textbf{Supervised Learning}: Algorithms are trained on labeled data, and the aim is to make predictions or map inputs to outputs.
            \item \textbf{Unsupervised Learning}: Algorithms are trained on unlabeled data, focusing on the underlying structure or distribution in the data.
            \item \textbf{Reinforcement Learning}: Algorithms learn to perform an action from state to state to maximize some type of reward or objective function.
        \end{itemize}

        In the following, we will focus on supervised and unsupervised learning, as they are the most relevant to our work.

        \subsubsection{Model Evaluation}
        Evaluating the performance of a machine learning model is crucial for understanding its effectiveness and suitability for a given task. It is not a trivial task as it depends on various factors like the type of data, the problem at hand, and the model itself \cite{EvaluatingQualityMLExplanations21}. In this section, we will discuss some common evaluation metrics for classification and regression models.
    
        \begin{minipage}{\dimexpr\linewidth-20pt}
            \paragraph{Precision}
            Precision is the ratio of correctly predicted positive observations to the total predicted positives. The formula for precision is:
            \par
            \vspace{2em}
            \begin{center}
                $
                \text{Precision} = \frac{\text{True Positives}}{\text{True Positives} + \text{False Positives}}
                $
            \end{center}
            \vspace{1em}
        \end{minipage}
    
        \begin{minipage}{\dimexpr\linewidth-20pt}
            \paragraph{Recall}
            Recall is the ratio of correctly predicted positive observations to all the observations in the actual class. The formula for recall is:
            \par
            \vspace{2em}
            \begin{center}
                $
                \text{Recall} = \frac{\text{True Positives}}{\text{True Positives} + \text{False Negatives}}
                $
            \end{center}
            \vspace{1em}
        \end{minipage}
    
        \begin{minipage}{\dimexpr\linewidth-20pt}
            \paragraph{F-1 Score}
            The F-1 Score is the weighted average of Precision and Recall, and it ranges from 0 to 1. The formula for the F-1 Score is:
            \par
            \vspace{2em}
            \begin{center}
                $
                \text{F-1 Score} = 2 \times \frac{\text{Precision} \times \text{Recall}}{\text{Precision} + \text{Recall}}
                $
            \end{center}
            \vspace{1em}
        \end{minipage}
    
        \begin{minipage}{\dimexpr\linewidth-20pt}
            \paragraph{Accuracy}
            Accuracy is the ratio of correctly predicted observations to the total observations. The formula for accuracy is:
            \par
            \vspace{2em}
            \begin{center}
                $
                \text{Accuracy} = \frac{\text{True Positives} + \text{True Negatives}}{\text{Total Observations}}
                $
            \end{center}
            \vspace{1em}
        \end{minipage}

        These are the most common metrics used for evaluating classification models. Other metrics like the Area Under the Curve (AUC) and the Receiver Operating Characteristic (ROC) curve are also used for evaluating binary classification models. For regression models, metrics like Mean Squared Error (MSE), Root Mean Squared Error (RMSE), and Mean Absolute Error (MAE) are commonly used \cite{ScikitLearn}.

        \subsection{Machine Learning Models for Binary Classification}
        Binary classification is a supervised learning task where the goal is to predict a binary outcome, i.e., one of two possible classes (0 or 1). This is a truly classical task for a \acrshort{ml} model, with a range of applicable algorithms. As such, we rely on Python's Scikit-learn library for implementation.

        \paragraph{Logistic Regression}
        A specialized form of regression tailored for predicting binary outcomes. It employs the logistic function to map predicted values between 0 and 1. While straightforward and interpretable, its performance may be limited on complex, non-linear data \cite{nick_logistic_2007}.

        \begin{minipage}{\dimexpr\linewidth-20pt}
            The logistic function is defined as:
            \par
            \vspace{2em}
            \begin{center}
                $
                P(y=1|x) = \frac{1}{1 + e^{-(\beta_0 + \beta_1 x)}}
                $
            \end{center}
            \vspace{1em}
            Where \( \beta_0 \) is the intercept and \( \beta_1 \) is the coefficient for the predictor variable \( x \).
        \end{minipage}

        \paragraph{Decision Trees}
        These are versatile models used for both classification and regression. They partition the feature space into regions, making decisions at each node. While easy to visualize, they are susceptible to overfitting \cite{kotsiantis_decision_2013}.

        \begin{minipage}{\dimexpr\linewidth-20pt}
            Decision Trees use metrics like Gini impurity or entropy to make splits:
            \par
            \vspace{2em}
            \begin{center}
                $
                \text{Gini}(T) = 1 - \sum_{i=1}^{C} p_i^2
                $
            \end{center}
            \vspace{1em}
            Where \( T \) is a node and \( p_i \) is the proportion of class \( i \) instances among the training instances in node \( T \).
        \end{minipage}

        \paragraph{Random Forest}
        An ensemble technique that aggregates predictions from multiple decision trees. Known for its robustness and ability to handle large, high-dimensional data \cite{probst_hyperparameters_2019}.

        \begin{minipage}{\dimexpr\linewidth-20pt}
            The final prediction is an average or majority vote from all trees:
            \par
            \vspace{2em}
            \begin{center}
                $
                \hat{y} = \frac{1}{n} \sum_{i=1}^{n} \hat{y_i}
                $
            \end{center}
            \vspace{1em}
            Where \( \hat{y} \) is the final prediction and \( \hat{y_i} \) is the prediction from the \( i^{th} \) tree.
        \end{minipage}

        \paragraph{Support Vector Machines (SVM)}
        Effective for both classification and regression, SVMs find the hyperplane that best separates the data into classes. They excel in high-dimensional spaces \cite{wu_analysis_2006}.

        \begin{minipage}{\dimexpr\linewidth-20pt}
            The objective is to maximize the margin between classes:
            \par
            \vspace{2em}
            \begin{center}
                $
                \max_{w, b} \frac{2}{\| w \|}
                $
            \end{center}
            \vspace{1em}
            Where \( w \) is the weight vector and \( b \) is the bias term.
        \end{minipage}

        \paragraph{k-Nearest Neighbors (k-NN)}
        An instance-based algorithm that classifies a new point based on the majority class among its 'k' nearest neighbors \cite{laaksonen_classification_1996}.

    \begin{minipage}{\dimexpr\linewidth-20pt}
        The distance between points is often calculated using Euclidean distance:
        \par
        \vspace{2em}
        \begin{center}
            $
            d(x, y) = \sqrt{\sum_{i=1}^{n} (x_i - y_i)^2}
            $
        \end{center}
        \vspace{1em}
        Where \( d(x, y) \) is the distance between points \( x \) and \( y \), and \( n \) is the number of dimensions.
    \end{minipage}

    \subsection{Deep Learning}
    Deep learning offers a powerful set of tools for automatically generating embeddings from the graph representation of heap dumps. Leveraging neural networks, we can build custom models using PyTorch to perform binary classification tasks, such as predicting key nodes in a heap dump. In this section, we will explore various neural network architectures that are well-suited for this task.

    Neural Networks serve as the backbone of deep learning models \cite{DeepLearningBook16}. They consist of interconnected nodes or "neurons" organized into layers. Deep learning extends this architecture by employing multiple hidden layers, enabling the model to learn complex representations from data. This is particularly useful when dealing with graph-based heap dump representations, where the relationships between nodes can be intricate \cite{KG22}. 
    
    Choosing Deep Learning models is dependent on the task. As such, we need to present the different types of neural networks and their advantages and disadvantages. Below are some common neural network architectures:

    \subsubsection{Recurrent Neural Networks (RNN)}
    \acrfull{rnn} are particularly effective for sequence-based data. In the context of heap dumps, if the memory addresses or keys exhibit some form of sequential pattern, RNNs can be employed to capture these temporal dependencies for binary classification \cite[10.2]{DeepLearningBook16}.

    \begin{itemize}
        \item \textbf{Advantages:}
        \begin{itemize}
            \item Good at capturing short-term dependencies in sequence data.
            \item Relatively simpler architecture.
        \end{itemize}
        \item \textbf{Disadvantages:}
        \begin{itemize}
            \item Struggles with long-term dependencies due to the vanishing gradient problem.
            \item May require more data for effective training.
        \end{itemize}
    \end{itemize}

    \subsubsection{Long Short-Term Memory (LSTM)}
    LSTMs are an extension of RNNs designed to capture long-term dependencies, making them suitable for more complex sequences \cite{hochreiter_long_1997}.

    \begin{itemize}
        \item \textbf{Advantages:}
        \begin{itemize}
            \item Effective in capturing long-term dependencies.
            \item Less susceptible to the vanishing gradient problem.
        \end{itemize}
        \item \textbf{Disadvantages:}
        \begin{itemize}
            \item More complex and computationally intensive than RNNs.
            \item May require fine-tuning of hyperparameters.
        \end{itemize}
    \end{itemize}

    \subsubsection{Gated Recurrent Units (GRU)}
    \acrfull{gru} offer a compromise between the simplicity of RNNs and the power of LSTMs. They are effective in capturing both short-term and long-term dependencies but with a less complex architecture \cite{chung_empirical_2014}.

    \begin{itemize}
        \item \textbf{Advantages:}
        \begin{itemize}
            \item Simpler architecture compared to LSTM.
            \item Efficient in capturing long-term dependencies.
        \end{itemize}
        \item \textbf{Disadvantages:}
        \begin{itemize}
            \item May not perform as well as LSTMs for very complex sequences.
            \item Still more computationally intensive than basic RNNs.
        \end{itemize}
    \end{itemize}

    \subsubsection{Convolutional Neural Networks (CNN)}
    Though traditionally used in image processing, CNNs can also be adapted for sequence data like heap dumps. They are excellent at identifying spatial hierarchies or patterns in the data \cite{DeepLearningBook16} \cite{lecun_gradient_based_1998}.

    \begin{itemize}
        \item \textbf{Advantages:}
        \begin{itemize}
            \item Highly effective in identifying local patterns.
            \item Less prone to overfitting due to pooling layers.
        \end{itemize}
        \item \textbf{Disadvantages:}
        \begin{itemize}
            \item May not capture global dependencies as effectively as RNNs or LSTMs.
            \item Requires a fixed-size input.
        \end{itemize}
    \end{itemize}

    \subsubsection{Graph Convolutional Networks (GCN)}
    Graph Convolutional Networks (GCNs) are a specialized form of neural networks designed to work directly with graphs \cite{KG22}. They are particularly useful for our task as they can automatically generate embeddings from the graph representation of heap dumps. These embeddings can then be used for binary classification to predict key nodes.

    \begin{itemize}
        \item \textbf{Advantages:}
        \begin{itemize}
            \item Capable of capturing both local and global graph structures.
            \item No need for manual feature extraction from graphs.
        \end{itemize}
        \item \textbf{Disadvantages:}
        \begin{itemize}
            \item May require fine-tuning and a well-defined graph structure.
            \item Computationally intensive for large graphs.
        \end{itemize}
    \end{itemize}

    Several Python libraries can be used to implement GCNs for our specific task. Some notable ones are PyTorch Geometric, Spektral, and DGL (Deep Graph Library). These libraries offer pre-built GCN layers and various utilities to facilitate the embedding and binary classification of heap dump graphs.

    \subsection{Graph Neural Networks}
    A \acrfull{gnn} constructs a neural network that mirrors the structure of the underlying data graph. In this architecture, nodes are linked according to their relationships in the data graph. The model is trained in a supervised fashion to map input features of nodes to their corresponding output features \cite{KG22}. These output features can either be manually annotated or sourced from a knowledge graph. 

    Unlike traditional knowledge graph embeddings, GNNs offer the advantage of end-to-end supervised learning tailored for specific tasks. 
    
    With a dataset of labeled examples, GNNs can classify individual nodes or even entire graphs. The challenge lies both in how to convert a given graph to a format that can be fed into a neural network (see \ref{sec:background:processing:embedding:graph}) and how to design a neural network that can effectively learn from the graph data. Note that the distinction between the two is not always clear since two problems are often tackled together. As such, the present classification is not exhaustive and arbitrary, even though it is based on the literature and meta-analysis of the field, like \citetitle{KG22} \cite{KG22} and \citetitle{GNNComprehensiveSurvey20} \cite{GNNComprehensiveSurvey20}.
    
    \acrshort{gnn}s have been employed in various applications, ranging from traffic prediction, recommender systems, and software verification \cite{GNNComprehensiveSurvey20}. Remarkably, GNNs can also serve as substitutes for conventional graph algorithms. For instance, they have been used to identify central nodes in knowledge graphs through supervised learning \cite{KG22}. The following sections delve into two specific types of GNNs: Recursive GNNs and Convolutional GNNs.

    \subsubsection{Recursive Graph Neural Networks (RecGNNs)}
    Recursive Graph Neural Networks (RecGNNs) serve as the foundational approach to graph neural networks \cite{KG22}. The model operates by passing messages between neighboring nodes to recursively compute results. The framework learns the functions that generate the expected output based on a training set of labeled nodes. \citeauthor{GNN08} \cite{GNN08} proposed a seminal \acrshort{gnn} model that uses feature vectors for nodes and edges, along with state vectors for nodes. Two parametric functions, the transition function and the output function, are used to update the state vectors and compute the final output for nodes, respectively. These functions are applied recursively until a fixpoint is reached. The model is highly flexible and can be adapted in various ways, such as defining neighboring nodes differently or using distinct parameters for each node \cite{GNN08}.

    \paragraph{Learning Process in \acrshort{gnn}s}
    The learning process in GNNs can be divided into three steps: input computation, node state update, and output computation. These steps are repeated until a fixpoint is reached, and the final output is computed for each node. Essentially, this involves finding the optimal parameters $ w $ such as for $ \varphi_{w} $ to best approximate the data in the learning dataset $ \mathcal{L} $. 

    \begin{minipage}{\dimexpr\linewidth-20pt}
        The learning task can be formulated as the minimization of a quadratic cost function $ e_{w} $ through iterative gradient descent \cite{GNN08}:
        \par
        \vspace{2em}
        \begin{center}
            \[
            e_{w} = \sum_{i=1}^{p}\sum_{j=1}^{q_{i}}(t_{i,j}-\varphi_{w}(G_{i},n_{i,j}) )^2
            \]
        \end{center}
        \vspace{1em}
    \end{minipage}

    Where:
    \begin{itemize}
        \item \( \mathcal{G} \): The set of graphs.
        \item \( \mathcal{N} \): The subset of nodes.
        \item \( \mathcal{D} = \mathcal{G} \times \mathcal{N} \): The set of pairs of a graph and nodes.
        \item \( G_{i} = (N_{i},E_{i}) \in \mathcal{G} \): The graph.
        \item \( N_{i} \): The set of nodes in graph \( G_{i} \).
        \item \( E_{i} \): The set of edges in graph \( G_{i} \).
        \item \( n_{i,j} \in N_{i} \): The \( j^{th} \) node of the graph \( G_{i} \).
        \item \( q_{i} \leq |N_{i}| \): The number of supervised nodes in the graph \( G_{i} \).
        \item \( p \leq |\mathcal{G}| \): The number of graphs in the learning set.
        \item \( \mathcal{L} = \{(G_{i},n_{i,j},t_{i,j})\} \): The learning set.
        \item \( t_{i,j} \in \mathbb{R}^m \): The target output for node \( n_{i,j} \).
        \item \( m \): The number of outputs.
        \item \( \mathbb{R}^m \): The \( m \)-dimensional Euclidean space.
        \item \( \varphi_{w} : \mathcal{D} \to \mathbb{R}^m \): The function that maps the input to the output.
        \item \( e_{w} \): Error function.
    \end{itemize}

    For graph-focused tasks, a special node is used for the target (\( t \)), whereas for node-focused tasks, supervision can be performed on every node. More information about the learning process can be found in \cite{GNN08}.

    \subsubsection{Convolutional Graph Neural Networks (ConvGNNs)}
    As introduced before, Graph Convolutional Networks extend the concept of convolution from images to graphs. \say{The core idea in the image setting is to apply small kernels (aka filters) over localized regions of an image using a convolution operator to extract features from that local region.} \cite{KG22}. They are designed to work with non-Euclidean data and are particularly useful for semi-supervised learning tasks on graphs. GCNs aim to learn a function that maps nodes to a low-dimensional space while considering their local neighborhood and features.

    \paragraph{Convolution filters}
    A Convolutional Neural Network (CNN) is a type of neural network that uses convolutional filters to extract features from images. These filters are applied to local regions of the image to capture spatial patterns \cite{CNNIntro15}. The same concept can be extended to graphs, where the convolution filters are applied to local regions of the graph to capture structural patterns. 

    \begin{minipage}{\dimexpr\linewidth-20pt}
        The convolution operation on an image is historically defined as follows \cite{ImageConvolution13} \cite{TutorialDeepLearningPart2}:
        \par
        \vspace{2em}
        \begin{center}
            \[
            y_{i,j} = \sum_{k=-m}^{m} \sum_{l=-n}^{n} x_{i+k,j+l} \times w_{kl}
            \]
        \end{center}
        \vspace{1em}
    \end{minipage}

    Where:
    \begin{itemize}
        \item \( y_{i,j} \): The output value at the coordinate $ (i,j) $ of the convolution output.
        \item \( x_{i,j} \): The input value at the coordinate $ (i,j) $ of the input image.
        \item \( w_{kl} \): The weight value (coefficient) of the kernel at the coordinate $ (k,l) $.
        \item \( m \): The kernel width.
        \item \( n \): The kernel height.
    \end{itemize}
    
    Similar to this, ConvGNNs implement the transition function using convolutions. One of the challenges in ConvGNNs is defining regions of a graph, as nodes in a graph may have varying numbers of neighbors. Solutions to this problem involve using spectral or spatial representations of graphs \cite{SemiSupervisedClassificationSpectralConvGCN16} \cite{KG22}. An alternative approach employs attention mechanisms to learn the most important features \cite{GAT17}. Unlike RecGNNs, ConvGNNs apply a fixed number of convolutional layers and can use different kernels/weights at each distinct step.


    Here are some common ConvGNN architectures. I recommend reading \citetitle{GCNAutoFiltering22} \cite{GCNAutoFiltering22} for a more in-depth discussion of these models:

    \begin{itemize}
        \item \textbf{Vanilla GCN}: The basic GCN model consists of an input layer, one or more hidden layers, and an output layer. Each layer is associated with a graph convolution operation that updates the node features based on their neighbors \cite{KG22}.
        
        \item \textbf{GraphSAGE (Graph Sample and Aggregation)}: This model generalizes the GCN framework by allowing various aggregation functions like mean, LSTM, and pooling to combine information from a node's neighbors \cite{GraphSAGE17}.
        
        \item \textbf{ChebNet}: ChebNet uses Chebyshev polynomials to generalize the convolution operation in the spectral domain. This allows the model to capture a broader range of graph structures \cite{GNNComprehensiveSurvey20}.
        
        \item \textbf{GAT (Graph Attention Networks)}: GAT introduces attention mechanisms into GCNs, enabling the model to weigh neighbors differently when aggregating information \cite{GAT17}.
        
        \item \textbf{MoNet}: This model employs a mixture model to generalize the convolution operation, making it capable of handling graphs with diverse structures \cite{MoNet17}.
        
        \item \textbf{Graph Isomorphism Network (GIN)}: GIN is designed to capture the isomorphism between different graphs, making it powerful for tasks like graph classification \cite{GCNAutoFiltering22}.
        
        \item Research on the topic is still ongoing, and a lot of new architectures are being proposed: Cluster-GCN, TAGCN, AutoGCN, and many more \cite{GCNAutoFiltering22}. It's worth noting that many of these models are not mutually exclusive and can be combined to create more powerful models.
    \end{itemize}

    The background chapter has introduced a lot of concepts and techniques that are relevant to our work. We have discussed the importance of data preprocessing and feature engineering, explored various machine learning algorithms, and looked at some common neural network architectures. Although not exhaustive, this overview should provide a solid foundation for the methods, design and implementation that will be discussed in the next chapters.

%% file: tex/chapters/methods.tex
\chapter{Methods}\label{chap:methods}


In the preceding chapters, all the necessary background knowledge to understand the methods have been introduced. In this chapter, we will present an overview of the challenges, the methods, and the tools we have developed for this thesis. We will first describe the dataset we have used. Then, we will describe the programs developed for this thesis. Finally, we will describe how we have packaged and deployed our programs with Nix.


\section{Hardware and software architecture}
    Throughout this thesis, we have used a variety of hardware and software architectures. 
    
    \subsection{Hardware development and testing environment}
    In this section, as a reference for the reader, we will describe shortly the hardware development environment. All environments are running some Linux \texttt{x86\_64} distribution.

    At the start of the project, around the end of 2022, the project started on an old laptop \textit{HP EliteBook Folio 1040 G2}, running \texttt{Ubuntu 22.04 LTS (Jammy Jellyfish)} with the following specifications:

    \begin{itemize}
        \item \textbf{CPU:} 5th Generation Intel Core i7-5600U 2.6 GHz (max turbo frequency 3.2-GHz), 4 MB L3 Cache, 15W
        \item \textbf{GPU:} Intel HD Graphics 5500
        \item \textbf{RAM:} 8GB DDR3L SDRAM (1600 MHz, 1.3v)
    \end{itemize}

    This device was used for the first experiments, and for the development of the first programs. However, it was not powerful enough to run the experiments on the whole dataset, and especially working on \acrshort{ml} part. As such, we have moved to a more powerful machine, a \textit{TUXEDO InfinityBook Pro 16 - Gen7} with the following specifications:

    \begin{itemize}
        \item \textbf{CPU:} 12th Gen Intel i7-12700H (20) @ 4.600GHz
        \item \textbf{GPU:} NVIDIA Geforce RTX 3070 Ti Laptop GPU
        \item \textbf{RAM:} 64GB DDR5 4800MHz Samsung
    \end{itemize}

    For the Operating System, we have switched from \texttt{Fedora 37} to \texttt{NixOS 23 (Tapir)}. This change was motivated by the fact that \texttt{NixOS} is a Linux distribution that uses a purely functional package management system \cite{NixOS08}. This means that the operating system is built by the Nix package manager, using a declarative configuration language. It allows to have a reproducible development environment, and to easily switch between different development environments. This has proved to be very useful in many areas like work environment isolation, on work collaboration with Clément Lahoche, and for software deployment to the server.
    
    Unfortunately, the \textit{TUXEDO InfinityBook Pro 16 - Gen7} laptop was not powerful enough to run the experiments on the whole dataset. Running the python script would have taken more than a week for some simple \acrshort{ml} experiments to run on the whole dataset. Small bash and python scripts have been run on this laptop, as well as tests of the different developed programs, but all the main experiments have been run on the server.
    
    In that context, we were provided 2 development servers towards the end of the thesis, around August 2023. The hardware server is a \textit{AS-4124GS-TNR} with the following specifications:

    \begin{itemize}
        \item \textbf{CPU:} 2x AMD EPYC 7662 (256) @ 2.000GHz
        \item \textbf{GPU:} NVIDIA Geforce RTX 3090 Ti
        \item \textbf{RAM:} 512GB DDR4 3200MHz
    \end{itemize}

    On this server, we have been given access to two docker instances running \texttt{Ubuntu 20.04.6 LTS}. The first instance is called \textbf{Deathstar} and the second one is called \textbf{Drogon}. For this Masterarbeit, I have mostly relied on Drogon for the final experiments, although Deathstar has been used for some preliminary experiments. 

    This server has be provided by the Department of Computer Science of Universität Passau, and especially the Chair of Data Science of Prof. Dr. Michael Granitzer. We would like to thank them for their support.

    \subsection{Software, languages and tools}
    In Computer Sciences, it doesn't take long to realize that testing hypotheses, diving deeper in problems and finding solutions to them is a very iterative process that requires a lot of experimentation. As such, the development of scripts and programs has been a substantial part of this thesis, from the very beginning to the very end. In this process, we have used a variety of tools and programming languages, such as Rust, Python, Bash, or Nix just to name the programming language used.

    In this section, as a reference for the reader, we will describe the software architectures, languages and tools that have been used throughout this thesis.

    Throughout the project, we have come to use a range of programming languages. Initial tests have been done using shell and bash command and simple scripts. However, as the project grew, we quickly moved to more powerful programming languages. 

    Python version 3.11 has been the main language for high level data science and \acrshort{ml} development. This new version of python features many improvements over the previous version, and especially in terms of performance. Better error messages, exception groups, improved support for f-strings, support for TOML configuration files, variadic generics, improved support for asyncio, and new modules for working with cryptography and machine learning are just some new features of this new version of python. While relatively new, this is why we have decided to use this version of python for the development of the \acrshort{ml} part of the project.

    Although Python is a popular and powerful language, it is not the most efficient language. As such, we have used Rust for some parts of the project, especially when no high level library is needed and when performances are critical to be able to parse efficiently the dataset. Rust is a systems programming language that runs blazingly fast, prevents segfaults, and guarantees thread safety. It is a very powerful language, and is especially useful for low-level programming. We have used it for the development of the algorithms that are used to extract the data from the dataset.

    \subsubsection{Packaging and deployment}
    We made an extensive use of git repositories for version control, with GitHub as a main platform for hosting the repositories. An ever-growing number of script and programs have been developed for this thesis. As such, we have needed a way to easily deploy those programs on different machines.

    Rust comes with a handful of tools for managing packages and dependencies. Cargo is Rust's build system and package manager. Cargo downloads your Rust project's dependencies, compiles your project, makes executables, and runs the tests. It is a powerful tool that allows to easily manage Rust projects. However, it is not the best tool for deploying programs on different machines. 

    On Python's side of things, things are a bit more complicated. For a long time, we have relied on virtual environments using the \texttt{conda} package manager. However, it is heavy to use, and it doesn't allow to easily export an environment from one Linux distribution to the other. 

    An example is the library \texttt{pygraphviz}. This library relies on third parties system libraries, that have different names depending on the Linux distribution:
    
    \begin{itemize}
        \item \textbf{Ubuntu:} \lstinline[language=bash]|sudo apt-get install graphviz graphviz-dev| is needed before a correct install of the Python \texttt{pygraphviz} library. 
        \item \textbf{Fedora:} \lstinline[language=bash]|sudo dnf install graphviz graphviz-devel| is needed before installing \texttt{pygraphviz}
    \end{itemize}

    Although it can be seen as a minor issue, this is just one example among dozens of libraries. This is real problem with \texttt{conda}. This is why we have decided to use Nix for managing python packages and dependencies. Nix is a purely functional package manager \cite{NixOriginalThesis06}. It allows to easily manage packages and dependencies, and to easily deploy programs on different machines as it guarantees reproducible builds. It is also very useful for development, as it allows to easily create isolated environments for development. This is why we have used Nix for managing the python packages and dependencies. Gradually, Nix has become a superset of other package managers like pip, conda, or cargo. 
    
    Any Nix project comes with either a \textit{shell.nix} or a more modern \textit{flake.nix}. Those files are used to describe the project, and to list all necessary dependencies. Since we are developing on NixOS, the integration of Nix with the operating system is very good, and can be easily setup.
    
    Nix is however really straightforward to install on any other distribution through the use of a single script available online. It can be installed in as little as one command.




\section{OpenSSH memory dumps dataset}\label{sec:background:kex:dataset}

    SmartKex has contributed to the research community by generating a comprehensive annotated dataset of OpenSSH heap memory dumps \cite{SmartKex22}. The dataset is publicly available on Zenodo \footnote{\url{https://zenodo.org/record/6537904}}. 

    \begin{minipage}{\dimexpr\linewidth-20pt}
        The dataset is organized into two top-level directories: $Training$ and $Validation$ with an additional $Performance\_Test$. The first two main directories are further divided based on the SSH scenario, such as immediate exit, port-forward, secure copy, and shared connection. Each of these subdirectories is then categorized by the software version that generated the memory dump. Within these, the heaps are organized based on their key lengths, providing a multi-layered structure that aids in specific research queries.

        \begin{figure}[H]
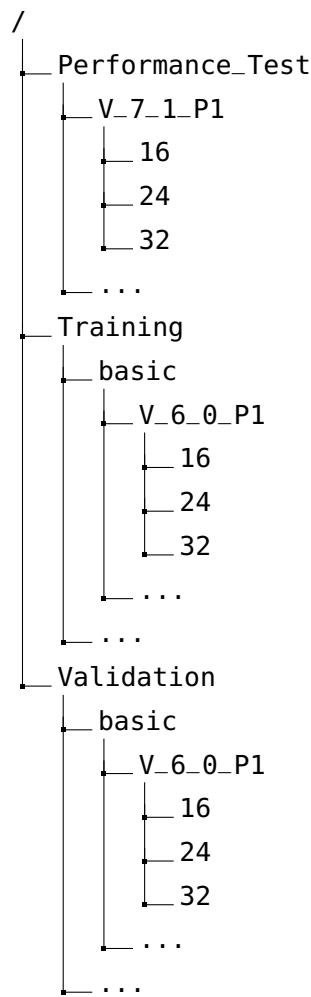

            \centering
            \caption{Illustration of the Dataset Directory Structure}
            \label{fig:dataset_structure}
            \begin{minipage}{0.6\textwidth}
            \dirtree{%
            .1 /.
            .2 Performance\_Test.
            .3 V\_7\_1\_P1.
            .4 16.
            .4 24.
            .4 32.
            .3 ....
            .2 Training.
            .3 basic.
            .4 V\_6\_0\_P1.
            .5 16.
            .5 24.
            .5 32.
            .4 ....
            .3 ....
            .2 Validation.
            .3 basic.
            .4 V\_6\_0\_P1.
            .5 16.
            .5 24.
            .5 32.
            .4 ....
            .3 ....
            }
        \end{minipage}
        \end{figure}
    \end{minipage}

    Two primary file formats are used to store the data: JSON and RAW. The JSON files contain meta-information like the encryption method, virtual memory address of the key, and the key's value in hexadecimal representation. The RAW files, on the other hand, contain the actual heap dump of the OpenSSH process.

    \begin{minipage}{\dimexpr\linewidth-20pt}
        Here is an example of content of a RAW memory dump file, displayed using \textit{vim} and \textit{xxd} commands:

        \begin{lstlisting}[style=hexdump, caption={16 bytes per line visualization of a Hex Dump from \textit{Training/basic/V\_7\_8\_P1/16/5070-1643978841-heap.raw}}]
            00000000: 0000 0000 0000 0000 5102 0000 0000 0000  ........Q.......
            00000010: 0607 0707 0707 0303 0200 0006 0401 0206  ................
            00000020: 0200 0001 0100 0107 0604 0100 0000 0203  ................
            00000030: 0103 0101 0000 0000 0000 0000 0000 0002  ................
            00000040: 0001 0000 0000 0000 0000 0100 0000 0001  ................
            00000050: 8022 1a3a 3456 0000 007f 1a3a 3456 0000  .".:4V.....:4V..
            00000060: f040 1a3a 3456 0000 9032 1a3a 3456 0000  .@.:4V...2.:4V..
            00000070: 608b 1a3a 3456 0000 9047 1a3a 3456 0000  `..:4V...G.:4V..
        \end{lstlisting}
    \end{minipage}

    The original file contains the raw byte content of the heap dump of a specific version of OpenSSH. It is a binary file, which means that it is not human-readable. However, it can be converted to a human-readable format using the \textit{xxd} command. The first column to the left represents the offset in hexadecimal. The last column represents the actual content of the bytes, in ASCII format. The columns in between represent the content of the bytes in hexadecimal format.

    Since hexadecimal is a base-16 number system, each byte is represented by two hexadecimal digits. The ASCII representation of the bytes is displayed on the right, and is only used for reference, as it is not always possible to convert the bytes to ASCII. For instance, the bytes at offset 0x10 are not printable characters, and thus cannot be converted to ASCII. Each line represents 16 bytes, and the offset is incremented by 16 for each line.

    For the purpose of this thesis, it will be more interesting to visualize the content of the heap dump as 8 bytes lines. This can be achieved by using the \textit{xxd} command with the \textit{-c} option, as shown in the following example:

    \begin{minipage}{\dimexpr\linewidth-20pt}
        The same example as before, a memory dump file, displayed using \textit{vim} and \textit{xdd -c 8} commands:

        \begin{lstlisting}[style=hexdump, caption={8 bytes per line visualization of a Hex Dump from \textit{Training/basic/V\_7\_8\_P1/16/5070-1643978841-heap.raw}}, label={lst:hexdump-8bytes}]
            00000000: 0000 0000 0000 0000  ........
            00000008: 5102 0000 0000 0000  Q.......
            00000010: 0607 0707 0707 0303  ........
            00000018: 0200 0006 0401 0206  ........
            00000020: 0200 0001 0100 0107  ........
            00000028: 0604 0100 0000 0203  ........
            00000030: 0103 0101 0000 0000  ........
            00000038: 0000 0000 0000 0002  ........
            00000040: 0001 0000 0000 0000  ........
            00000048: 0000 0100 0000 0001  ........
            00000050: 8022 1a3a 3456 0000  .".:4V..
            00000058: 007f 1a3a 3456 0000  ...:4V..
            00000060: f040 1a3a 3456 0000  .@.:4V..
            00000068: 9032 1a3a 3456 0000  .2.:4V..
            00000070: 608b 1a3a 3456 0000  `..:4V..
            00000078: 9047 1a3a 3456 0000  .G.:4V..
        \end{lstlisting}
    \end{minipage}

    This example shows the exact content of the preceding one. 

    To this RAW file is associated a JSON file, which contains its annotations.  

    \begin{minipage}{\dimexpr\linewidth-20pt}
         Here is an example of content of a JSON annotation file that comes with the previous RAW file:

        \begin{lstlisting}[style=json, caption={Complete JSON example, from \textit{Training/basic/V\_7\_8\_P1/16/5070-1643978841.json}}, label={lst:json-annotation-ex-1}]
            {
                "SSH_PID": "5070",
                "SSH_STRUCT_ADDR": "56343a1a4800",
                "session_state_OFFSET": "0",
                "SESSION_STATE_ADDR": "56343a1a8d30",
                "newkeys_OFFSET": "344",
                "NEWKEYS_1_ADDR": "56343a1aaa40",
                "NEWKEYS_2_ADDR": "56343a1aab40",
                "enc_KEY_OFFSET": "0",
                "mac_KEY_OFFSET": "48",
                "name_ENCRYPTION_KEY_OFFSET": "0",
                "ENCRYPTION_KEY_1_NAME_ADDR": "56343a1a9db0",
                "ENCRYPTION_KEY_1_NAME": "aes128-gcm@openssh.com",
                "ENCRYPTION_KEY_2_NAME_ADDR": "56343a1a3fb0",
                "ENCRYPTION_KEY_2_NAME": "aes128-gcm@openssh.com",
                "key_ENCRYPTION_KEY_OFFSET": "32",
                "key_len_ENCRYPTION_KEY_OFFSET": "20",
                "iv_ENCRYPTION_KEY_OFFSET": "40",
                "iv_len_ENCRYPTION_KEY_OFFSET": "24",
                "KEY_A_ADDR": "56343a1a3170",
                "KEY_A_LEN": "12",
                "KEY_A_REAL_LEN": "12",
                "KEY_A": "feb5fd4ef0759b034d69b858",
                "KEY_B_ADDR": "56343a1a33e0",
                "KEY_B_LEN": "12",
                "KEY_B_REAL_LEN": "12",
                "KEY_B": "f50b988297fa19709445c4ee",
                "KEY_C_ADDR": "56343a1aa1b0",
                "KEY_C_LEN": "16",
                "KEY_C_REAL_LEN": "16",
                "KEY_C": "f5b53280e944db0fe196668d877cd4c0",
                "KEY_D_ADDR": "56343a1a4010",
                "KEY_D_LEN": "16",
                "KEY_D_REAL_LEN": "16",
                "KEY_D": "ac4f18a963d9e72c857497b7dc9d088d",
                "KEY_E_ADDR": "56343a1a7d90",
                "KEY_E_LEN": "0",
                "KEY_E_REAL_LEN": "0",
                "KEY_E": "",
                "KEY_F_ADDR": "56343a1a2f60",
                "KEY_F_LEN": "0",
                "KEY_F_REAL_LEN": "0",
                "KEY_F": "",
                "HEAP_START": "56343a198000"
            }
        \end{lstlisting}
    \end{minipage}

    Those annotation files contain the meta-information about the heap dump, such as the encryption method, virtual memory address of the key, and the key's value in hexadecimal representation. Those annotations are invaluable for the development of machine learning models used for key prediction. 

    The dataset is not just limited to SSH key extraction; it also serves as a resource for identifying essential data structures that hold sensitive information. This makes it a versatile tool for various research applications. 

    \subsection{Assumptions}\label{sec:methods:dataset:assumptions}
    Before we dive in, let's make some assumptions about the dataset. We will use these assumptions to guide our exploration of the heap dump file. 

    \begin{itemize}
        \item \textbf{Heap dump file size:} We will assume that the heap dump file size is a multiple of 8 bytes. This is because the heap dump file is a memory dump, and memory is allocated in chunks that are multiples of 8 bytes. This means that we can expect the heap dump file to be composed of a sequence of 8 bytes blocks. If this assumption is not met, we will assume that padding the last block with 0s will not change the results of our exploration.
        \item \textbf{Chunk chaining:} We will assume that all the heap dump files have been generated using the same \lstinline[language=c]|malloc| implementation from GlibC. It means that we can expect to find the same patterns in all the heap dump files. Especially, we expect all the heap dump files to start by a first allocated in-use chunk. We can then follow the malloc header chaining to explore the heap dump file allocated memory chunks \cite{MallocInternalsWiki2023}.
        \item \textbf{Dataset key annotation format:} We will assume that the JSON annotation files have been generated using the same program. This means that we can expect the same format for all the JSON annotation files. This is important, as we will use the JSON annotation files to get the key addresses for annotating memory graphs used for the embedding step. If the format is not the same, we will assume that the JSON annotation file is corrupted, and we will skip it.
    \end{itemize}

    The \textbf{chunk chaining assumption} is absolutely crucial for the exploration of the heap dump file. It allows us to follow the malloc header chaining to explore the heap dump file allocated memory chunks. This assumption is supported by the code where we can find a comment stating that: \say{since chunks are adjacent to each other in memory, if you know the address of the first chunk (lowest address) in a heap, you can iterate through all the chunks in the heap by using the size information, but only by increasing address, although it may be difficult to detect when you've hit the last chunk in the heap} \cite{MallocGLIBC2001}.

    In the scripts and programs that have been developed for the following thesis, we have implemented a number of checks and tests to ensure that these assumptions are met. If not, the programs will raise an error, log the problem and generally skip the data. This behavior is implemented to ensure that the programs are robust to unexpected data, and to ensure that the results are reliable. These assumptions and related problems will be discussed and measured at several locations in the following sections.

    \subsection{Dataset production system information}\label{sec:methods:dataset:production_system_information}
    Neither the paper \citetitle{SmartKex22} nor the dataset itself provide information about the hardware and software used for its generation. This is important since we will be exploring allocated raw bytes which depend on the system and C library used. We obtained some information about the dataset generation by contacting the authors of the paper. 

    As specified in a mail from Reiser, Hans, the dataset was generated on a system with the following command output:

    \begin{minipage}{\dimexpr\linewidth-20pt}
    \begin{lstlisting}[language=bash, caption={Command and logs of the C-library version used for the dataset generation}]
        root@debian10:~# ldd --version
        ldd (Debian GLIBC 2.28-10) 2.28
        Copyright (C) 2018 Free Software Foundation, Inc.
        This is free software; see the source for copying conditions.  There is NO
        warranty; not even for MERCHANTABILITY or FITNESS FOR A PARTICULAR PURPOSE.
        Written by Roland McGrath and Ulrich Drepper.
    \end{lstlisting}
    \end{minipage}

    \begin{minipage}{\dimexpr\linewidth-20pt}
    \begin{lstlisting}[language=bash, caption={Command and logs of the Linux Standard Base Release used for the dataset generation}]
        root@debian10:~# lsb_release -a
        No LSB modules are available.
        Distributor ID:     Debian
        Description:        Debian GNU/Linux 10 (buster)
        Release:            10
        Codename:           buster
    \end{lstlisting}
    \end{minipage}

    \begin{minipage}{\dimexpr\linewidth-20pt}
    \begin{lstlisting}[language=bash, caption={Command and logs of the OS and kernel version used for the dataset generation}]
        root@debian10:~# uname -a
        Linux debian10 4.19.0-16-amd64 #1 SMP Debian 4.19.181-1 (2021-03-19) x86_64 GNU/Linux
    \end{lstlisting}
    \end{minipage}

    He also precise that the CPU used was an Intel Xeon CPU, either a E5-2609 or a E3-1230. From those commands, we can deduce the following crucial system related information:

    \begin{itemize}
        \item \textbf{CPU architecture:} \lstinline[language=bash]|x86_64|
        \item \textbf{OS version:} \lstinline[language=bash]|Debian GNU/Linux 10 (buster)|
        \item \textbf{Kernel version:} \lstinline[language=bash]|4.19.0-16-amd64|
        \item \textbf{C library version:} \lstinline[language=bash]|Debian GLIBC 2.28-10|
    \end{itemize}

    \subsection{Conventions and vocabulary}
    It's important to define some conventions and vocabulary that will be used in the following sections, since many concepts can be ambiguous depending on the context.

    \begin{itemize}
        \item \textbf{memory graph:} A memory graph is our particular case refers to a directed graph that represents the memory of a heap dump file. The memory graph is the main data structure used for the embedding step.
        \item \textbf{block:} In the following, we will refer to a block as a sequence of 8 bytes. This is because the heap dump file is a memory dump, and memory is allocated in chunks that are multiples of 8 bytes. This means that we can expect the heap dump file to be composed of a sequence of 8 bytes blocks.
        \item \textbf{chunk:} A chunk is a sequence of blocks bytes. This concepts directly comes from the \lstinline[language=c]|malloc| implementation. At its core, a chunk has a user data body composed of blocks and a malloc header block. A chunk can be in-use or free.
    \end{itemize}

    \subsection{Estimating the dataset balancing for key prediction}
    First, let's quickly estimate what the dataset is composed about. This will later be used to estimate the balancing of data for our key prediction goal. Some quick Linux commands can be used to get a general overview of the dataset.
    
    A first command can quickly give us an idea of the number of files in the dataset:
    \begin{lstlisting}[caption={Count all dataset files}, label=methods:code:count_all_dataset_files, language=bash]
        find /path/to/dataset -type f | wc -l
    \end{lstlisting}

    Another command can be used to get the total size of the dataset:
    \begin{lstlisting}[caption={Get the total size of the dataset}, label=methods:code:get_total_size_dataset, language=bash]
        du -sb /path/to/dataset
    \end{lstlisting}

    The first command indicates that the dataset contains $ 208749 $ files, which represents, according to second one, a total of $ 18203592048 $ bytes, or around 18 Gigabytes.

    We could just divide the number of files by the size of the dataset to get an average size of the files. However, this would not be accurate, as we are only interested in the size of the RAW files. Since JSON files are much smaller than RAW files, they would skew the average size of the files. Since we are only considering RAW files, we will use improved commands in order to determine the size of the RAW file only.

    The following command can be used to get a better understanding of the dataset, concerning the number of RAW files and their size:

    \begin{lstlisting}[caption={Find the number of RAW files in the dataset}, language=bash]
        find /path/to/dataset -type f -name "*.RAW" | wc -l
    \end{lstlisting}

    And the next one can be used to get the number of bytes of RAW files in the dataset:

    \begin{lstlisting}[caption={Find the number of bytes of RAW files in the dataset}, language=bash]
        find /path/to/dataset -type f -name "*.raw" -exec du -b {} + \
            | awk '{s+=$1} END {print s}'
    \end{lstlisting}

    Where:
    \begin{itemize}
        \item \lstinline[language=bash]!find phdtrack_data/ -type f -name "*.raw"! finds all the files in the dataset that have the extension \lstinline[language=bash]!.raw!.
        \item \lstinline[language=bash]!-exec du -b {} + | awk '{s+=$1} END {print s}'! executes the command \lstinline[language=bash]!du -b! on each file found by the previous command, and sums the size of each file.
    \end{itemize}

    These commands indicate that the dataset contains $ 103595 $ RAW files, which represents a total of $ 18067001344 $ bytes, or around 18 Gigabytes. This shows that the vast majority of the data is contained in RAW files, with JSON files representing less than a percent of the dataset in terms of byte size. As such, the average size for every RAW file is around 170 Kilobytes. 

    Now, considering that a given heap dump file is expecting to have only 6 keys (see \ref{sec:background:ssh:ssh_keys}), with keys maximal possible size being of 64 bytes, we can estimate that we have at maximum $ 39780480 $ or around 40 Megabytes of positively labeled samples. This, considering the total useful size of around 18 Gigabytes, means that our dataset is very imbalanced, with an expected upper-bounded ratio of $ 0.0022\% $ of positively labeled samples or around $ 2:1000 $.

    Considering that, a frontal \acrshort{ml} binary classification approach will not work. This is why the present report will discuss feature engineering and graph-based memory representation. The idea is to embed more information to our keys so as to be able to fight effectively the imbalanceness of the raw data.

    \subsection{Dataset validation}
    The dataset is merely a collection of heap dump RAW files for different use cases and versions of OpenSSH. Each heap dump file goes along a JSON annotation file that has been generated by the creators of the dataset to provide additional information about the heap dump, and especially encryption keys.
    
    However, it is worth noting that the dataset is not perfect. The use of the dataset for machine learning has revealed some issues. For instance, some JSON annotation files are not valid JSON files, and cannot be loaded as such. Some JSON annotation files are also not complete, with some crucial information missing. This is a problem, as we will use the JSON annotation files to get the key addresses for annotating memory graphs used for the embedding step. If the format is not the same, we will assume that the JSON annotation file is corrupted, and we will skip it.
    
    This is probably due to the fact that the annotations were generated automatically. For instance, in \textit{Training/basic/V\_7\_8\_P1/16/}, literally the first file of the dataset contains an incomplete annotation file, as some keys are missing. This is a limitation of the dataset that should be kept in mind when using it for research purposes.

    \begin{minipage}{\dimexpr\linewidth-20pt}
        Here is an example of content of a JSON annotation file with missing keys, and with missing annotations (like address or length) for the keys that are present:

        \begin{lstlisting}[style=json, caption={Missing keys in JSON annotation file \textit{Training/basic/V\_6\_0\_P1/16/24375-1644243522.json}}]
            {
                "ENCRYPTION_KEY_NAME": "aes128-ctr",
                "ENCRYPTION_KEY_LENGTH": "16",
                "KEY_C": "689e549a80ce4be95d8b742e36a229bf",
                "KEY_D": "76788e66a56d2b61eec294df37422fcb",
                "HEAP_START": "5589d41e0000"
            }
        \end{lstlisting}
    \end{minipage}

    \subsubsection{Automatic annotation validation}
    So as to determine really how much of the dataset can be used really for machine learned, we have developed a script that checks the validity of those annotations. This script called \texttt{check\_annotations.py}, is used to verify the quality, completeness, consistency and coherence of the annotations.

    Files are regrouped under the following categories:

    \begin{itemize}
        \item \textbf{Correct and Complete Files:} Files that have no missing keys, and that have all the keys with correct values.
        \item \textbf{Broken Files:} Files that are not valid JSON files, and cannot be loaded as such.
        \item \textbf{Incorrect Files:} Files that have contradictory information in their annotation file.
        \item \textbf{Missing key Files:} Files that have missing keys in their annotation file. A typical example is a JSON file with \lstinline[style=json]!"KEY_E": ""!. This means that the key E is missing, and that the key E address is not present in the annotation file, which is a problem for the machine learning since it means that we cannot label correctly the key E.
        \item \textbf{Incomplete key Files:} Files that have incomplete keys in their annotation file. A typical example is a JSON file with \lstinline[style=json]!"KEY_E": "689e549a80ce4be95d8b742e36a229bf"!. This means that the key E is present, but that the key E address is not present in the annotation file, which is a problem for the machine learning since it means that we cannot label correctly the key E.
    \end{itemize}

    The script is used to validate each JSON file using the following process:

    \begin{algorithm}[H]
        \caption{Json Annotation Validation}
        \begin{algorithmic}[1]
        \Procedure{ValidateJson}{$\text{json\_data}$}
            \State \textbf{Initialize} errors = Dictionnary\{\} \Comment{Serve as collection for counted errors}
            \State \textbf{Initialize} mandatory\_json\_keys = ['HEAP\_START', 'SSH\_STRUCT\_ADDR', 'SESSION\_STATE\_ADDR']
            \State \textbf{Initialize} key\_names = \{\}
            \State \textbf{Initialize} incorrect\_keys, missing\_keys, incomplete\_keys = 0
            
            \For{mandatory\_json\_key \textbf{in} mandatory\_json\_keys} \Comment{Check if some expected json keys are missing}
                \If{mandatory\_json\_key \textbf{not in} json\_data \textbf{or not} \text{correct hex address}}
                    \State errors[mandatory\_json\_key] = False
                \Else
                    \State errors[mandatory\_json\_key] = True
                \EndIf
            \EndFor
            
            \For{json\_key \textbf{in} json\_data.keys()} \Comment{Get all the keys names, like A, B, C, D, E, F}
                \If{json\_key.startswith("KEY\_")}
                    \State key\_name = GetLetterOfSSHKeyFromJSONKeyName(json\_key)
                    \State key\_names.add(key\_name)
                \EndIf
            \EndFor
            
            \For{key\_letter \textbf{in} key\_names}
                \State base\_key = "KEY\_" + key\_letter
                \State \textbf{PerformSSHKeyAnnotationValidationAndCompleteness}(base\_key, json\_data) that counts incorrect\_keys, missing\_keys, incomplete\_keys
            \EndFor
            
            \State Store counters in errors
            \State \Return errors
        \EndProcedure
        \end{algorithmic}
    \end{algorithm}

    The counting error algorithm done on each SSH key annotation by is described in the following:

    \begin{algorithm}[H]
        \caption{SSH Key Annotation Validation}
        \begin{algorithmic}[1]
        \Procedure{PerformSSHKeyAnnotationValidationAndCompleteness}{base\_key, json\_data}
            \State \textbf{Initialize} incorrect\_keys, missing\_keys, incomplete\_keys = 0
            \If{length(json\_data[base\_key]) == 0} 
                \State missing\_keys += 1 \Comment{missing key}
            \Else
                \State is\_key\_len\_present = exists(json\_data[base\_key\_LEN])
                \State is\_key\_addr\_present = exists(json\_data[base\_key\_ADDR])
                \State is\_key\_real\_len\_present = exists(json\_data[base\_key\_REAL\_LEN])
                
                \If{not is\_key\_len\_present \textbf{or} not is\_key\_addr\_present \textbf{or} not is\_key\_real\_len\_present} 
                    \State incomplete\_keys += 1 \Comment{Incomplete keys}
                    \State Generate and store error message about missing annotations
                \ElsIf{not is\_hex\_address\_correct(json\_data[base\_key\_ADDR])} 
                    \State incorrect\_keys += 1 \Comment{Incorrect address}
                    \State Generate and store error message about incorrect address
                \ElsIf{json\_data[base\_key\_LEN] is not a number or is negative} 
                    \State incorrect\_keys += 1 \Comment{Incorrect length}
                    \State Generate and store error message about incorrect length
                \Else
                    \State Validate key value length based on annotation length
                    \If{json\_data[base\_key\_LEN] == 0}
                        \State missing\_keys += 1 \Comment{missing key}
                    \ElsIf{length(json\_data[base\_key]) \textbf{!=} json\_data[base\_key\_LEN] * 2}
                        \State incorrect\_keys += 1 \Comment{contradictory length}
                        \State Generate and store error message about incorrect key value length
                    \EndIf
                \EndIf
            \EndIf
            \State \Return incorrect\_keys, missing\_keys, incomplete\_keys
        \EndProcedure
        \end{algorithmic}
    \end{algorithm}

    Note that I have simplified this algorithm. The \texttt{is\_hex\_address\_correct} function requires other manipulations to be called, since it checks that the given value is in the range of the heap dump addresses. To do so, it requires handling potentially missing \texttt{HEAP\_START} annotation, hexadecimal conversion with correct endianness, and other manipulations like determining the size of the heap dump. The full code is available in the \texttt{check\_annotations.py} file.

    The script runs in a few seconds on all the $103595$ JSON annotation files, and give the following results:

    \begin{itemize}
        \item \textbf{Number of Correct and Complete Files:} $ 26196 $ files 
        \item \textbf{Number of Broken Files:} $ 6 $ files are broken. A direct look at those files shows that they are empty files.
        \item \textbf{Number of Incorrect Files:} $ 0 $ files
        \item \textbf{Number of Missing key Files:} $ 58643 $ files have missing keys.
        \item \textbf{Number of Incomplete key Files:} $ 18750 $ files have incomplete keys.
    \end{itemize}

    We can also directly look at the keys in general:

    \begin{itemize}
        \item \textbf{Number of SSH keys:} $546534$ keys
        \item \textbf{Number of missing (empty) SSH keys:} $157244$ keys
        \item \textbf{Number of incompletely annotated SSH keys:} $37500$ keys
        \item \textbf{Number of incorrectly annotated SSH keys:} $0$ keys
    \end{itemize}

    \subsection{Dataset cleaning}\label{sec:methods:dataset:cleaning}
    We need to ensure that the subset of the original dataset that will be used for machine learning is correct and consistent. This means that we need to remove the broken files, and the files that have missing or incomplete keys. 

    In the new cleaned subset of the dataset, we kept only the files identified as correct and complete. This way, we are left with $26196$ RAW files.

    From this, we need to remove the raw files that do not respect the \textbf{Chunk chaining assumption} \ref{sec:methods:dataset:assumptions}. This cleaning process involves the chunk chaining algorithm that will be introduced later. During this process, out of the $26196$ RAW files, $ 5 $ of them have been detected to have 0 sized chunks. Those files have been removed from the cleaned dataset. This leaves us with $26191$ RAW files.

    \begin{lstlisting}[language=bash, caption={Command and logs of counting the number of RAW files in the cleaned dataset.}]
    $ find ~/code/phdtrack/phdtrack_data_clean/ -type f -name "*-heap.raw" | wc -l
    26191
    $ find ~/code/phdtrack/phdtrack_data_clean/ -type f -name "*.json" | wc -l
    26191
    \end{lstlisting}

    In total, this means that only $25.3\% $ of the RAW files with their JSON files are actually usable (correct, complete, with valid chunk chaining), and can be used for machine learning. This is because we don't have access to the packets that have been used to generate the dataset, and thus we cannot regenerate the annotations. Since the machine learning relies entirely on those annotations, we cannot afford to use partially annotated files. 
    
    This is a limitation of the dataset that should be kept in mind when using it for research purposes, and especially for supervised machine learning.

    \subsection{Exploring patterns in RAW heap dump files}
    Before diving into programming, we need to gain a better understanding of how to retrieve useful information from heap dump raw file. For that matter, we will continue to experiment with simple commands in RAW heap dump files. Note that in the following, number bases are indicated, since endianness and conversions can get confusing.

    Let's start by looking back at the RAW file we already presented in \ref{lst:hexdump-8bytes}.

    \subsubsection{Detecting potential pointers}
    The paper \citetitle{SmartKex22} indicates that the keys are 8-bytes aligned. In fact, this is the case for the whole heap dump file. This is why we have chosen to split the study of heap dump files in chunks or blocks of 8 bytes. The term \textit{block} in code is always referring to this, unless specified otherwise. The precision is important, since these blocks should not be confused with \textit{memory blocks} like the ones that are allocated by the \lstinline[language=c]|malloc()| function in C.

    \begin{minipage}{\dimexpr\linewidth-20pt}
        Let's re-open the heap dump file in vim, and let's use the following vim commands to explore the example heap dump file:

        \begin{itemize} 
            \item \lstinline[language=bash]!:%!xxd -c 8  5070-1643978841-heap.raw!: This vim command converts the opened file to a hex dump. The \lstinline[language=bash]!-c 8! option indicates that we want to display 8 bytes per line.
            \item \lstinline[language=bash]!:set hlsearch!: This vim command highlights the search results.
            \item \lstinline[language=bash]!:
            \item \lstinline[language=bash]!:%s/\v([0-9a-f]{8}:)/\1\ ! This vim command adds a whitespace after each 8 byte addresses.
            \item \lstinline[language=bash]!:%s/\v(([0-9a-f]{8}: )([0-9a-f]{16}))/\1\ ! This vim command adds a whitespace after each heap dump byte line.
        \end{itemize}
    \end{minipage}

    To find potential pointers, we can use the following command in vim:
    \begin{lstlisting}[language=bash, caption={Vim command to find potential pointers}]
        :/[0-9a-f]\{12}0\{4}
    \end{lstlisting}

    This is a search that looks for 12 hexadecimal digits followed by 4 zeros. This is because, the maximum possible addresses in the heap dump file have a size of around 12 hexadecimal digits, and because pointer addresses are in little-endian format, meaning that the last 4 bytes of the address are also the Most Significant Bytes (MSB) of the address. 
    
    The result is illustrated below \ref{fig:methods:dataset:pointer_examples_1010-1644391327-heap_potential_pointer_highlight}:

    \begin{figure}[H]\label{fig:methods:dataset:pointer_examples_1010-1644391327-heap_potential_pointer_highlight}
        \centering
        \includegraphics[width=16cm]{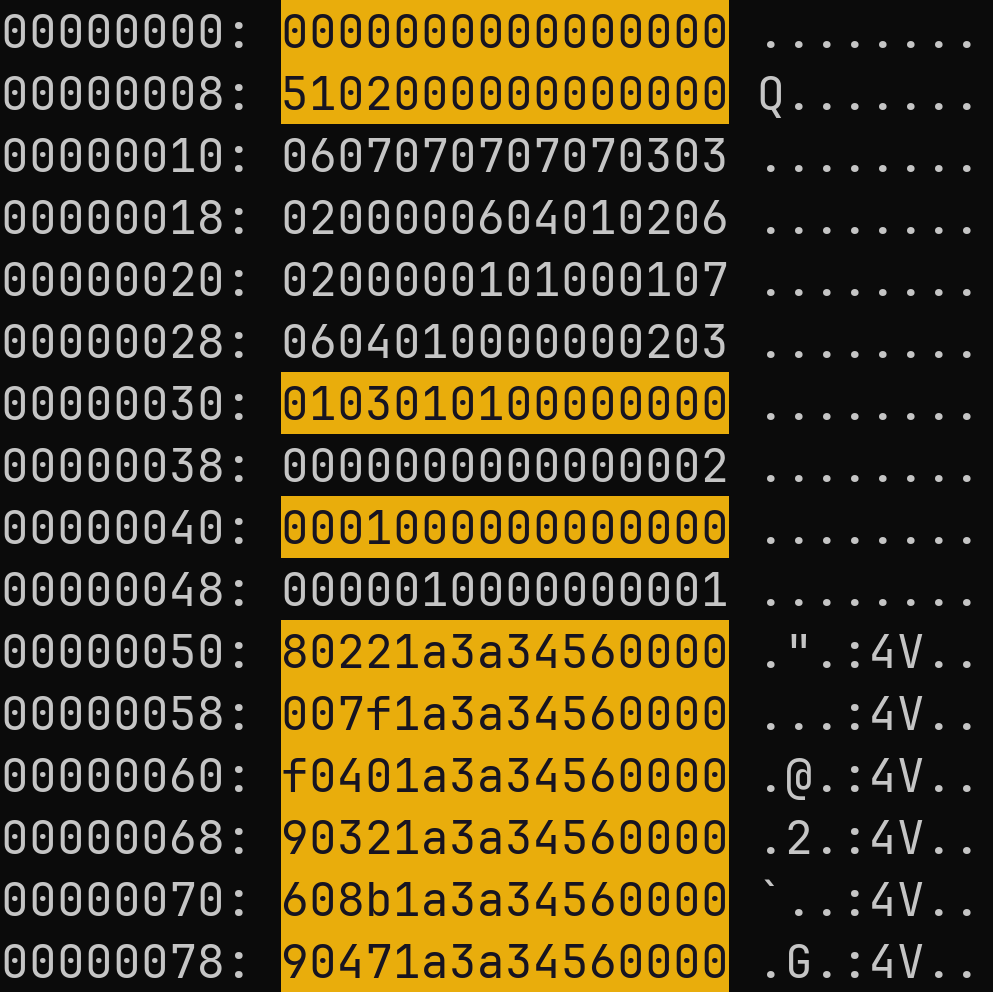}
        \caption{Binary RAW heap dump file loaded using \texttt{vim} and \texttt{xxd}, from \textit{/Training/Training/scp/V\_7\_8\_P1/16/1010-1644391327-heap.raw}, with highlight on rows with 12 hexadecimal digits followed by 4 zeros.}
    \end{figure}

    We have information about the starting address of the heap using \lstinline[style=json]!"HEAP_START": "56343a198000"!. Considering that the example heap dump file contains $ 135169 $ bytes, this means that for this given heap dump file, the pointer addresses range from value $ 94782313037824_{10} $ and $ 94782313172993_{10} $. Note that the little-endian hexadecimal representation of the heap end address is \lstinline[language=c]!0x01901b3a3456! which is 12 character long, or 6 bytes long.

    Note that conversions here can get confusing, since potential pointer strings extracted from the heap dump file are given in little-endian hexadecimal format, but the heap start address from the JSON annotation file is given in big-endian hexadecimal format.

    \begin{minipage}{\dimexpr\linewidth-20pt}
        That way, we can refine the detection of potential pointers by only considering the bytes that are in the range of the heap. Potential pointers are highlighted with "<<<" in the following hex dump:

        \begin{lstlisting}[language=python, caption={Conversions function from hex strings to decimal $ int $ values}.]
        # conversion from hex to decimal
        def hex_str_to_int(hex_str: str) -> int:
            """
            Convert a normal (big-endian) hex string to an int.
            WARNING: HEAP_START in JSON is big-endian.
            """
            bytes_from_str = bytes.fromhex(hex_str)
            return int.from_bytes(
                bytes_from_str, byteorder='big', signed=False
            )
        
        def pointer_str_to_int(hex_str: str) -> int:
            """
            Convert a pointer hex string to an int.
            WARNING: Pointer hex strings are little-endian.
            """
            bytes_from_str = bytes.fromhex(hex_str)
            return int.from_bytes(
                bytes_from_str, byteorder='little', signed=False
            )
        \end{lstlisting}
    \end{minipage}

    Using the functions above, we can check which potential pointers are indeed within the heap dump range.

    \begin{minipage}{\dimexpr\linewidth-20pt}
        That way, we can refine the detection of potential pointers. In the following, pointers are highlighted with \lstinline[style=hexdump]!<<<! in the following hex dump:

        \begin{lstlisting}[style=hexdump, caption={8 bytes per line visualization of a Hex Dump from \textit{Training/basic/V\_7\_8\_P1/16/5070-1643978841-heap.raw}}]
            00000000: 0000000000000000 ........
            00000008: 5102000000000000 Q.......
            00000010: 0607070707070303 ........
            00000018: 0200000604010206 ........
            00000020: 0200000101000107 ........
            00000028: 0604010000000203 ........
            00000030: 0103010100000000 ........
            00000038: 0000000000000002 ........
            00000040: 0001000000000000 ........
            00000048: 0000010000000001 ........
            00000050: 80221a3a34560000 .".:4V.. <<<
            00000058: 007f1a3a34560000 ...:4V.. 
            00000060: f0401a3a34560000 .@.:4V.. <<<
            00000068: 90321a3a34560000 .2.:4V.. <<<
            00000070: 608b1a3a34560000 `..:4V.. <<<
            00000078: 90471a3a34560000 .G.:4V.. <<<
        \end{lstlisting}
    \end{minipage}

    \begin{minipage}{\dimexpr\linewidth-20pt}
        One last check we can do, is verify if the potential pointers are  8-bytes aligned. This can be done by checking if the last 3 bits of the potential address are 0, or using a modulo 8 operation. A simple python function can be used to check that:

        \begin{lstlisting}[language=python, caption={Python function to check if a potential pointer is 8-bytes aligned}]
            def is_pointer_aligned(pointer: int) -> bool:
                """
                Check if a pointer is 8-bytes aligned.
                """
                return pointer % 8 == 0
        \end{lstlisting}
    \end{minipage}

    Using this function on the potential pointers we have found so far, we can see that all of them are indeed 8-bytes aligned. This is a good sign for pointer detection, as we now have a range of tests that can be used to detect potential pointers from other potentially random values.

    Here is the pseudocode for the pointer detection algorithm. This algorithm is presented for a full heap dump file:

    \begin{algorithm}
        \caption{Pointer Detection Algorithm}
        \begin{algorithmic}[1]
        \Procedure{PointerDetection}{$\text{heapDumpFile, HEAP\_START}$}
            \State $\text{heapStart} \gets \text{hex\_str\_to\_int}(HEAP\_START)$
            \State $\text{heapEnd} \gets \text{heapStart} + \text{FileSize}(\text{heapDumpFile})$
            \State $\text{position} \gets 0$
            \State $\text{potentialPointers} \gets []$
            \While{$\text{position} < \text{FileSize}(\text{heapDumpFile})$}
                \State $\text{block} \gets \text{Read8Bytes}(\text{heapDumpFile, position})$
                \If{$\text{block} \neq 0$}
                    \State $\text{pointer} \gets \text{pointer\_str\_to\_int}(\text{block})$
                    \If{$\text{heapStart} \leq \text{pointer} \leq \text{heapEnd}$}
                        \If{$\text{is\_pointer\_aligned}(\text{pointer})$}
                            \State $\text{Append}(\text{pointer}, \text{potentialPointers})$
                        \EndIf
                    \EndIf
                \EndIf
                \State $\text{position} \gets \text{position} + 8$
            \EndWhile
            \State \Return $\text{potentialPointers}$
        \EndProcedure
        \end{algorithmic}
    \end{algorithm}

    This pseudocode outlines the steps for detecting potential pointers in the heap dump file. It starts by reading the heap dump file 8 bytes at a time. For each 8-byte block, it checks if the block is non-zero and within the heap range. If so, it checks if the potential pointer is 8-bytes aligned using the \texttt{is\_pointer\_aligned} function we described before. If all conditions are met, the potential pointer is added to the list of potential pointers. The algorithm returns this list at the end.
    
    \subsubsection{Detecting potential keys}

    Encryption key prediction is the main focus of the present thesis. As such, we will now focus on how to detect potential keys in heap dump files. The paper \citetitle{SmartKex22} introduces 2 algorithms for key detection. The first one is a brute force approach that consists in trying all the possible keys in the heap dump file. The second one is a more sophisticated approach that uses a set of rules to detect potential keys.
    
    The first brute-force algorithm is given by:

    \begin{algorithm}[H]\label{alg:methods:dataset:smartkex22:brute_force}
    \caption{SSH keys brute-force algorithm from \citetitle{SmartKex22} \cite{SmartKex22}}
    \begin{algorithmic}[1]
    \Procedure{FindIVAndKey}{$\text{netPacket}, \text{heapDump}$}
        \State $\text{ivLen} \gets 16$ \Comment{Based on the encryption method}
        \State $\text{keyLen} \gets 24$ \Comment{Based on the encryption method}
        \State $i \gets \text{sizeof(cleanHeapDump)}$
        \State $r \gets 0$
        \While{$r < i$}
            \State $\text{pIV} \gets \text{heapDump}[r : r + \text{ivLen}]$
            \State $x \gets 0$
            \While{$x < i$}
                \State $\text{pKey} \gets \text{heapDump}[x : x + \text{keyLen}]$
                \State $f \gets \text{decrypt}(\text{netPacket}, \text{pIV}, \text{pKey})$
                \If{$f$ is TRUE}
                    \State \textbf{return} $\text{pIV}, \text{pKey}$
                \EndIf
                \State $x \gets x + 8$ \Comment{The IV is 8-bytes aligned}
            \EndWhile
            \State $r \gets x + 8$ \Comment{The key is 8-bytes aligned}
        \EndWhile
    \EndProcedure
    \end{algorithmic}
    \end{algorithm}
    
    This algorithm \ref{alg:methods:dataset:smartkex22:brute_force} outlines the brute-force approach for finding the Initialization Vector (IV) and the key. Initially, the lengths \(\text{ivLen}\) and \(\text{keyLen}\) are set based on the encryption method used for the heap. The algorithm then takes the first \(\text{ivLen}\) bytes from the heap dump to serve as the potential IV (\(pIV\)). Subsequently, \(\text{keyLen}\) bytes are extracted from the heap dump, starting from the first byte, to act as the potential key (\(pKey\)). The algorithm iterates through this potential key until it reaches the end of the heap dump. If decryption of the network packet is unsuccessful, the process is repeated by reading the next potential IV and the subsequent potential key \cite{SmartKex22}. 

    This algorithm is fairly straightforward, and can be implemented in a few lines of code. However, it is also very inefficient, as it tries all the possible keys in the heap dump file. It also needs some encrypted network packets to be able to test the keys, which are not included in the dataset. As such, we will not implement this algorithm.
    
    This is why the authors of the paper have also developed a more sophisticated algorithm that uses a set of rules to detect potential keys.

    No pseudocode is given for the second algorithm, but the paper \citetitle{SmartKex22} gives a description of the algorithm. It relies on the high-entropy assumption of encryption keys. The algorithm is inspired by image processing techniques, and can be described as follows:

    \begin{algorithm}
        \caption{Image-processing inspired Preprocessing Algorithm, as described in \citetitle{SmartKex22} \cite{SmartKex22}}
        \begin{algorithmic}[1]
        \Procedure{Preprocessing}{$\text{heapData}$}
            \State \textbf{Reshape} $\text{heapData}$ into $N \times 8$ matrix $X$
            \For{$i = 0$ to $N-1$}
                \For{$j = 0$ to $7$}
                    \State $Y[i][j] = |X[i][j] - X[i][j+1]| \& |X[i][j] - X[i+1][j]|$
                    \State $Z[i] = \text{count}(Y[i][k] == 0) \geq 4$
                    \If{$i < N-1$}
                        \State $R[i] = Z[i] \& Z[i+1]$
                    \EndIf
                \EndFor
            \EndFor
            \State \textbf{Extract} 128-byte slices from $R$ for training
        \EndProcedure
        \end{algorithmic}
    \end{algorithm}
    
    This Preprocessing Algorithm serves as a crucial step in adapting the heap data for machine learning models. The algorithm begins by reshaping the raw heap data into an \(N \times 8\) matrix \(X\), since keys are 8-bytes aligned \cite{SmartKex22}. Here, \(N \times 8\) is the size of the original heap data in bytes. It then calculates the discrete differences of the bytes in both vertical and horizontal directions, storing the results in matrix \(Y\). The algorithm employs a logical AND operation on these differences to identify high-entropy regions, which are likely candidates for encryption keys. Each 8-byte row in \(Y\) is examined for randomness, and if at least half of its bytes differ from adjacent bytes, it is marked as a potential part of an encryption key. The algorithm then filters out isolated rows that are unlikely to be part of an encryption key, resulting in an array \(R\). Finally, 128-byte slices are extracted from \(R\) for training the machine learning model. This preprocessing step not only adapts the data for machine learning but also narrows down the search space for potential encryption keys, thereby enhancing the efficiency of the subsequent steps. 

    However, this algorithm is not as efficient as it could be. It relies on using a kind of sliding window, which is not easily parallelizable. Also, the entropy-inspired computation is not as straightforward as it could be. That why we propose a new algorithm that is more efficient and more easily parallelizable.

    In order to perform some \acrshort{ml} techniques, and because the keys we are looking for can have a range of possible lengths (16, 24, 32, or 64 bytes), we shift the focus from detecting the full key, to just be able to predict the address of the key. That way, we can deal with keys of different sizes, and we can also use the same algorithm to detect the IV. This is why we will focus on detecting potential keys addresses, and not the full keys.

    We thus introduce a new algorithm for narrowing the search space for potential keys. This algorithm is inspired by the paper \citetitle{SmartKex22}, but is more efficient and more easily parallelizable, as it focuses on producing pairs of blocks of 8 bytes with high entropy. It uses directly the Shannon entropy formula, with each entropy computation being independent of the others.

    \begin{algorithm}
        \caption{Entropy Based Detection of Potential Key blocks}
        \begin{algorithmic}[1]
        \Procedure{EntropyDetection}{$\text{heapData}$}
            \State \textbf{Pad} $\text{heapData}$ with 0s to be 8-bytes aligned
            \State \textbf{Reshape} $\text{heapData}$ into $N \times 8$ matrix $X$
            \For{$i = 0$ to $N-1$,}
                \State $entropy[i] = \text{ShannonEntropy}(X[i])$ \Comment{Independents, compute in parallel.}
            \EndFor
            \State \textbf{Add} $entropy$ 2 by 2 pairs into $entropy\_pairs$ \Comment{Keep block indexes in resulting tuples.}
            \State \textbf{Sort} $entropy\_pairs$ by entropy as $sorted\_pairs$
            \State \Return $\text{SortedPairs}(\text{sorted\_pairs})$
        \EndProcedure
        \end{algorithmic}
    \end{algorithm}

    The \textit{Entropy Based Detection of Potential Key blocks} algorithm takes a raw heap dump, represented by the variable \texttt{heapData}, as input. The data is first padded with zeros to align it to 8-byte blocks and then reshaped into an $N \times 8$ matrix $X$. The Shannon entropy is computed for each 8-byte block in parallel, resulting in an array \texttt{entropy}. Subsequently, the entropy values of adjacent blocks are summed to form pairs, which are stored in \texttt{entropy\_pairs} along with their block indexes. These pairs are then sorted by their entropy sums to produce \texttt{sorted\_pairs}. The idea of using pairs of blocks instead of a single block or more than two blocks is related to the minimum key size, which is 16 bytes. This means that we need at least 2 blocks to be able to detect a potential key. The algorithm returns sorted pairs, so that we can easily extract the ones with the highest entropy sums. Given the index of a block, its actual memory address can be recomputed using the \texttt{HEAP\_START} address available in annotations.
    
    Using this algorithm, let's continue to explore our example heap dump file from \ref{lst:hexdump-8bytes}. We will use the following python function to compute the Shannon entropy of a given block of 8 bytes:

    \begin{minipage}{\dimexpr\linewidth-20pt}
    \begin{lstlisting}[language=python, caption={Python function to compute the Shannon entropy of a given block of 8 bytes}]
        def get_entropy(data: bytes):
            """
            Computes the entropy of a byte array, using Shannon's formula.
            """

            if len(data) == 0:
                return 0.0
            
            # Count the occurrences of each byte value
            _, counts = np.unique(data, return_counts=True)
            
            # Calculate the probabilities
            prob = counts / len(data)
            
            # Calculate the entropy using Shannon's formula
            entropy = -np.sum(prob * np.log2(prob))
            
            return entropy
    \end{lstlisting}
    \end{minipage}

    This function used NumPy array function for efficient computation. We can now use this function to compute the entropy of each block of 8 bytes in the heap dump file. We can then sort the pair of blocks by their entropy, and keep the ones with the highest entropy.
    
    When applied to the file \textit{Training/basic/V\_7\_8\_P1/16/5070-1643978841-heap.raw}, the algorithm produced $ 16896 $ entropy pairs, with $ 631 $ pairs having the maximum entropy sum. Another test using the index to address conversion and the JSON annotation file also indicate that all the 6 key addresses are within the $ 631 $ pairs with the highest entropy sum.
    
    This allows to reduce significantly the search space for potential keys, to already less that 4\% of the original heap dump file, which is significantly better that the 30\% reduction obtained by the preprocessing algorithm described in the paper SmartKex \cite{SmartKex22}, but less that the 2\% reduction obtained by the \acrshort{ml}-based processing algorithm described in the paper \cite{SmartKex22}. However, the same paper indicated that it was tested only for Key A and Key C, whereas this algorithm is tested for all the keys. Keep in mind that this is just an example for a single random file in the dataset, as a way to introduce the subject. In-depth experiments will be done in the dedicated chapter on \acrlong{ml}.

    Indeed, it is important to mention that we can rely on the JSON annotation files for providing labelling for key address prediction. Using this, we do not need to decrypt the network packets to be able to train our \acrshort{ml} models. This is a huge advantage, and is also required since we don't have the encrypted network packets in the dataset. Since we don't have those, and since the keys are already given, that is why we will focus on key address prediction, and not on key prediction.

    \subsection{Data structure exploration}
    Since the dataset contain whole heap dump file, we can also try to detect allocated chunks in those heap dumps. This can be done by looking for patterns in the heap dump file, in a similar fashion as we have done for potential pointers. However, for data structure, we can rely on our knowledge of the C library used to know exactly what to look for.
    
    Since OpenSSH is written in C, we can expect to find some C memory chunks in the heap dump files. C uses the \lstinline[language=c]|malloc| function to allocate memory. This function is used to allocate memory for a given data structure. It takes as input the size of the data structure to allocate, and returns a pointer to the allocated memory. We know that the dataset has been produced using \texttt{GLIBC 2.28} \ref{sec:methods:dataset:production_system_information}. Looking directly at the code for \lstinline[language=c]|malloc| in \texttt{GLIBC 2.28}, we can read in the comments that \say{Minimum overhead per allocated chunk: 4 or 8 bytes. Each malloc chunk has a hidden word of overhead holding size and status information} \cite{MallocGLIBC2001}. This is what we refer to as the \textit{malloc header}. This means that we can expect to find some 8-bytes aligned blocks in the heap dump file, that are not pointers, but that are the result of a \lstinline[language=c]|malloc| call. Detecting and using those \textit{malloc headers} is how we will try to detect memory chunks in heap dump files.

    In Linux on a \texttt{x86\_64} architecture, the malloc function typically uses a block (chunk) header to store metadata about each allocated block. This header is placed immediately before the block of memory returned to the user. The exact layout can vary depending on the implementation of the C library (e.g., glibc, musl), but generally, it contains the following:

    \begin{itemize}
        \item \textbf{Size of the Block}: The size of the allocated block, usually in bytes. This size often includes the size of the header itself and may be aligned to a multiple of 8 or 16 bytes.
        \item \textbf{Flags}: Various flags that indicate the status of the block, such as whether it is free or allocated, or whether the previous block is free or allocated. These flags are often stored in the least significant bits of the size field, taking advantage of the fact that the size is usually aligned, leaving the least significant bits zeroed.
    \end{itemize}

    \subsubsection{How \texttt{malloc} handles Heap Allocation}

    The \texttt{malloc} function in GLIBC 2.28 uses a boundary tag method to manage chunks of memory. Each chunk contains metadata that helps in the allocation and deallocation of memory \cite{MallocGLIBC2001} \cite{MallocInternalsWiki2023}. Below are the key components of a chunk:

    A chunk is a contiguous section of memory, in our case composed of several blocks of 8 bytes, that is handled by \texttt{malloc}. It contains the following components \cite{MallocInternalsWiki2023} \cite{StackExchangeMalloc2023}:

    \begin{enumerate}
        \item \textbf{Size of Previous Chunk}: This field exists only if the previous chunk is unallocated and its \texttt{P} (PREV\_INUSE) bit is clear. It helps in finding the front of the previous chunk.
        
        \item \textbf{Size of Chunk}: This field contains the size of the chunk in bytes along with three flags: \texttt{A} (NON\_MAIN\_ARENA), \texttt{M} (IS\_MMAPPED), and \texttt{P} (PREV\_INUSE). These flags are in the last 3 \acrshort{lsb}s of the size field. This precise block is considered in the following report as the \textit{malloc header} block. Note that the size of the chunk include the size of the \textit{malloc header}, chunk user data and \textit{footer} blocks.
        
        \item \textbf{User Data}: This is the actual memory space that is returned to the user.
        
        \item \textbf{Footer}: This is the same as the size of the chunk but is used for application data. Depending on how the chunk is represented, this is exactly the same as the \textbf{Size of Chunk} field. This is a more intuitive representation and is the one chosen in the schematic representation below.
        
        \item \textbf{Flags}:
        \begin{itemize}
            \item \texttt{A} (NON\_MAIN\_ARENA): Indicates if the chunk is in the main arena or a thread-specific arena.
            \item \texttt{M} (IS\_MMAPPED): Indicates if the chunk is allocated via \texttt{mmap}.
            \item \texttt{P} (PREV\_INUSE): Indicates if the previous chunk is in use. If false, it means the previous chunk is free.
        \end{itemize}
    \end{enumerate}

    The chunk allocation process involves the following concepts: 

    \begin{enumerate}
        \item \textbf{Initialization}: The very first chunk allocated always has the \texttt{P} bit set to prevent access to non-existent memory.
        
        \item \textbf{Free Chunks}: Free chunks are stored in circular doubly-linked lists. They contain forward and backward pointers to the next and previous chunks in the list.
        
        \item \textbf{Mmapped Chunks}: These chunks have the \texttt{M} bit set in their size fields and are allocated one-by-one.
        
        \item \textbf{Fastbins}: These are treated as allocated chunks and are consolidated only in bulk. These \textit{bins} are used to speed up the allocation process.
        
        \item \textbf{Top Chunk}: This is a special chunk that always exists. If it becomes less than \texttt{MINSIZE} bytes long, it is replenished.
    \end{enumerate}

    As explained directly in the code comments, an allocated chunk of 8 byte blocks can be described by the diagram below \cite{MallocGLIBC2001}. Note that is representation is personal to better suit the needs of our forensic analysis. The slight difference resides in the fact that the \texttt{footer} with the size of the considered chunk is represented as being part of the next chunk and not the current chunk. The footer of the previous chunk is actually the \texttt{mchunkptr} address. As stated in the GlicC wiki: \say{within the malloc library, a "chunk pointer" or \texttt{mchunkptr} does not point to the beginning of the chunk, but to the last word in the previous chunk - i.e. the first field in mchunkptr is not valid unless you know the previous chunk is free} \cite{MallocInternalsWiki2023}. Due to consideration of free/allocated chunks, it's actually easier to just consider the footer as being part of the next chunk, and not the current chunk. This is why the diagram below is slightly different from the one in the GLIBC wiki. This is just a difference in schematic representation, and does not change the actual data structure.

    \begin{figure}[H]
    \centering
    \begin{tikzpicture}[scale=0.99, every node/.style={scale=0.8}]
        
        \draw (4,1) rectangle (12,2);
        \draw (4,2) rectangle (12,3); 
        \draw (4,3) rectangle (12,4); 
        \draw (4,4) rectangle (12,5); 
        \draw (4,5) rectangle (12,6);
        \draw (4,6) rectangle (12,7);
        \draw (4,7) rectangle (12,8);
        \draw (4,8) rectangle (12,9); 
        \draw (4,9) rectangle (12,10); 
        \draw (4,10) rectangle (12,11);

        \draw[dashed] (3,3) rectangle (13,9);
        
        \draw[dotted] (0,8) -- (16,8);
        \draw[dotted] (0,4) -- (16,4);

        \draw[dashed] (4,0) -- (4,12);
        \draw[dashed] (12,0) -- (12,12);
        
        \node[anchor=west] at (0,9.5) {-last-footer-};
        \node[anchor=west] at (0,8.5) {-malloc-header-};
        \node[anchor=west] at (0,7.5) {-chunk-user-data-};
        \node[anchor=west] at (0,3.5) {-chunk-footer-};
        \node[anchor=west] at (0,2.5) {-next-malloc-header-};

        \node[anchor=west] at (14,8.5) {in-use chunk};
        
        \node[text width=14cm, align=center] at (8,9.5) {Size of previous chunk. If unallocated, next P=0};
        \node[text width=14cm, align=center] at (8,8.5) {Malloc header block: Size of chunk, in bytes |A|M|P};
        \node[text width=14cm, align=center] at (8,7.5) {User data starts here...};
        \node[text width=14cm, align=center] at (8,3.5) {(size of chunk, but used for application data)};
        \node[text width=14cm, align=center] at (8,2.5) {Size of next chunk, in bytes |A|0|1};
    \end{tikzpicture}
    \caption{Diagram of an allocated chunk in GLIBC 2.28 \cite{MallocGLIBC2001}.}
    \label{fig:allocated_chunk}
    \end{figure}
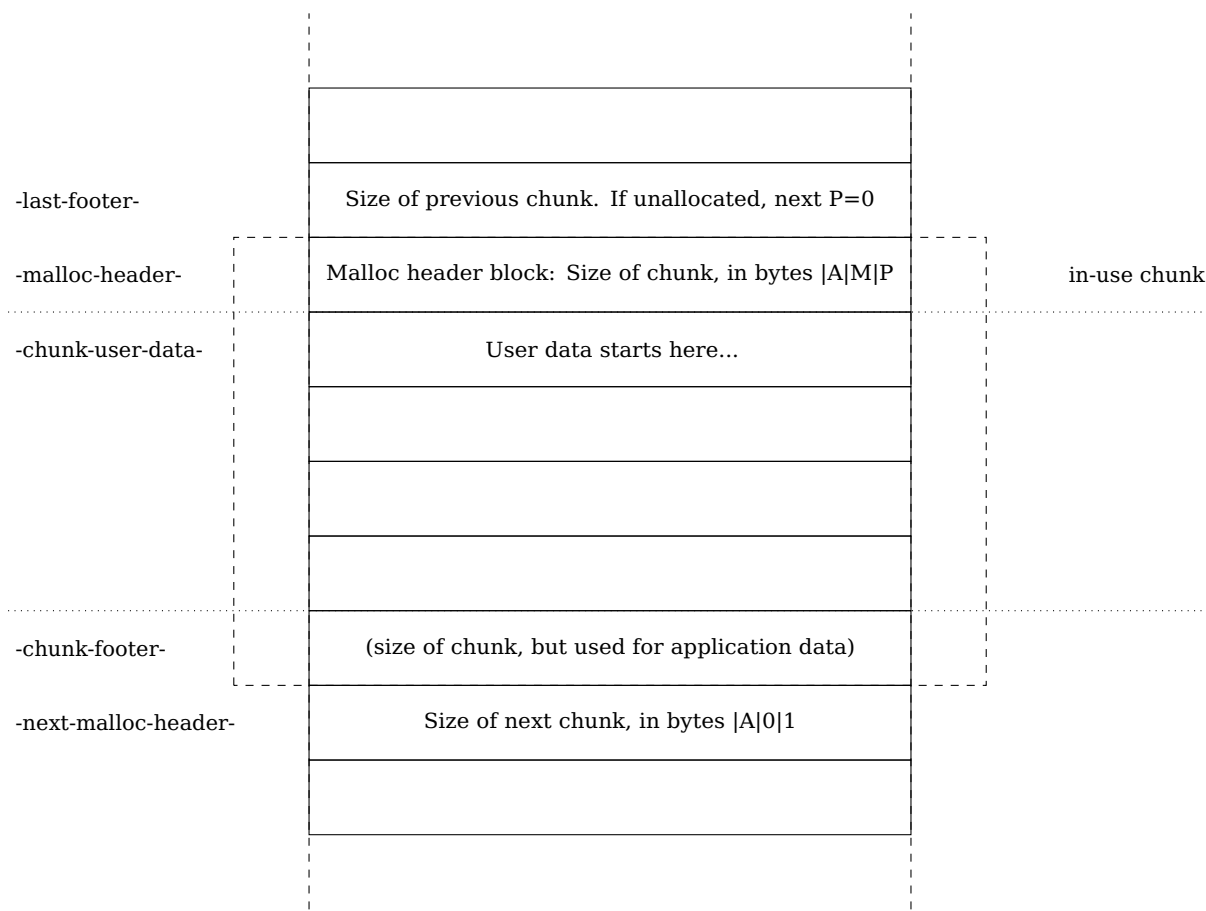

    The \texttt{malloc} function in GLIBC 2.28 uses a boundary tag method to manage chunks of memory. Each chunk contains metadata that helps in the allocation and deallocation of memory \cite{MallocGLIBC2001} \cite{MallocInternalsWiki2023}. The library stores available free chunks in circular doubly-linked lists called \say{bins}. This allows to quickly find a free chunk of memory of a given size. The problem is that we don't have access to those bins in the heap dump file. So to detect if a given chunk is in-use or free, we can rely on several methods. The first one is to look at the \texttt{P} bit of the malloc header. If it is set to 1, it means that the chunk is in use. If it is set to 0, it means that the chunk is free. 

    I also remarked that sometimes, the given heap dump file is cropped, and the last block is only composed of zeros and not complete. This is for instance the case with the last chunk of \textit{Training/basic/V\_7\_1\_P1/24/17016-1643962152-heap.raw}.

\begin{lstlisting}[language=bash, caption={Logs from chunk exploration script, related to the last chunk of the file \textit{Training/basic/V\_7\_1\_P1/24/17016-1643962152-heap.raw}. }]
WARN: Chunk [94022266975200] Chunk(block_index=10876, size=48176, flags=[A=False, M=False, P=True]) is out of bounds. Last block index: 16895 Iteration index: 16896 
WARN: Chunk [94022266975200] Chunk(block_index=10876, size=48176, flags=[A=False, M=False, P=True]) is out of bounds. Last block index: 16895 Iteration index: 16897
Chunk(block_index=10876, size=48176) is only composed of zeros.
\end{lstlisting}

    A free chunk contains the pointers of the next and previous free chunks in the heap, for its given bin. A representation of a free chunk, based directly on the code documentation \cite{MallocGLIBC2001}, is given below:

    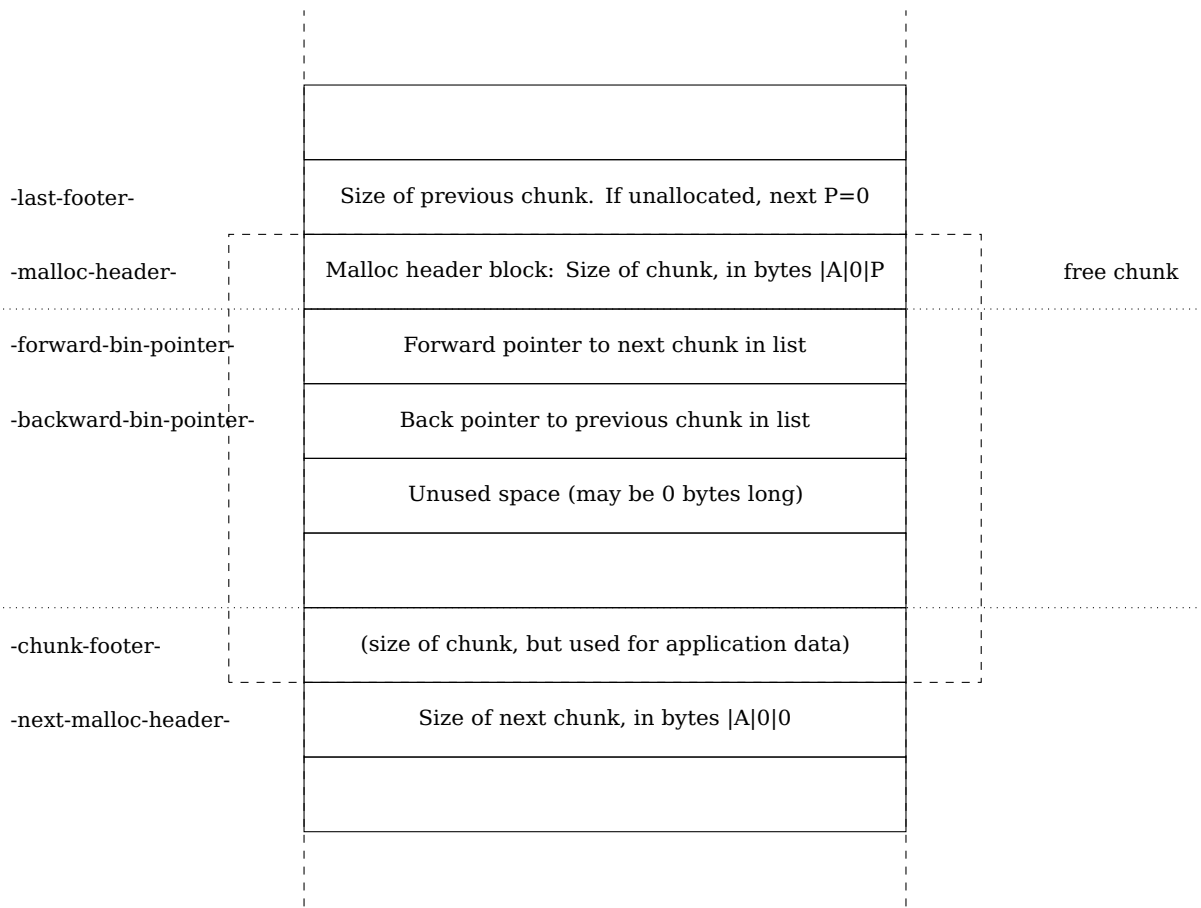
\begin{figure}[H]
        \centering
        \begin{tikzpicture}[scale=0.99, every node/.style={scale=0.8}]
            
            \draw (4,1) rectangle (12,2);
            \draw (4,2) rectangle (12,3); 
            \draw (4,3) rectangle (12,4); 
            \draw (4,4) rectangle (12,5); 
            \draw (4,5) rectangle (12,6);
            \draw (4,6) rectangle (12,7);
            \draw (4,7) rectangle (12,8);
            \draw (4,8) rectangle (12,9); 
            \draw (4,9) rectangle (12,10); 
            \draw (4,10) rectangle (12,11);
    
            \draw[dashed] (3,3) rectangle (13,9);
            
            \draw[dotted] (0,8) -- (16,8);
            \draw[dotted] (0,4) -- (16,4);
    
            \draw[dashed] (4,0) -- (4,12);
            \draw[dashed] (12,0) -- (12,12);
            
            \node[anchor=west] at (0,9.5) {-last-footer-};
            \node[anchor=west] at (0,8.5) {-malloc-header-};
            \node[anchor=west] at (0,3.5) {-chunk-footer-};
            \node[anchor=west] at (0,2.5) {-next-malloc-header-};

            \node[anchor=west] at (0,7.5) {-forward-bin-pointer-};
            \node[anchor=west] at (0,6.5) {-backward-bin-pointer-};
    
            \node[anchor=west] at (14,8.5) {free chunk};
            
            \node[text width=14cm, align=center] at (8,9.5) {Size of previous chunk. If unallocated, next P=0};
            \node[text width=14cm, align=center] at (8,8.5) {Malloc header block: Size of chunk, in bytes |A|0|P};
            \node[text width=14cm, align=center] at (8,7.5) {Forward pointer to next chunk in list};
            \node[text width=14cm, align=center] at (8,6.5) {Back pointer to previous chunk in list};
            \node[text width=14cm, align=center] at (8,5.5) {Unused space (may be 0 bytes long)};
            \node[text width=14cm, align=center] at (8,3.5) {(size of chunk, but used for application data)};
            \node[text width=14cm, align=center] at (8,2.5) {Size of next chunk, in bytes |A|0|0};
        \end{tikzpicture}
        \caption{Diagram of a free chunk in GLIBC 2.28 \cite{MallocGLIBC2001}.}
        \label{fig:free_chunk}
    \end{figure}

    \subsubsection{Chunk chaining}
    The chunk chaining algorithm relies on the \textbf{chunk chaining assumption} \ref{sec:methods:dataset:assumptions}. This assumption states that the allocator allocates chunks after chunks, and that the chunks are contiguous in memory. This means that we can expect to find the malloc header of the next chunk at the address $ \text{current\_malloc\_header\_chunk\_address} + \text{current\_chunk\_size} + 8 $, where 8 is the size of the malloc header block, or $ \text{current\_chunk\_user\_data\_address} + \text{current\_chunk\_size} $. It is the case for both free and allocated chunks. This is why we can use this assumption to detect chunks in the heap dump file. 
    
    This necessitates to understand malloc header blocks, and how they are represented in the heap dump file. In the specific case of \texttt{GLIBC 2.28}, the malloc header is defined as follows:

    \begin{minipage}{\dimexpr\linewidth-20pt}
        \begin{lstlisting}[language=c, caption={Malloc header definition in \texttt{GLIBC 2.28}}]
            #define SIZE_BITS (PREV_INUSE | IS_MMAPPED | NON_MAIN_ARENA)
        \end{lstlisting}
    \end{minipage}
    
    Since the malloc header respects the endianness of the system, we can expect to find the malloc header in little-endian format in the heap dump file. Using vim on \textit{Training/basic/V\_7\_8\_P1/16/5070-1643978841-heap.raw}, we can use the following command to find some potential malloc headers:

    \begin{lstlisting}[language=bash, caption={Vim command to find potential malloc headers}]
        :/[0-9a-f]\{4}0\{12}
    \end{lstlisting}
    
    This gives something like the following:

    \begin{figure}[H]
        \centering
        \includegraphics[width=16cm]{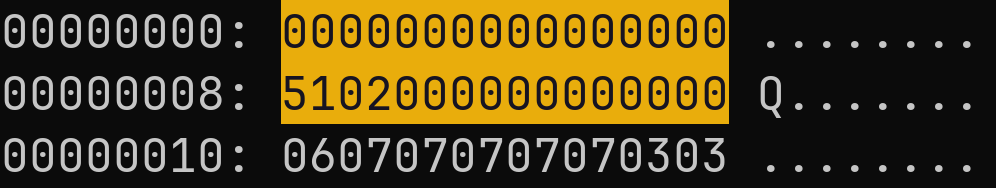}
        \caption{Attempt at malloc header detection in \textit{Training/basic/V\_7\_8\_P1/16/5070-1643978841-heap.raw}, at heap start.}
    \end{figure}

    Indeed, after a first zero block of 8 bytes (potential previous chunk footer), we expect a first data structure to be allocated at the start of the heap. Here this data structure is of size $ 5102000000000000_{16LE} $ (little-endian hex format) or $ 593_{10} $ bytes. The fact that it is an odd number is due to the \acrshort{lsb} being set to 1, to indicate that the preceding chunk is allocated (P flag). This means that the real size of the structure is actually $ 593_{10} - 1_{10} = 592_{10} $. This value is 8-byte aligned.

    Since we know that the allocator allocates chunks after chunks, we can expect the next chunk to be allocated at the address $ 5102000000000000_{16LE} + 592_{10} + 8_{10} = 5882193a34560000_{16LE} =  $. Note that we need to add 8 to the size to account for the malloc header block.
    
    In vim, since the address start at 0, we have to look at $ 592_{10} + 8_{10} = 258_{16} $. Let's have a look there:

    \begin{figure}[H]
        \centering
        \includegraphics[width=16cm]{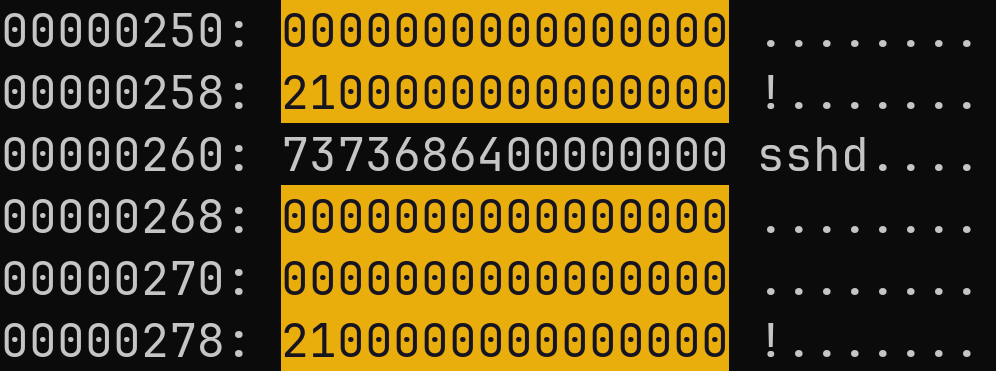}
        \caption{Attempt at malloc header detection in \textit{Training/basic/V\_7\_8\_P1/16/5070-1643978841-heap.raw}, at index $ 592_{10} = 250_{16} $.}
    \end{figure}

    There, we can see a zero block, followed by what we can expect to be another malloc header at address $ 258_{16} $. By doing the same process, we can thus propose an algorithm to detect the malloc headers, and thus the structures in the heap dump file.

    First, here is a simple algorithm to extract all the necessary information from a malloc header block:

    \begin{algorithm}[H]\label{alg:malloc_header_parsing}
        \caption{Malloc Header Parsing Algorithm}
        \begin{algorithmic}[1]
            \Procedure{MallocHeaderParsing}{$block$}
            \Require $block$ is a block of 8 bytes
            \Ensure $MallocHeader$ object
            \Ensure $Flags$ object
            \State \textbf{Note:} In this algorithm, $\&$ represents bitwise AND, and $\sim$ represents bitwise negation.
            \State $size\_and\_flags \leftarrow \text{ConvertBytesToInteger}(block, \text{'little-endian'})$
            \State $size \leftarrow size\_and\_flags \; \& \; (\sim 0x07)$ \Comment{Clear the last 3 bits to get the size}
            \State $Flags.a \leftarrow \text{bool}(size\_and\_flags \; \& \; 0x04)$
            \State $Flags.m \leftarrow \text{bool}(size\_and\_flags \; \& \; 0x02)$
            \State $Flags.p \leftarrow \text{bool}(size\_and\_flags \; \& \; 0x01)$
            \State \Return $MallocHeader\{size, Flags\}$
        \EndProcedure
        \end{algorithmic}
    \end{algorithm}

    We can also isolate the size parsing algorithm into a handy function:

    \begin{algorithm}[H]\label{alg:convert_to_size}
        \caption{Malloc Header block to size conversion Algorithm}
        \begin{algorithmic}[1]
            \Procedure{ConvertToSize}{$block$}
            \Require $block$ is a block of 8 bytes
            \State \textbf{Note:} In this algorithm, $\&$ represents bitwise AND, and $\sim$ represents bitwise negation.
            \State $size\_and\_flags \leftarrow \text{ConvertBytesToInteger}(block, \text{'little-endian'})$
            \State $size \leftarrow size\_and\_flags \; \& \; (\sim 0x07)$ \Comment{Clear the last 3 bits to get the size}
            \State \Return $size$
        \EndProcedure
        \end{algorithmic}
    \end{algorithm}

    Based on those algorithms, and in a similar fashion as what we have done manually by exploring the heap dump file with vim, we can propose the following algorithm to detect the malloc headers in a heap dump file:

    \begin{algorithm}[H]\label{alg:malloc_header_chaining}
        \caption{Malloc Header Chaining Algorithm}
        \begin{algorithmic}[1]
        \Procedure{MallocHeaderDetection}{$heapDumpFile$}
            \State \textbf{Note:} ConvertToSize is equivalent to MallocHeaderParsing($block$).size \Comment{See \ref{alg:convert_to_size}}
            \State \textbf{Initialize} malloc\_header\_list to empty list
            \State $position \gets 0$
            \While{$position < \text{FileSize}(heapDumpFile)$}
                \State $block \gets \text{Read8Bytes}(heapDumpFile, position)$
                \If{$\text{block} \neq 0$}
                    \State $size \gets \text{ConvertToSize}(block)$ \Comment{Be careful with flags}
                    \State \textbf{Assert} $size != 0$
                    \State \textbf{Assert} $size \mod 8 = 0$ \Comment{Check if the size is 8-bytes aligned}
                    \State $position \gets position + size$ \Comment{Leap over data structure.}
                \Else
                    \State $position \gets position + 8$
                \EndIf
            \EndWhile
            \State \Return $malloc\_header\_list$
        \EndProcedure
        \end{algorithmic}
    \end{algorithm}

    The idea behind the malloc header detection algorithm is simple. We start at the beginning of the heap dump file, and we look for the first non-zero block. Then we assume that the next block is a malloc header. We convert it to a size, and then leap over the user data and the footer up to the next chunk malloc header block index. The process is repeated until reaching the end of the heap dump file.

    Note that in case of a problem, like when the size obtained from malloc header parsing is equal to 0, this means that the heap dump chaining is broken. This has been handled in the dataset cleaning section \ref{sec:methods:dataset:cleaning}. 

    \subsubsection{Chunk chaining example}
    The program \texttt{chunk\_algorithms.py} has been developed specifically to test the chunk parsing and refine the associated algorithms.

    We can test our chunk parsing algorithm on a test file in the cleaned dataset.

    \begin{lstlisting}[language=bash, caption={Testing chunk parsing on \textit{Training/basic/V\_7\_1\_P1/24/17016-1643962152-heap.raw}. Partial log output. }]
$ python src/data_structure_detection.py --input /home/onyr/code/phdtrack/phdtrack_data_clean/Training/Training/basic/V_7_1_P1/24/17016-1643962152-heap.raw --debug
    Datetime: 2023_09_27_17_08_23_157209
    Chunk [1]: Chunk(block_index=2, size=592, flags=[A=False, M=False, P=True])
    Chunk [2]: Chunk(block_index=76, size=32, flags=[A=False, M=False, P=True])
    Chunk [3]: Chunk(block_index=80, size=32, flags=[A=False, M=False, P=True])
    Chunk [4]: Chunk(block_index=84, size=32, flags=[A=False, M=False, P=True])
    Chunk [5]: Chunk(block_index=88, size=32, flags=[A=False, M=False, P=True])
    Chunk [6]: Chunk(block_index=92, size=192, flags=[A=False, M=False, P=True])
    Chunk [7]: Chunk(block_index=116, size=32, flags=[A=False, M=False, P=True])
    Chunk [8]: Chunk(block_index=120, size=32, flags=[A=False, M=False, P=True])
    Chunk [...]: ...
    Chunk [911]: Chunk(block_index=10194, size=128, flags=[A=False, M=False, P=True])
    Chunk [912]: Chunk(block_index=10210, size=256, flags=[A=False, M=False, P=True])
    Chunk [913]: Chunk(block_index=10242, size=160, flags=[A=False, M=False, P=True])
    Chunk [914]: Chunk(block_index=10262, size=512, flags=[A=False, M=False, P=True])
    Chunk [915]: Chunk(block_index=10326, size=1296, flags=[A=False, M=False, P=True])
    Chunk [916]: Chunk(block_index=10488, size=1552, flags=[A=False, M=False, P=True])
    Chunk [917]: Chunk(block_index=10682, size=1552, flags=[A=False, M=False, P=True])
    Chunk [918]: Chunk(block_index=10876, size=48176, flags=[A=False, M=False, P=True])
    -----------> Statistics:
    Total number of files: 1
    Total number of chunks: 918
    Total number of blocks: 16896
    Total number of chunks with P=1: 903
    Total number of chunks with M=1: 0
    Total number of chunks with A=1: 0
    Total number of chunks only composed of zeros: 1
    \end{lstlisting}

    Looking at the first allocated chunks, we recognize what we had seen manually with vim for the file \textit{Training/basic/V\_7\_8\_P1/16/5070-1643978841-heap.raw}. The first chunk is of size 592, and the next one is of size 32. This is exactly what we had seen manually. It is a good sign that our algorithm is working as expected. We can also see that the last chunk is of size 48176, which is significantly bigger than the other chunks. This chunk is only composed of zeros, and is truncated, meaning that its size if bigger than the actual size of the heap dump file. 
    
    \subsubsection{Distinguishing between free and allocated chunks}
    The malloc header chaining algorithm allows to detect memory chunks in the heap dump file. However, it does not allow to distinguish between free and allocated chunks. This is a problem, since we want to be able to distinguish between free and allocated chunks, to be able to detect potential data structures and filter out useless blocks.
    
    Considering the structural differences between a free and in-use block, it's possible to try distinguishing free blocks by their \textit{forward} and \textit{backward} pointers. The issue is that the head dump raw file are not provided with any \textit{bins} information. As such, distinguishing between two normal pointers and the ones expected inside a free block is a non-trivial task. Hence, the tests performed on this idea are inconclusive. A more straightforward technique is to rely on the \texttt{P} malloc header flags. 
    
    \begin{figure}[H]
        \centering
        \begin{tikzpicture}[scale=0.99, every node/.style={scale=0.8}]
            \draw[dashed] (4,0) -- (4,15);
            \draw[dashed] (12,0) -- (12,15);

            \node[anchor=east] at (8.5,14.5) {...in-use...};

            \draw (4,13) rectangle (12,14);
            \node[anchor=west] at (0,13.5) {Chunk 0: In-use};
            \node[anchor=west] at (6.5,13.5) {header has P=1};

            \draw (4,12) rectangle (12,13);
            \node[anchor=west] at (0,12.5) {Chunk 1: In-use};
            \node[anchor=west] at (6.5,12.5) {header has P=1};
            
            \draw (4,11) rectangle (12,12);
            \node[anchor=west] at (0,11.5) {Chunk 2: Free};
            \node[anchor=west] at (6.5,11.5) {header has P=1};

            \draw (4,10) rectangle (12,11);
            \node[anchor=west] at (0,10.5) {Chunk 3: In-use};
            \node[anchor=west] at (6.5,10.5) {header has P=0};
    
            \node[anchor=east] at (8.5,9.5) {...in-use...};

            \draw (4,8) rectangle (12,9);
            \node[anchor=west] at (0,8.5) {Chunk 100: In-use};
            \node[anchor=west] at (6.5,8.5) {header has P=1};
            
            \draw (4,7) rectangle (12,8);
            \node[anchor=west] at (0,7.5) {Chunk 101: Free};
            \node[anchor=west] at (6.5,7.5) {header has P=1};
            
            \draw (4,6) rectangle (12,7);
            \node[anchor=west] at (0,6.5) {Chunk 103: In-use};
            \node[anchor=west] at (6.5,6.5) {header has P=0};
            
            \node[anchor=east] at (8.5,5.5) {...in-use...};
            
            \draw (4,4) rectangle (12,5);
            \node[anchor=west] at (0,4.5) {Chunk 1000: In-use};
            \node[anchor=west] at (6.5,4.5) {header has P=1};
            
            \draw (4,3) rectangle (12,4);
            \node[anchor=west] at (0,3.5) {Chunk 1001: Free};
            \node[anchor=west] at (6.5,3.5) {header has P=1};
            
            \draw (4,2) rectangle (12,3);
            \node[anchor=west] at (0,2.5) {Chunk 1002: In-use};
            \node[anchor=west] at (6.5,2.5) {header has P=0};

            \draw (4,1) rectangle (12,2);
            \node[anchor=west] at (0,1.5) {Chunk 1002: In-use};
            \node[anchor=west] at (6.5,1.5) {header has P=1};
            
            \node[anchor=east] at (8.5,0.5) {...in-use...};
    
        \end{tikzpicture}
        \caption{Heap dump showing a mix of free and in-use chunks. Note: each chunk immediately after a free chunk has a P flag set to 0. Each rectangle represents a chunk.}
        \label{fig:heap_dump}
    \end{figure}
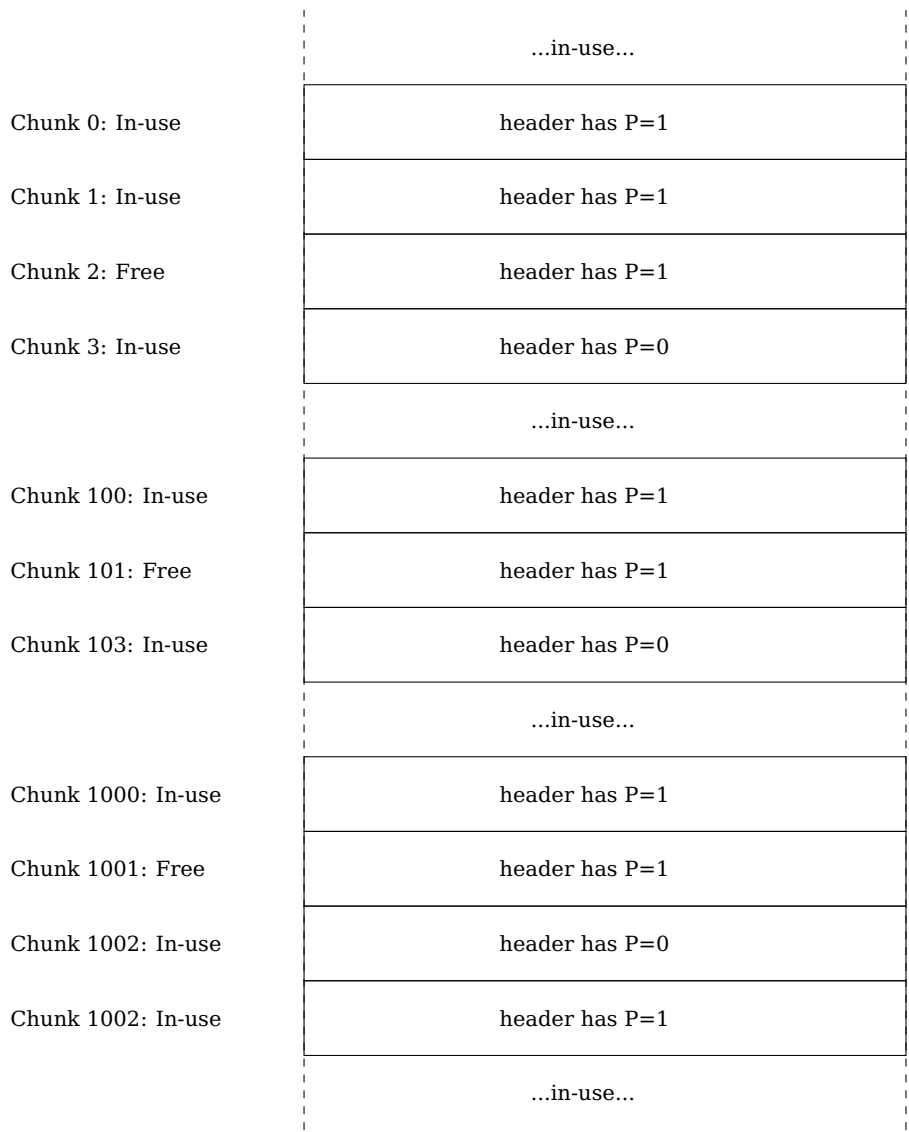
    
    For a given chunk, the follow-up chunk in ascending address number, contains such a flag in its header block. If the flag value is 0, then the current chunk is free. If the flag value is 1, then the current chunk is in use by the program. This is the technique that has been used in the final implementation of the chunk chaining algorithm. 

    \begin{algorithm}[H]
        \caption{Chunk Parsing Algorithm}
        \begin{algorithmic}[1]
        \Procedure{ChunkParsing}{$heapDumpFile, HEAP\_START$}
            \State \textbf{Note:} ConvertToSize is equivalent to MallocHeaderParsing($block$).size
            \State \textbf{Note:} Get8BytesBlocks returns a list of 8 bytes blocks from the heap dump file.
            \Ensure $MallocHeader$ object
            \Ensure $Flags$ object
            \Ensure $Chunk$ object
            \Ensure $HEAP\_START$ provided from annotation file is a correct address.
            \State \textbf{Note:} In this algorithm, $\&$ represents bitwise AND, and $\sim$ represents bitwise negation.
            \State \textbf{Initialize} $chunk\_list$ to empty list
            \State $blocks \gets \text{Get8BytesBlocks}(heapDumpFile)$
            \State \textbf{Initialize} $index \gets 0$
            \While{$index < lenght(blocks)$}
                \State $block \gets blocks[index]$
                \State \textbf{Initialize} $Chunk$ to empty object
                \If{$\text{block} \neq 0$}
                    \State $Chunk.header : \{size, Flags\} \gets \text{MallocHeaderParsing}(block)$ \Comment{See \ref{alg:malloc_header_parsing}}
                    \State \textbf{Assert} $Chunk.header.size \geq 2 $ \Comment{Must contains at least header and footer}
                    \State \textbf{Assert} $Chunk.header.size \mod 8 = 0$ \Comment{Check if the size is 8-bytes aligned}
                    \State $Chunk.block\_index \gets index$ \Comment{Index of the first block of the chunk after header}
                    \State $Chunk.address \gets HEAP\_START + (index * 8)$ \Comment{Address of $block\_index$}
                    \State $footer\_index \gets index + Chunk.header.size - 1$ \Comment{Index of the footer block}
                    \If{$footer\_index < lenght(blocks)$}
                        \State $footer \gets blocks[footer\_index]$
                        \If{$\text{ConvertToSize}(footer) = Chunk.footer.size$}
                            \State $Chunk.correct_footer \gets True$
                        \Else
                            \State $Chunk.correct_footer \gets False$
                        \EndIf
                    \Else
                        \State $Chunk.correct_footer \gets False$
                    \EndIf
                    \State $next\_chunk\_header\_index \gets index + Chunk.header.size$ \Comment{Index of the next chunk header block}
                    \If{$next\_chunk\_header\_index < lenght(blocks)$}
                        \State $next\_chunk\_header \gets blocks[next\_chunk\_header\_index]$
                        \State $Chunk.is\_in\_use \gets \text{MallocHeaderParsing}(next\_chunk\_header).flags.p$
                    \Else
                        \State $Chunk.is\_in\_use \gets False$ \Comment{See \footnote{Experiments show that last chunk can be cropped and in that case, is only composed of zeros. We can thus consider it as free.}}
                    \EndIf
                    \State $index \gets index + Chunk.header.size$ \Comment{Leap over chunk.}
                \Else
                    \State $index \gets index + 8$ \Comment{Leap over zero block.}
                \EndIf
            \EndWhile
            \State \Return $malloc\_header\_list$
        \EndProcedure
        \end{algorithmic}
    \end{algorithm}

    Note that this algorithm is based on the malloc header chaining algorithm \ref{alg:malloc_header_chaining}. The main difference is that we now have access to the malloc header flags from the following chunk, and that we can thus distinguish between free and allocated chunks. The algorithm also includes the footer parsing technique discussed briefly in the following section.

    \subsubsection{Chunk footer}
    The documentation of the \texttt{malloc} function of GLIBC  states that the footer of a chunk is the same as the size of the chunk considered. In the current report, we represent the footer as being part of the chunk itself.

    Below are two chunks content of similar size:

    \begin{lstlisting}[language=bash, caption={Printing some free and in-use chunks from \textit{Training/basic/V\_7\_1\_P1/24/17016-1643962152-heap.raw}.}]
Printing Chunk [addr:0x80a2d1438355] [status:in-use] [footer:incorrect] Chunk(block_index=80, size=32, flags=[A=False, M=False, P=True])
        Block [79]: 	b'!\x00\x00\x00\x00\x00\x00\x00' 		33 		-malloc-header-
        Block [80]: 	b'\xa0\xa2\xd1C\x83U\x00\x00' 		94022266888864 		
        Block [81]: 	b'\xc0\xa2\xd1C\x83U\x00\x00' 		94022266888896 		
        Block [82]: 	b'\x00\x00\x00\x00\x00\x00\x00\x00' 		0 		-footer-
Printing Chunk [addr:0xa09fd2438355] [status:free] [footer:correct] Chunk(block_index=8180, size=32, flags=[A=False, M=False, P=True])
        Block [8179]: 	b'!\x00\x00\x00\x00\x00\x00\x00' 		33 		-malloc-header-
        Block [8180]: 	b'\xb0\xbc\xe1\xeeS\x7f\x00\x00' 		139998466784432 		
        Block [8181]: 	b'\xb0\xbc\xe1\xeeS\x7f\x00\x00' 		139998466784432 		
        Block [8182]: 	b' \x00\x00\x00\x00\x00\x00\x00' 		32 		-footer-
    \end{lstlisting}

    Here, the status of the chunk has been detected using the \texttt{P} flag technique. At first sight, those two blocks seems similar. The first 2 blocks in the user data space of the chunks both seems to contain what looks like pointers. As one can see, the first chunk in this example, with a \texttt{block\_index=80} has clearly a malloc header and footer as expected. Note that here, the value $33$ represents the size of the block ($32$ bytes which correspond to 4 blocks) with the \acrshort{lsb} being set to 1 meaning the preceding chunk is in use. However, the in-use block footer doesn't correspond to the value we expect. This difference of behavior is observed throughout the cleaned dataset. 

    \subsection{Discussing chunk parsing for problem scale reduction}
    Now that we have presented all the necessary knowledge and algorithms used to be able to parse the RAW heap dump files, we can discuss the results of those algorithms and their uses and limitations. Many tests have been needed in order to develop the final algorithms. This testing process has also unveiled some interesting properties of the dataset that will be used as basis for the semantic embedding of the memory graph representation and subsequent machine learning steps.

    The program \texttt{chunk\_algorithms.py} has been developed specifically to test the chunk parsing and refine the associated algorithms on the cleaned dataset. Below are presented the global statistics produced by the final version of the program:

    \begin{lstlisting}[language=bash, caption={Printing cleaned dataset chunk parsing global statistics.}]
        Input is directory: /home/onyr/code/phdtrack/phdtrack_data_clean/
        Found 26191 files in /home/onyr/code/phdtrack/phdtrack_data_clean/.
        Processing files: 100%|\blacksquare \blacksquare \blacksquare \blacksquare \blacksquare | 26191/26191 [12:11<00:00, 35.81it/s, file=7091-1650972335]
        ------> Statistics:
        Total number of parsed files: 26191
        Total number of skipped files: 0
        Total number of chunks: 37682063
        Total number of blocks: 674232832
        Total number of chunks with P=1: 37346373
        Total number of chunks with M=1: 0
        Total number of chunks with A=1: 0
        Total number of free chunks: 354410
        Total number of chunks only composed of zeros: 18720
        Total number of blocks in free chunks: 183331224
        Total number of chunks with correct footer value: 1009522
        Total number of chunks both free and with correct footer value: 335690
        Total number of chunks free and annotated: 0
        Total number of potential footers with annotations (should be 0): 0
        Total number of annotated chunks: 209528
        Total number of chunks in use, with correct footer, and annotated: 7668
        Total number of chunks in use, with correct footer, and key annotated: 7668
        Percentage of free chunks: 0.9405270619074121%
        Percentage of blocks in free chunks: 27.19108523033183%
        Percentage of free chunks with correct footer value: 94.71798199825061%
        Percentage of in-use chunks with correct footer value: 1.8051818044922352%
        Average number of annoted chunks per file: 8.0
        Average number of chunks in use with correct footer and annotated per file: 0.2927723263716544
        Set of sizes of key chunks: {32, 48, 64}
        Sizes of key chunks with their number of occurences:
        Size: 32  Number of occurences: 34366
        Size: 48  Number of occurences: 109346
        Size: 64  Number of occurences: 13434
        Number of sizes: 157146
        Number of unique sizes: {32, 48, 64}
    \end{lstlisting}

    The cleaned dataset contains 26191 RAW files and their corresponding annotation files. The program has been able to parse all those files, and has been able to detect 37682063 chunks, which represents 674232832 blocks. This is a huge number of blocks. The goal being to be able to predict which of those blocks are first key blocks, we need to be able to filter out the useless blocks as much as possible to both optimize computations and scale down the problem. 

    Using the \texttt{P} flag technique, we can see that 37346373 chunks are in use, and 354410 chunks are free. Although the proportion of free chunks is only 0.94\%, there are 27.19\% of the blocks that are in free chunks. More importantly, we can see that no free chunk is annotated. This means we can filter out all free chunks and their blocks. This allows a huge reduction of the scale of the problem. 
    
    The average number of annotated chunks per file being a perfect value of 8, this means that all the parsed files indeed contains the 6 key annotations with the additional \texttt{SSH\_STRUCT} and \texttt{SSH\_KEY} annotations. The dataset is very imbalanced since we have only 6 keys times the number of RAW files as positive labels and the rest as negative, thus the need for advanced reduction techniques. 

    \subsubsection{From a block-based to a chunk-based approach}
    
    The exact code to annotate the chunks can be as simple as the following:

    \begin{algorithm}[H]
        \caption{Annotate Chunk Algorithm}
        \begin{algorithmic}[1]
        \Procedure{AnnotateChunk}{$chunk, keys\_addresses, ssh\_struct\_addr, session\_state\_addr$}
            \Ensure $chunk$ object
            \Ensure $keys\_addresses$ list of integers
            \Ensure $ssh\_struct\_addr$ integer
            \Ensure $session\_state\_addr$ integer
            \State \textbf{Note:} Annotations should be done after free chunk detection.
            
            \Procedure{AssertChunkUsedThenAnnotate}{$chunk, annotation$}
                \State \textbf{Assert} $chunk.is\_in\_use$ \Comment{Make sure we don't annotate free chunks}
                \State $chunk.annotations.append(annotation)$
            \EndProcedure
            
            \If{$chunk.address \in keys\_addresses$}
                \State \textsc{AssertChunkUsedThenAnnotate}($chunk, ChunkAnnotation.ChunkContainsKey$)
            \ElsIf{$chunk.address = ssh\_struct\_addr$}
                \State \textsc{AssertChunkUsedThenAnnotate}($chunk, ChunkAnnotation.ChunkContainsSSHStruct$)
            \ElsIf{$chunk.address = session\_state\_addr$}
                \State \textsc{AssertChunkUsedThenAnnotate}($chunk, ChunkAnnotation.ChunkContainsSessionState$)
            \EndIf
        \EndProcedure
        \end{algorithmic}
    \end{algorithm}
    
    This algorithm in itself and the results observed is an important discovery. The annotations are actually always given for the \texttt{chunk.address} which corresponds to the address of the first block after the malloc header block. This means that the annotations are actually given for the beginning of the user data space of a chunk. This is crucial discovery, since it means that we can filter out the malloc header and footer blocks, and only keep the first block of the user data space of the chunks we want to embed. There are $674232832 - 183331224 = 490901608$ blocks in use. But there is only $37346373$ chunks in use. This means that we can reduce the number of blocks to embed from $490901608$ to $37346373$ which is an additional reduction applied after the previous filtering that reduces the scale of the problem by a factor of 13. 

    \subsubsection{Using chunk footer for filtering is not possible}
    
    Now let's look at the footer parsing. We can see in the logs that 94.72\% of free chunks are said to have a correct footer value. But this value is misleading. Since the last chunk of a heap dump is often cropped, it means it has no footer. But we consider those special last chunks as free chunks. In fact, in the $354410$ free chunks, we have $18720$ or around 5.28\% of them that are those special last cropped chunks only composed of zeros. With this perspective, we understand that 100\% of the free chunks should be considered with correct footer value. In contrast, only 1.81\% of in-use chunks have a correct footer value. It's tempting to think that maybe, those few chunks could maybe be actually empty too and removed. But this is not the case since a few chunks are actually both in-use, with a key annotation and a correct footer value. This means that we need to keep those chunks.

    \subsubsection{Chunk filtering}
    Based on the previous observations, we can propose different ways of filtering some chunks out. The objective is to reduce the number of chunks before any further processing to reduce the imbalanceness.

    Since we have seen that free chunks are never annotated, we can filter them out. This filtering technique allows to reduce the number of chunks from $37682063$ to $37346373$, which is a small reduction of 0.89\%. It is not a huge reduction of chunks, but is a much more significant reduction of the number of blocks since 27.2\% of the blocks are in free chunks.

    We can also filter the chunks whose size is not 32, 48 or 64 bytes. This is based on the observation that the key chunks are of those sizes. Such a filtering technique allows to reduce the number of chunks significantly, since there is a cumulated $109346+34366+13434 = 157146$ chunks of those size, which represents, compare to the original number of chunks, a diminution of 99.6\%. It is indeed a huge reduction of the scale of the problem.

    A last approach to chunk filtering for key prediction consists in measuring the entropy of the first bytes of a chunk. Since we have previously discovered that keys are always located at the beginning of a chunk, and since the keys are composed of random bytes, we can expect the entropy of the first bytes of a chunk to be high.

    The following graph illustrates this phenomenon:

    \begin{figure}[H]\label{methods:entropy_of_all_chunks}
        \centering
        \includegraphics[width=16cm]{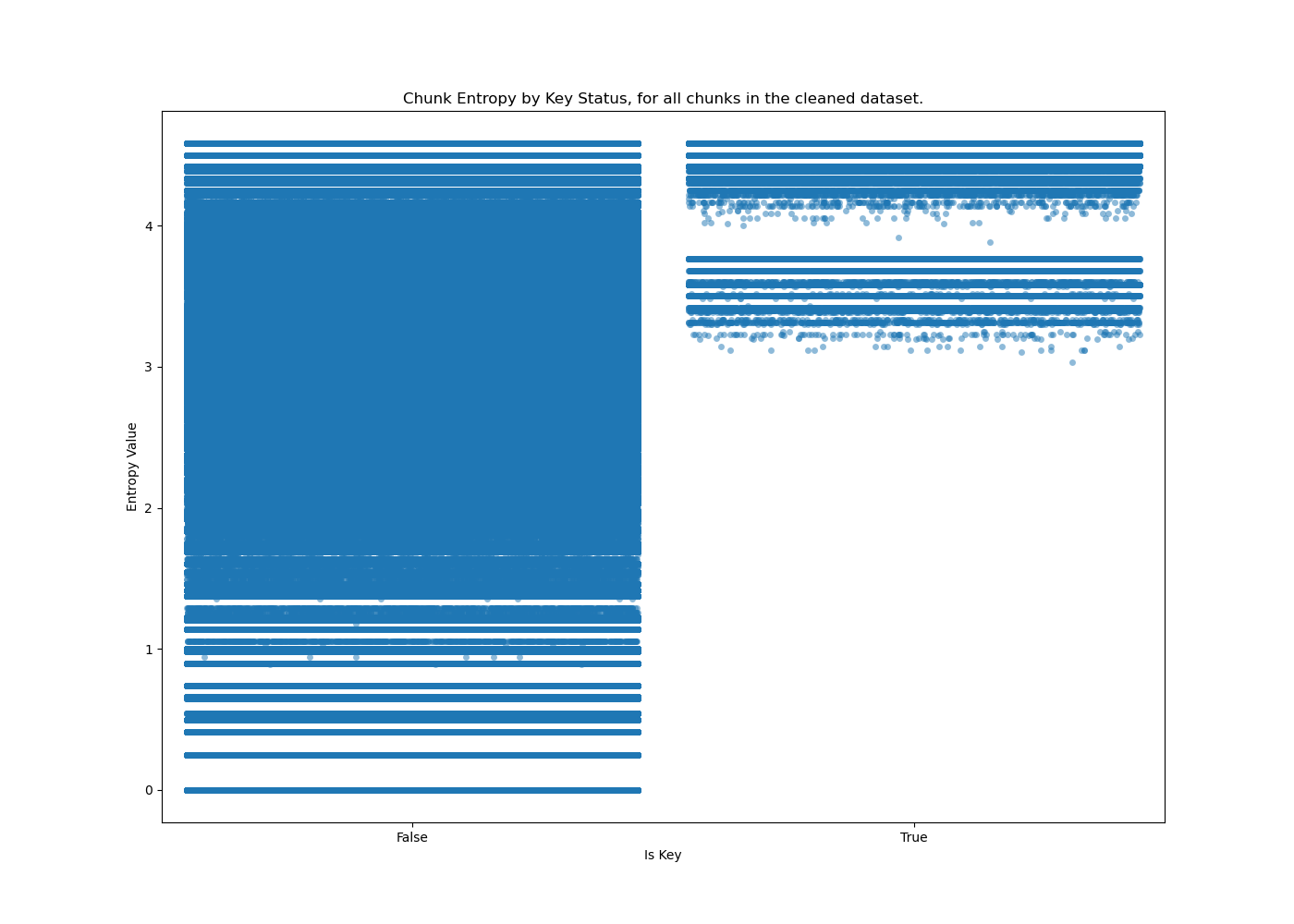}
        \caption{Visualization of the entropy distribution for all chunks of the \textit{phdtrack\_data\_clean/} RAW heap dump dataset.}
    \end{figure}

    Using a script to perform some counting, we realize that the number of chunks whose entropy is less than the minimum entropy of the chunks that contains a key (is key chunk) is 19690826, which represents 52.3\% of the total number of chunks. It's not entirely clear why the key chunk entropy values seem spread across 2 strips of values, but is probably an effect of the different distributions of chunks across the input dataset, depending on the version, use can and number of chunks in the considered input files.

    After all those extensive analysis and tests, we have gained invaluable knowledge about how we can reduce the scale of the problem and parse the files. Now, we need a way to create meaningful embeddings for the blocks we want to perform machine learning on. This is the goal of the next section.

\section{Graph-based memory dumps embedding}\label{chap:mem_2_graph}
Now that we have a decent understanding of the dataset as well as the low level memory dump format, we can start to think about how to convert the memory dumps into graphs. As a recall, we want to be able to convert a memory dump into a graph representation that can be used for machine learning, since we want to be able to create a memory modelization as a basis for efficient embedding and feature engineering later. This is inherently due to the imbalanceness of the dataset, as we want to add more information to each memory block that just its raw bytes. The goal is to have a graph representation of the memory dump that can be used for efficient machine learning.

\subsection{Initial work from Python to Rust}

Initially, we have been working and manipulating the code provided by SmartKex\footnote{SmartKex GitHub repository: \url{https://github.com/smartvmi/Smart-and-Naive-SSH-Key-Extraction}} for key detection. Our first explorations of the dataset quickly gave birth to some Jupyter Notebooks, which were used to explore the dataset and to understand the code, like \texttt{search\_in\_heap\_mem.ipynb}. Rapidly, we decided to rebuild a complete Python 3.11 version of the code. This was done for several reasons:

\begin{itemize}
    \item The provided code had no type hinting, which makes it hard to read and understand.
    \item We wanted to explore the dataset and learn by doing.
    \item The original code was not designed to be used as a library, but rather as a standalone script.
    \item The original code was just a few hundred lines of code and was not designed to be easily extensible, nor to be able to handle a large number of memory dumps.
    \item We wanted to modernize code by using the latest stable version of Python.
\end{itemize}

We decided to build a memory graph representation at that moment because we wanted to be able to add more information to the memory blocks than just their raw bytes. This new program was called \texttt{ssh\_key\_discover}, and relied on a number of Python libraries to work, like \texttt{graphviz}. This was a all-in-one library, composed of 2 sections, \texttt{mem\_graph} and \texttt{ml\_discovery}. The first one was devoted to build memory graphs, while the second one was dedicated to the data science and machine learning part.

This initial program was already capable of handling several data processing pipelines, including machine learning pipelines with models like Random Forest, a grid search for hyperparameter optimization, a cross validation pipeline, several balancing strategies and of course, a memory graph representation with a semantic embedding. As an early development version, this program was not optimized for performance, and just loading a given heap dump file and its annotation, then building the memory graph representation could take from 30 seconds to a minute (on the TUXEDO machine), depending on the size of the heap dump file. As the original dataset comprises more than $ 10^{6} $ files, a rapid estimation of the time needed just for the semantic embedding of the memory graph representation was above a month. In this regard, this initial program was just used on a bunch of files as a way to develop the semantic embedding model, parsing algorithms and start working on feature engineering and machine learning. But it could not be used to produce final results on the whole dataset due to the performance issues described above.

Such an optimization issue was clearly not acceptable, and we decided to rewrite the graph part in Rust. This is a compiled language that leverages zero-cost abstractions, and thus, is several order of magnitude faster than Python. It was also a good opportunity to learn Rust, which is a language that is gaining more and more popularity, especially in the security community. This new program was called \texttt{mem2graph}. Switching from Rust to Python and doing a proper use of multithreading allowed us to reduce the time needed to build the memory graph representation from 30 seconds to less than 1 second. In out case, and comparing using only the TUXEDO laptop, this represents an estimated minimum of a 130x speedup. But this is even much better on the server, where the multithreading can really be leverage. This was a significant improvement which allowed us to build the memory graph representation for the whole original dataset in just a few hours.

\subsection{Memory Graph Representation}
Now, let's describe the memory graph representation. The goal is to be able to represent a memory dump as a graph. This modelization makes sense since the heap dump can be considered as having memory chunks as nodes, being connected by pointers acting as arrows. This is a very natural way to represent a memory dump. However, in our cases, and since the goal is to make predictions on raw bytes, we will not use the chunks as nodes, but rather the memory blocks directly. This is because we want to be able to make predictions on raw blocks of bytes, and not on chunks. 

Our memory graph representation is composed of a directed graph, where each node is a memory block of bytes, and each edge is either indicative of a pointer link or a chunk membership relationship. This second representation is directly inspired by collection representation in \acrlong{kg} ontologies. In the case of \acrshort{rdf}, this could be equivalent to a \textbf{rdf:Bag}, which is an unordered container \cite{OrderedDataInRDF20} (see \ref{sec:background:ontology}). The graph is directed because the pointers are directed. We will also consider the relationship of belonging to a chunk as oriented from the data structure header block to the data structure member blocks.

Our memory graph representation is inherently a property graph. Each node and edge can have properties. The properties of an edge are the type of the edge, which can be either a pointer or a structure membership relationship.

\begin{itemize}
    \item \textbf{dts:} Data Structure Membership Relationship
    \item \textbf{ptr:} Pointer Relationship (direction is from the source to the target)
\end{itemize}

In our case, the properties of a node are at minimum the address and the byte block. The graph is also heterogeneous since our nodes can have different types corresponding to their inferred characteristics. 

\begin{itemize}
    \item \textbf{PN:} Pointer Node. This is a node whose bytes have been identified as a pointer.
    \item \textbf{CHN:} Chunk Header Node. This is a node whose bytes have been identified as a memory (malloc) chunk header. In the graph, this node is the root node of a memory structure managed by the C library responsible for memory management and allocation.
    \item \textbf{KN:} Key Node. This is a node whose bytes have been identified as a key. This identification relies both on the annotations and some verification checks.
    \item \textbf{VN:} Value Node. These are all blocks that have not been identified. It is the default node type.
\end{itemize}

These nodes and edges form the base of the memory graph representation. Below is a simplified (truncated) example of a memory graph representation. The full example is available in \ref{appendix:mem_graph:17016-1643962152:full}. For clarity, the addresses are not displayed in this simplified version. Another version of this graph with real addresses is available in \ref{appendix:mem_graph:17016-1643962152:truncated}.

\begin{figure}[H]\label{methods:mem_graph:17016-1643962152:simplified}
    \centering
    \includegraphics[width=16cm]{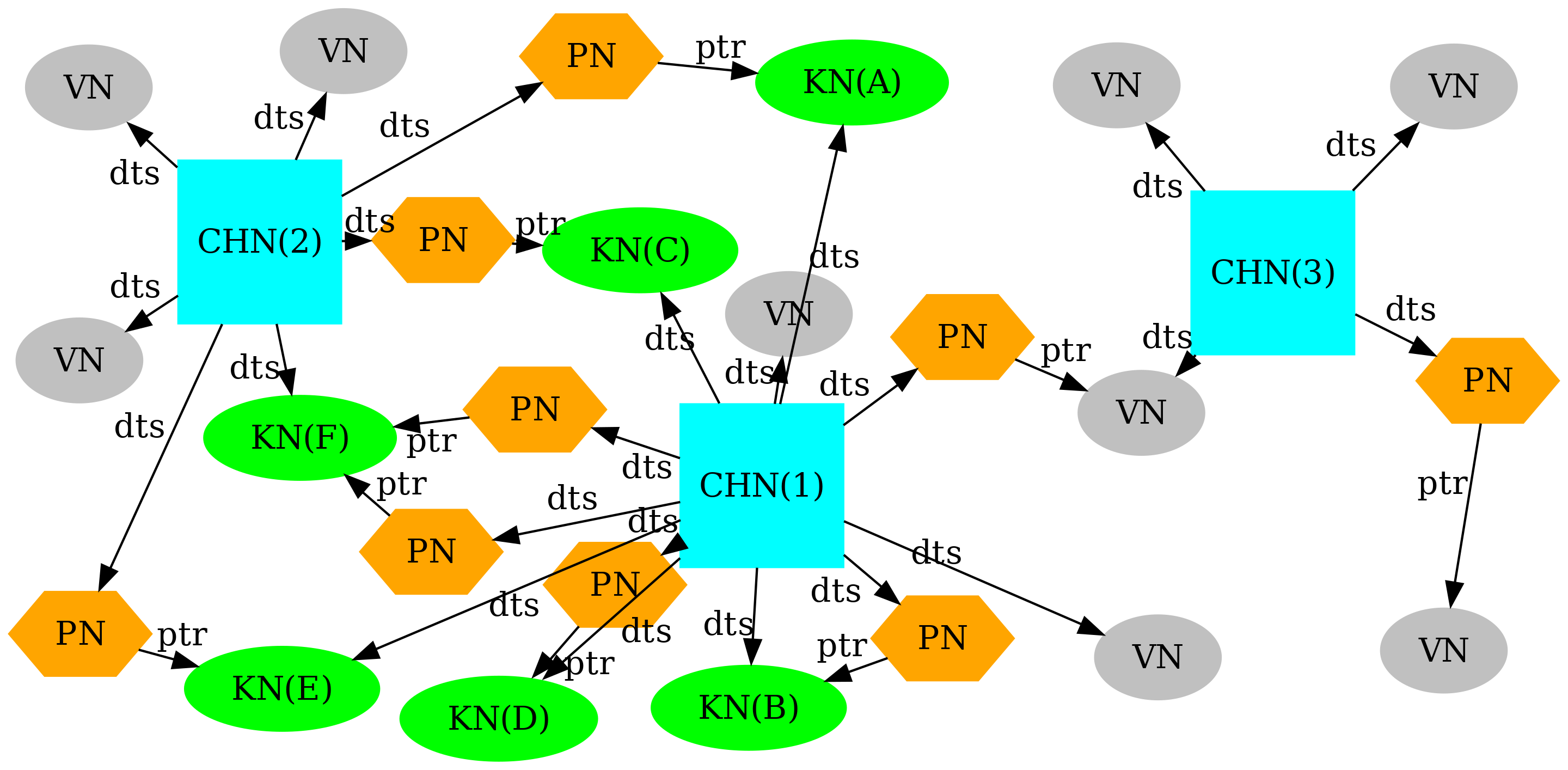}
    \caption{Visualization of a truncated memory graph generated from \textit{Training/basic/V\_7\_1\_P1/24/17016-1643962152-heap.raw}. The addresses are not displayed for improved readability. Version with addresses here \ref{appendix:mem_graph:17016-1643962152:truncated}.}
\end{figure}

The given graph represents a memory layout with various types of nodes, each serving a specific purpose. The graph contains \gls{chn} nodes, which act as the root nodes for allocated structures and are colored in cyan. These DTN nodes are connected to \gls{kn} nodes, which are identified as keys and are colored in green. The \gls{pn} nodes, colored in orange, are pointers and can be connected to value nodes or key nodes. Finally, the graph includes \gls{vn} nodes, which are the default node types and are colored in gray. These nodes have not been identified as any specific type and may contain arbitrary values.

The idea behind this representation will be to try to make predictions on the \gls{kn} nodes, which are the nodes that have been identified as keys. Using the graph, we can build an embedding of the nodes and as such, add more information to a given byte block than just its raw bytes. This is the basis of the semantic embedding, which will be discussed later.

This example is based on the heap dump file \textit{Training/basic/V\_7\_1\_P1/24/17016-1643962152-heap.raw} and has been generated using \texttt{mem2graph}, and the \textbf{sfdp} layout algorithm from \texttt{graphviz} using the following command:

\begin{lstlisting}[language=bash, caption={Command used to generate the memory graph visualization of \textit{Training/basic/V\_7\_1\_P1/24/17016-1643962152-heap.raw}}]
    sfdp -Gsize=30! -Goverlap=voronoi -Tpng 17016-1643962152_truncated_no_addresses.gv > 17016-1643962152_truncated_no_addresses.png
\end{lstlisting}

\section{From heap dump to memory graph embeddings}
Now that the basis of the memory graph representation has been described, let's dive in the different phases involved in transforming a raw heap dump file into a memory graph file with some custom embeddings that can later be loaded and used by some data analysis and \acrshort{ml} programs.

\subsection{Initialization and data checking}

The graph construction process begins with the initialization phase. In this phase, the first step is graph initialization, which involves loading a given heap dump file and its associated annotation file. Once the files are loaded, several checks are performed on the annotation file to ensure its validity. These checks include verifying that all annotations are present and formatted correctly, as well as ensuring that the annotation file is neither empty nor contains errors.

\subsubsection{Graph Construction steps}

The second major step in the process is graph building. This involves constructing the graph from heap dump byte blocks. The first part of this step is the data structure detection, where blocks are parsed from start to finish. The parsing process leaps over blocks by using chunk sizes that are stored in chained chunk headers. Each chunk is then verified for its size, alignment, and the presence of a potential footer. Following this, the pointer detection step is carried out. In this phase, potential pointers are identified using the previously introduced pointer detection algorithm and are added to the graph. 

An optional step that can be performed is the chunk pointer reduction. This step removes any blocks that are not Chunk Header Nodes, effectively transforming the graph from a block-based graph to a chunk-based graph. While this step is not mandatory, it can be useful for reducing the scale of the problem. This approach will be extensively used in the machine learning section, as it has been shown that the key block prediction problem is equivalent to a key chunk prediction problem.

\subsubsection{Graph Annotation}

The third major step in the process is graph annotation. In this phase, Value Nodes in the graph are replaced by Key Nodes, utilizing the annotations provided in the JSON annotation file. Additional annotations, such as SSH\_STRUCT, can also be added at this stage. Following the completion of the annotation step, it becomes possible to export the graph for various purposes. The graph can be exported to file formats like \texttt{.dot} or \texttt{.gv}, which are suitable for visualization or other analytical tasks.

At this step, the graph is looking similar to the example shown before \ref{methods:mem_graph:17016-1643962152:simplified}. Using the pointer reduction to a chunk-only \gls{memgraph}, we can obtain something like the following:

\begin{figure}[H]\label{methods:mem_graph:585-1644391327:chunk_only}
    \centering
    \includegraphics[width=16cm]{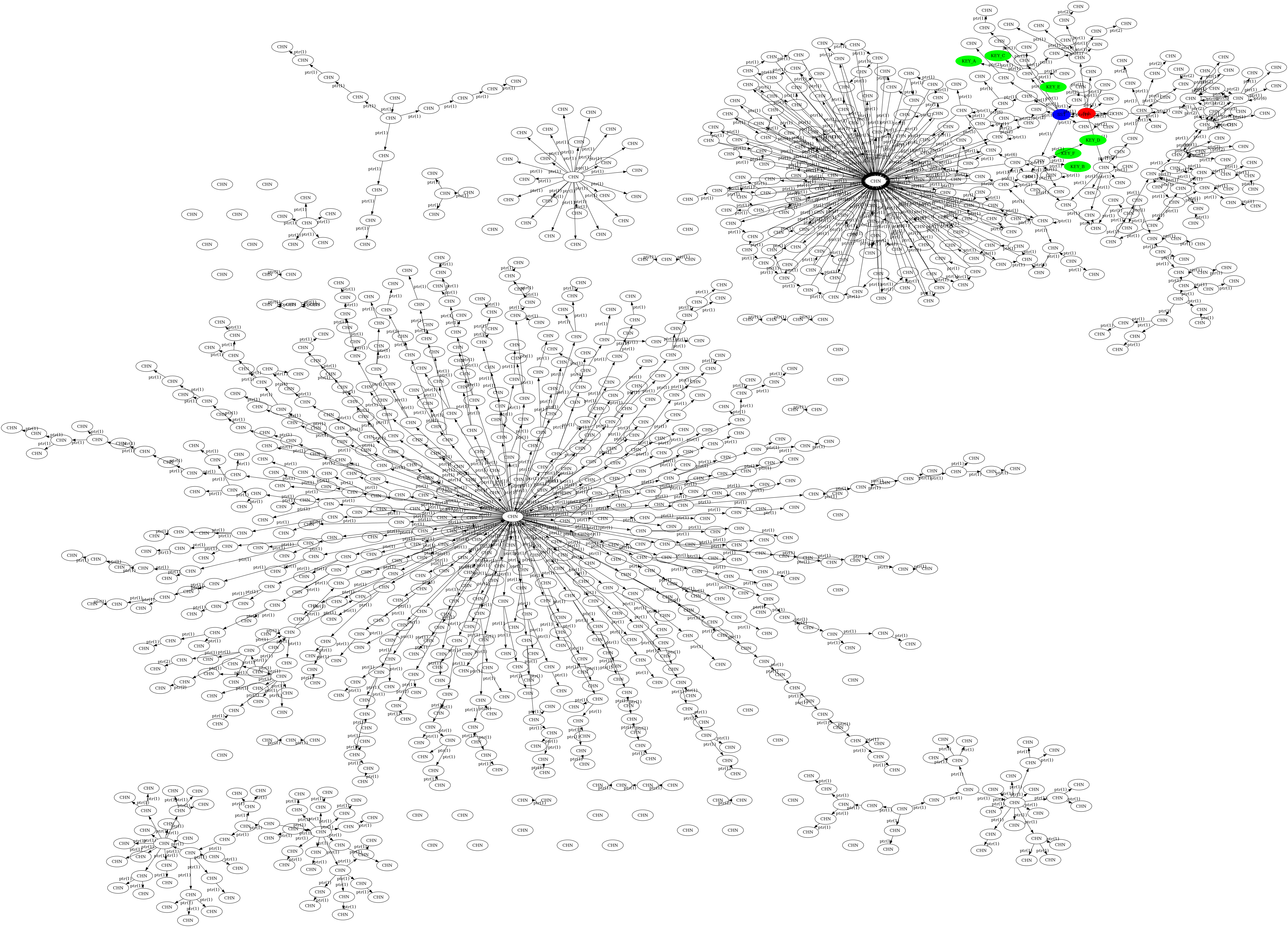}
    \caption{Visualization of the chunk memory graph, with only Chunk Header Nodes representing chunks, generated from \textit{Training/scp/V\_7\_8\_P1/16/
    585-1644391327-heap.raw}.}
\end{figure}

Note how we can identify some data structures formed by pointer-connected chunks. This is a very interesting property of the memory graph representation, since it allows identifying data structures and their members based only on the shape of connections. This is a very important property that will be used later for feature engineering and embeddings. 

\subsubsection{Custom Graph-Based Embeddings}

The fourth step in the workflow involves generating embeddings from the graph. Multiple types of embeddings can be generated, each serving a unique purpose and offering different insights into the graph structure. One such embedding is the semantic embedding. This is a general approach that enriches each node with information related to its graph structure vicinity. It captures the essence of the node's position and relationships within the graph, making it useful for various machine learning and analytics tasks.

In addition to semantic embeddings, other types of embeddings can also be generated. For instance, statistical embeddings focus on capturing the statistical properties of the graph. Another interesting type of embedding is the random walk embedding and related version called Node2Vec. This method leverages the random walk algorithm to generate embeddings, capturing the local and global structure of the graph by simulating random paths through it.

Each of these embeddings offers a unique lens through which to analyze and interpret the graph, and the choice of embedding can be tailored to the specific requirements of the task at hand. It's also possible to add additional features like entropy of the chunk start bytes and filtering information.

\subsubsection{Exporting the Graph}

The fifth step in the process involves exporting the graph. The graph is exported to a \texttt{.gv} DOT file format. Custom embeddings are integrated into the graph by utilizing the comment fields in a slightly modified stringified JSON format. This approach allows for easy reading of the embeddings associated with each node while maintaining the DOT file as a valid format that can be used with tools and libraries supporting the DOT graph formal.

The DOT (Directed Orthogonal Text) format is a plain text graph description language that is widely used for representing structured information. An example of a \gls{memgraph} in DOT format without embeddings is provided for reference. In this example, nodes and edges are represented along with their attributes such as color, shape, and labels.

When embeddings are added to the graph, additional comment fields are included in the DOT file nodes. These comment fields contain a JSON string that holds the embedding information for each node. Moreover, the graph starts with a pseudo JSON comment field that contains a serialized JSON object specifying the embedding type and feature names. 

Here is an example of a comment field in a node:

\begin{lstlisting}[style=text, caption={A comment field example for a node with embedding. Output is cropped.}]
    comment="[0,94918015119368,1,108,108,108,108,108,108,108,1,67,82,139,175,204,...
    0,0,2.355388542207534]"
\end{lstlisting}

Below is an example of a \gls{memgraph} comment field containing a JSON serialized object with embedding type and feature names.

\begin{lstlisting}[style=text, caption={A memgraph comment field example containing JSON serialized object with embedding type and feature names. Output is cropped.}]
    comment="{ 'embedding-type': 'chunk-semantic-embedding', 'embedding-fields': ['block_position_in_chunk','chn_addr','chns_ancestor_1',...'chns_children_8','chunk_byte_size','chunk_number_in_heap','chunk_ptrs','chunk_vns','ptrs_ancestor_1','ptrs_ancestor_2',...,'ptrs_children_8','entropy'] }"
\end{lstlisting}

The inclusion of these comment fields serves multiple purposes. First, it allows for the storage of embeddings along with additional information, making the graph more informative. Second, this format can be easily integrated into Python machine learning pipelines, facilitating the use of the graph in various machine learning tasks. Lastly, the DOT format serves as a standard format for graph representation, making it a versatile choice for both visualization and computational tasks.

\section{A wide range of features and embeddings}
In the following, we will explore the features and embeddings developed and used in this thesis. We will start by the features and embeddings based on the memory graph characteristics, then we will explore the graph-agnostic embeddings, and finally, we will discuss the machine learning models used for the binary classification task.

\subsection{Embeddings based on custom features}
While doing the construction of the graph, we can add some custom features to the nodes. Those features have been developed to embed the unique characteristics of each node of the graph, depending on its type, parent chunk characteristics as well as its vicinity in the memory graph. Those features have been regrouped in several custom embeddings that will be discussed in the following.

\subsubsection{Remark on the collaborative work}
This section focuses on the embeddings developed during a Masterarbeit project around OpenSSH heap dump analysis. Clément Lahoche and the author of the present report have done a collaborative effort on the matter. Since Clément has focused on the embeddings, the following section will not discuss in too many details the embeddings that are already described and analyzed in details by Clément's work. The reader is invited to read his Masterarbeit report \cite{ClementEmbeddingsMasterarbeit23} for more information. This section includes some elements that are clearly identified as coming from Clément's work.

As a notable difference, the Node2Vec embedding, which is a graph-agnostic embedding, will also be discussed in the following, which is not the case in Clément's thesis. This is because the present report focuses on the machine learning part, and especially on graph representation learning.

\subsubsection{Semantic graph embedding}\label{sec:mem_2_graph:semantic_embedding}

The focus of this stage is on semantic embedding, a technique that transforms the graph into a low-dimensional vector space. Each vector encapsulates the local neighborhood of a graph chunk, enabling the application of advanced machine learning methods. The embedding process is intricate, considering both direct and indirect connections to and from each chunk. It starts by counting the number of pointers and chunks directly linked to a specific chunk, and then extends this by recursively exploring deeper layers of connections. A parallel reverse analysis is also conducted to capture child nodes. The outcome is a compact vector that richly represents the chunk's contextual relationships within the graph.

The following algorithm describes the process of generating the semantic embedding for a given chunk:

\begin{algorithm}[H]
    \caption{Generate Ancestor/Children Embedding.}
    \label{algo:embedding:generate_ancestor_children_embedding}
    \begin{algorithmic}
        \Function{GenerateNeighborsCHN}{$chunk\_node, dir$}
            \State $ancestor\_nodes \gets$ an empty set
            \State $children \gets$ graph.neighbors\_directed($chunk\_node, OUT$) \Comment{Get members of the chunk}
            \For{$child$ \textbf{in} $children$}
                \State $ancestor\_nodes$.insert($child$)
            \EndFor
            \State $result \gets$ an empty list
            \State $current\_nodes \gets$ an empty set
            \For{$\_$ \textbf{in} $0$ \textbf{to} $DEPTH$}
                \State $current\_nodes \gets$ $ancestor\_nodes$ \Comment{switch ancestor nodes and current nodes}
                \State $ancestor\_nodes \gets$ an empty set
                \State $nb\_chn \gets 0$
                \State $nb\_ptr \gets 0$
                \For{$current\_node$ \textbf{in} $current\_nodes$}
                    \If{$node$ is ChunkHeaderNode} \Comment{Update number of chunks and pointers}
                        \State $nb\_chn \gets nb\_dtn + 1$
                    \ElsIf{$node$ is PointerNode}
                        \State $nb\_ptr \gets nb\_ptr + 1$
                    \EndIf
                    \Comment{Get neighbors of the current node}
                    \For{$neighbor$ \textbf{in} graph.neighbors\_directed($current\_node, dir$)}
                        \State $ancestor\_nodes$.insert($neighbor$) \Comment{Add neighbors to the next ancestor nodes}
                    \EndFor
                \EndFor
                \State $result$.append($nb\_chn$) \Comment{Add number of data structures}
                \State $result$.append($nb\_ptr$) \Comment{Add number of pointers}
            \EndFor
            \State \textbf{return} $result$
        \EndFunction
    \end{algorithmic}
\end{algorithm}

Note that this algorithm is taken from \citeauthor{ClementEmbeddingsMasterarbeit23}, from his Masterarbeit report \cite{ClementEmbeddingsMasterarbeit23}. It has been developed and implemented as a collaborative effort on this project.

The embedding algorithm is applied to each chunk in the graph, exploring up to a predefined depth, generally 8, which is a hyperparameter of this embedding. This results in a 32-unit embedding, broken down into 8 units each for ancestor pointers, ancestor chunks, child pointers, and child chunks. Basic chunk attributes like block position, byte size, and number of pointers and value nodes are also included, bringing the total embedding size to 37 units. Despite its comprehensiveness, the embedding has limitations, such as the potential for noise from value nodes and the complexity of capturing intricate relationships.

A way that have been used extensively, to both reduce the number of nodes and to improve the quality of the embeddings, is to reduce the graph to a chunk-only graph. This is done by removing all the nodes that are not chunk header nodes. It is a very interesting approach, since it allows to reduce the scale of the problem by a factor of 10, and also allows focusing on the data structures, which are the most interesting nodes to embed. This approach will be used in the machine learning section, since we have shifted the focus from block to chunk prediction.

\subsubsection{Semantic features from essential chunks attributes}

Every chunk in the heap dump comes with fundamental attributes that provide insights on its structure and content. These attributes are not limited to the primary chunk nodes but are also inherited by value and pointer nodes, which are subcomponents of a chunk. The key attributes include the block's position within the chunk, the chunk's byte size, the total number of pointers and value nodes in the chunk, and the chunk's index in the heap. These details collectively offer a thorough understanding of each chunk's makeup and its relative position in the heap.

\subsubsection{Statistical Embedding}

Statistical embeddings serve as a powerful tool for reducing high-dimensional data while preserving essential patterns and probabilistic relationships. One key technique employed is the use of n-gram values, specifically focusing on bit combinations to manage dimensionality. This approach aligns with the primary goal of identifying SSH keys, which inherently display a wide range of frequencies. Various n-gram sizes are utilized, including 1-gram, 2-gram, 3-gram, up to 8-gram, with the latter contributing significantly to capturing broader contextual patterns. 

In addition to n-grams, other statistical metrics like mean, standard deviation, MAD, skewness, kurtosis, and Shannon entropy are incorporated. These metrics offer a multi-faceted view of the data, aiding in the identification of SSH keys. However, chunks with a standard deviation of zero are excluded from the analysis, as they are unlikely to contribute to the identification of random patterns like SSH keys. These skipped values are replaced with \texttt{NaN} values in embedding comment of nodes, that needs to be handled by the machine learning pipeline. It has been chosen to replace those values with zeros. 

Finally, the statistical embedding vector for each chunk is constructed by combining n-gram values and these additional statistical metrics. The vector also includes basic chunk information, resulting in a comprehensive vector that encapsulates the chunk's characteristics.

\subsubsection{Start-bytes Embedding}
In addition to the aforementioned embeddings, a simpler approach was implemented to serve as a baseline for comparison. This method focuses solely on the initial bytes of each chunk for vectorization. The sample vector is initialized with basic chunk information and then populated with the first bytes of the chunk, up to a predefined limit. If the chunk has fewer starting bytes than the predefined limit, zeros are added to fill the remaining positions. This straightforward approach provides a straightforward embedding, suitable for comparative evaluations with more intricate embeddings.

\subsection{Embedding transformations depending on the model}

When it comes to feeding embeddings into machine learning models, the shape and size of the embeddings need to be tailored to fit the model's requirements. For classic machine learning algorithms like Random Forest, the requirement is that the embedding matrices must be of a fixed, predefined size. This presents a unique challenge when working with graph embeddings, as graphs usually have a variable number of nodes. To address this, padding is added to the embedding matrices to ensure they all match the size of the largest graph. In contrast, Graph Convolutional Networks (GCN) offer more flexibility in this aspect. GCNs are capable of handling variable-size embedding matrices as long as the number of features is fixed. This eliminates the need for padding the matrices, which is advantageous as it simplifies the preprocessing steps.

\subsubsection{Node filtering to feature}

On the topic of node filtering in the context of chunk memory graphs, it's important to note that active rebalancing isn't performed, despite the number of positive nodes (key chunks) being substantially lower than the number of negative nodes (non-key chunks). The rationale behind this choice is to enable models to learn from complete graphs. This is particularly relevant for Graph Convolutional Networks, which are capable of handling graphs of variable sizes. The goal is for the models to be able to identify key chunks even in completely unlabeled memory graphs. The ability of GCNs to process variable-size graphs makes them especially suitable for this kind of task, as it allows the model to learn from the full structure of the graph without the need for compromising the integrity of the data through techniques like rebalancing.

\subsection{Graph-agnostic Embeddings}

Unlike our previous embeddings, which were developed manually to suit the intricacies of chunk graphs, there are also pre-existing, generalized graph embeddings. These graph-agnostic embeddings offer the benefit of being applicable to a wide range of graphs without requiring specific customization based on the characteristics of the underlying data.

\subsubsection{RandomWalk}

The RandomWalk algorithm offers a straightforward approach to graph embedding. It simulates random walks starting from each node in the graph and uses these walks to create vector representations of the nodes. One of the advantages of RandomWalk is its simplicity, both in terms of implementation and interpretation. The algorithm excels at capturing local structures within the graph, making it particularly effective for tasks such as community detection and link prediction. However, its focus on local characteristics means that it might not capture global properties of the graph as effectively.

\subsubsection{Node2Vec}

Node2Vec extends the capabilities of the RandomWalk algorithm by introducing additional parameters that allow for a more nuanced exploration of the graph. This makes Node2Vec more versatile than RandomWalk, enabling it to capture both local and global graph structures. Because of its flexibility, Node2Vec is well-suited for a range of applications including the specific task of providing an embedding for the node of memory graphs. While its versatility is a strong point, it comes at the cost of increased computational complexity due to the introduction of several hyperparameters.

\textbf{Hyperparameters:}
\begin{itemize}
  \item \textbf{p:} The return parameter, which controls the likelihood of the walk returning to the node it just left.
  \item \textbf{q:} The in-out parameter, which differentiates between inward and outward nodes in the walk.
  \item \textbf{Length of Walk:} Determines the length of each random walk.
  \item \textbf{Number of Walks:} Specifies the number of walks to initiate from each node.
\end{itemize}

Note that the two last hyperparameters have a huge impact on the performance of the algorithm. These hyperparameters play a crucial role in shaping the behavior of the Node2Vec algorithm. Specifically, the return and in-out parameters help guide the random walks in a way that allows the algorithm to capture different types of structural information from the graph. The length and number of walks, meanwhile, impact the granularity and quality of the embeddings generated.

\section{Machine Learning Binary Classification}

Binary classification is a type of machine learning task where the model is trained to differentiate between two classes. In the context of key chunk prediction, binary classification serves to identify whether a given chunk is a "key chunk" or not. Successfully predicting key chunks is crucial as it leads to a 100\% successful key retrieval rate. This is because, in our case, all keys are situated at the beginning of a chunk, and no chunk contains more than one key. Various machine learning models, ranging from classic approaches to more modern methods like Graph Convolutional Networks (GCNs), have been employed for this task.

\subsection{Classic Models of Machine Learning}

For baseline comparisons, we have experimented with classic machine learning models including Random Forest, Logistic Regression, and the SGD Classifier. These models serve as a well-studied and understood starting point for our classification problem, providing a frame of reference against which more complex models can be compared.

\subsubsection{Random Forest}

Random Forest is an ensemble learning method that operates by constructing multiple decision trees during training and outputs the class that is the mode of the classes of the individual trees for classification tasks. It is highly flexible and can handle a wide range of data types, making it a strong candidate for various use cases, including key chunk prediction.

\textbf{Strong and Weak Points:}
\begin{itemize}
  \item \textbf{Strong:} The model is robust to overfitting and can handle high dimensional data well.
  \item \textbf{Weak:} Random Forest models can be computationally expensive and may require a long training time, especially for larger datasets.
\end{itemize}

As often with models, Random Forest has a bunch of hyperparameters:

\textbf{Hyperparameters:}
\begin{itemize}
  \item \textbf{n\_estimators:} Number of trees in the forest.
  \item \textbf{max\_features:} The number of features to consider when looking for the best split.
  \item \textbf{max\_depth:} The maximum depth of the tree.
  \item \textbf{min\_samples\_split:} The minimum number of samples required to split an internal node.
  \item \textbf{min\_samples\_leaf:} The minimum number of samples required to be at a leaf node.
\end{itemize}

We have used the implementation of Random Forest available in the Scikit-learn library \cite{ScikitLearn}.

\subsubsection{Logistic Regression}

Logistic Regression is a statistical model commonly used for binary classification tasks. It models the log-odds of the probability of the event occurring as a linear combination of the predictor variables. The model is particularly effective when the probability of the outcome (dependent variable) can be expressed as a logistic function of the predictor (independent variable).

\textbf{Strong and Weak Points:}
Logistic Regression is straightforward to implement and understand, making it a good starting point for many classification problems. However, its simplicity is both a strength and a weakness; it might not perform well when the relationship between the variables is not log-linear or when the dataset has high dimensionality.

\textbf{Hyperparameters:}
\begin{itemize}
  \item \textbf{C:} Inverse of regularization strength; smaller values specify stronger regularization.
  \item \textbf{solver:} Algorithm to use for optimization, such as 'liblinear' or 'saga'.
  \item \textbf{max\_iter:} Maximum number of iterations for the solver to converge.
\end{itemize}

We have relied on the Logistic Regression implementation available in the Scikit-learn library for our experiments\cite{ScikitLearn}, using the default hyperparameters.

\subsubsection{SGD Classifier}

The Stochastic Gradient Descent (SGD) Classifier is a linear classifier optimized by stochastic gradient descent. It is especially useful for large-scale and sparse machine learning problems. The SGD Classifier can approximate other types of linear classifiers like Logistic Regression and Support Vector Machines.

\textbf{Strong and Weak Points:}
SGD Classifier is computationally efficient, making it well-suited for large datasets. However, it requires careful tuning of its hyperparameters and might be sensitive to feature scaling.

\textbf{Hyperparameters:}
\begin{itemize}
  \item \textbf{alpha:} Regularization term that discourages large coefficients to prevent overfitting.
  \item \textbf{loss:} Specifies the loss function to be used, such as 'hinge' for SVM or 'log' for logistic regression.
  \item \textbf{max\_iter:} Maximum number of passes over the training data.
  \item \textbf{learning\_rate:} The learning rate schedule, could be 'constant', 'optimal', 'invscaling', or 'adaptive'.
\end{itemize}

For the SGD Classifier as well, we used the Scikit-learn implementation\cite{ScikitLearn}.

\subsection{Graph Convolutional Networks (GCN)}

As GCNs have already been introduced in the background section, this part will primarily focus on our specific implementation and the variants of GCN models employed for the task of key chunk prediction.

GCNs are a specialized form of neural networks designed to operate directly on graphs. One of their main characteristics is their ability to capture the graph's structural information. They accomplish this by using edge connectivity information, either from an adjacency matrix or an edge list, as part of their input. This makes them particularly effective for tasks involving irregular data structures like the memory graphs discussed previously.

We used the PyTorch Geometric library for the development of our GCN models. This library offers a robust set of tools and abstractions, making it easier to construct custom graph-based neural networks \cite{PyTorchGeometric19}.

GCNs excel at capturing the topological features of graphs, making them a strong candidate for our task of key chunk prediction. However, they can be computationally intensive, especially for large graphs, and also require careful tuning of hyperparameters for optimal performance.

\subsubsection{Very Simple GCN}

The Very Simple GCN model is a minimalist approach, capturing the essential features of a Graph Convolutional Network. This model consists of just one graph convolution layer followed by a single fully connected layer. The convolution layer takes the input features and transforms them into a 16-dimensional space.

After the graph convolution operation, a ReLU (Rectified Linear Unit) activation function is applied to the output. ReLU is defined as \( f(x) = \max(0, x) \), replacing all negative values in the tensor with zeros. In the context of GCNs, ReLU is commonly used to introduce non-linearity into the model. It enables the network to learn from the error and make adjustments, making it more capable of handling complex, non-linear relationships in the graph data.

Finally, the output from the ReLU activation is passed through a fully connected layer to produce the final output for classification. Due to its simplicity, this model is computationally efficient but may not capture complex graph structures effectively.

\subsubsection{Simple GCN Models}

The Simple GCN model, or \texttt{LessSimplifiedGNN}, is a step-up in complexity from the very simplified version. It incorporates two graph convolutional layers, doubling the depth of the network. The first convolution transforms the input features to a 12-dimensional space, and the second one further transforms these 12-dimensional vectors into 24 dimensions. Two fully connected layers follow, ultimately producing the final output. This model is capable of capturing more complex features in the graph but is computationally more demanding than the simpler model.

\subsubsection{First GCN Model}

The First GCN model is the first version that was actually implemented and tested for the task. It's a more complex architecture optimized for higher performance. It consists of two graph convolution layers that transform the input features first into a 16-dimensional and then into a 32-dimensional space. Following these, the network contains three fully connected layers. These layers are intended to capture more intricate patterns in the data. Failed attempts to scale the model computation speed by delegating the fully connected layers to the GPU were made, but ultimately, the model was too memory intensive to be trained on the GPU. This model is designed for capturing more nuanced relationships in the graph but at the cost of higher computational requirements.

\subsection{The Impact of Complexity}

The complexity of a model is a double-edged sword when it comes to performance metrics like precision and recall. On one hand, more complex models with additional layers or more neurons can capture intricate patterns in the graph data, potentially improving precision by accurately identifying key chunks. On the other hand, the complexity often leads to overfitting, especially when the training dataset is small or lacks diversity. Overfitting can adversely affect recall, as the model might become too specialized in recognizing training graphs and fail to generalize well to unseen data.

In the case of Graph Convolutional Networks (GCNs), the non-linear transformations and multiple layers can allow the network to learn highly specialized features, which is excellent for achieving high precision. However, these can be detrimental to recall if the model becomes so tailored to the training data that it fails to identify key chunks in new, unseen graphs.

The impact of the model's complexity also varies with the number of input graphs. For datasets with few graphs, a simpler model is often more appropriate to prevent overfitting. In contrast, when numerous diverse graphs are available for training, a more complex model can be employed to capture the rich set of features inherent in the data, thereby potentially improving both precision and recall.

Therefore, the choice of model complexity should be carefully considered, weighing its impact on performance metrics and the size and diversity of the available training data. This is why more complex GCN models have also been implemented and tested.

\subsection{Understanding Dropout in GCNs}

Dropout is a regularization technique used in neural networks, including Graph Convolutional Networks (GCNs) and Convolutional Neural Networks (CNNs), to prevent overfitting. The dropout layer randomly sets a fraction of the input units to 0 during training, which helps to make the model more robust and improves generalization. In the \acrshort{ml} pipelines developed in Python, a dropout rate of 0.5 is used, meaning approximately 50\% of the neurons will be turned off during each forward pass. This section will discuss the effect of dropout on the model's performance and the intuition behind it.

\subsubsection{Effect on Binary Classification}

In a binary classification problem, dropout can have several effects:

\begin{itemize}
    \item \textbf{Reduced Overfitting}: By randomly deactivating neurons, the model is less likely to rely on any single feature, making it generalize better to unseen data. This is expected to reduce overfitting and thus improve the model's performance, especially the recall of the positive class.
    \item \textbf{Increased Robustness}: Dropout can make the model more robust to noise in the training data.
    \item \textbf{Variable Performance}: While dropout can improve generalization, it may also lead to increased variance in the model's predictions, especially if the dropout rate is too high.
\end{itemize}

In the code of some GCN models implemented, dropout is applied after the activation functions of the GCN layers and the first fully connected layer:

\begin{verbatim}
    x = F.relu(x)
    x = self.dropout(x)
\end{verbatim}

This is a common practice as it allows the model to learn more robust features. However, the dropout rate and its placement in the architecture are yet other hyperparameters that may need to be tuned for optimal performance. Since we already have a large number of hyperparameters, we will not tune the dropout rate in this thesis.

\subsubsection{Batch Normalization}

The code also includes Batch Normalization layers, denoted by \texttt{self.bn1} and \texttt{self.bn2}. These layers normalize the features to have zero mean and unit variance, which can accelerate training and provide some regularization effect, complementing the dropout layers.

\subsection{More Complex GCN Models}

In an effort to enhance the performance of our initial GCN models, we explored more complex architectures. The motivation behind increasing the complexity was to evaluate whether additional layers or techniques could lead to improved performance metrics such as precision and recall. These advanced models have been tested against the simpler GCN models to determine their effectiveness.

\subsubsection{GCN with Dropout and Batch Normalization}

Building upon the First GCN model, we incorporated dropout and batch normalization layers. Dropout is employed to mitigate the issue of overfitting, especially relevant for complex models trained on limited data. With a dropout rate of 0.5 after each ReLU activation, we were able to regulate the model's complexity during training. 

Batch normalization, on the other hand, aims to accelerate training and stabilize the learning process. By normalizing the output features of each layer, batch normalization helps in alleviating internal covariance shift, making the model training more resilient to the choice of hyperparameters.

This version of the model aligns closely with the First GCN model, but the addition of dropout and batch normalization layers offer greater robustness, especially when training on imbalanced or smaller datasets.

\subsubsection{Advanced GCN}

The Advanced GCN model represents the most intricate architecture we have experimented with. This model introduces several additional components compared to the simpler models.

The Advanced GCN consists of three Graph Convolution layers, followed by Batch Normalization layers. The initial Graph Convolution layer transforms the input features into a 32-dimensional space, which is then normalized using \texttt{BatchNorm1d}. The second and third Graph Convolution layers further increase the dimensions to 64 and 128, respectively, also followed by batch normalization steps. These layers aim to capture more complex features from the graph structure.

ReLU (Rectified Linear Unit) activation functions are applied after each batch normalization, introducing non-linearity to the model and helping to capture intricate relationships in the data. Additionally, dropout layers with a rate of 0.5 are placed after each ReLU activation to mitigate the risk of overfitting.

After the Graph Convolution layers, the architecture includes a series of fully connected layers that transform the 128-dimensional feature vector into a 256-dimensional space, which is further compressed into 128 and then 64 dimensions. Finally, the model outputs a vector of dimensions corresponding to the number of classes. Each of these fully connected layers is followed by ReLU activations, except for the final output layer.

\begin{itemize}
    \item \textbf{Graph Conv Layers}: Three layers with dimensions 32, 64, and 128
    \item \textbf{Batch Normalization}: Applied after each Graph Conv layer
    \item \textbf{Fully Connected Layers}: Layers with dimensions 256, 128, and 64
    \item \textbf{Dropout}: Applied after each ReLU activation with a rate of 0.5
\end{itemize}

In summary, the model is engineered to capture more complex relationships in the data, at the cost of increased computational requirements and a higher risk of overfitting, especially when trained on small or less diverse datasets. The dropout and batch normalization layers are integrated to combat overfitting to some extent.

The current chapter has been an overview of the dataset, development environment, and tools used for this thesis. In the next chapter, we will dive deeper into developed programs, experimentations conducted and subsequent results.

%% file: tex/chapters/results.tex
\chapter{Results}\label{chap:results}

The following section describes the experimental setup, the used and generated datasets as well as parameters to conclude with the experimental results achieved.

\section{Developed programs}
Many programs have been developed for the need of the present thesis. Early data exploration scripts have paved the way towards efficient highly parallel programs in both Rust and Python for data analysis, graph and embedding generation of ML and DL tested on different models and contexts. All the necessary concepts and methods have been introduced previously, so it is now time to present those main program in details.

\subsection{Mem2Graph}
The present report has already presented the \textit{Mem2Graph} program throughout the heap dump memory parsing algorithm, graph construction and embeddings. This program has been developed in Rust and is an active collaboration between the author and Clément Lahoche, another PhDTrack student at Passau. The program is available on GitHub at \url{https://github.com/passau-masterarbeit-2023/mem2graph}. 

The program is composed of several layers that build on top of each other. The first layer is dedicated into loading a RAW heap dump file with its annotation JSON file, performs some checks and prepare the data for further analysis. The next one performs the graph construction following the algorithms introduced in the Methods section. Another layer performs some annotations of the nodes. The final layer is more versatile and dependent on the input program parameters. For the need of this report, several pipelines of memgraph with and without embeddings have been added, namely \texttt{graph} and \texttt{graph-with-embedding-comments}. The first pipeline doesn't perform any embedding and export the memory graph to a text file following the DOT format \cite{DotFormat15}. The second pipeline performs the same operations, but also exports the memory graph with the embeddings as comments in the DOT file. 

Since the prediction effort is focused on memory chunks, this embedding is generally called with the \texttt{--no-value-node} parameter, which transform the memory graph from a graph of blocks of 8 bytes, into a memory graph of memory chunks, connected by their pointers inside their respective user data space. Several chunk node embeddings have been implemented, and the user can choose which one to use. The chunk node embeddings are the following: \texttt{chunk-semantic-embedding}, \texttt{chunk-statistic-embedding}, and \texttt{chunk-start-bytes-embedding}.

\subsection{Machine Learning Pipelines Runner}

When working with machine learning, Python is a dominant language, benefiting from a rich ecosystem of libraries and frameworks.

The project leverages a wide range of Python libraries to build comprehensive machine learning pipelines:
\begin{itemize}
    \item \textit{NetworkX} for graph-based data structures and algorithms.
    \item \textit{PyGraphviz} for graph visualization.
    \item \textit{Torch} for tensor computations and building neural networks.
    \item \textit{Scikit-learn} for classical machine learning algorithms.
    \item \textit{Pandas} for data manipulation and analysis.
    \item \textit{NumPy} for numerical computations.
    \item \textit{Matplotlib} for data visualization.
    \item \textit{Torch-Geometric} for Geometric Deep Learning extensions for PyTorch.
\end{itemize}

The program encompasses a variety of machine learning models to compare how different models react to the embeddings and representations developed earlier. It includes classical machine learning models from the Scikit-learn library, such as Random Forest, Stochastic Gradient Descent (SGD), and Logistic Regression. These models serve a point of comparison since they don't rely on graph-based data structures and algorithms. 

The program also includes Graph Convolutional Network (GCN) models built on top of PyTorch and Torch-Geometric. These models are more complex and powerful, leveraging the graph structure and embeddings to achieve better results. Those models specifically leverage graph-based embeddings and input data, generated using the Node2Vec algorithm to create dense and continuous node features that can be used for subsequent analysis or machine learning tasks.

\textbf{Main Pipelines:}
The program is organized around three main pipelines:
\begin{enumerate}
    \item \textit{GCN Pipeline:} For Graph Convolutional Network models, built using libraries such as Torch and Torch-Geometric.
    \item \textit{Classical ML Pipeline:} Leveraging algorithms like Random Forest, SGD, and Logistic Regression from Scikit-learn.
    \item \textit{Feature Evaluation Pipeline:} Primarily aimed at evaluating the quality and importance of generated features or graph embeddings.
\end{enumerate}

\subsection{Other programs}
A lot of other programs have been developed for the need of this thesis, but they are not as important as the ones presented above. Those scripts and short program have been developed for several purposes:

\begin{itemize}
    \item \textbf{Data exploration:} Several scripts have been developed to explore the data, and to understand the structure of the heap dump files, the annotations, and the memory graphs.
    \item \textbf{Heap Dump parsing algorithm testers:} Several scripts have been developed to test the heap dump parsing algorithms, and to ensure that the algorithms are working as intended on all possible situations.
    \item \textbf{Graph and embedding generation testers:} Others have been developed to test the graph and embedding generation algorithms, and to ensure that the algorithms are working as intended on all possible situations.
    \item \textbf{Result visualization, analysis and latex generators:} Later in the report, the results will be presented and discussed. Several scripts have been developed to generate the latex tables and plots, and to analyze the results.
\end{itemize}

All those programs represent a consequent amount of work, and have been invaluable to the success of this thesis. Using CLOC (Count Lines of Code),  a command-line utility that can count lines of code in various languages, the following statistics have been obtained:

\begin{lstlisting}[language=bash, caption={Command used to count the number of lines of code in the \textit{phdtrack} directory, containing the reposiroties of the present thesis.}]
    cloc mem2graph research-base predicting_ssh_key_masterarbeit_report phdtrack_server_scripts phdtrack_project_3 memory_graph_gcn data_processing_masterarbeit --exclude-dir=.venv
\end{lstlisting}

The following table shows the number of lines of code for each programming language used in the present thesis:

\begin{table}[h]
    \centering
    \caption{Code Statistics for Masterarbeit}
    \label{tab:cloc_output}
    \begin{tabular}{|l|r|r|r|r|}
        \hline
        Language & Files & Blank Lines & Comments & Code Lines \\
        \hline
        CSV & 867 & 0 & 0 & 158305697 \\
        Text & 25 & 272 & 0 & 50073 \\
        Python & 132 & 2326 & 2682 & 8050 \\
        Rust & 50 & 1343 & 1149 & 6823 \\
        TeX & 28 & 963 & 141 & 6636 \\
        Markdown & 33 & 1016 & 0 & 2375 \\
        JSON & 1345 & 1 & 0 & 1677 \\
        Jupyter Notebook & 2 & 0 & 1811 & 381 \\
        Nix & 12 & 50 & 31 & 290 \\
        TOML & 1 & 2 & 1 & 22 \\
        Bourne Shell & 3 & 8 & 8 & 18 \\
        make & 1 & 8 & 3 & 18 \\
        Dockerfile & 1 & 1 & 0 & 2 \\
        \hline
    \end{tabular}
\end{table}

In the context of this thesis, three programming languages stand out for their specialized roles: Python, Rust, and Nix. Python is predominantly used for the machine learning pipeline, offering ease of use and a rich ecosystem for data science tasks. Rust serves as the backbone for the Mem2Graph program, providing the efficiency required for graph construction and manipulation. Nix is employed for package management and building, ensuring reproducibility across different computing environments. These languages complement each other well, with Python offering high-level abstractions for machine learning, and Rust providing low-level control for performance-critical tasks. Additionally, CSV files are utilized to store model results. TeX is used for generating this report, highlighting the diverse yet complementary set of tools and languages employed in this research. 

\section{Experimental Setup}

The experimental setup serves as the backbone of this research, providing a structured framework for conducting large-scale experiments on the server. This section delves into the intricacies of the setup, detailing the steps involved and the tools used. It also includes selected program output logs to offer a granular view of the program parameters, environment, and usage.

The elements discussed in this section are not merely illustrative; they offer invaluable insights into the challenges encountered during the experiments. These details are particularly crucial when discussing the large-scale experiments, as they provide a comprehensive understanding of the various facets involved.

The final experiments were conducted in a systematic manner, following these steps:

\begin{enumerate}
    \item \textbf{Data Cleaning:} The original Zenodo dataset was cleaned to produce a RAW heap dump dataset, serving as the foundational data for the experiments.
    
    \item \textbf{Graph and Embedding Generation:} The Mem2Graph Rust program was employed to generate graphs along with their embeddings. A Python launcher script facilitated the generation of multiple memgraph datasets, each with varying combinations of program parameters.
    
    \item \textbf{Data Preloading and Validation:} A sanity check Python program was used for data preloading and validation, ensuring the integrity and consistency of the data before proceeding to the experimental phase.
    
    \item \textbf{ML and DL GCN Pipelines:} The main Python program was responsible for the seamless launching of data science tasks, machine learning training, and model evaluation. It was designed to cover a predefined range of parameters and model hyperparameters.
    
    \item \textbf{Result Collection and Evaluation:} The final step involved the collection, evaluation, and visualization of the results. Error handling mechanisms were in place to make necessary corrections and prepare for the next iteration of experiments. This step also facilitated the confrontation of hypotheses and research questions.
\end{enumerate}

Initial small-scale tests were conducted on a laptop to validate the programs and their results. The final, large-scale experiments were carried out on the Drogon server, equipped with 80 threads and 256 GB of RAM. The computational resources provided by the University of Passau have been invaluable for conducting these experiments, sometimes running for several days straight.

\subsection{Generation of the memgraph datasets}
Using Mem2Graph powerful features, it is possible to generate several memgraph datasets with different parameters. The following command has been used to generate 6 memgraph datasets, each containing 26202 graphs, for a total of 157212 graphs. Those datasets account for different chunk node embeddings and a potential additional filtering feature. The command is run on the Drogon server, with 80 threads.

\begin{lstlisting}[language=bash, caption={Sample of final logs of the \texttt{Mem2Graph} program, after generating 6 memgraph datasets from the cleaned heap dump dataset.}, literate={_}{\_}{1}]
    [2023-10-24T20:42:44 UTC][INFO mem_to_graph::exe_pipeline::pipeline] OK [t: worker-63] [N*202 / 26202 files] [fid: 8683-1650977906]    (Nb samples: 0)
    [2023-10-24T20:42:44 UTC][INFO mem_to_graph::graph_data::heap_dump_data] FILE heap dump raw file path: "/root/phdtrack/phdtrack_data_clean/Performance_Test/V_8_1_P1/32/8794-1650977906-heap.raw"
    [2023-10-24T20:42:44 UTC][INFO mem_to_graph::exe_pipeline::pipeline] OK [t: worker-63] [N*203 / 26202 files] [fid: 8794-1650977906]    (Nb samples: 0)
    [2023-10-24T20:42:44 UTC][INFO mem_to_graph::exe_pipeline::pipeline] TIME [total pipeline time: 114.84s]
    100%|##############################| 6/6 [1:07:15<00:00, 672.62s/it]
    Finished! Total time: hours: 1, minutes: 7, seconds: 15 (Drogon Server)
\end{lstlisting}

It took a little more than an hour to generate 6 memgraph datasets, each containing 26202 graphs. The total number of graphs generated is 157212. The generated memgraph datasets are stored in the \texttt{data/} directory, as follows:

\begin{figure}[H]
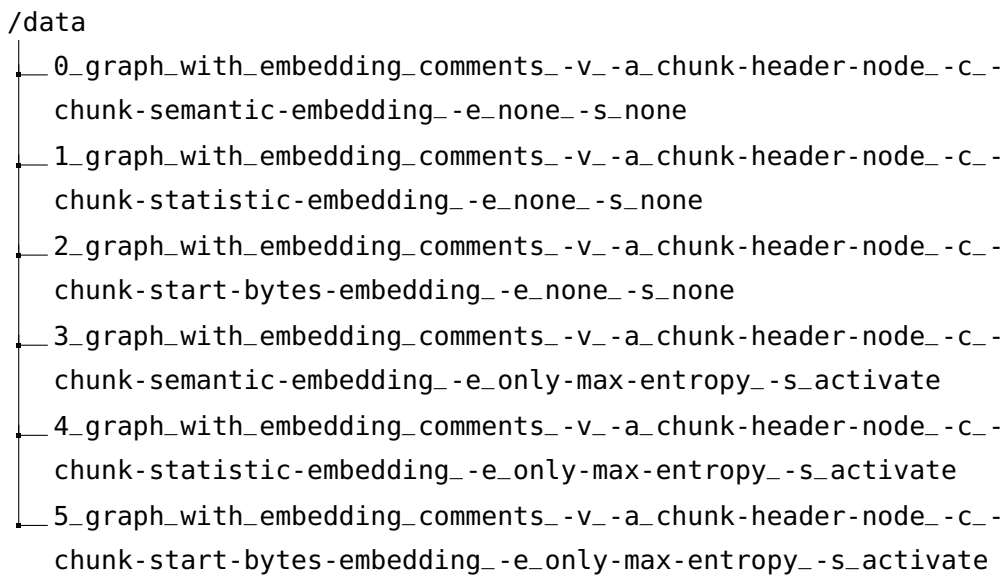

    \centering
    \caption{Illustration of the Data Directory Structure}
    \label{fig:data_structure}
    \begin{minipage}{0.8\textwidth}
        \dirtree{%
            .1 /data.
            .2 0\_graph\_with\_embedding\_comments\_-v\_-a\_chunk-header-node\_-c\_chunk-semantic-embedding\_-e\_none\_-s\_none.
            .2 1\_graph\_with\_embedding\_comments\_-v\_-a\_chunk-header-node\_-c\_chunk-statistic-embedding\_-e\_none\_-s\_none.
            .2 2\_graph\_with\_embedding\_comments\_-v\_-a\_chunk-header-node\_-c\_chunk-start-bytes-embedding\_-e\_none\_-s\_none.
            .2 3\_graph\_with\_embedding\_comments\_-v\_-a\_chunk-header-node\_-c\_chunk-semantic-embedding\_-e\_only-max-entropy\_-s\_activate.
            .2 4\_graph\_with\_embedding\_comments\_-v\_-a\_chunk-header-node\_-c\_chunk-statistic-embedding\_-e\_only-max-entropy\_-s\_activate.
            .2 5\_graph\_with\_embedding\_comments\_-v\_-a\_chunk-header-node\_-c\_chunk-start-bytes-embedding\_-e\_only-max-entropy\_-s\_activate.
        }
    \end{minipage}
\end{figure}

As one can see, the dataset directory names are composed with the most important \texttt{Mem2Graph} program parameters, responsible for some feature and embedding generations. The \texttt{-e} flag, short for \texttt{--entropy-filter} is responsible for the filtering using the Shannon entropy, the \texttt{-s} for \texttt{--chunk-byte-size-filter} for chunk byte size filtering. The \texttt{-c}, \texttt{--graph-comment-embedding-type} controls the custom embedding being save alongside each node in the generated .GV memgraph files.

\subsection{Sanity checking and preloading the generated memgraph datasets}
Loading the graph from DOT files into NetworkX graph in Python is a resource intensive operation that consumes all the available computing power on all tested platforms (laptop, servers). It takes several dozens of seconds up to a minute to load a memory graph containing only around 1000 chunk nodes. It has been experimented that saving those loaded graphs using the \textit{pickle} python library allows to perform checks while loading the graph, add more information about each graph. The subsequent retrieval of the graph is much faster, and allows to perform the sanity checks before any further processing. So to verify the memory graph dataset generation as well as transforming DOT files into pre-saved NetworkX graph, a sanity checking script has been developed.

Below is a sample of the logs generated by the sanity check script:

\begin{lstlisting}[language=bash, caption={Result logs of the \texttt{memory\_graph\_gcn/src/sanity\_check\_gv\_files.py} program.}, literate={_}{\_}{1}]
    * Running program...
   Passed program params:
   param[0]: src/sanity_check_gv_files.py
   param[1]: -k
   Parsed program params:
   keep_old_output: True
   skip_dir_starting_with_number: None
   dry_run: False
    * Now, performing data loading and sanity checks...
    * Looking for Mem2Graph dataset directories in /root/phdtrack/mem2graph/data...
    * Skipping .gitignore...
    * Found 6 Mem2Graph dataset directories.
   [...]
   Loading graphs: 100%|##########| 26201/26202 [3:22:17<00:00,  1.17it/s]
   Loading graphs: 100%|##########| 26202/26202 [3:22:25<00:00,  3.19s/it]
   Loading graphs: 100%|##########| 26202/26202 [3:22:25<00:00,  2.16it/s]
    * Checking embedding length of graphs in /root/phdtrack/mem2graph/data/5_graph_with_embedding_comments_-v_-a_chunk-header-node_-c_chunk-start-bytes-embedding_-e_only-max-entropy_-s_activate...
   -> [x] 26202 graphs in /root/phdtrack/mem2graph/data/5_graph_with_embedding_comments_-v_-a_chunk-header-node_-c_chunk-start-bytes-embedding_-e_only-max-entropy_-s_activate have been loaded and checked.
   -> [_] 0 graphs in /root/phdtrack/mem2graph/data/5_graph_with_embedding_comments_-v_-a_chunk-header-node_-c_chunk-start-bytes-embedding_-e_only-max-entropy_-s_activate have been skipped (deleted).
   [x] 157212 total graphs in the input mem2graph dataset dir paths have been loaded and checked.
   [_] 0 total graphs in the input mem2graph dataset dir paths have been skipped (deleted).
   <END> Program took: 42626.062309 total sec (11h 50m 26s)
\end{lstlisting}

The sanity checking file can be considered very long to run, having taken almost 12 hours straight, but it is a necessary step to ensure the quality of the generated memgraph datasets. It is also a good way to check the validity of the generated embeddings, and to ensure that the graphs are re-exported as \texttt{NetworkX} graphs that can be loaded much faster than the original DOT files.

\subsection{Launching the experiments}
Two pass of experiments have been conducted, with the exact same parameters and input memgraph dataset. Contrary to expectations, the experiments were much faster on the laptop that on the Drogon server. 

\begin{itemize}
    \item \textbf{Time take for the experiments on the laptop:} 12h 31m 53s
    \item \textbf{Time take for the experiments on the Drogon server:} 29h 17m 57s
\end{itemize}

The experiments have been launched using the following command (here, on Drogon server):

\begin{lstlisting}[language=bash, caption={Command used to launch final experiments, on Drogon server.}, label={results:final-launching:run-experiments:command}, literate={_}{\_}{1}]
    nohup python src/main_gcn.py -i /root/phdtrack/phdtrack\_data\_clean/ -p gcn-pipeline classic-ml-pipeline feature-evaluation-pipeline -b 6 -a -q -n 16  > output\_ml\_2023\_10\_27\_16h\_35.log 2>&1 &
\end{lstlisting}

Using \texttt{nohup} and redirecting the output to a log file allows the experiments to run in the background and enables the retrieval of the output at a later time. The command specifies the input directory containing the annotated DOT (.gv) graph directory with the \texttt{-i} flag. All three pipelines are chosen: \texttt{gcn-pipeline}, \texttt{classic-ml-pipeline}, and \texttt{feature-evaluation-pipeline}, as indicated by the \texttt{-p} flag. The batch size for parallel processing is set to 6 using the \texttt{-b} flag, and the \texttt{-a} flag indicates the use of all available Mem2Graph datasets. The \texttt{-q} flag enables quiet mode for Node2Vec, and the \texttt{-n} flag specifies the use of 16 input graphs.

\subsection{Dealing with hyperparameter tuning}

In the quest to optimize the performance of both \acrshort{ml} and \acrshort{dl} models, hyperparameter tuning plays a crucial role. This section elaborates on the various strategies and tools employed for hyperparameter tuning in this research.

\begin{itemize}
    \item \textbf{Precise Command Lines for Tuning:} The compute instances and experiment parameters can be finely tuned using precise command-line options. This flexibility allows for a more targeted approach to model optimization, enabling the user to specify various hyperparameters and settings right from the terminal.
    
    \item \textbf{Python Program for Experiment Launch:} A dedicated Python program has been developed to automate the launching of \acrshort{ml}, \acrshort{gcn} and \acrshort{fe} pipelines. This program takes different hyperparameters as input and initiates the corresponding experiments, thereby streamlining the entire process.
    
    \item \textbf{Automatic Logging in CSV:} All the results from each experiment, along with the hyperparameters used, are automatically logged into a CSV file. This facilitates easy tracking and comparison of different model configurations and their corresponding performances.
    
    \item \textbf{Timestamps and Duration Steps:} Each experiment is meticulously logged with precise timestamps and duration steps. This includes the time taken for generating embeddings, as well as the time required for the training and testing phases. Such detailed logging aids in identifying bottlenecks and optimizing the pipeline further.
    
    \item \textbf{Extensive Experimentation:} Over the course of this research, thousands of experiments have been conducted. These experiments span a wide range of parameters, models, and embeddings, providing a comprehensive understanding of the model behaviors and their sensitivities to different hyperparameters.
    
    \item \textbf{Automated Result Analysis and Visualization:} To aid in the interpretation of the extensive experimental results, automated scripts have been developed for result analysis and visualization. These scripts generate various plots and metrics that provide insights into the performance and reliability of the different models and configurations.
\end{itemize}

Of all the parameters, the model types, their respective hyperparameters, the combinations of possible embeddings with their own parameters, and the number of input graphs are the most important. All those parameters explains the large number of experiments that have been conducted, and the need for a precise and automated way to launch them.

\subsubsection{Limited number of input graphs due to compute time and memory usage}

It also explains why the input number of memgraphs has been limited to 16, as it is already a very large number of experiments to run. Each file taking several dozens of seconds to be transformed into an embedding, this represents around 10 minutes just for the embedding generation phase, for each pipeline. With more than 600 pipelines to run, this already represents dozens of hours of compute time. Depending on the context and model, the training and evaluation phases are also time-consuming, and the more input graphs there are, the longer it takes to train and test the model. 

The compute time is not the only issue with dealing with a large number of input graphs. The memory usage is also a problem, as the memory usage increases linearly with the number of input graphs. The parallel processing of only 6 pipelines having 16 memgraphs as inputs generally represents between 16 and 50 GB of RAM usage, depending on the model and the embedding used. Due to this, all tests consisting of trying to run this already limited number of input graphs on the GPU have failed, as the GPU memory is not sufficient to handle the memory usage of the program. The slowness of the CPU alongside memory bandwidth limitations are the main bottlenecks of the program, and the main reasons why the number of input graphs has been limited to 16.

\subsubsection{Parallel launch of experiments for hyperparameters tuning}

To maximize efficiency and expedite the research process, a Python program was developed to launch multiple experiments in parallel. This approach allows for simultaneous testing of various input graphs, machine learning and deep learning models, embedding techniques, and hyperparameters. By leveraging parallel computing, the program significantly reduces the time required for extensive experimentation, thereby accelerating the overall research and development cycle.

Below are the hyperparameters used during the main experiments, as stored in the \texttt{hyperparams.json} file:

\begin{lstlisting}[style=json, caption={JSON hyperparameters used during experiments}, label={results:json-hyperparams-range}, literate={_}{\_}{1}]
        {
            "node2vec_dimensions_range": [128],
            "node2vec_walk_length_range": [16],
            "node2vec_num_walks_range": [50],
            "node2vec_p_range": [0.5, 1.0, 1.5],
            "node2vec_q_range": [0.5, 1.0, 1.5],
            "node2vec_window_range": [10],
            "node2vec_batch_words_range": [8],
            "node2vec_workers_range": [16],
            "randomforest_trees_range": [100, 500, 1000],
            "gcn_training_epochs_range": [20]
        }
\end{lstlisting}

Even though the above JSON file shows the hyperparameters selected for the large scale final experiments, pre-experiments have been previously conducted to find some usable values. Due to compute time limitations, the ranges of hyperparameters have been limited to a few values, and the number of input graphs has been limited to 16. The following JSON file shows the hyperparameters used during the pre-experiments:

\begin{lstlisting}[style=json, caption={JSON hyperparameters used during experiments}, label={results:json-hyperparams-range}, literate={_}{\_}{1}]
    {
        "node2vec_dimensions_range": [16, 32, 128],
        "node2vec_walk_length_range": [16, 32],
        "node2vec_num_walks_range": [50, 100],
        "node2vec_p_range": [0.5, 1.0, 1.5],
        "node2vec_q_range": [0.5, 1.0, 1.5],
        "node2vec_window_range": [10, 20],
        "node2vec_batch_words_range": [4, 8, 16],
        "node2vec_workers_range": [16, 32],
        "randomforest_trees_range": [100, 500, 1000],
        "gcn_training_epochs_range": [5, 10, 20]
    }
\end{lstlisting}

Several observations were made during the early experiments concerning the impact of different hyperparameters on the model's performance, especially concerning the Node2Vec parameters. The following observations were made:

\begin{itemize}
    \item \textbf{Walk Length and Number of Walks:} Increasing the number of walks and the walk length generally improved the model's performance. However, this came at the cost of significantly increased computational time. The benefits plateaued after reaching a certain threshold.
    
    \item \textbf{Number of Dimensions:} A higher number of dimensions generally led to better results. Lower values were found to be detrimental to the model's performance, indicating the importance of this parameter.
    
    \item \textbf{Impact of \( p \) and \( q \):} The parameters \( p \) and \( q \) had an unpredictable impact on the model's performance. While some combinations seemed to yield better results, no clear pattern was observed, making these parameters challenging to optimize.
    
    \item \textbf{Batch Word Range:} The range of batch words had a minimal impact on the model's performance. As a result, an intermediate value was selected for this parameter to balance computational efficiency and performance.
\end{itemize}

These observations provide valuable insights into the behavior of the models under different hyperparameter settings and serve as a guide for future experiments.

\subsubsection{Description of the \texttt{results.csv} File}

The \texttt{*results.csv} files are used to store the results of machine learning and deep learning classifiers, including Graph Convolutional Networks (GCNs) and classical classifiers like Random Forest. The files are regular CSVs organized with the following headers:

\begin{itemize}
    \item \textbf{system}: The operating system on which the experiment was run. Here, \textit{Linux}.
    \item \textbf{node\_name}: The name of the node in the cluster. Here, either \textit{nixos} (local machine) or \textit{rascoussie} (lab server).
    \item \textbf{release, version}: OS release and version information.
    \item \textbf{machine, processor}: Hardware details.
    \item \textbf{physical\_cores, total\_cores}: Number of physical and total cores.
    \item \textbf{total\_memory, available\_memory}: Total and available memory in bytes.
    \item \textbf{start\_time}: The start time of the experiment.
    \item \textbf{nb\_input\_graphs}: Number of input graphs.
    \item \textbf{duration\_embedding}: Time taken for embedding.
    \item \textbf{subpipeline\_name, index, pipeline\_name}: Information about the pipeline model used, with values like \textit{sgd-classifier} or \textit{gcn-with-dropout}.
    \item \textbf{input\_mem2graph\_dataset\_dir\_path}: Path to the dataset directory.
    \item \textbf{node\_embedding}: Type of node embedding used. Several values are possible, like solo embeddings like \textit{chunk-header-node} or \textit{node2vec}, or combined embeddings like \textit{node2vec-chunk-semantic-embedding}.
    \item \textbf{node2vec\_*}: Parameters for the Node2Vec algorithm, like the number of walks, the walk length, the window size, the number of dimensions, the number of epochs, and the p and q parameters.
    \item \textbf{random\_forest\_*}: Parameters for the Random Forest algorithm, like the number of estimators (trees) or the number of parallel jobs.
    \item \textbf{imbalance\_ratio}: Ratio of the classes in the dataset. For instance, a value of 496.33 means that the dataset contains 496.33 times more negative samples than positive samples. Since no filtering is applied, the ratio is always greater than 1, and generally very high.
    \item \textbf{precision\_class\_*, recall\_class\_*, f1\_score\_class\_*, support\_class\_*}: Metrics for each class.
    \item \textbf{true\_positives, true\_negatives, false\_positives, false\_negatives}: Confusion matrix elements.
    \item \textbf{AUC}: Area under the ROC curve.
    \item \textbf{duration\_train\_test}: Time taken for training and testing.
    \item \textbf{nb\_node\_features}: Number of node features. This value directly depends on the node embedding used, and the input memory graph dataset parameters used during generation.
    \item \textbf{first\_gcn\_training\_epochs}: Number of epochs in GCN training phase.
\end{itemize}

Each row in the \texttt{*results.csv} files represents a single experiment run with a specific configuration and its corresponding results. Keeping a precise track of the parameters used for each experiment is crucial for reproducibility and traceability. They also form the basis for the analysis and visualization of the results.

\section{Obtained Results}

This section presents a comprehensive overview of the results obtained with the final experimentation. The outcomes are detailed in multiple formats, including correlation matrices, performance metrics tables, and visualizations, to provide a precise understanding of the model performances. The aim is to elucidate the effectiveness of different features and embeddings in the context of our machine learning pipelines. In-depth discussions on these results will be reserved for the following "Discussions" part.

\subsection{Feature Engineering results}

This section delves into the results obtained from the feature engineering efforts carried out during the experiments. The results are presented in multiple correlation matrices.

\begin{figure}[H]\label{results:corr_matrices:kendall}
    \centering
    \includegraphics[width=16cm]{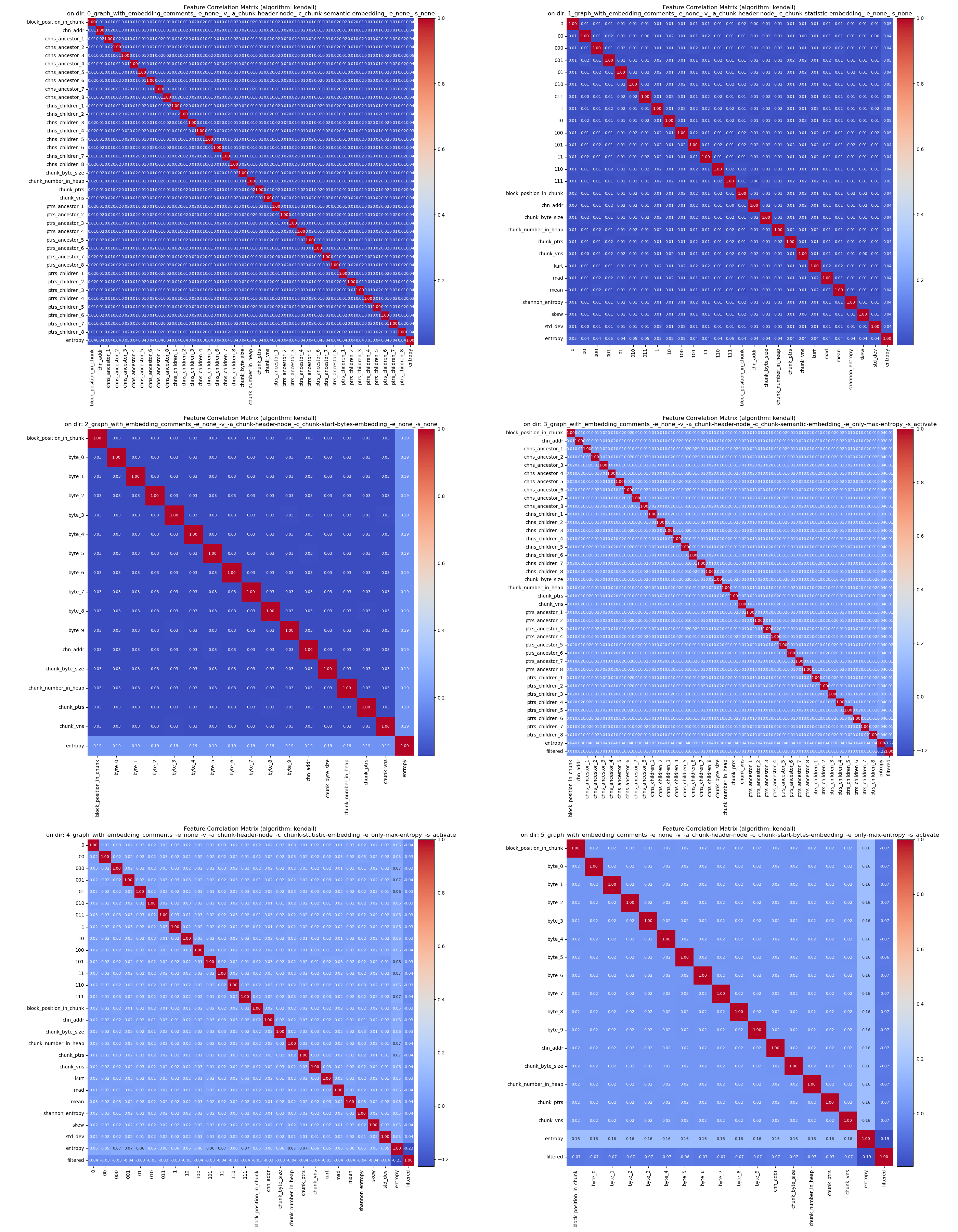}
    \caption{Feature correlation matrices on the different Mem2Graph-generated datasets. Used algorithm: Kendall.}
\end{figure}

\begin{figure}[H]\label{results:corr_matrices:pearson}
    \centering
    \includegraphics[width=16cm]{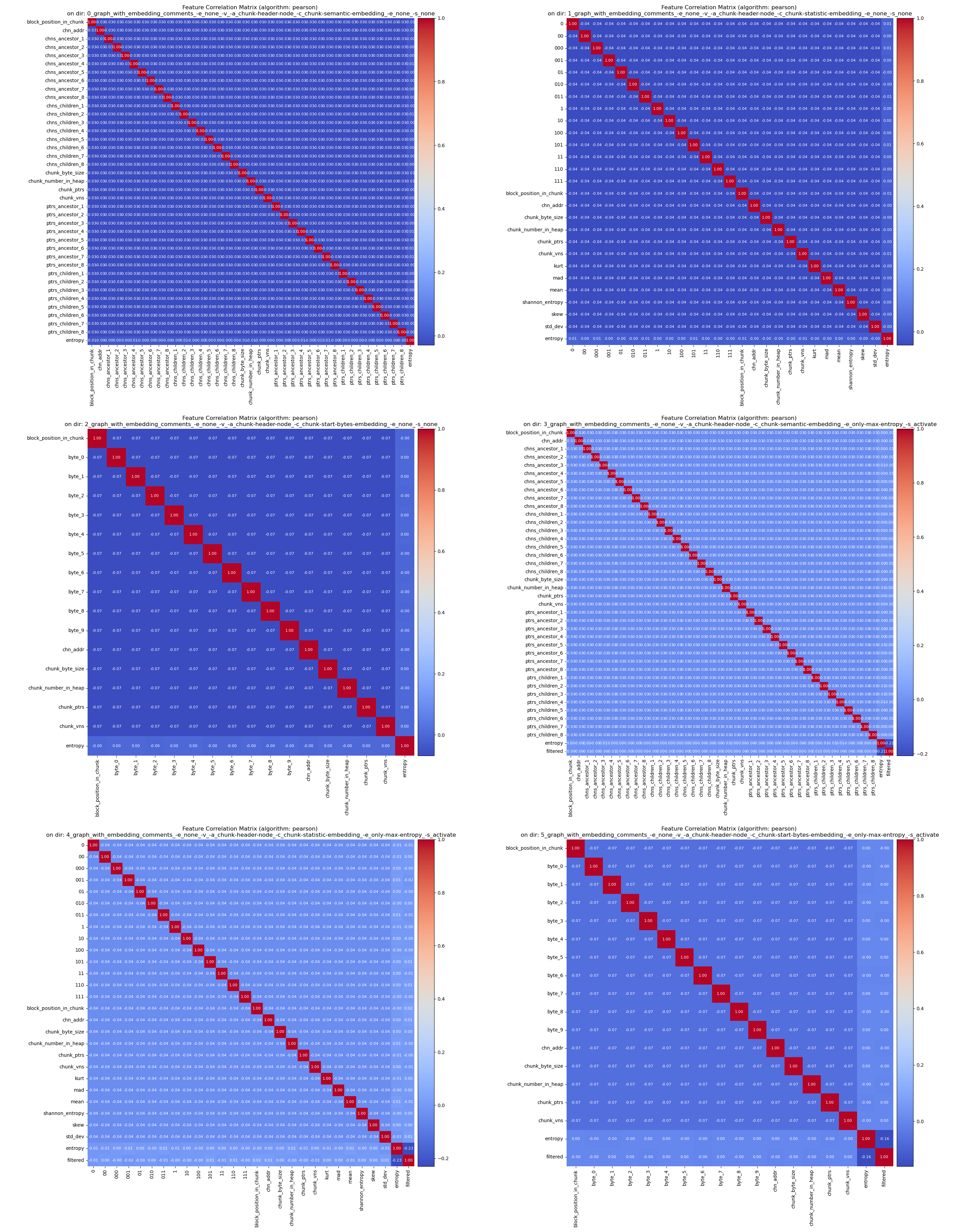}
    \caption{Feature correlation matrices on the different Mem2Graph-generated datasets. Used algorithm: Pearson.}
\end{figure}

\begin{figure}[H]\label{results:corr_matrices:spearman}
    \centering
    \includegraphics[width=16cm]{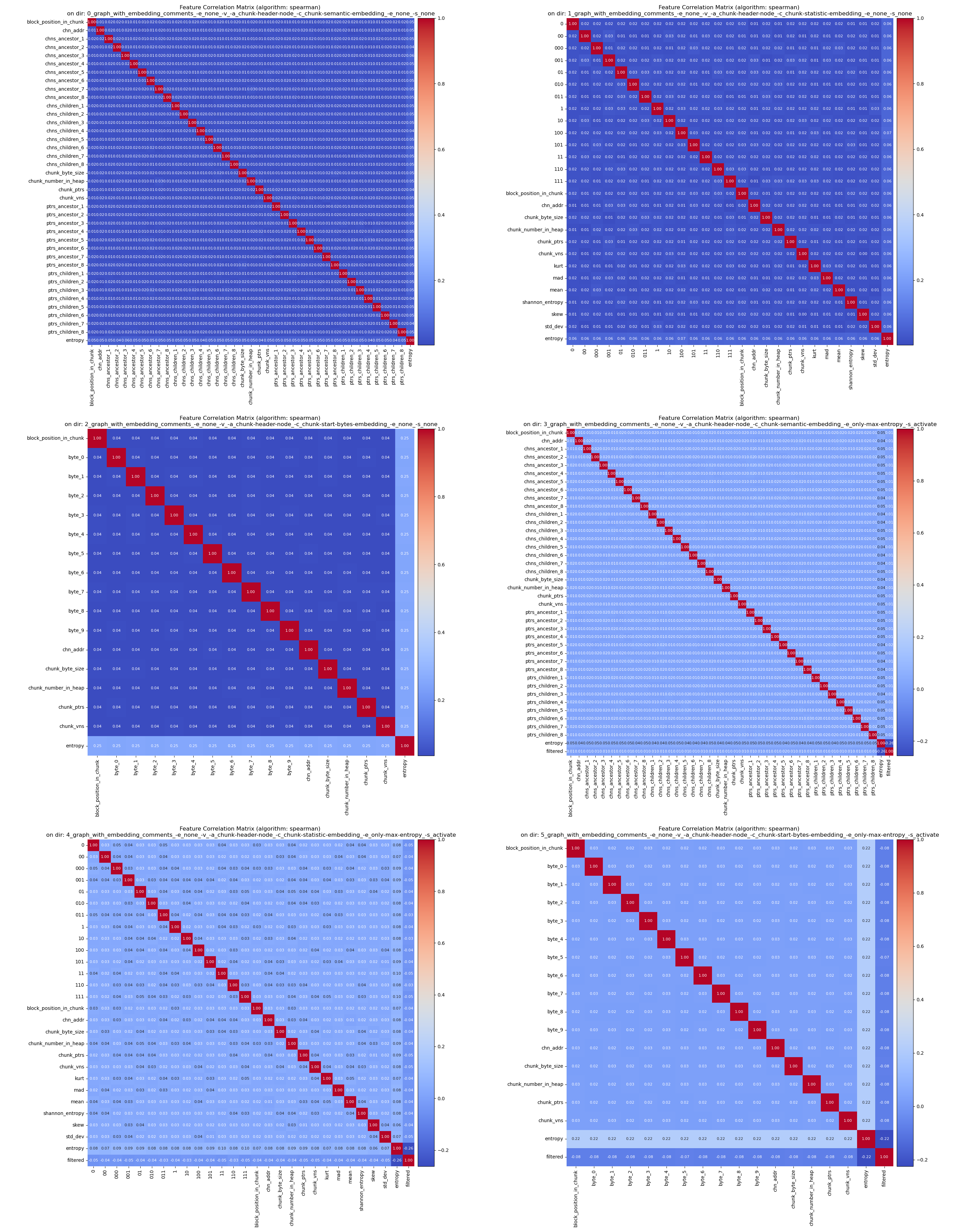}
    \caption{Feature correlation matrices on the different Mem2Graph-generated datasets. Used algorithm: Spearman.}
\end{figure}

\subsection{Classic Model results}

Tables are provided to summarize the performance of different pipelines and models. These tables include four classical machine learning metrics: precision, recall, F1 score, and the Area Under the Curve (AUC). Each table offers a snapshot of how well each model performs on the key chunk prediction task.

\begin{table}[H]
    \centering
    \caption{Best instances of model: logistic-regression.}
    \begin{tabular}{lcccccc}
      \textbf{Best at}  & \textbf{Precision} & \textbf{Recall} & \textbf{F1 Score} & \textbf{AUC} \\
        precision & 1.0000 & 0.0417 & 0.0800 & 0.5208 \\
        recall & 0.3333 & 0.5000 & 0.4000 & 0.7471 \\
        f1 score & 0.3333 & 0.5000 & 0.4000 & 0.7471 \\
        AUC & 0.2449 & 0.5000 & 0.3288 & 0.7486 \\
    \end{tabular}
\end{table}

\begin{table}[H]
    \centering
    \caption{Best instances of model: random-forest.}
    \begin{tabular}{lcccccc}
      \textbf{Best at}  & \textbf{Precision} & \textbf{Recall} & \textbf{F1 Score} & \textbf{AUC} \\
        precision & 1.0000 & 0.0417 & 0.0800 & 0.5208 \\
        recall & 1.0000 & 0.0833 & 0.1538 & 0.5417 \\
        f1 score & 1.0000 & 0.0833 & 0.1538 & 0.5417 \\
        AUC & 1.0000 & 0.0833 & 0.1538 & 0.5417 \\
    \end{tabular}
\end{table}

\begin{table}[H]
    \centering
    \caption{Best instances of model: sgd-classifier.}
    \begin{tabular}{lcccccc}
      \textbf{Best at}  & \textbf{Precision} & \textbf{Recall} & \textbf{F1 Score} & \textbf{AUC} \\
        precision & 1.0000 & 0.0417 & 0.0800 & 0.5208 \\
        recall & 0.4615 & 1.0000 & 0.6316 & 0.9962 \\
        f1 score & 0.4615 & 1.0000 & 0.6316 & 0.9962 \\
        AUC & 0.4615 & 1.0000 & 0.6316 & 0.9962 \\
    \end{tabular}
\end{table}

\subsection{Deep Learning GCN Model results}
Best models obtained after the hyperparameter search, on a range of embeddings and models, this time focusing on the \acrshort{gcn} models.

\begin{table}[H]
    \centering
    \caption{Best instances of model: very-simple-gcn.}
    \begin{tabular}{lcccccc}
      \textbf{Best at}  & \textbf{Precision} & \textbf{Recall} & \textbf{F1 Score} & \textbf{AUC} \\
        precision & 0.6000 & 0.5000 & 0.5455 & 0.7489 \\
        recall & 0.2609 & 1.0000 & 0.4138 & 0.9907 \\
        f1 score & 0.6000 & 0.5000 & 0.5455 & 0.7489 \\
        AUC & 0.2609 & 1.0000 & 0.4138 & 0.9907 \\
    \end{tabular}
\end{table}

\begin{table}[H]
    \centering
    \caption{Best instances of model: simple-gcn.}
    \begin{tabular}{lcccccc}
      \textbf{Best at}  & \textbf{Precision} & \textbf{Recall} & \textbf{F1 Score} & \textbf{AUC} \\
        precision & 0.5000 & 0.5000 & 0.5000 & 0.7484 \\
        recall & 0.2308 & 1.0000 & 0.3750 & 0.9891 \\
        f1 score & 0.5000 & 0.5000 & 0.5000 & 0.7484 \\
        AUC & 0.2609 & 1.0000 & 0.4138 & 0.9907 \\
    \end{tabular}
\end{table}

\begin{table}[H]
    \centering
    \caption{Best instances of model: first-gcn.}
    \begin{tabular}{lcccccc}
      \textbf{Best at}  & \textbf{Precision} & \textbf{Recall} & \textbf{F1 Score} & \textbf{AUC} \\
        precision & 0.5000 & 0.5000 & 0.5000 & 0.7484 \\
        recall & 0.2727 & 1.0000 & 0.4286 & 0.9913 \\
        f1 score & 0.5000 & 0.5000 & 0.5000 & 0.7484 \\
        AUC & 0.2727 & 1.0000 & 0.4286 & 0.9913 \\
    \end{tabular}
\end{table}

\begin{table}[H]
    \centering
    \caption{Best instances of model: gcn-with-dropout.}
    \begin{tabular}{lcccccc}
      \textbf{Best at}  & \textbf{Precision} & \textbf{Recall} & \textbf{F1 Score} & \textbf{AUC} \\
        precision & 0.3500 & 0.2333 & 0.2800 & 0.6152 \\
        recall & 0.0863 & 0.8000 & 0.1558 & 0.8810 \\
        f1 score & 0.2110 & 0.7667 & 0.3309 & 0.8767 \\
        AUC & 0.0863 & 0.8000 & 0.1558 & 0.8810 \\
    \end{tabular}
\end{table}

\begin{table}[H]
    \centering
    \caption{Best instances of model: advanced-gcn.}
    \begin{tabular}{lcccccc}
      \textbf{Best at}  & \textbf{Precision} & \textbf{Recall} & \textbf{F1 Score} & \textbf{AUC} \\
        precision & 0.2097 & 0.4333 & 0.2826 & 0.7129 \\
        recall & 0.0533 & 0.9000 & 0.1006 & 0.8943 \\
        f1 score & 0.2097 & 0.4333 & 0.2826 & 0.7129 \\
        AUC & 0.1552 & 0.9000 & 0.2647 & 0.9390 \\
    \end{tabular}
\end{table}

\section{Compared Performances of models and embeddings}

In our experiments, we limited the analysis to a maximum of 16 input graphs, which may appear to be a relatively low number. However, this limitation was necessary due to the extensive range of hyperparameters, embeddings, and models that we aimed to evaluate. Despite this constraint, we were able to run a total of 7976 machine learning pipelines, accumulating over 100 hours of compute time. This extensive computational effort underscores the complexity and depth of the feature engineering and model evaluation processes undertaken in this study. The following tables and visualizations provide a comprehensive overview of the results obtained from these experiments.

\begin{table}[H]
    \centering
    \caption{Results for the model very-simple-gcn}
    \begin{tabular}{lcccccc}
      \textbf{Model}  & \textbf{Best Precision} & \textbf{Best Recall} & \textbf{Best F1 Score} & \textbf{Best AUC} \\
        advanced-gcn & 0.2097 & 0.9000 & 0.2826 & 0.9390 \\
        first-gcn & 0.5000 & 1.0000 & 0.5000 & 0.9913 \\
        gcn-with-dropout & 0.3500 & 0.8000 & 0.3309 & 0.8810 \\
        logistic-regression & 1.0000 & 0.5000 & 0.4000 & 0.7486 \\
        random-forest & 1.0000 & 0.0833 & 0.1538 & 0.5417 \\
        sgd-classifier & 1.0000 & 1.0000 & 0.6316 & 0.9962 \\
        simple-gcn & 0.5000 & 1.0000 & 0.5000 & 0.9907 \\
        very-simple-gcn & 0.6000 & 1.0000 & 0.5455 & 0.9907 \\
    \end{tabular}
\end{table}

In addition to the tabular data, we also offer visualizations to facilitate a more intuitive comparison between models and embeddings. These graphical representations aim to make the complex data more digestible and provide insights that may not be immediately obvious from the tables alone.

\begin{figure}[H]\label{results:compare:models:full}
    \centering
    \includegraphics[width=16cm]{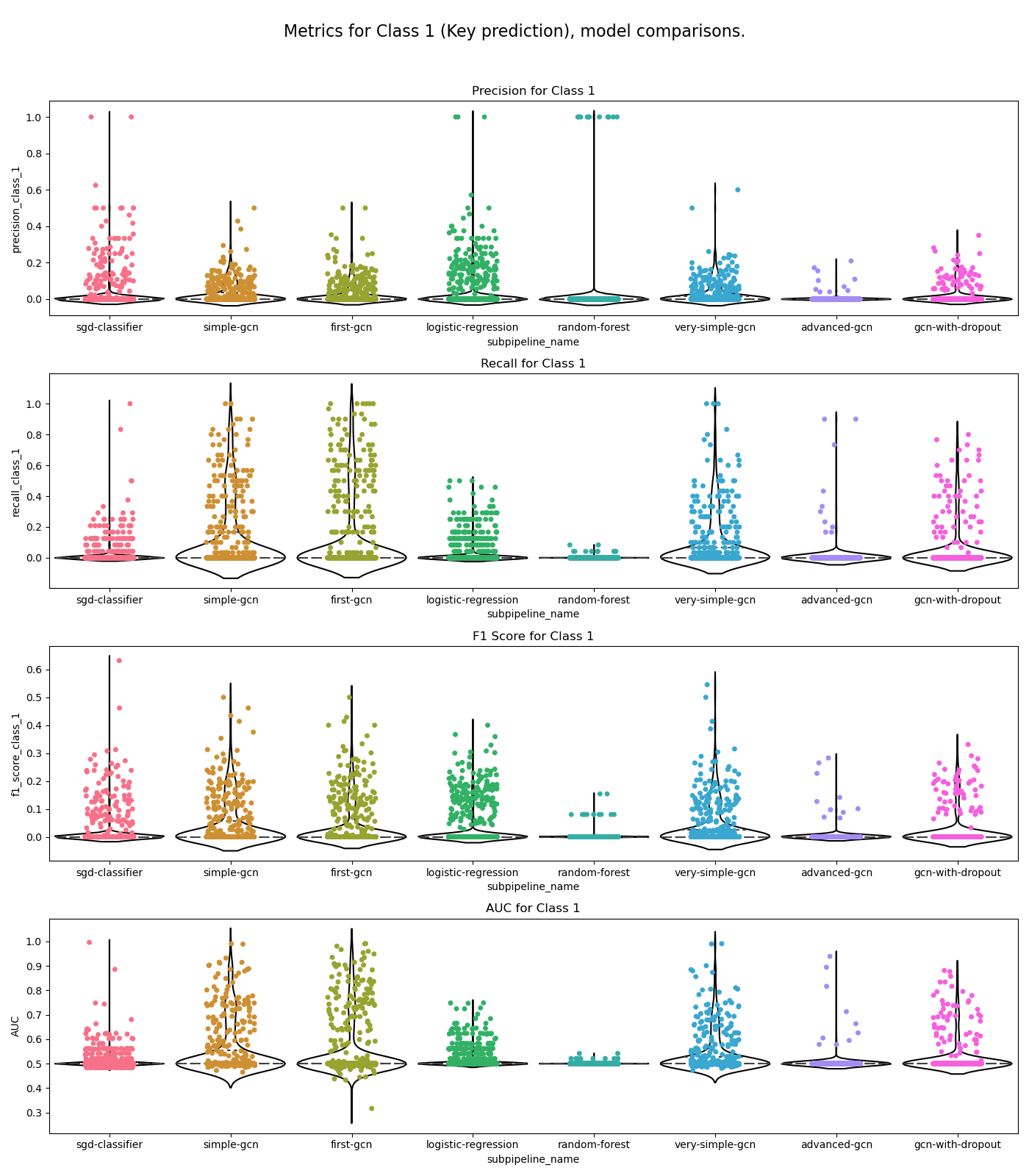}
    \caption{Visualization of the result metrics use to compare model performance on memory graphs, for different embeddings and hyperparameters.}
\end{figure}

\begin{figure}[H]\label{results:compare:embeddings:full}
    \centering
    \includegraphics[width=16cm]{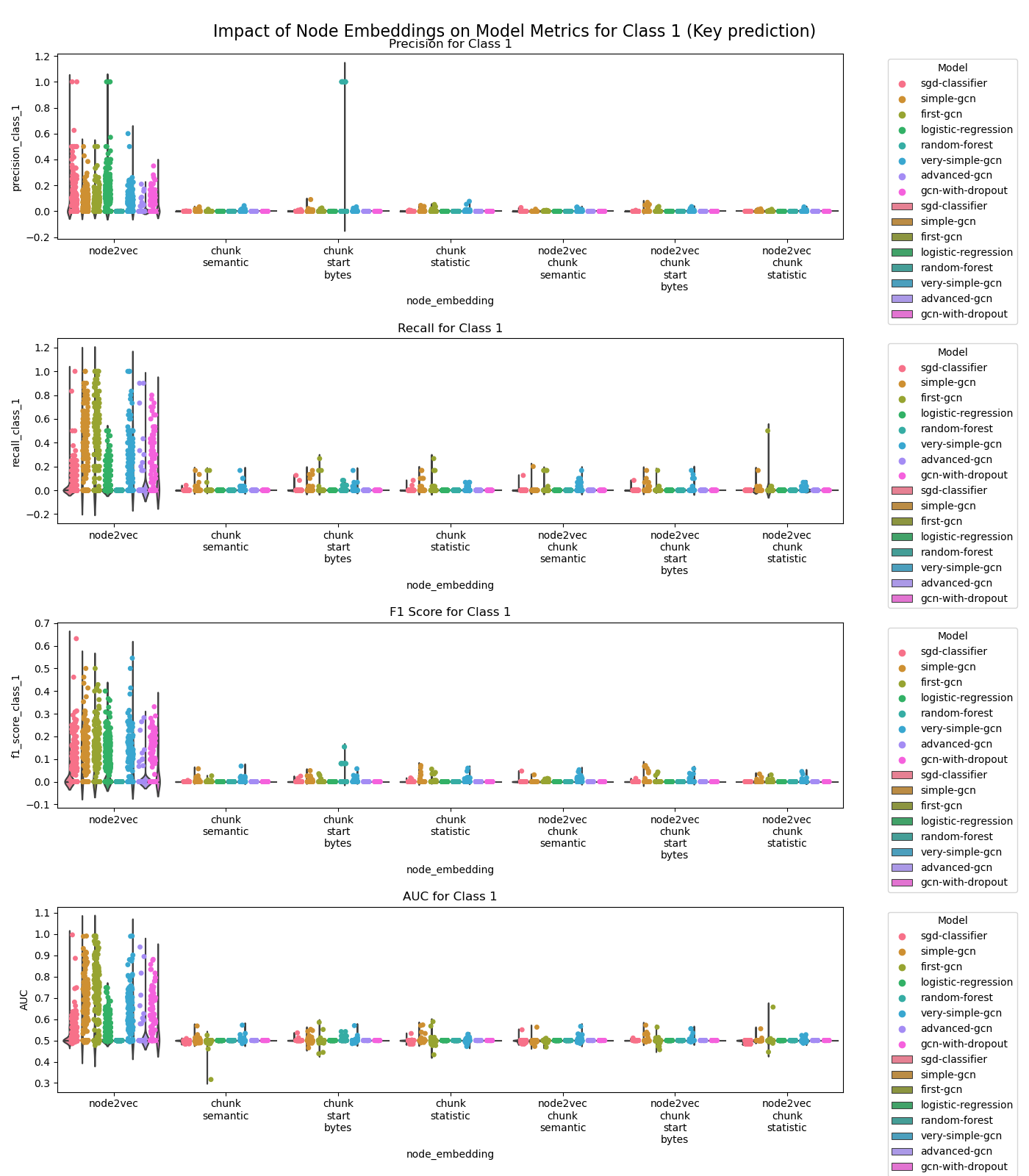}
    \caption{Visualization of the result metrics use to compare model performance per memory graph node embedding strategies.}
\end{figure}

It's worth noting that while this section provides a detailed account of the results, an in-depth discussion about these findings, their implications, and potential future work will be covered in the following "Discussions" section.

%% file: tex/chapters/discussion.tex
\chapter{Discussion}\label{chap:discussion}
In the previous chapter, the results of the experiments were presented. This chapter aims to provide an in-depth discussion of those results, as well as to identify the limitations of the experiments and to propose avenues for future research.

\section{Discussion of the results}
The following subsections will discuss the results obtained in the experiments, and will provide insights into the performance of the different models, as well as the impact of the different features and embeddings on the performance of the models.

\subsection{Objectives of the experiments}

The primary objectives of the experiments conducted in this study are multi-faceted. First and foremost, we aim to demonstrate the feasibility of utilizing machine learning and deep learning models to predict chunks with keys in the OpenSSH program based on a graph-like representation of the heap dump files provided in the original dataset. To achieve this, we employ a range of algorithms to extract features from memory graphs, or 'memgraphs'. These algorithms include not only custom solutions tailored to our specific needs but also well-established, powerful algorithms like Node2Vec. Furthermore, we seek to evaluate the impact of these diverse features on the performance metrics of the predictive models. Lastly, we compare the performances of various models to identify the most effective approaches for our specific use-case.

\subsection{Discussing features and embeddings}

In this section, we delve into the intricacies of the features and embeddings used in our experiments, focusing particularly on their interrelationships as revealed by correlation matrices. Correlation matrices provide a quantitative measure of how different features of custom embeddings relate to each other. Each cell in the matrix represents the correlation coefficient between two features, which ranges from -1 to 1. A high positive value indicates a strong positive correlation, meaning that as one feature increases, the other tends to also increase. A negative value would indicate the opposite.

It's worth noting that performing this analysis on Node2Vec embeddings is generally considered irrelevant. Node2Vec embeddings are designed to capture the neighborhood structure of nodes in a way that is optimized for machine learning tasks, and their dimensions do not have an easily interpretable meaning. Therefore, analyzing the correlation between different dimensions of a Node2Vec embedding is unlikely to provide insights that are useful for feature engineering.

To interpret the correlation matrices, we use a color-coded system where red signifies high correlation and blue signifies low or no correlation. In the context of machine learning, understanding feature correlation is crucial for several reasons:

\begin{itemize}
    \item \textbf{Feature Selection:} Highly correlated features carry redundant information, which may not only be unnecessary but can also lead to overfitting and poor generalization.
    \item \textbf{Interpretability:} Understanding how features are correlated can provide insights into the underlying structure of the data and the problem being solved.
    \item \textbf{Computational Efficiency:} Eliminating correlated features can reduce the dimensionality of the problem, making the model simpler and faster to train.
\end{itemize}

Therefore, the correlation matrices serve as a valuable tool for both feature selection and model interpretation. In that context, and looking at the correlation matrices provided for the different algorithms like Pearson, Spearman, and Kendall correlation, we can see that the correlation between the different features is generally very low, meaning that the features are not correlated to each other. This is a good thing, as it means that the features are not redundant, and that they are all bringing new information to the model. No matter the correlation algorithm used, the matrices look very similar, and the correlation between the features is very low. The only features that stands a bit from the rest are the features corresponding to the filtering and entropy. This is actually just a sign that the entropy was indeed used in the filtering algorithm, since key chunks are generally more entropic than non-key chunks. In practice, that's also why the experiments have been run with and without this filtering feature.

\subsection{Classic ML models performances}

In this subsection, we discuss the performance of three tested classical binary classification machine learning models, namely Logistic Regression, Random Forest, and SGD Classifier, in the context of key chunk prediction. The models were evaluated based on four key metrics: Precision, Recall, F1 Score, and AUC (Area Under the Curve).

\subsubsection{Logistic Regression}
The Logistic Regression model excels in precision with a perfect score of 1.0000 but falls short in recall, F1 score, and AUC. The model is highly precise but fails to capture the majority of the positive instances, as indicated by the low recall of 0.0417. This suggests that while the model makes very few false-positive errors, it misses a large number of true positives.

\subsubsection{Random Forest}
Random Forest shows excellent precision at the expense of recall. It has a high precision of 1.0000 but a very low recall of 0.0833, indicating that it is precise but not sensitive. The AUC of 0.5417 suggests that the model's ability to distinguish between the classes is slightly better than random guessing.

\subsubsection{SGD Classifier}
The SGD Classifier stands out in terms of recall and AUC, both scoring close to a perfect score This indicates that the model is excellent at identifying all the positive instances and distinguishing between the two classes. However, its precision isn't always perfect, suggesting a higher rate of false positives.

\subsubsection{Comparison}
Upon comparing the three models, it's evident that each has its own strengths and weaknesses. Logistic Regression has a much better overall recall than Random Forest which is very precise but fails to find a lot of cases. However, SGD Classifier is clearly the best overall model here, since it is highly sensitive and excellent in class separation but lacks in precision.

In summary, the SGD Classifier would be more appropriate in the binary classification task of predicting is a chunk is a key chunk or not. Random Forest offers a balanced but mediocre performance and could serve as a baseline model. Those models merely serve as a comparison point for the deep learning models.

\subsection{GCN models performances}

In this subsection, we delve into the performance metrics of five different \acrfull{gcn} models: Very Simple GCN, Simple GCN, First GCN, GCN with Dropout, and Advanced GCN. These models were evaluated on the same four key metrics as the classical models: Precision, Recall, F1 Score, and AUC.

\subsubsection{Very Simple GCN}
The Very Simple GCN model shows generally balanced performances. It's worth noting that the best instance of the model reached a perfect recall of 1.0000, indicating excellent sensitivity. However, the precision of the model is at best only of 0.6000, suggesting a higher rate of false positives. Its best instance having an AUC score of 0.9907 suggests excellent class separation capabilities.

\subsubsection{Simple GCN}
This model has similar or slightly lower performance metrics to the Very Simple GCN, with a precision and recall of 0.5000. The AUC score is slightly lower at 0.9891 but still indicates excellent class separation.

\subsubsection{First GCN}
The First GCN model also looks very similar to the two previous models. Its best AUC instance has a high AUC indicating excellent sensitivity and class separation. 

\subsubsection{GCN with Dropout}
This model has the lowest precision among the GCN models at 0.3500 but shows a decent AUC score of 0.8810. The model overall seems to be more sensitive, but less precise compare to the simpler models introduced before. This was indeed expected due to the use of dropout, which is known to increase sensitivity at the expense of precision.

\subsubsection{Advanced GCN}
The Advanced GCN model, despite its complexity, does not outperform the simpler models in this limited dataset. It displays the worst performances in precision, f1 score and AUC, but has the best recall of all the GCN models, at 0.9000. This suggests that the model is very sensitive but not very precise, which is not surprising considering the complexity of the model, and the limited number of input memgraphs. This trend is likely to change if tested with a larger number of training graphs.

\subsubsection{Layer Complexity}
It's worth noting that the simpler models (Very Simple and Simple GCN) with only 2 layers tend to perform better in this limited training dataset. This could be due to the low number of example graphs, which might not be sufficient to train the more complex, 4-layer Advanced GCN model effectively. Yet, the Advanced GCN model has the best recall of all the GCN models, suggesting that it is more sensitive than the simpler models.

In summary, each GCN model has its own set of strengths and weaknesses. While some excel in precision, others are more sensitive or better at class separation. The choice of model would depend on the specific requirements of the task at hand.

\subsection{Comparing the embeddings impact on the models}
Based on the results of the experiments, it's evident that the choice of embedding has a significant impact on the performance of the models. As this can be seen in \ref{results:compare:embeddings:full}, and surprisingly, it seems that the custom embeddings developed specifically for this research are actually not the best performing embeddings. As such, adding the features from those custom embeddings to the Node2Vec embeddings does not improve the performances of the models, but actually degrade them. This is a very surprising result. The Node2Vec embedding is actually the best performing embedding, no matter the model used, which is even more surprising, since the GCN models were actively using the graph structure, which is not the case for the classic ML models. This means that the Node2Vec embedding, which is purely based on the graph structure, is actually the best performing embedding, and that the custom embeddings are not bringing any additional information to the models. Thus, this demonstrates that the memgraph structure is in and on itself sufficiently meaningful to perform the classification task. 

It's worth mentioning the master thesis of Clément Lahoche that actually dives deeper in the question of embedding quality and impacts. His work displays significantly different results than the ones obtained in this research about embeddings. It shows that the question of embeddings is indeed a complex topic, and that the results can vary a lot depending on the dataset, the model, and the approach used.

The big difference between this present work and his thesis is that he used a much larger number of input memgraphs, but also that he actually perform a real rebalancing on the dataset before training. This means that whereas the experiments perform in this work have been dealing with a very high imbalance rate, his classifiers have been trained on a much better balanced dataset. This is a huge difference, and it's not surprising that the results are so different. The reason why this thesis did not use active rebalancing is that we wanted to explore the impact of learning on a full memgraph dataset, without alteration of those memgraphs. This is due to the fact that this work has been focused around those graphs and their equivalent in the classification phase with GCNs that rely on the graph structure. The goal was to see if the graph structure was enough to perform the classification, and the results are very promising. Results show that the imbalance ratio is not a problem for the models, as they are able to perform well despite the very high imbalance ratio.

\subsection{Comparing GCN and classical ML models}

In this subsection, we aim to compare the performance of classical machine learning models with Graph Convolutional Networks (GCNs) for the task of key chunk prediction. The comparison is based the results of the experiments, the best instances of each model on each metric, but also the distribution of the metrics as illustrated in \ref{results:compare:models:full}.

\subsubsection{Overall Performance}

For overall balanced performance, the SGD Classifier from the classical models and the Very Simple GCN from the GCN models stand out. The SGD Classifier excels in recall and AUC, making it highly sensitive and excellent in class separation, although it lacks in precision. On the other hand, the Very Simple GCN shows balanced metrics across the board. It has a decent recall and AUC, but its precision is not perfect. However, it is the best performing GCN model in terms of precision.

\subsubsection{Precision}

If precision is the primary concern, then Logistic Regression and Random Forest from the classical models and the Very Simple GCN from the GCN models are the best choices. Logistic Regression and Random Forest both have a perfect precision score of 1.0000, while the Very Simple GCN has a precision of 0.6000, which is the highest among the GCN models.

\subsubsection{Recall}

For applications where high recall is crucial, the SGD Classifier from the classical models and the Advanced GCN from the GCN models are the most suitable. The SGD Classifier has a near-perfect recall, while the Advanced GCN has the highest recall among the GCN models at 0.9000.

\subsubsection{Class Separation (AUC)}

If the ability to distinguish between classes is of utmost importance, then the SGD Classifier from the classical models and the Very Simple GCN from the GCN models are the best options. Both models have AUC scores close to 1, indicating excellent class separation capabilities, between key and non-key chunks.

\subsubsection{Considerations for Small Datasets}

It's worth noting that the simpler GCN models (Very Simple and Simple GCN) tend to perform better on the limited dataset used for training. This suggests that for small datasets, simpler models may be more effective. The Advanced GCN model, despite its complexity, does not perform as well but shows promise in terms of high recall, which could potentially improve with more training data.

\subsubsection{Summary}

In summary, the choice of model would depend on the specific metric that is most important for the task. For balanced performance, the SGD Classifier and Very Simple GCN are recommended. For high precision, Logistic Regression or Very Simple GCN should be considered. For high recall, the SGD Classifier or Advanced GCN would be the most appropriate. Finally, for excellent class separation, the SGD Classifier and Very Simple GCN are the best choices according to the experiments.

\section{Limitations of the Experiments}

While the experiments conducted offer valuable insights into the performance and capabilities of the machine learning and deep learning classifiers, it is crucial to acknowledge the limitations that were inherent in the experimental setup. These limitations range from computational resources to the scale and duration of the experiments. Understanding these constraints is essential for interpreting the results accurately and for identifying avenues for future research. This section aims to discuss the following limitations in detail:

\begin{itemize}
    \item \textbf{Number of Compute Instances:} The experiments were constrained by the available number of compute instances, affecting the scale at which they could be conducted.
    
    \item \textbf{Number of Input Graphs:} The quantity of input graphs used in the experiments was rather limited, which could impact the generalizability of the results.
    
    \item \textbf{Duration of the Experiments:} The time allocated for each experiment was finite, potentially affecting the depth of the analysis.
    
    \item \textbf{CPU-only Environment:} The experiments were conducted in a CPU-only setting, without the acceleration benefits that a GPU could offer, due to problems of memory consumption being too high for the GPU. Additional work on this aspect could significantly improve the performance of the experiments, especially the Node2Vec embedding generation.
    
    \item \textbf{High Memory Bandwidth Usage:} The experiments were characterized by high memory bandwidth usage, which could have implications for performance.
\end{itemize}

The subsequent subsections will delve into each of these limitations, providing a comprehensive understanding of their impact on the experiments.

\subsection{Limited number of input graphs}
The number of input graphs is a parameter that can be easily changed, and the experiments can be run again with a higher number of input graphs, but it would take a very long time to run, and the results would be similar although improved to the ones obtained with 16 input graphs. Improving the performances could be done essentially by recoding the program Node2Vec embedding part and adding those results directly inside the \textit{Mem2Graph} program. Leveraging the Rust zero-abstraction costs philosophy, it would be possible to improve the performances by a probable factor of 100 to 1000 times, and to run the experiments with a higher number of input graphs. For ease of development and handling of the results, I would still recommend to perform the machine learning related experiments using the powerful Python dedicated libraries.

That being said, it's remarkable that the models can perform so well considering the very small number of training memgraph and the very high imbalance ratio of the dataset. The imbalance ratio is the ratio of the number of negative samples over the number of positive samples. In the case of the dataset used in this research, the imbalance ratio is very high, ranging generally at several hundred times more negative samples than positive samples. No rebalancing has been performed on the dataset since we wanted to explore the impact of learning on a full memgraph dataset, without alteration of those memgraphs.

The results obtained in this research are already very promising, and the imbalance ratio is not a problem for the models, as they are able to perform very well despite the very high imbalance ratio.

\subsection{Duration of the experiments}

Although a lot of efforts have been put to deal both with dataset reduction, for instance transforming the initial block address prediction into a chunk address prediction problem, then optimizing a lot of computing through the use of a dedicated Rust parallel and optimized program, then using techniques like file preloading, the sheer number of hyperparameters and the number of experiments to run, as well as the compute time for the Node2Vec embedding generation, make the experiments very long to run. 

Below are the duration times for the different steps of the experiments.

\begin{table}[H]
    \centering
    \caption{Duration times for duration of embedding generation in ML/DL/FE pipeline (in seconds).}
    \begin{tabular}{lcccccc}
      \textbf{Model}  & \textbf{Min duration} & \textbf{Max duration} & \textbf{Min duration} \\
        advanced-gcn & 506.5721079074733 & 1548.909129 & 0.129933 \\
        first-gcn & 503.49931116140345 & 1548.909129 & 0.129933 \\
        gcn-with-dropout & 506.5721079074733 & 1548.909129 & 0.129933 \\
        logistic-regression & 505.3690870955711 & 1565.660571 & 0.06828 \\
        random-forest & 505.3690870955711 & 1565.660571 & 0.06828 \\
        sgd-classifier & 505.3690870955711 & 1565.660571 & 0.06828 \\
        simple-gcn & 506.5721079074733 & 1548.909129 & 0.129933 \\
        very-simple-gcn & 506.5721079074733 & 1548.909129 & 0.129933 \\
    \end{tabular}
\end{table}

Considering the tested models are not especially complex, and since the number of input memgraph stays limited, the duration of the training and testing steps is actually quite small:

\begin{table}[H]
    \centering
    \caption{Duration times for duration of training and testing in ML/DL/FE pipeline (in seconds).}
    \begin{tabular}{lcccccc}
      \textbf{Model}  & \textbf{Min duration} & \textbf{Max duration} & \textbf{Min duration} \\
        advanced-gcn & 10.843279690391459 & 496.664757 & 2.247229 \\
        first-gcn & 5.5063934491228075 & 279.007307 & 0.583914 \\
        gcn-with-dropout & 8.272014035587189 & 418.445809 & 0.905496 \\
        logistic-regression & 1.3495811305361307 & 4.165722 & 0.362695 \\
        random-forest & 11.72722453030303 & 48.723031 & 0.315739 \\
        sgd-classifier & 0.6751382750582751 & 6.405859 & 0.020952 \\
        simple-gcn & 4.337255024911032 & 163.587265 & 0.509536 \\
        very-simple-gcn & 8.188304871886121 & 58.307836 & 0.242535 \\
    \end{tabular}
\end{table}

All those values have been produced only by the python pipeline program. The embedding time is actually mostly accounting for the Node2Vec generation, since the other embeddings are already loaded in memory as they are produced and included in the outputs of Mem2Graph. The Node2Vec generation is the most time-consuming part of the pipeline, and it is the reason why the experiments take so long to run. Transferring this algorithm into Mem2Graph would be a huge improvement, and would allow to run the experiments with a much higher number of input memgraphs.

Throughout this report, we have introduced a lot of concepts, diverse algorithms, and different models. The experiments conducted in this research were limited in scope due to the focus around a large set of models, embeddings and hyperparameters which already represented a consequent amount of work and computational resources. However, there are many more avenues for future research, which will be introduced with the conclusion of this report.

%% file: tex/chapters/conclusion.tex
\chapter{Conclusion}\label{chap:conclusion}


The evolving landscape of cybersecurity necessitates robust techniques for safeguarding digital communications. OpenSSH, a pivotal element in this landscape, is a popular implementation of the Secure Shell (\acrshort{ssh}) protocol, which enables secure communication between two networked devices. The protocol is widely used in the industry, particularly in the context of remote access to servers. Using digital forensic techniques, it is possible to extract the SSH keys from memory dumps, which can then be used to decode encrypted communications thus allowing the monitoring of controlled systems. At the crux of this Masterarbeit is the development of algorithms and machine learning models to predict SSH keys within these heap dumps, focusing on using graph-like-structures and vectorization for custom embeddings. With an interdisciplinary approach that fuses traditional feature engineering with graph-based methods as well as memory modelization for inductive reasoning and learning inspired by recent developments in \acrfull{kg}s, this research not only leverages existing machine learning paradigms but also explores new avenues, such as \acrfull{gcn} applied to memory forensics. The present work also introduces a new memory forensics tool, \textit{mem2graph}, which is designed to be modular and extensible, and which can be used to generate memory graphs from memory dumps. 

\section{Summary of Results}
Below is a summary of the results achieved in the present work.

\subsection{Dataset Exploration}
A careful exploration of the dataset, and deep understanding of the original heap dumps have been invaluable in discovering patterns in the raw data. This exploration has allowed the development of a range of parsing algorithm able to extract information like structure and content of a given heap dump.

It has been discovered that the problem of finding the address of keys in the heap dump can be reduced to identifying the chunks that contain those keys. This allows to reduce the size of the problems from around 100 000 of blocks per heap dump, to around 1000 chunks per file. This also allows to concentrate the heap dump memory graph representation around the chunks.

It has also been demonstrated that two powerful chunk filtering techniques can drastically reduce the number of chunks to consider. The first filter criterion consists in the Shannon's entropy value of a chunk user data start bytes. This is because the keys are expected to have a high entropy compared to other raw data types. The second important criterion is the chunk byte size. It has been shown that key chunks actually have a small size in the range of possible key size. If filtering is not possible, as it is the case with \acrshort{gcn} models, those filters can actually be converted in powerful float and boolean features.

However, its worth noting that instead of doing active node filtering and data rebalancing, we have run the experiments and model training and evaluation on full memory graphs with high imbalance ratio. This is because we wanted to test GCN that are able to handle imbalanced data with graphs of varying size, and because we wanted to test the ability of the GCN to learn from the imbalanced data. 

\subsection{Memory Graph Generation}
This Masterarbeit has introduced a range of algorithms able to generate memory graphs from memory dumps. The algorithms are designed to be as generic as possible, and can be applied to any memory dump dataset. The algorithms are mostly implemented in the \textit{mem2graph} program, and many exploration and sanity checking scripts are also available in Python.

With those algorithms, it is possible to parse a RAW heap dump file, and transform it into a memory graph, or memgraph. This graph is a data structure, where each node represents a memory block with a precise address in the heap. Each edge represents either a pointer pointing to another block, or materialize the fact that a block belongs to a specific chunk. In order to reduce the size of the graph, it is possible to compact the block graph into a chunk graph, where each node represents a chunk, and chunks are connected through their pointers. Those kinds of graphs are only composed of Chunk Header Nodes whose address is considered the address of the related chunk. This allows to reduce the size of the graph by a factor of 10 to 100.

\subsection{Feature Engineering and Embeddings}
The memory graph can be used to extract features from the memory dump, and to apply machine learning algorithms to the memory dump. It can also be used for direct graph visualization. The memory graph serves as a direct source of embedding whether they are made manually or using readily available and tested techniques like RandomWalks or Node2Vec.

All those embeddings can be combined. The feature evaluation has shown that those features are very lowly correlated, meaning that their quality is high. However, all those different embeddings doesn't have the same results on the \acrshort{ml} and \acrshort{gcn} models, depending on the strengths and weaknesses of the different models. 

However, it's worth noting that the best results are always obtained when using the Node2Vec embedding, no matter the 
type of model used. These observations are likely to be due to the fact that Node2Vec is able to capture the structure of the graph, and that the structure of the graph is the most important feature for the models, given a relatively small number of input memgraph and considering that those memgraphs are highly imbalanced.

\subsection{Conclusion on Model Performances}

In this study, we compared the efficiency of classical machine learning models and \acrfull{gcn} in the task of key chunk prediction. Our findings indicate that the choice of model largely depends on the specific metric of interest. For a balanced performance encompassing recall, precision, and AUC, the SGD Classifier from the classical models and the Very Simple GCN from the GCN models are the most promising. 

If precision is the primary metric of concern, Logistic Regression and Random Forest from the classical models excel with perfect scores, while the Very Simple GCN leads among the GCNs. For scenarios where high recall is crucial, the SGD Classifier and the Advanced GCN model stand out. Both models also perform exceptionally well in class separation, as indicated by their high AUC scores.

It's also worth noting that simpler GCN models like the Very Simple and Simple GCN tend to perform better on limited datasets, suggesting their suitability for tasks with constrained data availability. In contrast, more complex models like the Advanced GCN show promise in terms of high recall but require more extensive training data for optimal performance.

In summary, the optimal model selection is contingent upon the specific requirements of the task, whether it be balanced performance, high precision, high recall, or excellent class separation. The choice of model also depends on the availability of training data. For instance, if the dataset is limited, a simpler GCN model like the Very Simple GCN is preferable. However, if the dataset is extensive, a more complex model like the Advanced GCN is more suitable. 

\section{Outlook on Future Work}\label{conclusion:sec:future_work}


The current report, in conjunction with the associated Masterarbeit, has introduced numerous novel algorithms and implementations. These have been instrumental in addressing the initial research questions. However, as with most research endeavors, new queries and potential avenues for enhancement have emerged, paving the way towards further exploration.

The methodologies and algorithms introduced for the OpenSSH memory dump dataset are versatile and can be extended to other memory dump datasets utilizing the GLIBC library. Given that this library is the default for Linux, adapting the methods from this Masterarbeit to other applications requires minimal effort. The \textit{mem2graph} program is inherently modular and built for extensibility. Furthermore, this tool can be employed to produce memory graphs for diverse datasets. Thanks to the universal character of the generated embeddings and memory graphs, new datasets can be readily integrated into the \acrshort{ml} and \acrshort{dl} pipelines crafted in Python. While an extensive array of features and embedding techniques have been explored in this report, there remains ample opportunity for innovative experimentation.

For a seamless fusion of machine learning into the \textit{mem2graph} program, further effort is required. Embedding machine learning immediately post-memory graph creation can substantially boost efficiency, particularly when aiming to craft a real-time OpenSSH memory forensics utility. However, this integration is challenging due to the current limited \acrshort{ml} support within Rust.

Another avenue for enhancement involves analyzing the effects of different C libraries on allocated chunks and the layout of heap dump memory. Investigating various languages could also be insightful. Depending on the level of variation encountered, modifications to the algorithms might be required, especially concerning the architecture involved in generating or extracting heap dump configurations. Pursuing this direction could significantly advance the development of a universal machine learning-assisted memory forensics tool for key extraction.

While the background section underscores the vast array of \acrshort{ml} architectures available, it's clear that not all can be thoroughly explored. This research has primarily addressed the most common and promising ones, yet numerous others await investigation. The tools crafted to bolster \acrshort{ml} pipelines present a solid foundation for such endeavors. Another dimension to consider is hyperparameter optimization. Given the constraints of time and resources, only certain parameter ranges were tested. Expanding these tests, incorporating larger datasets, and harnessing increased computational capacity can directly enhance performance.

%% file: tex/appendix.tex
\chapter*{Appendix}

\section{Code and files}

\begin{lstlisting}[style=text, caption={The DOT file of uncompressed block memgraph, here \textit{Training/basic/V\_7\_1\_P1/24/17016-1643962152-heap.raw}, with real addresses. Output is cropped.}, label={appendix:dot:17016-1643962152:cropped}]
    strict digraph "17016-1643962152" {
        "CHN(0x558343d21d40)" [label="CHN(1)" color="cyan" style=filled shape=square];
        "CHN(0x558343d1a448)" [label="CHN(2)" color="cyan" style=filled shape=square];
        "VN(0x558343d1a450)" [label="VN" color="grey" style=filled];
        "VN(0x558343d1a458)" [label="VN" color="grey" style=filled];
        "PN(0x558343d24ae8)" [label="PN" color="orange" style=filled shape=hexagon];
        "KN_KEY_A(0x558343d29460)" [label="KN(A)" color="green" style=filled];
        "KN_KEY_B(0x558343d2b960)" [label="KN(B)" color="green" style=filled];
        "CHN(0x558343d21d40)" -> "KN_KEY_A(0x558343d29460)" [label="dts" weight=1]
        "PN(0x558343d204e8)" -> "KN_KEY_A(0x558343d29460)" [label="ptr" weight=1]
        "CHN(0x558343d21d40)" -> "KN_KEY_B(0x558343d2b960)" [label="dts" weight=1]
        "PN(0x558343d2deb8)" -> "KN_KEY_B(0x558343d2b960)" [label="ptr" weight=1]
        "CHN(0x558343d21d40)" -> "KN_KEY_C(0x558343d29080)" [label="dts" weight=1]
        "PN(0x558343d204e0)" -> "KN_KEY_C(0x558343d29080)" [label="ptr" weight=1]
        "PN(0x558343d24ae8)" -> "VN(0x558343d1a010)" [label="ptr" weight=1]
        "PN(0x558343d1a240)" -> "VN(0x558343d20680)" [label="ptr" weight=1]
    }
\end{lstlisting}

\section{Memory Graphs}

\begin{figure}[H]\label{appendix:mem_graph:302-1644391327:full}
    \centering
    \includegraphics[width=16cm]{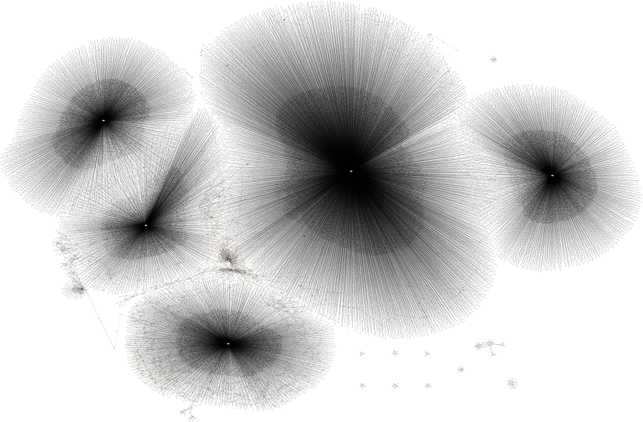}
    \caption{Visualization of the full memory graph generated from \textit{Training/scp/V\_7\_8\_P1/16/302-1644391327-heap.raw}.}
\end{figure}

\begin{figure}[H]\label{appendix:mem_graph:17016-1643962152:full}
    \centering
    \includegraphics[width=16cm]{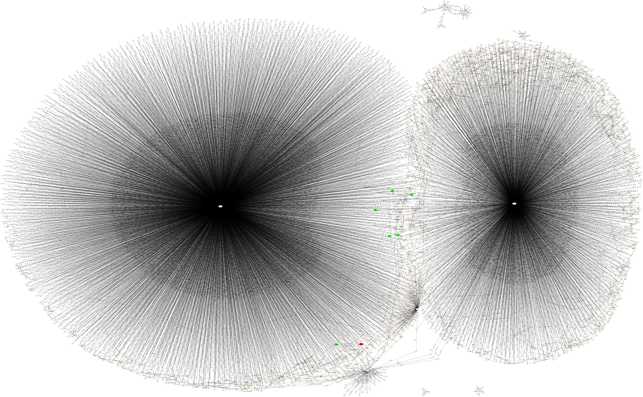}
    \caption{Visualization of the full memory graph generated from \textit{Training/basic/V\_7\_1\_P1/24/17016-1643962152-heap.raw}.}
\end{figure}

\begin{figure}[H]\label{appendix:mem_graph:17016-1643962152:truncated}
    \centering
    \includegraphics[width=16cm]{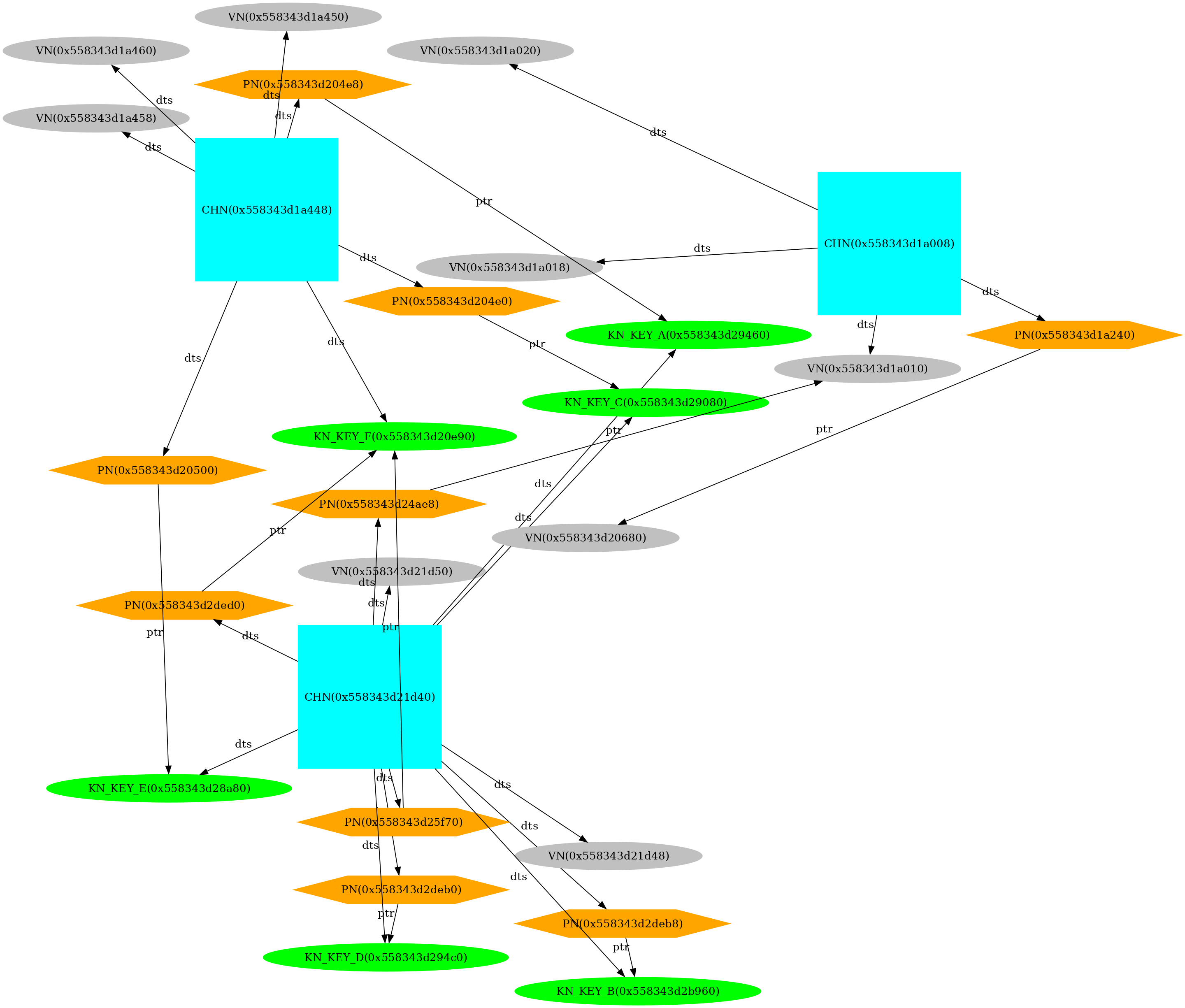}
    \caption{Visualization of a truncated memory graph generated from \textit{Training/basic/V\_7\_1\_P1/24/17016-1643962152-heap.raw}. Here with real addresses.}
\end{figure}

Generated using a slightly different command, for better layout of the nodes:

\begin{lstlisting}[language=bash, caption={Command used to generate the memory graph visualization of \textit{Training/basic/V\_7\_1\_P1/24/17016-1643962152-heap.raw} here using real addresses.}]
    sfdp -Gsize=30! -Goverlap=ortho -Tpng 17016-1643962152_truncated.gv > 17016-1643962152_truncated.png
\end{lstlisting}

\begin{figure}[H]
    \centering
    \includegraphics[width=16cm]{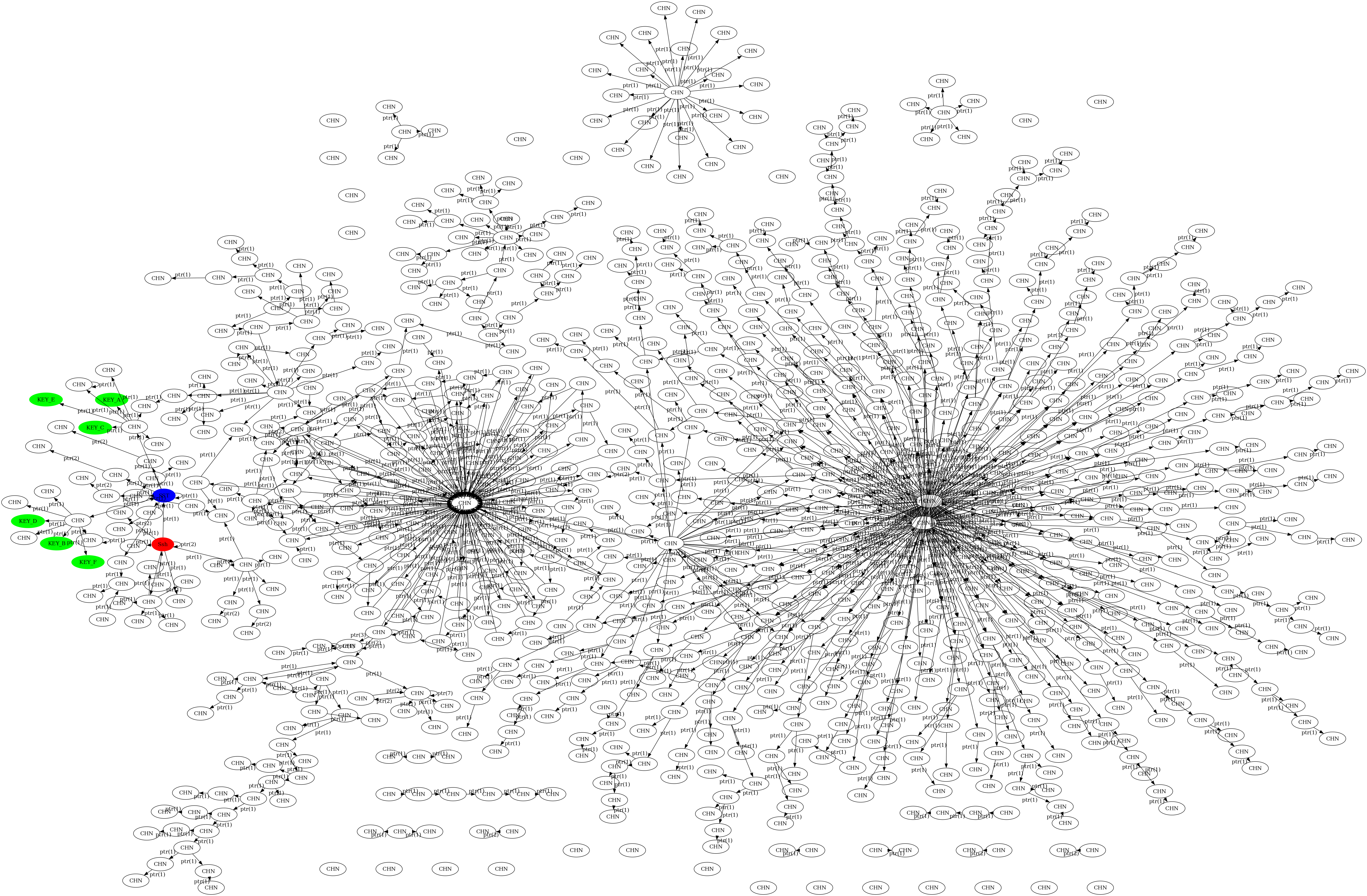}
    \caption{Visualization of a chunk memory graph generated from \textit{Validation/Validation/basic/V\_7\_8\_P1/24/8708-1643979488-heap.raw}.}
\end{figure}

\begin{figure}[H]
    \centering
    \includegraphics[width=16cm]{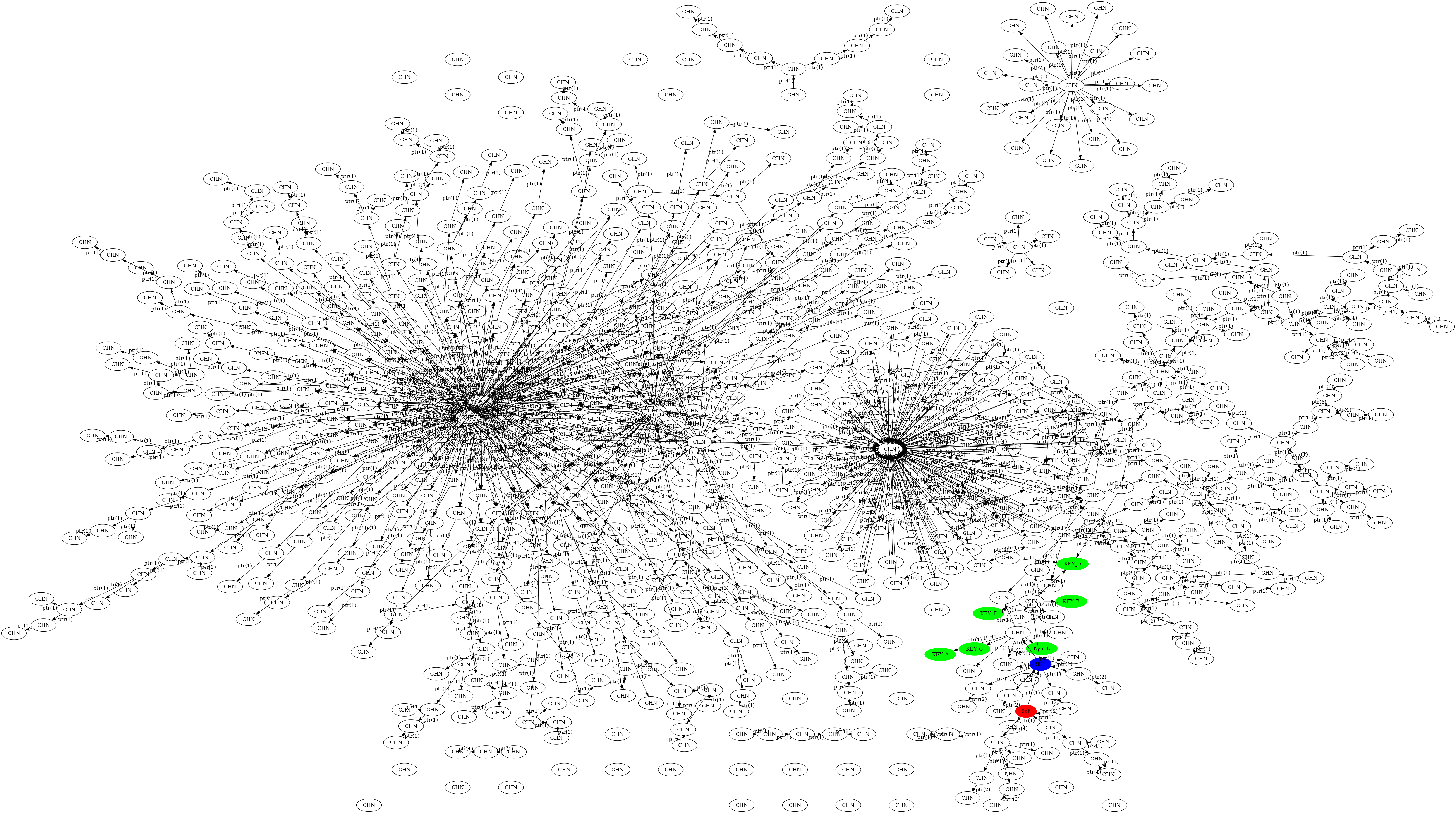}
    \caption{Visualization of a chunk memory graph generated from \textit{Training/Training/basic/V\_6\_8\_P1/24/28621-1643890740-heap.raw}.}
\end{figure}

\section{Latest ML Experiment results}
The following are the latest results for ML experiments. Those tables were generated after the submission of the thesis.

\begin{table}[H]
    \centering
    \caption{Best instance for each model, with respect to accuracy.}
    \begin{tabular}{|l|c|c|c|c|c|c|} \hline 
      \textbf{Model}  & \textbf{Accuracy} & \textbf{Precision} & \textbf{Recall} & \textbf{F1 Score} & \textbf{Embedding}  \\ \hline 
        random-forest & 0.9984 & 1.0000 & 0.0833 & 0.1538 & chunk-start-bytes \\ \hline 
        sgd-classifier & 0.9984 & 0.6250 & 0.2083 & 0.3125 & node2vec \\ \hline 
        logistic-regression & 0.9983 & 1.0000 & 0.0417 & 0.0800 & node2vec \\ \hline 
        advanced-gcn & 0.9969 & 0.0000 & 0.0000 & 0.0000 & node2vec \\ \hline 
        gcn-with-dropout & 0.9969 & 0.0000 & 0.0000 & 0.0000 & node2vec \\ \hline 
        first-gcn & 0.9963 & 0.0000 & 0.0000 & 0.0000 & node2vec-chunk-statistic \\ \hline 
        simple-gcn & 0.9962 & 0.0000 & 0.0000 & 0.0000 & chunk-statistic \\ \hline 
        very-simple-gcn & 0.9959 & 0.0000 & 0.0000 & 0.0000 & node2vec \\ \hline 
    \end{tabular}
\end{table}

\begin{table}[H]
    \centering
    \caption{Best instance for each model, with respect to precision.}
    \begin{tabular}{|l|c|c|c|c|c|c|} \hline 
      \textbf{Model}  & \textbf{Accuracy} & \textbf{Precision} & \textbf{Recall} & \textbf{F1 Score} & \textbf{Embedding}  \\ \hline 
        logistic-regression & 0.9944 & 1.0000 & 0.0417 & 0.0800 & node2vec \\ \hline 
        random-forest & 0.9983 & 1.0000 & 0.0417 & 0.0800 & chunk-start-bytes \\ \hline 
        sgd-classifier & 0.9983 & 1.0000 & 0.0417 & 0.0800 & node2vec \\ \hline 
        very-simple-gcn & 0.9946 & 0.6000 & 0.5000 & 0.5455 & node2vec \\ \hline 
        first-gcn & 0.9935 & 0.5000 & 0.5000 & 0.5000 & node2vec \\ \hline 
        simple-gcn & 0.9935 & 0.5000 & 0.5000 & 0.5000 & node2vec \\ \hline 
        gcn-with-dropout & 0.9920 & 0.3500 & 0.2333 & 0.2800 & node2vec \\ \hline 
        advanced-gcn & 0.9898 & 0.2097 & 0.4333 & 0.2826 & node2vec \\ \hline 
    \end{tabular}
\end{table}

\begin{table}[H]
    \centering
    \caption{Best instance for each model, with respect to recall.}
    \begin{tabular}{|l|c|c|c|c|c|c|} \hline 
      \textbf{Model}  & \textbf{Accuracy} & \textbf{Precision} & \textbf{Recall} & \textbf{F1 Score} & \textbf{Embedding}  \\ \hline 
        first-gcn & 0.9826 & 0.2727 & 1.0000 & 0.4286 & node2vec \\ \hline 
        sgd-classifier & 0.9924 & 0.4615 & 1.0000 & 0.6316 & node2vec \\ \hline 
        simple-gcn & 0.9783 & 0.2308 & 1.0000 & 0.3750 & node2vec \\ \hline 
        very-simple-gcn & 0.9815 & 0.2609 & 1.0000 & 0.4138 & node2vec \\ \hline 
        advanced-gcn & 0.8887 & 0.0533 & 0.9000 & 0.1006 & node2vec \\ \hline 
        gcn-with-dropout & 0.9613 & 0.0863 & 0.8000 & 0.1558 & node2vec \\ \hline 
        logistic-regression & 0.9912 & 0.3333 & 0.5000 & 0.4000 & node2vec \\ \hline 
        random-forest & 0.9984 & 1.0000 & 0.0833 & 0.1538 & chunk-start-bytes \\ \hline 
    \end{tabular}
\end{table}

\begin{table}[H]
    \centering
    \caption{Best instance for each model, with respect to f1 score.}
    \begin{tabular}{|l|c|c|c|c|c|c|} \hline 
      \textbf{Model}  & \textbf{Accuracy} & \textbf{Precision} & \textbf{Recall} & \textbf{F1 Score} & \textbf{Embedding}  \\ \hline 
        sgd-classifier & 0.9924 & 0.4615 & 1.0000 & 0.6316 & node2vec \\ \hline 
        very-simple-gcn & 0.9946 & 0.6000 & 0.5000 & 0.5455 & node2vec \\ \hline 
        first-gcn & 0.9935 & 0.5000 & 0.5000 & 0.5000 & node2vec \\ \hline 
        simple-gcn & 0.9935 & 0.5000 & 0.5000 & 0.5000 & node2vec \\ \hline 
        logistic-regression & 0.9912 & 0.3333 & 0.5000 & 0.4000 & node2vec \\ \hline 
        gcn-with-dropout & 0.9858 & 0.2110 & 0.7667 & 0.3309 & node2vec \\ \hline 
        advanced-gcn & 0.9898 & 0.2097 & 0.4333 & 0.2826 & node2vec \\ \hline 
        random-forest & 0.9984 & 1.0000 & 0.0833 & 0.1538 & chunk-start-bytes \\ \hline 
    \end{tabular}
\end{table}

\begin{table}
    \centering
    \caption{Best model instance for each metric.}
    \begin{tabular}{|l|c|c|c|c|c|c|} \hline 
      \textbf{Metric} & \textbf{Model}  & \textbf{Accuracy} & \textbf{Precision} & \textbf{Recall} & \textbf{F1 Score} & \textbf{Embedding}  \\ \hline 
        accuracy & random-forest & 0.9984 & 1.0000 & 0.0833 & 0.1538 & chunk-start-bytes \\ \hline 
        precision & logistic-regression & 0.9944 & 1.0000 & 0.0417 & 0.0800 & node2vec \\ \hline 
        recall & first-gcn & 0.9826 & 0.2727 & 1.0000 & 0.4286 & node2vec \\ \hline 
        f1 score & sgd-classifier & 0.9924 & 0.4615 & 1.0000 & 0.6316 & node2vec \\ \hline 
    \end{tabular}
\end{table}

%% file: tex/german_affidavit.tex

\section*{Eidesstattliche Erkl\"arung}

	Hiermit versichere ich, dass ich diese Masterarbreit selbstst\"andig und ohne Benutzung anderer als der angegebenen Quellen und Hilfsmittel angefertigt habe und alle Ausf\"uhrungen, die w\"ortlich oder sinngem\"a\ss{} übernommen wurden, als solche gekennzeichnet sind, sowie, dass ich die Masterarbreit ~in gleicher oder \"ahnlicher Form noch keiner anderen Pr\"ufungsbeh\"orde vorgelegt habe.

	\vspace{3cm}

	Passau, \today

	\vspace{2cm}

	\parbox{8cm}{
		\hrule \strut \theauthor
	}

%% file: biblio.bib
@article{SSHkex22, 
  title={SSHkex: Leveraging virtual machine introspection for extracting SSH keys and decrypting SSH network traffic}, 
  volume={40}, 
  url={https://linkinghub.elsevier.com/retrieve/pii/S2666281722000063}, DOI={10.1016/j.fsidi.2022.301337}, 
  journal={Forensic Science International: Digital Investigation}, 
  author={Sentanoe, Stewart and Reiser, Hans P.}, 
  year={2022}, 
  pages={301337}, 
  language={en} 
}

@online{ClementEmbeddingsMasterarbeit23,
  title={Structure embeddings for OpenSSH heap dump analysis},
  author={Clement Lahoche},
  year={2023},
  url={https://github.com/passau-masterarbeit-2023/masterarbeit-report-clement},
  language={en},
  note={Available online},
  urldate={2023-11-01}
}

@dataset{CleanedZenodoDataset2022,
  author       = {Rascoussier, Florian O. and Lahoche, C.},
  title        = {{Machine Learning Assisted SSH Keys Extraction From The Heap Dump | Cleaned for Embedding}},
  version      = {0.1.1},
  publisher    = {Zenodo},
  year         = {2022},
  doi          = {10.5281/zenodo.10514199},
  url          = {https://doi.org/10.5281/zenodo.10514199}
}

@misc{PyTorchGeometric19,
      title={Fast Graph Representation Learning with PyTorch Geometric}, 
      author={Matthias Fey and Jan Eric Lenssen},
      year={2019},
      eprint={1903.02428},
      archivePrefix={arXiv},
      primaryClass={cs.LG}
}

@online{DotFormat15,
  title={Drawing graphs with dot},
  author={Gansner, Emden R. and Koutsofios, Eleftherios and North, Stephen},
  year={2015},
  month={1},
  day={5},
  url={https://graphviz.org/pdf/dotguide.pdf},
  note={Accessed: 2023-10-27}
}

@online{MallocGLIBC2001,
  author       = {Wolfram Gloger and Doug Lea},
  title        = {Malloc implementation for multiple threads without lock contention},
  year         = {2001},
  note         = {Version ptmalloc2-20011215, based on VERSION 2.7.0, Sun Mar 11 14:14:06 2001 by Doug Lea},
  url          = {https://elixir.bootlin.com/glibc/glibc-2.28/source/malloc/malloc.c},
  urldate      = {2023-09-22},
  organization = {Free Software Foundation, Inc.},
  copyright    = {Copyright (C) 1996-2018 Free Software Foundation, Inc.},
  note         = {Accessed: 2023-09-22}
}

@online{MallocInternalsWiki2023,
  author       = {DJ Delorie and Florian Weimer and Carlos O'Donell and Andreas Schwab},
  title        = {Malloc Internals},
  year         = {2023},
  url          = {https://sourceware.org/glibc/wiki/MallocInternals},
  urldate      = {2023-09-25},
  note         = {Commit: b39f275c, Accessed: 2023-09-25},
  organization = {Sourceware},
  note         = {Accessed: 2023-09-22}
}

@online{StackExchangeMalloc2023,
  author       = {Unknown},
  title        = {How does glibc malloc work?},
  year         = {2023},
  url          = {https://reverseengineering.stackexchange.com/questions/15033/how-does-glibc-malloc-work/15038#15038},
  note         = {Asked 6 years, 6 months ago; Modified 3 years, 6 months ago; Viewed 11k times; Accessed: 2023-09-25}
}

@online{OrderedDataInRDF20,
  author       = {Joep Meindertma},
  title        = {Ordered data in RDF: About Arrays, Lists, Collections, Sequences and Pagination},
  year         = {2020},
  date         = {2020-02-07},
  url          = {https://ontola.io/blog/ordered-data-in-rdf},
  urldate      = {2023-09-22},
  note         = {Accessed: 2023-09-22}
}

@article{SmartKex22,
  title        = {SmartKex: Machine Learning Assisted SSH Keys Extraction From The Heap Dump},
  url          = {http://arxiv.org/abs/2209.05243},
  DOI          = {10.48550/arXiv.2209.05243},
  abstractNote = {Digital forensics is the process of extracting, preserving, and documenting evidence in digital devices. A commonly used method in digital forensics is to extract data from the main memory of a digital device. However, the main challenge is identifying the important data to be extracted. Several pieces of crucial information reside in the main memory, like usernames, passwords, and cryptographic keys such as SSH session keys. In this paper, we propose SmartKex, a machine-learning assisted method to extract session keys from heap memory snapshots of an OpenSSH process. In addition, we release an openly available dataset and the corresponding toolchain for creating additional data. Finally, we compare SmartKex with naive brute-force methods and empirically show that SmartKex can extract the session keys with high accuracy and high throughput. With the provided resources, we intend to strengthen the research on the intersection between digital forensics, cybersecurity, and machine learning.},
  note         = {arXiv:2209.05243 [cs]},
  number       = {arXiv:2209.05243},
  publisher    = {arXiv},
  author       = {Fellicious, Christofer and Sentanoe, Stewart and Granitzer, Michael and Reiser, Hans P.},
  year         = {2022},
  month        = {09}
}

@techreport{RFC9142,
  title={Key Exchange (KEX) Method Updates and Recommendations for Secure Shell (SSH)},
  author={M. Baushke},
  year={2022},
  month={01},
  institution={Internet Engineering Task Force (IETF)},
  type={RFC},
  number={9142},
  note={Updates: 4250, 4253, 4432, 4462; Errata exist},
  category={Standards Track},
  issn={2070-1721}
}

@article{NSAFoilSafeguards2013,
  title={N.S.A. Able to Foil Basic Safeguards of Privacy on Web},
  author={Nicole Perlroth and Jeff Larson and Scott Shane},
  journal={The New York Times},
  year={2013},
  month={09},
  day={5},
  url={https://www.nytimes.com/2013/09/06/us/nsa-foils-much-internet-encryption.html}
}

@article{GuardianEncryption2013,
  title={Revealed: how US and UK spy agencies defeat internet privacy and security},
  author={James Ball and Julian Borger and Glenn Greenwald},
  journal={The Guardian},
  year={2013},
  month={09},
  day={6},
  url={https://www.theguardian.com/world/2013/sep/05/nsa-gchq-encryption-codes-security},
  note={Published at 11.24 BST}
}

@online{Adamantiadis2013,
  title={OpenSSH introduces curve25519-sha256@libssh.org key exchange !},
  author={Aris Adamantiadis},
  year={2013},
  month={11},
  day={3},
  url={https://www.libssh.org/2013/11/03/openssh-introduces-curve25519-sha256libssh-org-key-exchange/},
  note={Retrieved 2023-09-05},
  publisher={libssh.org}
}

@online{OpenSSHReleaseNotes5-7,
  title={OpenSSH 5.7 release notes},
  author={OpenSSH},
  year={2011},
  month={01},
  day={24},
  note={Retrieved 2022-11-13},
  publisher={OpenSSH},
  url={https://www.openssh.com/txt/release-5.7}
}

@inproceedings{SarraceniaSSHHoneypot18,
  title={Sarracenia: Enhancing the Performance and Stealthiness of SSH Honeypots Using Virtual Machine Introspection},
  author={Stewart Sentanoe and Benjamin Taubmann and Hans P. Reiser},
  booktitle={Lecture Notes in Computer Science},
  volume={11252},
  year={2018},
  month={11},
  day={02},
  url={https://link.springer.com/chapter/10.1007/978-3-030-03638-6_16},
  note= {Conference paper}
}

@online{OpenSSHReleaseNotes6-5,
  title={OpenSSH 6.5 release notes},
  author={OpenSSH},
  year={2014},
  month={01},
  day={29},
  note={Retrieved 2022-11-13},
  publisher={OpenSSH},
  url={https://www.openssh.com/txt/release-6.5}
}

@online{OpenSSHReleaseNotes7-0,
  title={OpenSSH 7.0 release notes},
  author={OpenSSH},
  year={2015},
  month={08},
  day={11},
  note={Retrieved 2022-11-13},
  publisher={OpenSSH},
  url={https://www.openssh.com/txt/release-7.0}
}

@article{GraphTheorySolnon,
  title={Th{\'e}orie des graphes et optimisation dans les graphes},
  author={Solnon, Christine},
  journal={INSA de Lyon}
}

@book{GraphTheoryIntro01,
  title={Introduction to graph theory},
  author={West, Douglas Brent and others},
  volume={2},
  year={2001},
  publisher={Prentice hall Upper Saddle River}
}

@online{OpenSSHReleaseNotes7-2,
  title={OpenSSH 7.2 release notes},
  author={OpenSSH},
  year={2016},
  month={01},
  day={29},
  note={Retrieved 2022-11-13},
  publisher={OpenSSH},
  url={https://www.openssh.com/txt/release-7.2}
}

@online{OpenSSHReleaseNotes8-2,
  title={OpenSSH 8.2 release notes},
  author={OpenSSH},
  year={2020},
  month={02},
  day={14},
  note={Retrieved 2022-11-13},
  publisher={OpenSSH},
  url={https://www.openssh.com/txt/release-8.2}
}

@online{OpenSSHReleaseNotes8-8,
  title={OpenSSH 8.8 release notes},
  author={OpenSSH},
  year={2021},
  month={09},
  day={26},
  note={Retrieved 2022-11-13},
  publisher={OpenSSH},
  url={https://www.openssh.com/txt/release-8.2}
}

@inproceedings{FeatureSelecExtract14,
	title        = {A survey of feature selection and feature extraction techniques in machine learning},
	author       = {Khalid, Samina and Khalil, Tehmina and Nasreen, Shamila},
	year         = 2014,
	month        = {08},
	booktitle    = {2014 Science and Information Conference},
	pages        = {372-378},
	doi          = {10.1109/SAI.2014.6918213},
	abstractnote = {Dimensionality reduction as a preprocessing step to machine learning is effective in removing irrelevant and redundant data, increasing learning accuracy, and improving result comprehensibility. However, the recent increase of dimensionality of data poses a severe challenge to many existing feature selection and feature extraction methods with respect to efficiency and effectiveness. In the field of machine learning and pattern recognition, dimensionality reduction is important area, where many approaches have been proposed. In this paper, some widely used feature selection and feature extraction techniques have analyzed with the purpose of how effectively these techniques can be used to achieve high performance of learning algorithms that ultimately improves predictive accuracy of classifier. An endeavor to analyze dimensionality reduction techniques briefly with the purpose to investigate strengths and weaknesses of some widely used dimensionality reduction methods is presented.}
}

@book{FeatureEngineeringMadeEasy18,
  title={Feature Engineering Made Easy: Identify unique features from your dataset in order to build powerful machine learning systems},
  author={Ozdemir, Sinan and Susarla, Divya},
  year={2018},
  publisher={Packt Publishing Ltd}
}

@article{TheoryOfCommunicationShannon1948,
	title        = {A Mathematical Theory of Communication},
	author       = {Shannon, C E},
	year         = 1948,
	month        = {10},
	journal      = {The Bell System Technical Journal},
	volume       = 27,
	pages        = {379-423},
	language     = {en}
}

@article{ScikitLearn,
  title={Scikit-learn: Machine Learning in {P}ython},
  author={Pedregosa, F. and Varoquaux, G. and Gramfort, A. and Michel, V.
          and Thirion, B. and Grisel, O. and Blondel, M. and Prettenhofer, P.
          and Weiss, R. and Dubourg, V. and Vanderplas, J. and Passos, A. and
          Cournapeau, D. and Brucher, M. and Perrot, M. and Duchesnay, E.},
  journal={Journal of Machine Learning Research},
  volume={12},
  pages={2825--2830},
  year={2011},
  note={Software available at https://scikit-learn.org/stable/index.html}
}

@book{StatisticalMethodsInPractice09,
  title={Statistical methods in practice: for scientists and technologists},
  author={Boddy, Richard and Smith, Gordon},
  year={2009},
  publisher={John Wiley \& Sons}
}

@article{ScienceMachineLearning15,
  title={Machine learning: Trends, perspectives, and prospects},
  author={Jordan, Michael I and Mitchell, Tom M},
  journal={Science},
  volume={349},
  number={6245},
  pages={255--260},
  year={2015},
  publisher={American Association for the Advancement of Science},
  url={https://www.science.org/doi/full/10.1126/science.aaa8415}
}

@article{nick_logistic_2007,
	title = {Logistic regression},
	pages = {273--301},
	journaltitle = {Topics in biostatistics},
	author = {Nick, Todd G and Campbell, Kathleen M},
	date = {2007},
	note = {Publisher: Springer}
}

@article{kotsiantis_decision_2013,
	title = {Decision trees: a recent overview},
	volume = {39},
	issn = {0269-2821, 1573-7462},
	url = {http://link.springer.com/10.1007/s10462-011-9272-4},
	doi = {10.1007/s10462-011-9272-4},
	shorttitle = {Decision trees},
	pages = {261--283},
	number = {4},
	journaltitle = {Artificial Intelligence Review},
	shortjournal = {Artif Intell Rev},
	author = {Kotsiantis, S. B.},
	urldate = {2023-08-30},
	date = {2013-04},
	langid = {english}
}

@article{probst_hyperparameters_2019,
	title = {Hyperparameters and Tuning Strategies for Random Forest},
	volume = {9},
	issn = {1942-4787, 1942-4795},
	url = {http://arxiv.org/abs/1804.03515},
	doi = {10.1002/widm.1301},
	pages = {e1301},
	number = {3},
	journaltitle = {{WIREs} Data Mining and Knowledge Discovery},
	shortjournal = {{WIREs} Data Min \& Knowl},
	author = {Probst, Philipp and Wright, Marvin and Boulesteix, Anne-Laure},
	urldate = {2023-08-30},
	date = {2019-05},
	langid = {english},
	eprinttype = {arxiv},
	eprint = {1804.03515 [cs, stat]}
}

@article{lecun_gradient_based_1998,
	title = {Gradient-Based Learning Applied to Document Recognition},
	journaltitle = {proc of the {IEEE}},
	author = {{LeCun}, Yann and Bottou, Leon and Bengio, Yoshua and Ha, Patrick},
	date = {1998},
	langid = {english}
}

@inproceedings{ConvKB18,
	title        = {A Novel Embedding Model for Knowledge Base Completion Based on Convolutional Neural Network},
	author       = {Dai Quoc Nguyen and Tu Dinh Nguyen and Dat Quoc Nguyen and Dinh Phung},
	year         = {2018},
	booktitle    = {Proceedings of the 2018 Conference of the North American Chapter of the Association for Computational Linguistics: Human Language Technologies, Volume 2 (Short Papers)},
	publisher    = {Association for Computational Linguistics},
	doi          = {10.18653/v1/n18-2053},
	url          = {https://doi.org/10.18653%2Fv1%2Fn18-2053}
}

@inproceedings{SDNE16,
	title        = {Structural Deep Network Embedding},
	author       = {Wang, Daixin and Cui, Peng and Zhu, Wenwu},
	year         = {2016},
	booktitle    = {Proceedings of the 22nd ACM SIGKDD International Conference on Knowledge Discovery and Data Mining},
	location     = {San Francisco, California, USA},
	publisher    = {Association for Computing Machinery},
	address      = {New York, NY, USA},
	series       = {KDD '16},
	pages        = {1225-1234},
	doi          = {10.1145/2939672.2939753},
	isbn         = 9781450342322,
	url          = {https://doi.org/10.1145/2939672.2939753},
	numpages     = 10
}

@inproceedings{RGCN18,
  title={Modeling relational data with graph convolutional networks},
  author={Schlichtkrull, Michael and Kipf, Thomas N and Bloem, Peter and Van Den Berg, Rianne and Titov, Ivan and Welling, Max},
  booktitle={The Semantic Web: 15th International Conference, ESWC 2018, Heraklion, Crete, Greece, June 3--7, 2018, Proceedings 15},
  pages={593--607},
  year={2018},
  organization={Springer},
  url={https://arxiv.org/pdf/1703.06103.pdf}
}

@article{KGBERT19,
  title={KG-BERT: BERT for knowledge graph completion},
  author={Yao, Liang and Mao, Chengsheng and Luo, Yuan},
  journal={arXiv preprint arXiv:1909.03193},
  year={2019},
  url={https://arxiv.org/pdf/1909.03193.pdf}
}

@article{ComOverInsecureChannels78,
  author = {Merkle, Ralph C.},
  title = {Secure Communications over Insecure Channels},
  year = {1978},
  issue_date = {April 1978},
  publisher = {Association for Computing Machinery},
  address = {New York, NY, USA},
  volume = {21},
  number = {4},
  issn = {0001-0782},
  url = {https://doi.org/10.1145/359460.359473},
  doi = {10.1145/359460.359473},
  journal = {Commun. ACM},
  month = {04},
  pages = {294-299},
  numpages = {6}
}

@inproceedings{KGBART21,
  title={Kg-bart: Knowledge graph-augmented bart for generative commonsense reasoning},
  author={Liu, Ye and Wan, Yao and He, Lifang and Peng, Hao and Philip, S Yu},
  booktitle={Proceedings of the AAAI Conference on Artificial Intelligence},
  volume={35},
  number={7},
  pages={6418--6425},
  year={2021},
  url={file:///home/onyr/Downloads/16796-Article%20Text-20290-1-2-20210518.pdf}
}

@inproceedings{Node2vec16,
  title={node2vec: Scalable feature learning for networks},
  author={Grover, Aditya and Leskovec, Jure},
  booktitle={Proceedings of the 22nd ACM SIGKDD international conference on Knowledge discovery and data mining},
  pages={855--864},
  year={2016},
  url={https://dl.acm.org/doi/pdf/10.1145/2939672.2939754}
}

@inproceedings{LINEEmbedding15,
  title={Line: Large-scale information network embedding},
  author={Tang, Jian and Qu, Meng and Wang, Mingzhe and Zhang, Ming and Yan, Jun and Mei, Qiaozhu},
  booktitle={Proceedings of the 24th international conference on world wide web},
  pages={1067--1077},
  year={2015},
  url={https://dl.acm.org/doi/pdf/10.1145/2736277.2741093}
}

@article{GAT17,
  title={Graph attention networks},
  author={Velickovic, Petar and Cucurull, Guillem and Casanova, Arantxa and Romero, Adriana and Lio, Pietro and Bengio, Yoshua and others},
  journal={stat},
  volume={1050},
  number={20},
  pages={10--48550},
  year={2017},
  url={https://personal.utdallas.edu/~fxc190007/courses/20S-7301/GAT-questions.pdf}
}

@article{CNNIntro15,
  author       = {Keiron O'Shea and
                  Ryan Nash},
  title        = {An Introduction to Convolutional Neural Networks},
  journal      = {CoRR},
  volume       = {abs/1511.08458},
  year         = {2015},
  url          = {http://arxiv.org/abs/1511.08458},
  eprinttype    = {arXiv},
  eprint       = {1511.08458},
  bibsource    = {dblp computer science bibliography, https://dblp.org}
}

@book{NixOriginalThesis06,
  title={The purely functional software deployment model},
  author={Dolstra, Eelco},
  year={2006},
  publisher={Utrecht University},
  url={https://edolstra.github.io/pubs/phd-thesis.pdf}
}

@online{WasIstOpenScience23,
  author       = {OpenScience ASAP},
  title        = {Was ist Open Science?},
  year         = {2023},
  url          = {http://openscienceasap.org/open-science/},
  urldate      = {2023-09-19},
  note         = {Accessed: 2023/09/19}
}

@article{WhyNotShareData22,
  author      = {Gomes, DGE and Pottier, P and Crystal-Ornelas, R and Hudgins, EJ and Foroughirad, V and Sánchez-Reyes, LL and Turba, R and Martinez, PA and Moreau, D and Bertram, MG and Smout, CA and Gaynor, KM},
  title       = {Why don't we share data and code? Perceived barriers and benefits to public archiving practices},
  journal     = {Proc Biol Sci},
  year        = {2022},
  volume      = {289},
  number      = {1987},
  pages       = {20221113},
  doi         = {10.1098/rspb.2022.1113},
  date        = {2022-11-30},
  eprint      = {Epub 2022 Nov 23},
  pmid        = {36416041},
  pmcid       = {PMC9682438}
}

@article{NixOS08,
  author = {Dolstra, Eelco and L\"{o}h, Andres},
  title = {NixOS: A Purely Functional Linux Distribution},
  year = {2008},
  issue_date = {September 2008},
  publisher = {Association for Computing Machinery},
  address = {New York, NY, USA},
  volume = {43},
  number = {9},
  issn = {0362-1340},
  url = {https://doi.org/10.1145/1411203.1411255},
  doi = {10.1145/1411203.1411255},
  journal = {SIGPLAN Not.},
  month = {09},
  pages = {367-378},
  numpages = {12}
}

@article{ImageConvolution13,
  title={Image convolution},
  author={Ludwig, Jamie},
  journal={Portland State University},
  year={2013},
  url={https://web.pdx.edu/~jduh/courses/Archive/geog481w07/Students/Ludwig_ImageConvolution.pdf}
}

@article{RGCN22,
  title={R-GCN: the R could stand for random},
  author={Degraeve, Vic and Vandewiele, Gilles and Ongenae, Femke and Van Hoecke, Sofie},
  journal={arXiv:2203.02424 preprint},
  year={2022},
  url={https://arxiv.org/pdf/2203.02424.pdf}
}

@inproceedings{Deepwalk14,
  title={Deepwalk: Online learning of social representations},
  author={Perozzi, Bryan and Al-Rfou, Rami and Skiena, Steven},
  booktitle={Proceedings of the 20th ACM SIGKDD international conference on Knowledge discovery and data mining},
  pages={701--710},
  year={2014},
  url={https://dl.acm.org/doi/pdf/10.1145/2623330.2623732}
}

@article{GraphSAGE17,
  title={Inductive representation learning on large graphs},
  author={Hamilton, Will and Ying, Zhitao and Leskovec, Jure},
  journal={Advances in neural information processing systems},
  editor = {I. Guyon and U. Von Luxburg and S. Bengio and H. Wallach and R. Fergus and S. Vishwanathan and R. Garnett},
  volume={30},
  year={2017},
  url={https://proceedings.neurips.cc/paper_files/paper/2017/file/5dd9db5e033da9c6fb5ba83c7a7ebea9-Paper.pdf},
  url_short = {https://proceedings.neurips.cc/paper_files/paper/2017/file/5dd9db5e033da9c6fb5ba83c7a7ebea9-Paper.pdf}
}

@article{GNNComprehensiveSurvey20,
  title={A comprehensive survey on graph neural networks},
  author={Wu, Zonghan and Pan, Shirui and Chen, Fengwen and Long, Guodong and Zhang, Chengqi and Philip, S Yu},
  journal={IEEE transactions on neural networks and learning systems},
  volume={32},
  number={1},
  pages={4--24},
  year={2020},
  publisher={IEEE},
  url={https://ieeexplore.ieee.org/stamp/stamp.jsp?tp=&arnumber=9046288}
}

@article{GCNAutoFiltering22,
  title={Beyond low-pass filtering: Graph convolutional networks with automatic filtering},
  author={Wu, Zonghan and Pan, Shirui and Long, Guodong and Jiang, Jing and Zhang, Chengqi},
  journal={IEEE Transactions on Knowledge and Data Engineering},
  year={2022},
  publisher={IEEE},
  url={https://ieeexplore.ieee.org/stamp/stamp.jsp?tp=&arnumber=9806316}
}

@inproceedings{MoNet17,
  title={Geometric deep learning on graphs and manifolds using mixture model cnns},
  author={Monti, Federico and Boscaini, Davide and Masci, Jonathan and Rodola, Emanuele and Svoboda, Jan and Bronstein, Michael M},
  booktitle={Proceedings of the IEEE conference on computer vision and pattern recognition},
  pages={5115--5124},
  year={2017},
  url={https://openaccess.thecvf.com/content_cvpr_2017/papers/Monti_Geometric_Deep_Learning_CVPR_2017_paper.pdf}
}

@article{GNN08,
  title={The graph neural network model},
  author={Scarselli, Franco and Gori, Marco and Tsoi, Ah Chung and Hagenbuchner, Markus and Monfardini, Gabriele},
  journal={IEEE transactions on neural networks},
  volume={20},
  number={1},
  pages={61--80},
  year={2008},
  publisher={IEEE},
  url={https://ieeexplore.ieee.org/stamp/stamp.jsp?tp=&arnumber=4700287}
}

@article{SpectralNetworks13,
  title={Spectral networks and locally connected networks on graphs},
  author={Bruna, Joan and Zaremba, Wojciech and Szlam, Arthur and LeCun, Yann},
  journal={arXiv preprint arXiv:1312.6203},
  year={2013},
  url={https://arxiv.org/pdf/1312.6203.pdf}
}

@article{TutorialDeepLearningPart2,
  title={A tutorial on deep learning part 2: Autoencoders, convolutional neural networks and recurrent neural networks},
  author={Le, Quoc V and others},
  journal={Google Brain},
  volume={20},
  pages={1--20},
  year={2015},
  url={https://ai.stanford.edu/~quocle/tutorial2.pdf}
}

@article{SemiSupervisedClassificationSpectralConvGCN16,
  title={Semi-supervised classification with graph convolutional networks},
  author={Kipf, Thomas N and Welling, Max},
  journal={arXiv preprint arXiv:1609.02907},
  year={2016},
  url={https://arxiv.org/pdf/1609.02907.pdf}
}

@misc{chung_empirical_2014,
	title = {Empirical Evaluation of Gated Recurrent Neural Networks on Sequence Modeling},
	url = {http://arxiv.org/abs/1412.3555},
	number = {{arXiv}:1412.3555},
	publisher = {{arXiv}},
	author = {Chung, Junyoung and Gulcehre, Caglar and Cho, {KyungHyun} and Bengio, Yoshua},
	urldate = {2023-08-23},
	date = {2014-12-11},
	langid = {english},
	eprinttype = {arxiv},
	eprint = {1412.3555 [cs]}
}

@inproceedings{laaksonen_classification_1996,
	location = {Washington, {DC}, {USA}},
	title = {Classification with learning k-nearest neighbors},
	volume = {3},
	isbn = {978-0-7803-3210-2},
	url = {http://ieeexplore.ieee.org/document/549118/},
	doi = {10.1109/ICNN.1996.549118},
	eventtitle = {International Conference on Neural Networks ({ICNN}'96)},
	pages = {1480--1483},
	booktitle = {Proceedings of International Conference on Neural Networks ({ICNN}'96)},
	publisher = {{IEEE}},
	author = {Laaksonen, J. and Oja, E.},
	urldate = {2023-08-30},
	date = {1996},
	langid = {english}
}

@article{hochreiter_long_1997,
	title = {Long short-term memory},
	volume = {9},
	pages = {1735--1780},
	number = {8},
	journaltitle = {Neural computation},
	author = {Hochreiter, Sepp and Schmidhuber, Jürgen},
	urldate = {2023-08-23},
	date = {1997},
	note = {Publisher: {MIT} Press}
}

@book{DeepLearningBook16,
  title={Deep learning},
  author={Goodfellow, Ian and Bengio, Yoshua and Courville, Aaron},
  year={2016},
  publisher={MIT press},
  url={https://books.google.de/books?hl=en&lr=&id=omivDQAAQBAJ&oi=fnd&pg=PR5&dq=deep+learning&ots=MNV2eosBRS&sig=jN2QwFikq3g_YqU3hJVPEP0XIJ4&redir_esc=y#v=onepage&q=deep%20learning&f=false}
}

@article{wu_analysis_2006,
	title = {Analysis of Support Vector Machine Classiﬁcation},
	volume = {8},
	number = {2},
	journaltitle = {Journal of Computational Analysis \& Applications},
	author = {Wu, Qiang and Zhou, Ding-Xuan},
	date = {2006},
	langid = {english}
}

@article{EvaluatingQualityMLExplanations21,
	title = {Evaluating the Quality of Machine Learning Explanations: A Survey on Methods and Metrics},
	volume = {10},
	issn = {2079-9292},
	url = {https://www.mdpi.com/2079-9292/10/5/593},
	doi = {10.3390/electronics10050593},
	shorttitle = {Evaluating the Quality of Machine Learning Explanations},
	pages = {593},
	number = {5},
	journaltitle = {Electronics},
	shortjournal = {Electronics},
	author = {Zhou, Jianlong and Gandomi, Amir H. and Chen, Fang and Holzinger, Andreas},
	urldate = {2023-09-11},
	date = {2021-03-04},
	langid = {english}
}

@Inbook{UnderstandingWordEmbeddingsAndLM20,
  author="Gomez-Perez, Jose Manuel
  and Denaux, Ronald
  and Garcia-Silva, Andres",
  title="Understanding Word Embeddings and Language Models",
  bookTitle="A Practical Guide to Hybrid Natural Language Processing: Combining Neural Models and Knowledge Graphs for NLP",
  year="2020",
  publisher="Springer International Publishing",
  address="Cham",
  pages="17--31",
  isbn="978-3-030-44830-1",
  doi="10.1007/978-3-030-44830-1_3",
  url="https://doi.org/10.1007/978-3-030-44830-1_3"
}

@online{OpenSSHReleaseNotes9-0,
  title={OpenSSH 9.0 release notes},
  author={OpenSSH},
  year={2022},
  month={04},
  day={08},
  note={Retrieved 2022-11-13},
  publisher={OpenSSH},
  url={https://www.openssh.com/txt/release-9.0}
}

@online{StackExchangeSSHuseRASandDH23,
  author       = {Unknown},
  title        = {How does SSH use both RSA and Diffie-Hellman?},
  year         = {2023},
  url          = {https://security.stackexchange.com/questions/76894/how-does-ssh-use-both-rsa-and-diffie-hellman},
  note         = {Asked 8 years, 9 months ago; Modified 8 years, 9 months ago; Viewed 22k times; Accessed: 2023-09-21}
}

@techreport{RFC4251,
  title={The Secure Shell (SSH) Protocol Architecture},
  author={T. Ylonen and C. Lonvick},
  year={2006},
  month={01},
  institution={Network Working Group},
  type={RFC},
  number={4251},
  note={Updated by: 8308, 9141},
  category={Standards Track},
  publisher={SSH Communications Security Corp; Cisco Systems, Inc.}
}

@techreport{RFC4253,
  title={The Secure Shell (SSH) Transport Layer Protocol},
  author={T. Ylonen and C. Lonvick},
  year={2006},
  month={01},
  institution={Network Working Group},
  type={RFC},
  number={4253},
  note={Updated by: 6668, 8268, 8308, 8332, 8709, 8758, 9142; Errata Exist},
  category={Standards Track},
  publisher={SSH Communications Security Corp; Cisco Systems, Inc.}
}

@article{OpenSSHUnderHood07,
  title={The OpenSSH Protocol under the Hood},
  author={Girish Venkatachalam},
  journal={Linux Journal},
  year={2007},
  month={04},
  number={156},
  url={https://www.ecb.torontomu.ca/~courses/coe518/LinuxJournal/elj2007-156-OpenSSH.pdf}
}

@INPROCEEDINGS{NTRUPostQuantum17,
  author={Guillen, Oscar M. and Pöppelmann, Thomas and Bermudo Mera, Jose M. and Bongenaar, Elena Fuentes and Sigl, Georg and Sepulveda, Johanna},
  booktitle={Design, Automation \& Test in Europe Conference \& Exhibition (DATE), 2017}, 
  title={Towards post-quantum security for IoT endpoints with NTRU}, 
  year={2017},
  volume={},
  number={},
  pages={698-703},
  doi={10.23919/DATE.2017.7927079}
}

@inproceedings{InferenceEndianness17,
  title={Inference of Endianness and Wordsize From Memory Dumps},
  author={de Souza, Paulo Nunes and Gladyshev, Pavel},
  booktitle={European Conference on Cyber Warfare and Security},
  pages={619--627},
  year={2017},
  organization={Academic Conferences International Limited}
}

@article{McLaren2019,
  title={Decrypting live SSH traffic in virtual environments},
  author={P. McLaren and G. Russell and W.J. Buchanan and Z. Tan},
  journal={Digital Investigation},
  year={2019},
  volume={29},
  pages={109--117},
  url={https://www.sciencedirect.com/science/article/abs/pii/S1742287619300647}
}

@misc{PortableOpenSSHGitHub,
  title={Portable OpenSSH},
  author={Tatu Ylonen},
  year={1995},
  url={https://github.com/openssh/openssh-portable/},
  note={Github repositor. Accessed on 25.08.2023}
}

@article{Kerckhoffs1883,
  title={La cryptographic militaire},
  author={Auguste Kerckhoffs},
  journal={Journal des sciences militaires},
  year={1883},
  pages={5--38}
}

@article{SSHBotnetInfect21,
  title        = {A novel Machine Learning-based approach for the detection of SSH botnet infection},
  author       = {Martínez Garre, José Tomás and Gil Pérez, Manuel and Ruiz-Martínez, Antonio},
  year         = 2021,
  month        = {02},
  journal      = {Future Generation Computer Systems},
  volume       = 115,
  pages        = {387-396},
  doi          = {10.1016/j.future.2020.09.004},
  url          = {https://www.sciencedirect.com/science/article/pii/S0167739X20303265},
  abstractnote = {Botnets are causing severe damages to users, companies, and governments through information theft, abuse of online services, DDoS attacks, etc. Although significant research is being made to detect them and mitigate their effect, they are exponentially increasing due to new zero-day attacks, a variation of their behavior, and obfuscation techniques. High Interaction Honeypots (HIH) are the only honeypots able to capture attacks and log all the information generated by attackers when setting up a botnet. The data generated is being processed using Machine Learning (ML) techniques for detection since they can detect hidden patterns. However, so far, research has been focused on intermediate phases of the botnet’s life cycle during operation, underestimating the initial phase of infection. To the best of our knowledge, this is the first solution in the infection phase of SSH-based botnets. Therefore, we have designed an approach based on an SSH-based HIH to generate a dataset consisting of executed commands and network information. Herein, we have applied ML techniques for the development of a real-time detection model. This approach reached a very high level of prediction and zero false negatives. Indeed, our system detected all known and unknown SSH sessions intended to infect our honeypots. Thus, our research has demonstrated that new SSH infections can be detected through ML techniques.}
}

@techreport{SSHReport18,
  title={SSH Annual Report 2018},
  author={SSH Communications Security},
  year={2018},
  url={https://info.ssh.com/hubfs/2021%20Investor%20documents/SSH_Annual_Report_2018_final.pdf},
  institution={SSH Communications Security},
  type={Annual Report},
  note={Accessed: 2023-08-30}
}

@inproceedings{SSHIdentityTheft05,
	title        = {A first step toward detecting SSH identity theft in HPC cluster environments: discriminating masqueraders based on command behavior},
	author       = {Yurcik, W. and Liu, Chao},
	year         = 2005,
	month        = {05},
	booktitle    = {CCGrid 2005. IEEE International Symposium on Cluster Computing and the Grid, 2005.},
	volume       = 1,
	pages        = {111--120 Vol. 1},
	doi          = {10.1109/CCGRID.2005.1558542},
	abstractnote = {Recent attacks enabled by stolen authentication passwords and keys have allowed intruders to masquerade as legitimate users on high performance computing clusters. With the motivation of detecting masqueraders on clusters, this work seeks to discriminate different types of users based on their command behavior - in particular, user command behavior on a multi-user public machine versus user command behavior on a high performance computing cluster. Our intuition is that these users act differently and the unique high performance cluster environment is constrained such that command behavior discrimination is enhanced versus enterprise environments. We formalize this into a classification problem to be solved by a support vector machine with TF-IDF feature construction techniques from the field of Information Retrieval. We present results showing the effectiveness of this approach exhibiting high precision depending on the length of monitoring in both time and number of commands. In particular we show that as few as 10 commands may be enough to recognize a masquerading attacker on a high performance computing cluster.}
}

@misc{SSH1Vulnerability01,
	url          = {https://www.kb.cert.org},
  title        = {Weak CRC allows packet injection into SSH sessions encrypted with block ciphers},
	year         = 2001,
  month        = {11},
  note = {Accessed: 2023-08-30}
}

@misc{CoreSecurity23,
  author = {Core Security Technologies},
  title = {SSH Insertion Attack},
  year = {2023},
  note = {Archived from the original on 2011-07-08},
  howpublished = {\url{https://www.coresecurity.com/core-labs/advisories/ssh-insertion-attack}},
  accessed = {2023-08-30}
}

@misc{USCERT2011,
  author = {US CERT},
  title = {SSH CBC vulnerability},
  subtitle = {Vulnerability Note VU\#958563 - SSH CBC vulnerability},
  year = {2011},
  note = {Archived from the original on 2011-06-22},
  howpublished = {\url{https://www.kb.cert.org/vuls/id/958563}},
  accessed = {2023-08-30}
}

@misc{Spiegel14,
  author = {Spiegel Online},
  title = {Prying Eyes: Inside the NSA's War on Internet Security},
  year = {2014},
  note = {Archived from the original on January 24, 2015},
  howpublished = {Spiegel Online},
  url = {https://www.spiegel.de/international/germany/inside-the-nsa-s-war-on-internet-security-a-1010361.html},
  accessed = {2023-08-30}
}

@book{ESETWindigo14,
	title        = {Operation WINDIGO},
	author       = {Bilodeau, Olivier and Bureau, Pierre-Marc and Calvet, Joan and Dorais-Joncas, Alexis and Léveillé, Marc-Étienne and Vanheuverzwijn, Benjamin},
	year         = 2014,
	month        = {03},
	pages        = 69,
	url          = {https://web-assets.esetstatic.com/wls/2014/03/operation_windigo.pdf},
	abstractnote = {The vivisection of a large Linux server-side credential stealing malware campaign},
	institution  = {ESET},
	language     = {en}
}

@article{APTTactics19,
	title        = {Cyber Kill Chain-Based Taxonomy of Advanced Persistent Threat Actors: Analogy of Tactics, Techniques, and Procedures},
	author       = {Bahrami*, Pooneh Nikkhah and Dehghantanha**, Ali and Dargahi***, Tooska and Parizi****, Reza M. and Choo*****, Kim-Kwang Raymond and Javadi******, Hamid H. S.},
	year         = 2019,
	month        = {11},
	journal      = {Journal of Information Processing Systems},
	volume       = 15,
	number       = 4,
	pages        = {865-889},
	doi          = {10.3745/JIPS.03.0126},
	url          = {http://xml.jips-k.org/full-text/view?doi=10.3745/JIPS.03.0126},
	abstractnote = {Journal of Information Processing Systems Full-text XML}
}

@inproceedings{ClassificationMalware21,
	title        = {Classification of IOT-Malware using Machine Learning},
	author       = {Madan, Sanjay and Singh, Monika},
	year         = 2021,
	month        = {11},
	booktitle    = {2021 International Conference on Technological Advancements and Innovations (ICTAI)},
	pages        = {599-605},
	doi          = {10.1109/ICTAI53825.2021.9673185},
	abstractnote = {Every day, attackers target embedded IoT devices, causing damage to key cyber-infrastructure, obtaining users’ personal information, and misusing it to a greater extent. Data confidentiality, authentication, and privacy, denial of service, nonrepudiation, and digital content protection are only a few of the difficult security challenges that must be handled. To exploit these resource-constrained Cyber-physical systems, attackers use brute-force assaults, man-in-the-middle attacks, injecting malicious code, eavesdropping, and backdoors, among other methods. In this research, we present a hybrid analysis method for analyzing Linux-based IoT malware and event correlation for incident management using anomaly detection. For malware classification, the machine learning model is built using information from both static and dynamic analysis of harmful programs. For anomaly identification and event correlation, the anomalous DDoS traffic detection approach is also proposed. The F1-score is maximized for various DDoS attacks using the threshold selection approach, and the results are compared to the state-of-the-art literature.}
}

@inproceedings{SSHHoneypotEffectiveness23,
	title        = {Analysis of SSH Honeypot Effectiveness},
	author       = {Hetzler, Connor and Chen, Zachary and Khan, Tahir M.},
	year         = 2023,
	month        = {03},
	booktitle    = {Advances in Information and Communication},
	publisher    = {Springer Nature Switzerland},
	address      = {Cham},
	series       = {Lecture Notes in Networks and Systems},
	pages        = {759-782},
	doi          = {10.1007/978-3-031-28073-3_51},
	isbn         = 9783031280733,
	abstractnote = {The number of cyberattacks has increased in the twenty-first century, with the FBI receiving 791,790 individual internet crime complaints in the United States in 2020. As attackers become more sophisticated with their ransomware and malware campaigns, there is a significant need for security researchers to assist the greater community by running vulnerable honeypot machines to collect malicious software. The ability to collect meaningful malware from attackers depends on how the attackers receive the honeypot. Most attackers fingerprint targets before they launch their attack, so it would be very beneficial for security researchers to understand how to hide honeypots from fingerprinting and trick the attackers into depositing malware. This study investigated the use of a cloaked and uncloaked SSH honeypot to learn how attackers fingerprint SSH honeypots and the efficacy of cloaking the honeypot by changing features that attackers fingerprint. This paper compares the number of logins and commands run by the attackers captured on default and cloaked Cowrie honeypots. The project lasted just over a month, and the cloaked honeypot received 74.5\% of the total login attempts and 53\% of the commands executed on the honeypot systems, as it was more believable than the uncloaked. This paper reports that modifying SSH honeypot systems improves malware collection effectiveness. This project focuses on monitoring attacker tactics on a standard SSH honeypot to understand their fingerprinting commands, integrate those that are missing into the cloaked honeypot, as well as adding/modifying files with which the attackers interact. The attackers utilized a variety of UNIX commands and files on the honeypots. The attackers who successfully logged into the SSH honeypot downloaded malware, including IRC bots, Mirai, Xmrig cryptominers, and many others. What is certain is that if a cautious attacker believes they are in a honeypot, they will leave without depositing malware onto the system, which reduces the effectiveness of the honeypot for security research.},
	editor       = {Arai, Kohei},
	collection   = {Lecture Notes in Networks and Systems},
	language     = {en},
  url          = {}
}

@misc{honeyssh17,
  author = {ppacher},
  title = {honeyssh},
  year = {2017},  
  howpublished = {\url{https://github.com/ppacher/honeyssh}},
  note = {GitHub repository},
  accessed = {2023-08-30}
}

@article{BotnetBusinessModels23,
	title        = {Botnet Business Models, Takedown Attempts, and the Darkweb Market: A Survey},
	author       = {Georgoulias, Dimitrios and Pedersen, Jens Myrup and Falch, Morten and Vasilomanolakis, Emmanouil},
	year         = 2023,
	month        = {11},
	journal      = {ACM Computing Surveys},
	volume       = 55,
	number       = 11,
	pages        = {1-39},
	doi          = {10.1145/3575808},
	url          = {https://dl.acm.org/doi/10.1145/3575808},
	abstractnote = {Botnets account for a substantial portion of cybercrime. Botmasters utilize darkweb marketplaces to promote and provide their services, which can vary from renting or buying a botnet (or parts of it) to hiring services (e.g., distributed denial of service attacks). At the same time, botnet takedown attempts have proven to be challenging, demanding a combination of technical and legal methods, and often requiring the collaboration of a plethora of entities with varying jurisdictions. In this article, we map the elements associated with the business aspect of botnets and utilize them to develop adaptations of two widely used business models. Furthermore, we analyze the 28 most notable botnet takedown operations carried out from 2008 to 2021, in regard to the methods employed, and illustrate the correlation between these methods and the segments of our adapted business models. Our analysis suggests that the botnet takedown methods have been mainly focused on the technical side, but not on the botnet economic components. We aim to shed light on new takedown vectors and incentivize takedown actors to expand their efforts to methods oriented more toward the business side of botnets, which could contribute toward eliminating some of the challenges that surround takedown operations.},
	language     = {en}
}

@article{KG21,
  author = {Aidan Hogan and Eva Blomqvist and Michael Cochez and Claudia D'Amato and Gerard De Melo and Claudio Gutierrez and Sabrina Kirrane and José Emilio Labra Gayo and Roberto Navigli and Sebastian Neumaier and Axel-Cyrille Ngonga Ngomo and Axel Polleres and Sabbir M. Rashid and Anisa Rula and Lukas Schmelzeisen and Juan Sequeda and Steffen Staab and Antoine Zimmermann},
  title = {Knowledge Graphs},
  year = {2021},
  issue_date = {May 2022},
  publisher = {Association for Computing Machinery},
  address = {New York, NY, USA},
  volume = {54},
  number = {4},
  issn = {0360-0300},
  url = {https://doi.org/10.1145/3447772},
  doi = {10.1145/3447772},
  journal = {ACM Comput. Surv.},
  month = {7},
  articleno = {71},
  numpages = {37},
  bibsource = {dblp computer science bibliography, https://dblp.org}
}

@article{KG22,
	title        = {Knowledge Graphs (Extended)},
	author       = {Hogan, Aidan and Blomqvist, Eva and Cochez, Michael and d'Amato, Claudia and de Melo, Gerard and Gutierrez, Claudio and Gayo, José Emilio Labra and Kirrane, Sabrina and Neumaier, Sebastian and Polleres, Axel and Navigli, Roberto and Ngomo, Axel-Cyrille Ngonga and Rashid, Sabbir M. and Rula, Anisa and Schmelzeisen, Lukas and Sequeda, Juan and Staab, Steffen and Zimmermann, Antoine},
	year         = 2022,
	month        = {05},
	journal      = {ACM Computing Surveys},
	volume       = 54,
	number       = 4,
	pages        = {1-37},
	doi          = {10.1145/3447772},
	url          = {http://arxiv.org/abs/2003.02320},
	note         = {arXiv:2003.02320 [cs]},
	abstractnote = {In this paper we provide a comprehensive introduction to knowledge graphs, which have recently garnered significant attention from both industry and academia in scenarios that require exploiting diverse, dynamic, large-scale collections of data. After some opening remarks, we motivate and contrast various graph-based data models and query languages that are used for knowledge graphs. We discuss the roles of schema, identity, and context in knowledge graphs. We explain how knowledge can be represented and extracted using a combination of deductive and inductive techniques. We summarise methods for the creation, enrichment, quality assessment, refinement, and publication of knowledge graphs. We provide an overview of prominent open knowledge graphs and enterprise knowledge graphs, their applications, and how they use the aforementioned techniques. We conclude with high-level future research directions for knowledge graphs.}
}

@article{KGKE22,
  title={Knowledge Graphs and their Role in the Knowledge Engineering of the 21st Century},
  author = {Paul Groth and Elena Simperl and Marieke van Erp and Denny Vrandečić},
  journal={Dagstuhl Reports},
  volume={12},
  number={9},
  pages={60--120},
  year={2022},
  publisher={Dagstuhl Seminar},
  doi={10.4230/DagRep.12.9.60},
  note={Report from Dagstuhl Seminar 22372. Specific usage: pp. 60-72, Subsection "3.2 A Brief History of Knowledge Engineering: A Practitioner's Perspective"}
}

@article{CKG23,
  title={Construction of Knowledge Graphs: State and Challenges},
  author={Marvin Hofer and Daniel Obraczka and Alieh Saeedi and Hanna Köpcke and Erhard Rahm},
  journal={arXiv preprint arXiv:2302.11509},
  year={2023},
  url={https://doi.org/10.48550/arXiv.2302.11509}
}

@article{TDKG16,
  title={Towards a Definition of Knowledge Graphs},
  author={Ehrlinger, Lisa and Wöß, Wolfram},
  booktitle={SEMANTiCS (Posters, Demos, SuCCESS)},
  year={2016},
  pages={1--4}
}

@book{hulme1902proverb,
  title={Proverb Lore: Many Sayings, Wise Or Otherwise, on Many Subjects, Gleaned from Many Sources},
  author={Hulme, Frederick Edward},
  year={1902},
  publisher={E. Stock},
  pages={188}
}

@article{googleblog2023knowledgegraph,
  author = {Google},
  title = {Introducing the Knowledge Graph: Things, not strings},
  journal = {Google Blog},
  year = {2012},
  month = {05},
  day = {16},
  url = {https://blog.google/products/search/introducing-knowledge-graph-things-not/},
  note = {Accessed: 2023-06-16}
}

@online{venables2019,
  author = "Michelle Venables",
  title = "An Introduction to Graph Theory",
  year = "2019",
  url = "https://towardsdatascience.com/an-introduction-to-graph-theory-24b41746fabe",
  publisher = "Towards Data Science",
  note = "Accessed: 2023-06-12"
}

@online{fiorelli2013,
    author = "Gianluca Fiorelli",
    title = "Best of 2013: No 13 - Search in the Knowledge Graph era",
    year = "2013",
    url = "https://www.stateofdigital.com/search-in-the-knowledge-graph-era/",
    publisher = "State of Digital",
    note = "Accessed: 2023-06-12"
}

@online{gilkey2019,
    author = "Jackson Gilkey",
    title = "Graph Theory and Data Science",
    year = "2019",
    url = "https://towardsdatascience.com/graph-theory-and-data-science-ec95fe2f31d8",
    publisher = "Towards Data Science",
    note = "Accessed: 2023-05-25"
}

@article{JAWAD2023102124,
  title = {Adoption of knowledge-graph best development practices for scalable and optimized manufacturing processes},
  journal = {MethodsX},
  volume = {10},
  pages = {102124},
  year = {2023},
  issn = {2215-0161},
  doi = {https://doi.org/10.1016/j.mex.2023.102124},
  url = {https://www.sciencedirect.com/science/article/pii/S2215016123001255},
  author = {M.S. Jawad and Chitra Dhawale and Azizul Azhar Bin Ramli and Hairulnizam Mahdin},
}
